\documentclass[aps,pre,superscriptaddress,amsmath,amssymb,reprint]{revtex4-1}
\usepackage[utf8]{inputenc}
\usepackage{mfirstuc}
\usepackage{graphicx}
\usepackage{subfigure}
\usepackage{float}
\usepackage{gensymb}
\usepackage{amsmath}
\usepackage{color}
\usepackage{multirow}
\usepackage{comment}
\usepackage{enumitem}
\usepackage[colorlinks=true, linkcolor=blue, citecolor=blue, urlcolor=blue]{hyperref}
\usepackage{url}

\newcommand{\SK}[1]{\textcolor{black}{{#1}}}

\newcommand{\rev}[1]{\textcolor{black}{{#1}}}
\newcommand{\SKrev}[1]{\textcolor{black}{{#1}}}
\newcommand{\SKReRev}[1]{\textcolor{black}{{#1}}}
\newcommand{\SKfinal}[1]{\textcolor{black}{{#1}}}

\begin{document}


\title{Role of Fragility of the Glass Formers in the Yielding Transition under Oscillatory Shear}

\author{Roni Chatterjee}
\author{Monoj Adhikari}
\author{Smarajit Karmakar}
\email{smarajit@tifrh.res.in}

\affiliation{Tata Institute of Fundamental Research, 36/P, Gopanpally Village, Serilingampally Mandal, Ranga Reddy District, Hyderabad 500046, Telangana, India}

\begin{abstract}
\noindent\textbf{Abstract:}
Amorphous materials, especially metallic glasses, are known for their exceptional mechanical properties, such as high yield strength and large yield strain. Understanding the microscopic mechanisms behind their failure, particularly the yielding transition, remains an active area of research. Previous studies have shown that yielding behavior depends on the initial age of the sample. Through extensive computer simulations, we demonstrate that this age dependence varies across different materials and is influenced by the specific characteristics of the initial glass former, particularly its fragility. Both strong and fragile glass formers exhibit similar yielding behaviour in poorly annealed conditions with a critical yield strain, $\gamma_c$ that does not depend on the initial conditions. However, below a critical degree of annealing, the yield point increases significantly with further annealing for fragile glasses, while it remains relatively constant for strong glasses. The results are found to be universal across a wide variety of model glassy systems with varying fragility, including metallic glasses, molecular glasses, model granular glasses, and network-forming glasses like Silica. We rationalise these findings by introducing a modified mean-field elastoplastic model that explicitly incorporates the crucial role of changing energy barrier with increasing annealing in the yielding process. This simple model reproduces all the simulation results and provides critical insights into how energy barriers influence the physics of the yielding transition including the critical yield strain under oscillatory shear deformation.
\end{abstract}

\maketitle


\noindent{\bf Introduction:}
Understanding the mechanical behaviour of amorphous solids is crucial due to their enormous applications ranging from bulk to nanoscales \cite{fielding2000aging,falk2011deformation,schuh2007mechanical,bonn2017yield} in our day-to-day life. Specifically, the study of amorphous solids under external shear deformation has drawn significant interest in designing better materials for future applications in recent years. Among these, a large number of studies focus on comprehending the response of amorphous solids to uniform shear deformation \cite{shi2005strain,shi2007evaluation,karmakar2010statistical,karmakar2010predicting,keim2013yielding,zaccone2014microscopic,denisovSR15,lin2014scaling,bhowmik2019effect,barbot2020rejuvenation,zaccone2020rheology}. At small amplitude deformations, the response closely resembles that of an elastic solids; however, under large deformation, numerous plastic deformations occur, eventually leading to yielding and the system starts to flow. Although the nature of the yielding transition is a subject of debate \cite{kawasaki2016macroscopic,procaccia2017mechanical,jaiswal2016mechanical,shrivastav2016yielding}, energy, or stress shows a discontinuity when driven by oscillatory shear deformation at the yielding transition \cite{PREFiocco,regev2013reversibility,nagamanasa2014experimental,PKetal,parmar2019strain}.

\SK{Recent investigations using cyclic shear predominantly under athermal quasistatic (AQS) \cite{maloney2006amorphous} conditions indicate that the yielding transition depends on the initial sample's degree of annealing or inherent structure (IS) energy. In poorly annealed systems, stroboscopic energy decreases with increasing amplitude of deformation \rev{($\gamma_{max}$) until a critical yield strain amplitude ($\gamma_c$),} after which it increases again. Near yielding, the system undergoes an absorbing to diffusing transition: for smaller amplitudes below yielding, the system reaches a steady state called the absorbing state with an invariant stroboscopic configuration; for amplitudes larger than yielding, \rev{$\gamma_{max} > \gamma_c$,} the steady states are diffusing states where the configuration keeps changing with the number of cycles. This scenario changes significantly if the systems are sufficiently annealed so that the initial energy is below the threshold energy, shown to be the energy at the Mode Coupling Transition temperature \rev{$T_{MCT}$} \cite{yeh2020glass,bhaumik2021role}. The system's energy remains constant until it reaches the yielding amplitude \rev{$\gamma_Y$ at those annealing conditions}, where it undergoes a sudden increase and reaches the diffusive states. \rev{Note that $\gamma_Y \ge \gamma_c$ in all systems that are studied}. A similar discontinuous jump in energy is observed under uniform shear as well \cite{ozawa2018random,bhaumik2021role} when a large system size is considered. This effect of annealing on yielding is robust across various model systems \cite{divoux2023ductile,ozawa2018random,barlow2020ductile,ozawa2023creating,keim2022mechanical,pollard2022yielding,mutneja2023yielding}, different driving protocols \cite{krishnan2023annealing,adhikari2022yielding}, finite temperature, finite shear rate \cite{singh2020brittle,lamp2022brittle}, and different dimensions \cite{singh2020brittle,ozawa2020role,bhaumik2022yielding,chatterjee2026memory}.
}

\SK{For poorly annealed glasses, the stress vs. strain curve exhibits a monotonic trend, and the transition to a flowing state is smooth or gradual, particularly under uniform shear conditions. Under cyclic shear, the yielding point is more precisely defined. However, in both cases, material failure resembles ductile behavior for poorly annealed glass. As the degree of annealing increases, a large discontinuous stress drop is observed at yielding in both uniform and cyclic shear, indicating an abrupt and catastrophic transition, characteristic of brittle behavior. Observing two distinct yielding behaviors, solely by adjusting the degree of annealing, which correlates with the temperature at which the liquid is prepared, raises questions about the existence of a potential phase transition. To answer these questions, various coarse-grained models have been introduced \cite{ozawa2018random,sastry2021models,barlow2020ductile,parley2022mean,Maloney2021,liu2022fate,kumar2022mapping,rossi2022finite,parley2023towards} which successfully reproduce the observed behavior in simulation and also attempt to answer the existence of a possible non-equilibrium phase transition belonging to the driven Random Field Ising Model (RFIM) universality class\cite{rossi2022finite,mutneja2025RandomPinning}. Another recent work using random particle pinning showed strong evidence of such a non-equilibrium phase transition from brittle to ductile states at a finite critical disorder \cite{mutneja2025RandomPinning}. These results suggest a renewed interest in the community in understanding the nature of the yielding transition in amorphous solids. While many of these studies support the notion of a phase transition distinguishing between brittle and ductile yielding, the question of its persistence in the thermodynamic limit remains a topic of ongoing debate.}

All these studies on yielding described above primarily focused on the initial energy of the system, providing limited insights into the nature of the initial glass former, notably its fragility, which characterises how rapidly the viscosity of the glass former increases as it approaches the putative glass transition point. Fragility is a crucial parameter for glass-forming liquids as it connects to various dynamical and structural properties such as dynamical heterogeneity, configurational entropy, violation of the Stokes-Einstein relation, glass-forming ability and so on \cite{sastry2001relationship, scopigno2003fragility, ruocco2004landscapes, tanaka2005relationship,yu2015strain,adhikari2021spatial,alba2022perspective}. In \cite{bhaumik2021role}, the authors present the oscillatory shear responses of two model glasses: Silica, a well-known strong glass that forms networks, and the Kob-Andersen Model, which is recognised as a fragile glass former. One can clearly see that the oscillatory shear yielding diagrams for these two models are very different, notably for well-annealed states below the critical energy. \SKfinal{The yield strain increases significantly with increasing annealing for the Kob-Andersen model ($\approx 46.6\%$), whereas for silica, the increase is modest ($\approx 4.3\%$).} The results inspired us to study the fundamental role of fragility in the yielding transition under oscillatory shear.

\SK{In this article, we investigate whether yielding behavior is similar among different glass formers \rev{with varying degrees of fragility}. Through extensive computer simulations, we demonstrate that the nature of the yielding transition can vary depending on the fragility of the initial glass former. \rev{We show that our observation is universal across a wide variety of glassy systems with a large difference in their fragilities. This includes metallic glass, molecular glass, model granular glass, and network-forming glass such as silica glass. Details of the models and simulation protocols are provided in the Methods section and the Supplementary Information (SI).} Additionally, we introduce a \rev{mean-field} elastoplastic model \rev{that incorporates the critical role of energy barrier in the yielding transition}, successfully reproducing the simulation results. The analysis of this model allows us to rationalise the observed changes in yielding behaviour, attributing them to changes in energy barriers with the annealing history.
}
\begin{figure*}[htpb]
  \centering
  \includegraphics[width=0.99\textwidth]{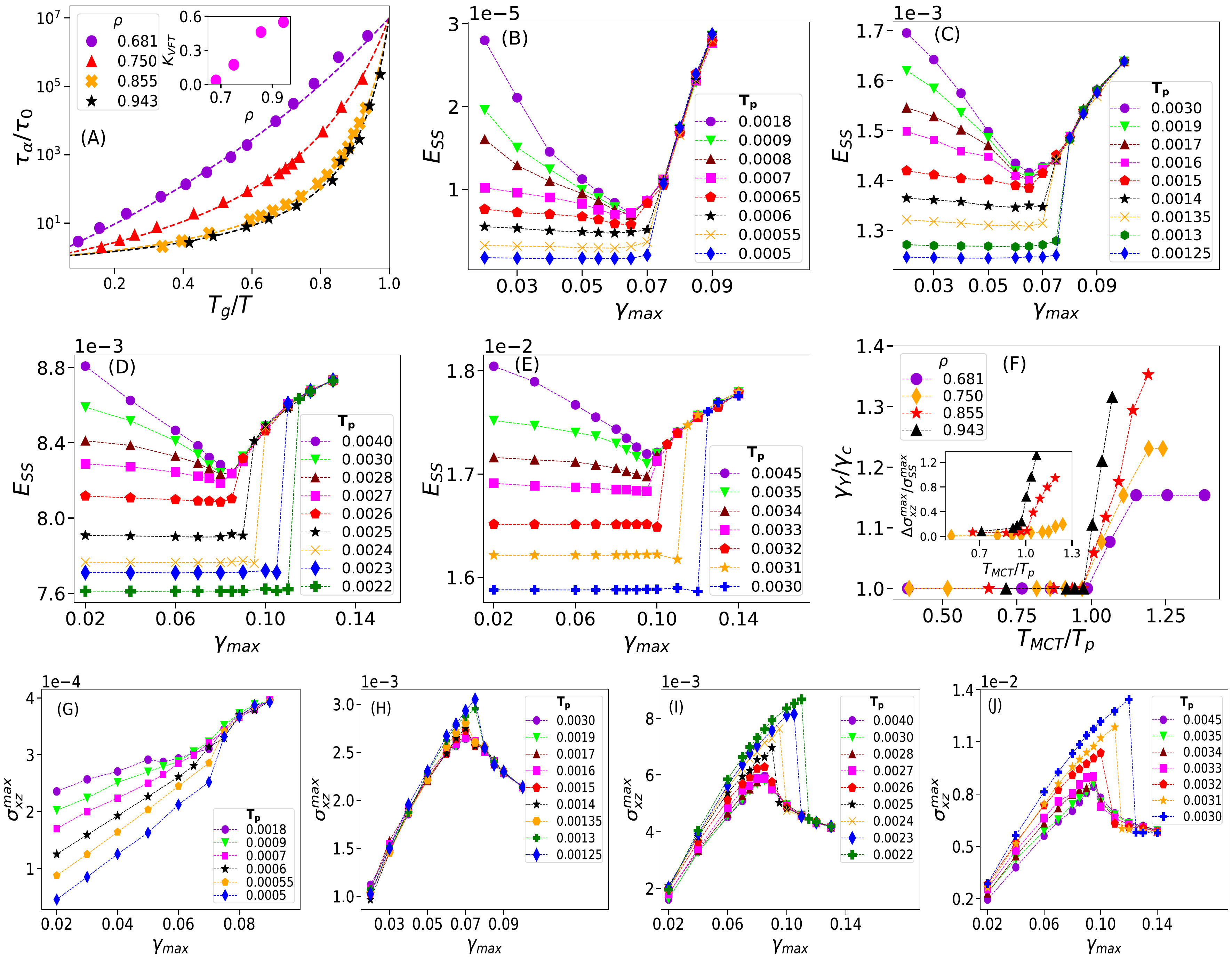}
  \caption{{\bf Oscillatory Shear Yielding Diagram \rev{of 3D HP Model}:} (A) Relaxation time, $\tau_\alpha$ (scaled with $\tau_0$ from Eq. 1) is plotted as a function of scaled temperature, $T_g/T$, for all the four studied densities. $T_g$ is the calorimetric glass transition temperature (see text for definition). The marked variation in the shape of the relaxation curve with changing density highlights the large change in fragility in these systems with increasing density. The inset shows the kinetic fragility as a function of density. The dashed lines represent fits to the data. (B) Steady-state energy $E_{SS}$ vs cyclic shear amplitude $\gamma_{max}$ is plotted for differently annealed samples as indicated by their parent temperatures ($T_p$) for $\rho = 0.681$. Notice that the main characteristic of the absence of any mechanical annealing below a critical energy scale, along with the nearly no shift in the yield strain ($\gamma_Y$) with increasing annealing in this strong glass former. Similar phase diagrams for $\rho = 0.750$ (C), $\rho = 0.855$ (D), and $\rho = 0.943$ (E) are shown to highlight how the yielding point shifts to higher strain amplitude as the fragility of the glass formers is systematically changed. For the most fragile glass former (panel E), the shift in the yield strain at low temperature is very large (nearly $30\%$). (F) Shows the shifting of $\gamma_Y$ with increasing annealing for all four densities. For better comparison, the x-axis has been scaled by their respective MCT temperature, $T_{MCT}$. Inset shows the stress drop relative to steady state stress (after yielding) with annealing. Maximum steady state stress, $\sigma^{max}_{xz}$, is plotted against $\gamma_{max}$ for $\rho = 0.681$ (G) $\rho = 0.750$ (H), $\rho = 0.855$ (I) and $\rho = 0.943$ (J), to highlight the stark difference in the yielding transition between strong and fragile glass formers. Strong glass formers continued to show ductile-like yielding with increasing annealing, whereas fragile glass formers have shown sharp, brittle yielding with increasing annealing, as clear from the sharp stress drops across the yielding strain. \SKrev{Results for $\rho = 0.681$ shows some interesting difference compared to other densities. Jamming transition seems to be playing an important role in these systems (see text for details). The dashed lines are guides to the eye.}
  }
\label{fig:1}
\end{figure*}

\vskip +0.1in
\noindent{\bf \large Results:}
\SK{\rev{First, we show results for the binary harmonic sphere model in 3D (referred to as 3D HP) with changing density, and then we discuss results from other models.} In Fig.\ref{fig:1}A, we show the Angell plot to highlight the large variation of fragility \rev{for the 3D HP model} with densities by plotting the relaxation time, $\tau_{\alpha}$ against scaled temperature ($T_g/T$). We obtain the Angell plot by defining the calorimetric glass transition temperature $T_g$ such that $\tau_{\alpha} (T_g)=10^{7}$ \rev{in appropriate reduced unit}. Subsequently, we fit this relaxation time using the Vogel Fulcher Tamannam (VFT) formula: 
\begin{equation}
 \tau_{\alpha}=\tau_0 \exp\left[{\frac{1}{K_{VFT}(T/T_{VFT}-1)}}\right],   
\end{equation}
where $T_{VFT}$ denotes the temperature at which the relaxation time might diverge \rev{upon extrapolation} and $K_{VFT}$ represents the kinetic fragility, \rev{the rate at which the relaxation time diverges at $T_{VFT}$}.  In the inset of Fig.\ref{fig:1}A, we show $K_{VFT}$ as a function of density. Our observations corroborate with the previous studies that fragility in this soft sphere glass model increases significantly with increasing density \cite{berthier2009compressing, adhikari2021spatial,tah2022fragility}. Specifically, the system demonstrates behavior reminiscent of a strong liquid at a density $\rho = 0.681$ with $K_{VFT} \simeq 0.034$. \rev{Note that the zero temperature jamming density for this model is $\rho \simeq 0.654-0.674$ depending on the protocol and initial configurations \cite{chaudhuri2010jamming,ozawa2012jamming,das2020unified}.} However, as the density increases, the system becomes more fragile. At the highest studied density ($\rho = 0.943$), $K_{VFT} \simeq 0.549$. Thus, the change in fragility is around $16$, corresponding to a density change by a factor of $1.38$. We then perform cyclic shear on these four glasses at various degrees of annealing as represented by the parent temperatures ($T_p$). Parent temperature refers to the temperature at which the liquid is equilibrated before quenching to zero temperature to prepare the amorphous solid samples. For each chosen cyclic shear amplitude ($\gamma_{max}$), we perform a couple of hundred to a thousand cycles of shear deformation (\rev{depending on the fragility}) until the energy reaches a steady state value $E_{SS}$. Using these extensive cyclic shear simulations, we obtain the yielding diagrams of the systems by plotting $E_{SS}$ as a function of $\gamma_{max}$ \rev{for all the four densities} and the studied parent temperatures.} 

\SK{Variations of energy with cycles are shown in the SI. Fig.\ref{fig:1}(B), (C), (D), and (E) show the yielding diagrams for densities $\rho = 0.681$, $0.750$, $0.855$, and $0.943$, respectively, \rev{for 3D HP model with large fragility change}. In poorly annealed systems, all glass formers behave similarly: the steady-state energy decreases as $\gamma_{max}$ increases until reaching the critical yielding point ($\gamma_c$), after which $E_{SS}$ rises again. For all poorly annealed systems ($T_p > T_{MCT}$), the systems yield at the same point. We call $\gamma_c$ critical yield strain because this strain demarcates the difference between the poorly annealed and well-annealed samples. However, stark differences emerge among different glass formers in the well-annealed regime. Although $E_{SS}$ remains unchanged with increasing $\gamma_{max}$, the yielding point, $\gamma_Y$, increases significantly with increasing annealing for fragile liquids, whereas it remains nearly the same for strong liquids. The yielding transition for strong liquid (Fig. \ref{fig:1}(B)) occurs more or less at the same point irrespective of the degree of annealing, whereas for the most fragile liquid (Fig. \ref{fig:1}(E)), the yielding point increases very rapidly with increasing annealing for temperatures $T_p < T_{MCT}$. To highlight the changes in the yield strain qualitatively, we estimated the percentage changes in the yield point $\gamma_Y$ with respect to the critical yield point $\gamma_c$, and for $\rho = 0.681$, the percentage change in yield point is around $15\%$ whereas, for the fragile glass formers, the changes are as large as $35\%$. These results provide important insights into observations reported in previous studies, which we discuss below.} 

\SK{In Fig.\ref{fig:1}(F), we illustrate $\gamma_Y/\gamma_c$ as a function of $T_{MCT}/T_p$, where $\gamma_Y$ denotes the yielding point \rev{of the system at the corresponding $T_p$} and $\gamma_c$ denotes the yielding point of poorly annealed glasses \rev{for all $T_p > T_{MCT}$}. The yielding point does not change for different glasses until $T_{MCT}$; however, below $T_{MCT}$, it increases rapidly with increasing fragility, whereas for the strong glasses, it remains almost constant. \rev{The inset of the same figure panel shows that changes in stress drop, $\sigma^{max}_{xz}$, across the yielding transition normalised by the steady state stress, $\sigma^{max}_{SS}$. It is clear that for fragile glasses, the relative magnitude of stress drop at the yielding transition is much larger than for strong glass formers, and it becomes even larger with increasing annealing.}}
\begin{figure*}[htpb]
  \centering
  \includegraphics[width=0.99\textwidth]{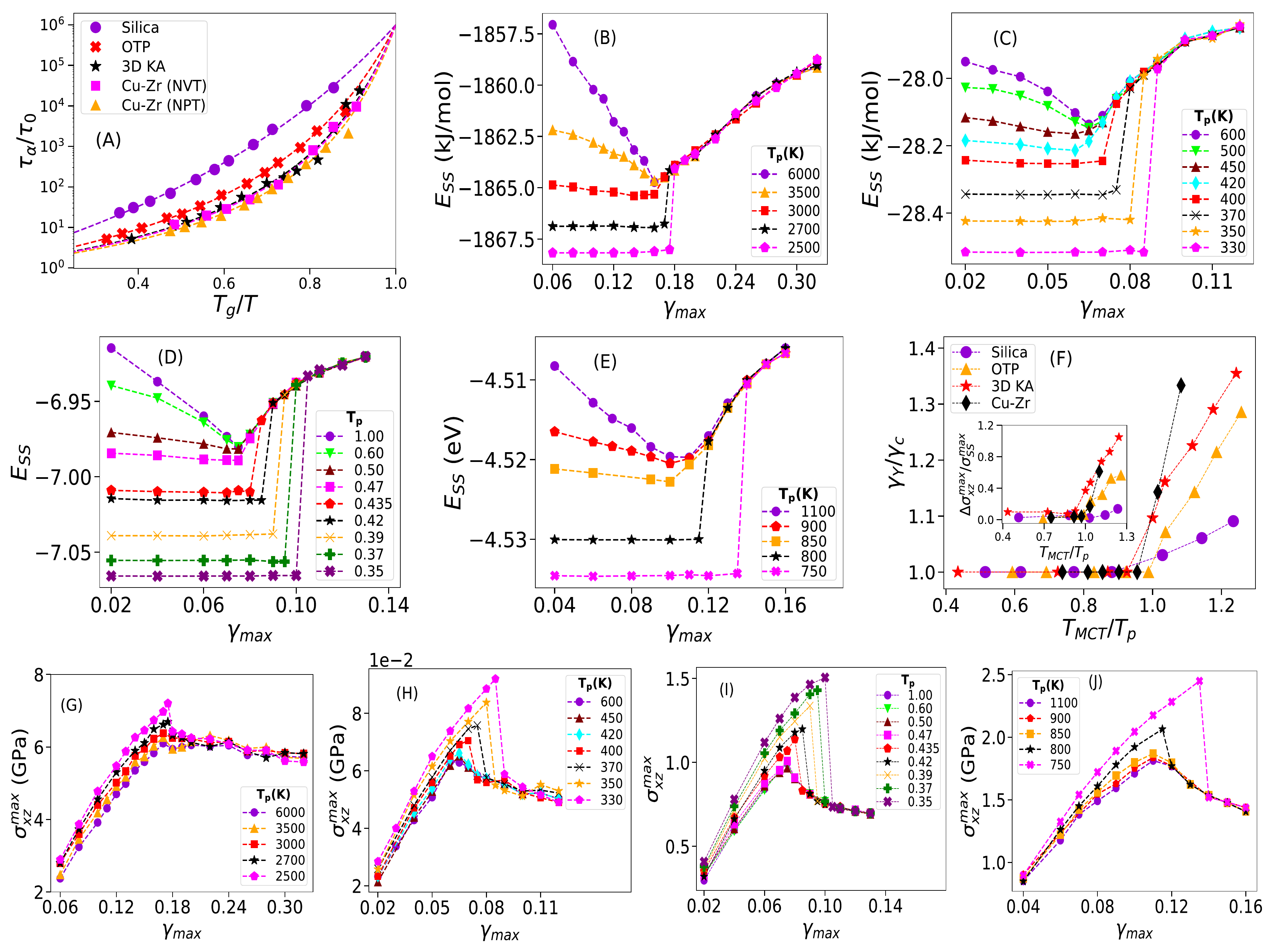}
  \caption{\rev{{\bf Universality: Oscillatory Shear Yielding Diagram Across Models with Varying Fragility.} (A) Angell plot showing the changes in fragility in four different model glassy systems - BKS Silica, OTP, 3D KA and Cu-Zr metallic glass. Note that the origin of fragility in these models is very different. The dashed lines represent fits to the data. (B) The steady state energy diagram as a function of $\gamma_{max}$ for BKS Silica Model. Note that within the accessible temperature range, BKS Silica does not show significant changes in the yield strain below $T = 3000 K$. (C) The steady state energy diagram for a model ortho-terphenyl (OTP) glass former. This model is more fragile than Silica in the studied temperature range and shows an increase in $\gamma_Y$ below $T = 420 K$. Similar plot for 3D KA model (D) and Cu-Zr metallic Glass (E). (F) Changes in yield strain $\gamma_Y$ normalised by $\gamma_c$ as a function of $T_{MCT}/T_p$ for all four models. Note that within the studied temperature range, Cu-Zr shows the largest change in yield strain, in complete agreement with results reported for the Harmonic sphere model. Inset shows the stress drop relative to steady state stress (after yielding) with annealing. (G-J) Shows the stress-strain curve $\sigma_{max}$ vs $\gamma_{max}$ for BKS Silica (G), OTP (H), 3D KA (I) and Cu-Zr (J). Again, note that the stress jump in the Cu-Zr model is largest at the yield strain as compared to Silica.} The dashed lines are guides to the eye.}
\label{allModel}
\end{figure*}

\SK{To further elucidate the yielding transition under cyclic shear, in Fig.\ref{fig:1}(G)-(J), we show the maximum steady state stress $\sigma^{max}_{xz}$ as a function of $\gamma_{max}$ for different parent temperatures. \SKrev{We first discuss $\rho = 0.750, 0.855$ and $0.943$ cases before discussing the $\rho = 0.681$ case. We see very interesting behaviour for $\rho = 0.681$ which seems to be linked to a nearby unjamming transition at $\rho\simeq 0.645$. We discuss this case in detail in the subsequent paragraph}. In the case of one of the strong glasses (Fig.\ref{fig:1}(H)), we observe that for various parent temperatures, $\sigma^{max}_{xz}$ increases as $\gamma_{max}$ increases until it reaches the yield point and subsequently stress drops to a lower value. Interestingly, the yield points appear to be similar across different parent temperatures for strong glass. This stress-strain behaviour not only reaffirms that the yielding point does not change with the degree of annealing but also suggests that strong glasses remain ductile even in the most annealed samples.} 

\SK{However, the stress-strain behaviour changes significantly for fragile glass (Fig. \ref{fig:1}(J)). With increasing annealing, the yielding point increases very rapidly and becomes brittle for well-annealed glasses with sharp stress drops at the yielding strain. Thus, this result, combined with the yielding diagram described above, reveals that the nature of yielding is highly dependent on the fragility of the glass formers. We have seen similar phenomena in our recent investigation, where we have tuned the fragility by pinning a fraction of particles chosen randomly in the 3D KA model \cite{chatterjee2024effect}. \rev{As particle pinning can be challenging to implement in experiments, particularly with molecular glass formers, we have investigated a pinning-like effect using heavy mass particles as soft pinning sites. We found that increasing the mass of the soft-pinned particles leads to a decrease in fragility, similar to the effects observed with traditional particle pinning. By conducting finite-rate oscillatory shear simulations for this model, we demonstrated that $\gamma_Y$ increases more rapidly with respect to $\gamma_c$ in fragile glass formers compared to strong ones. These findings suggest that this phenomenon is observed in different driving conditions and is very likely to be universal in nature.}}

\rev{In the 3D HP model, fragility increases with density or packing fraction. One could attempt to understand this increase by following the argument of \cite{krausser2015interatomic}, which suggests that the softness of the repulsive part of the interatomic potential dictates a system's fragility. At higher densities, particles experience the steep repulsive part of the potential, leading to more fragile behavior. The jamming-unjamming transition, which this model exhibits at zero temperature, could provide an alternative explanation of changes in fragility. For the model system studied here the unjamming-to-jamming transition occurs a range of densities ( $\rho \simeq 0.654-0.674$) depending on the protocol and initial configurations \cite{chaudhuri2010jamming,ozawa2012jamming,das2020unified}. One could argue that the origin of fragility is closely related to this underlying jamming transition, suggesting that the strong case at low density (\( \rho = 0.681 \)) may be significantly influenced by this transition. Previous study \cite{scalliet2019nature} has shown that for a repulsive model system with a large polydispersity (around \( 23\% \)), some low-density states near the jamming density can undergo an unjamming transition due to annealing under simple shear. It is important to note that with high polydispersity, the jamming density or packing fraction also increases to larger values.} 

\SK{In our binary model system, we are working at a density slightly greater than the largest jamming density for this model \cite{das2020unified}. However, upon closer examination of the case where $\rho = 0.681$, we observe some anomalies. For instance, the curves of $\sigma_{max}^{xz}$ versus $\gamma_{max}$ for various parent temperatures do not resemble those observed at higher densities \SKrev{as shown in Fig.\ref{fig:1}(G). Stress for small amplitudes is significantly lower in well-annealed glasses compared to poorly annealed ones. This contrasts with higher densities, where the difference in stress for small amplitudes is negligible.} Despite this, the shear modulus increases systematically with higher annealing temperatures or lower parent temperatures. Additionally, the oscillatory shear results clearly indicate that even after multiple cycles of annealing, the system does not undergo an unjamming transition at low temperatures, as evidenced by the non-zero inherent structure energy. \SKrev{The noticeable reduction in stress for a smaller amplitude may be related to the behavior associated with annealing-induced unjamming \cite{scalliet2019nature}. Future research exploring the impact of this unjamming transition on the oscillatory shear yielding diagram would be a fascinating direction to pursue.}} 

\rev{Nonetheless, to understand whether our observation of a direct relation between fragility and the yielding transition is universal or not, we have studied four more different model glassy systems with varying fragility. These models are (1) the BKS (van Beest-Kramer-van Santen) Silica (SiO$_2$) model \cite{van1990force,saika2004free}, (2) the ortho-terphenyl (OTP) model \cite{lewis1993relaxation,lewis1994molecular,lewis1994rotational,mossa2002dynamics}, (3) the 3D Kob-Andersen model (3D KA) \cite{kob1994scaling,kob1995testing}, and (4) the Copper-Zirconium (Cu-Zr) metallic glass model \cite{mendelev2019development}. We have performed extensive molecular dynamics simulations of all four models (see the Methods section for the model details). In Fig.\ref{allModel}(A), we show the Angell plot of all these models to highlight the significant changes in fragility across these models. Note that all these models are fundamentally different in their microscopic details; for example, BKS Silica is a network-forming glass, whereas OTP is a molecular glass made of three bonded monomers. The Kob-Andersen (KA) model is a model of metallic glass to model Ni$_{80}$P$_{20}$ interacting via the simple Lenard-Jones potential. The Copper-Zirconium (Cu-Zr) binary metallic glass is a well-studied model of metallic glass former interacting via an Embedded Atom Model (EAM) potential \cite{mendelev2019development}. The overall change in the fragility from Silica to Cu-Zr glass is around $5$ times, which is similar to the shift in fragility between $\rho = 0.750$ and $\rho = 0.943$ in the 3D HP model. If the effect of fragility is universal in nature, then one would expect to see similar relative changes in the yield strain relative to their poorly annealed critical yield strain with increasing degree of annealing. 
}


\begin{figure}
    \centering
    \includegraphics[width=0.47\textwidth,height=0.70\textwidth]{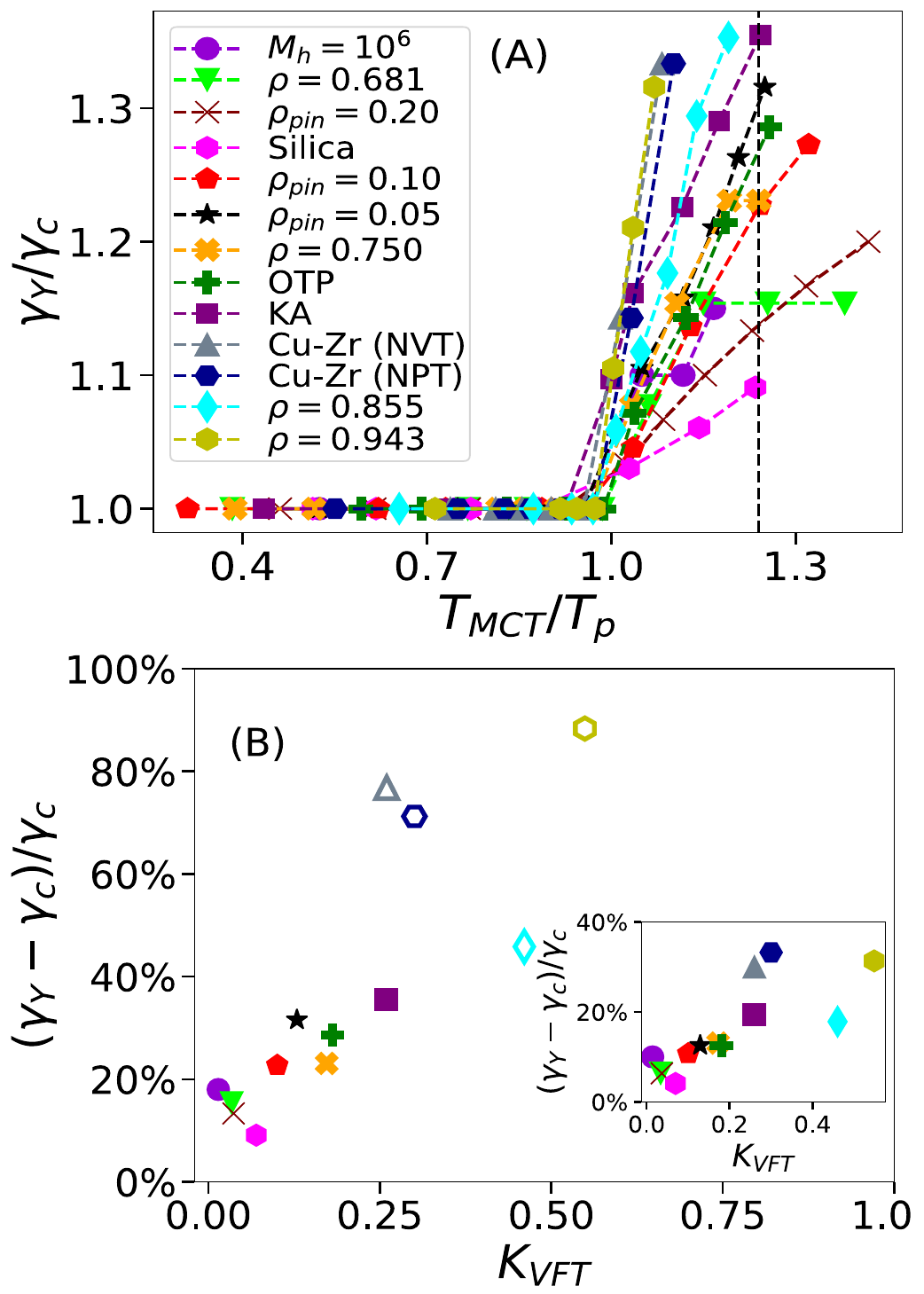}
    \caption{\rev{{\bf Universal Relation between Shift in $\gamma_Y$ and Kinetic Fragility $K_{VFT}$.} (A) Shows the normalised yield strain $\gamma_Y/\gamma_c$ as a function of normalised parent temperature, $T_{MCT}/T_p$ for 3D HP model (for all densities), Silica, OTP, 3D KA, Cu-Zr metallic glass model in NVT and NPT ensembles, 3D KA model with random pinning and soft pinning (see Ref.\cite{chatterjee2024effect}). The dashed lines are guides to the eye. (B) The normalised shift in yield strain, $(\gamma_Y - \gamma_c)/\gamma_c$, is first computed at a fixed degree of annealing ($T_p = 0.75 T_{MCT}$) across different model systems and plotted as a function of kinetic fragility parameter, $K_{VFT}$. The strong correlation observed confirms the fundamental role of fragility in the yielding transition in amorphous solids.} \SKrev{For a few fragile glass-formers we linearly extrapolated the data to obtain the shift in yield strain and they are shown using open symbols (see text for details). The inset shows the same correlation by taking $T_{MCT}/T_p = 1.07$. It shows the same correlation without any extrapolation.} }
    \label{GammaKvft}
\end{figure} 
\rev{In Fig. \ref{allModel}(B), we show the yielding diagram for the steady state energy of the BKS Silica model. This is the strongest (lowest fragility) glass-former out of the four models. The critical energy scale for this model is $T\simeq 3000K$, which is close to the MCT transition temperature, and the shift in the yield strain below this temperature is very small. Note that the absolute value of the yield strain for Silica is $\gamma_c\simeq 0.165$, which is much larger than the other models studied. We reiterate that we are not making a direct comparison between the absolute value of the yield strain and the fragility parameter, but rather we are making a direct connection between the shift in yield strain from its critical value $\gamma_c$ and fragility. Although, as suggested in \cite{lunkenheimer2023thermal,zaccone2014microscopic,zaccone2020rheology}, there is a possibility of a strong correlation between the yield strain and the fragility, which we have discussed later. In Fig. \ref{allModel}(C-E), we show a similar yielding diagram for OTP (C), 3D KA model (D) and Cu-Zr metallic glass (E). One can clearly see that the yield strain increases significantly with increasing annealing for OTP, 3D KA and metallic glass models. Fig. \ref{allModel}(F) shows a normalised plot of changes in yield strain ($\gamma_Y/\gamma_c$) with respect to the scaled parent temperature ($T_{MCT}/T_p$) for all four models. This figure clearly demonstrates that the largest fragile liquid (Cu-Zr glass in this case) shows the largest change in relative yield strain compared to the less fragile liquid (Silica glass) at a comparable degree of annealing. This result is in complete agreement with the results obtained for the 3D HP model. Thus, we can conclude that fragility solely controls the shift in the yield strain in well-annealed glass, irrespective of the origin of fragility in these model systems. This establishes beyond doubt that our results are universally valid across a wide range of glass-forming systems, irrespective of their microscopic details and the origin of their kinetic fragility.}

\begin{figure*}[htpb]
  \centering
  \includegraphics[width=0.97\textwidth]{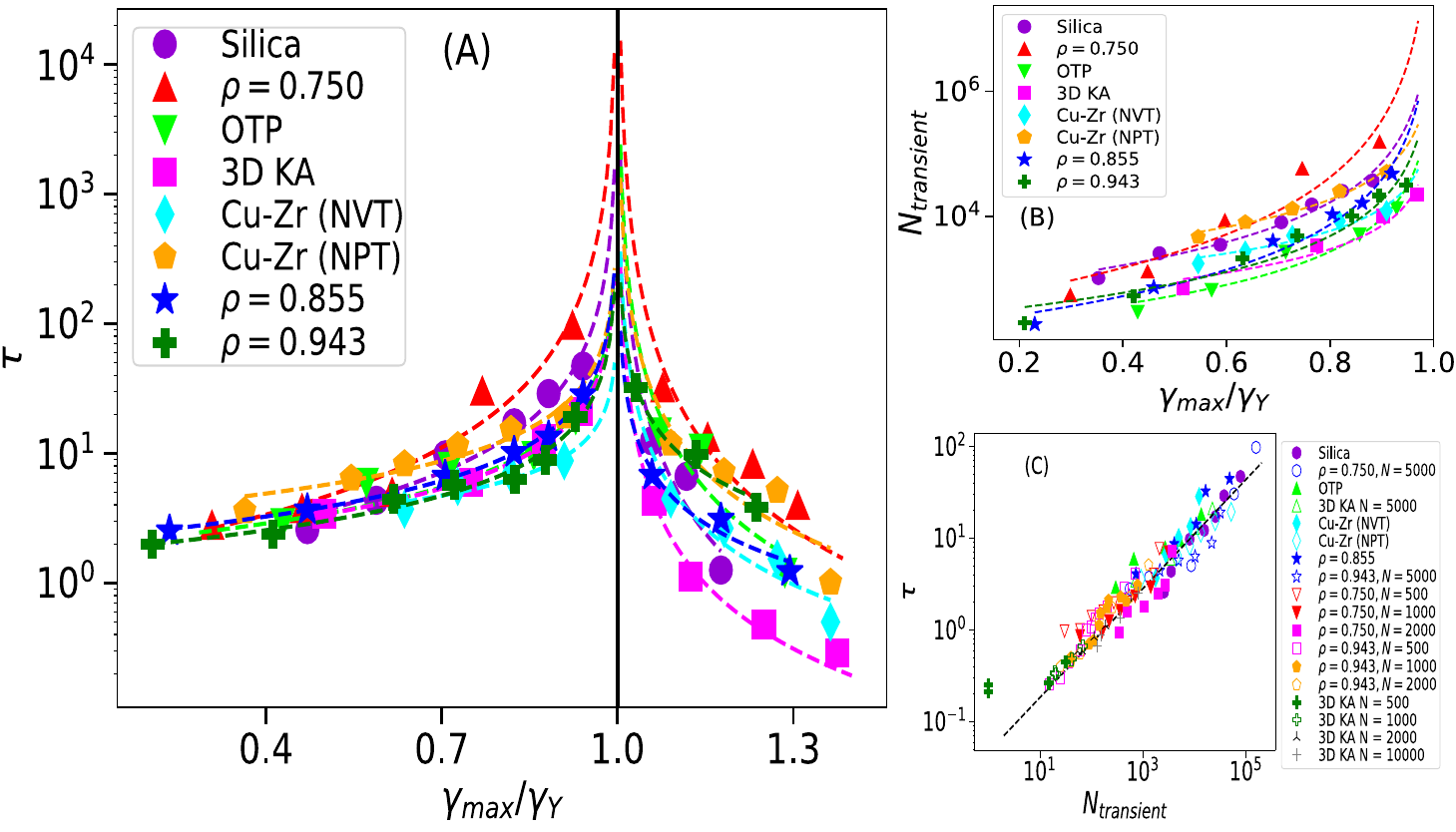}
  \caption{{\bf Diverging Timescales and Accumulated Plastic Events.} \rev{(A) The characteristic timescale $\tau$ related to the number of cycles required to reach an absorbing state for $\gamma_{max} < \gamma_Y$ or steady state for $\gamma_{max} > \gamma_Y$ is plotted as a function of $\gamma_{max}/\gamma_Y$. The divergence of $\tau$ follows a power-law behaviour (dashed lines) with varying exponents for different models. Strong glass-formers show statistically longer times than fragile glass-formers.} (B) Number of plastic drops $N_{transient}$ during transient cycles (before reaching the absorbing/steady state) is plotted against $\gamma_{max}/\gamma_Y$ for all models studied. \rev{$N_{transient}$ shows similar power-law divergence (dashed lines) with $\gamma_{max}/\gamma_Y$.} (C) The correlation between timescale $\tau$ and $N_{transient}$ for all studied models are shown. It is clear that $\rho = 0.750$ \rev{and Silica} go through a significantly larger number of plastic drops when $\gamma_{max}$ is close to $\gamma_Y$, and the corresponding timescale is also much higher than the others. \rev{It is very interesting to see a universal power-law relation between $\tau$ and $N_{transient}$ \SKReRev{for all the studied models irrespective of system sizes}. The dotted line represents a power-law fit with an exponent close to $0.594 \pm 0.051$.} }
\label{fig:4}
\end{figure*}
\begin{figure*}[htpb]
     \centering
     \includegraphics[width=0.94\textwidth]{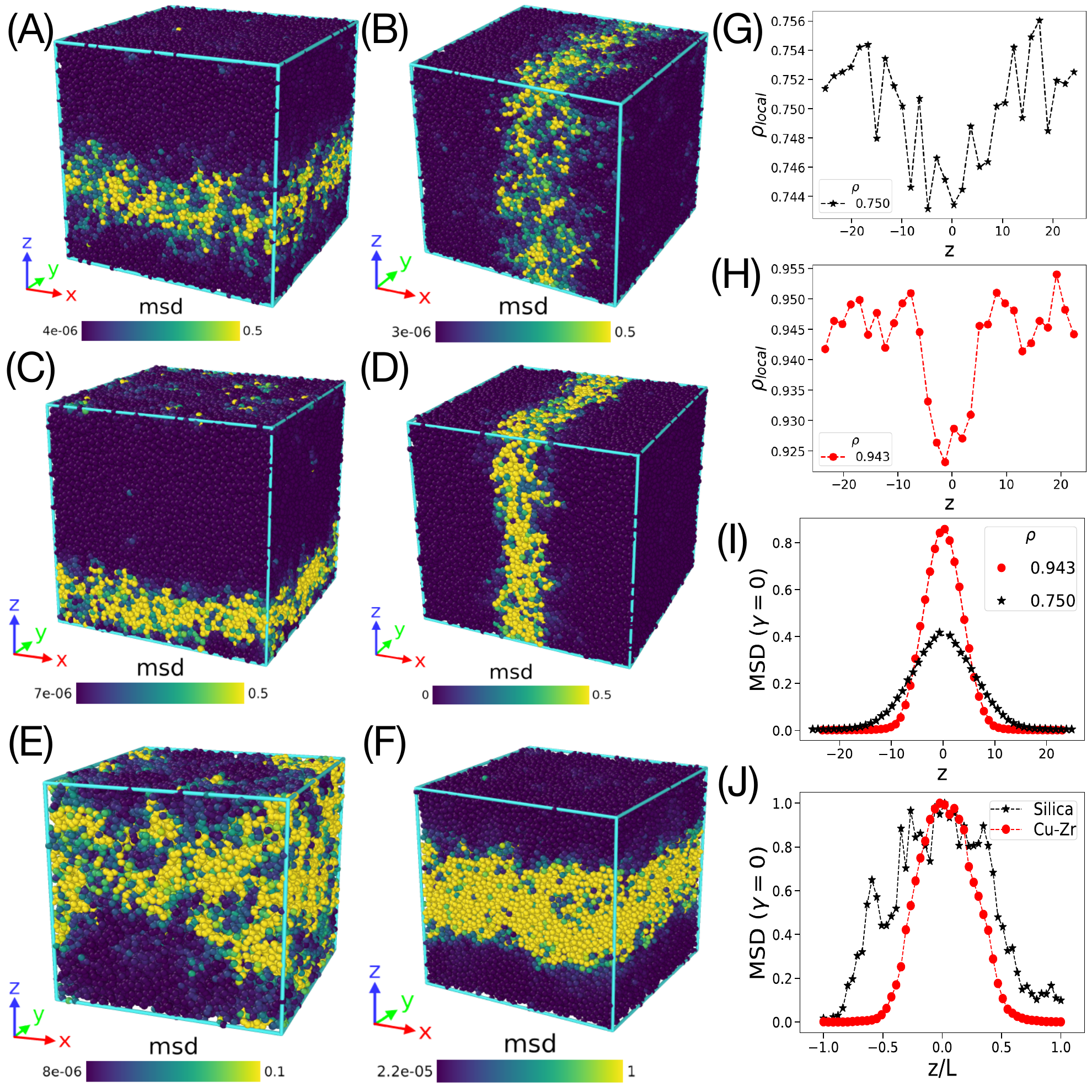}
     \caption{{\bf Structure of Shear Bands.} Typical horizontal (A) and vertical (B) shear bands for the strong glass-former ($T_p = 0.0015$ sheared with $\gamma_{max} = 0.08$) at the density $\rho = 0.750$. Similar horizontal (C) and vertical (D) shear bands for the most fragile glass-former ($T_p = 0.0032$ sheared with $\gamma_{max} = 0.10$) studied in this work at the density $\rho = 0.943$. \SKrev{(E) and (F) Typical horizontal shear bands for BKS Silica at $T_p = 2700$K and Cu-Zr metallic glass at $T_p = 750$K. Samples sheared at $\gamma_{max} = 0.18$ for Silica and $\gamma_{max} = 0.14$ in the case of metallic glass.} The plot of local density $\rho_{local}$ averaged in the shear plane across the shear band (z-direction) for configurations ($\gamma = 0$) sheared with $\gamma_{max} \approx \gamma_Y$ for two different densities $\rho = 0.750$ (G) and $\rho = 0.943$ (H). (I) Plot of MSD ($\gamma = 0$) across shear band along z-direction for configurations ($\gamma = 0$) sheared at $\gamma_{max} \approx \gamma_Y$ for two different densities $\rho = 0.750$ and $\rho = 0.943$. Dashed lines are fitted data via a Gaussian distribution. Width of the fitted curve $\sigma = 5.226$ for $\rho = 0.943$ and $\sigma = 8.351$ for $\rho = 0.750$. \SKrev{(J) Shows the distribution of MSD($\gamma = 0$) for these two glasses to highlight that Silica has relatively more diffuse shear band than metallic glass, as evidenced by the width of the distributions.} The dashed lines are guides to the eye. The shear is performed in the xz-plane with `x' being the shear direction and `z' being the perpendicular direction. `L' represents the simulation box length.}
     \label{shearband}
 \end{figure*}
\rev{In Fig. \ref{allModel}(G-J), we present the stress-strain curves ($\sigma_{max}$ vs. $\gamma_{max}$) for the Silica (G), OTP (H), 3D KA (I) models, and Cu-Zr metallic glass (J). It is evident that, within the studied temperature range, Silica glass exhibits less brittle behaviour than the Cu-Zr metallic glass model. In the OTP, 3D KA and Cu-Zr models, we observe a significantly larger stress drop, the magnitude of which increases with higher annealing, indicating brittle yielding via sharp shear band formation. The responses of these model systems under simple shear are detailed in the SI. In Fig. \ref{GammaKvft}(A), we have presented the shift in $\gamma_Y$ for all the models studied in this work and additional results for randomly pinned systems as well as systems with soft pinning using heavy masses (from \cite{chatterjee2024effect}).} 

\SK{To establish a quantitative correlation between shift in $\gamma_Y$ with fragility, we have computed the relative shift $(\gamma_Y - \gamma_c)/\gamma_c$ at a fixed degree of annealing by setting $T_p = 0.75 T_{MCT}$ for all the model systems and then plotted them against obtained kinetic fragility, $K_{VFT}$ in Fig.\ref{GammaKvft}(B). We had to extrapolate the shift in yield strain for the most fragile case ($\rho = 0.943$ for 3D HP model and Cu-Zr model) to get the change at $0.75 T_{MCT}$. Across a wide range of model glassy systems, including granular glass, molecular glasses, network glass, and metallic glasses, we observe a strong correlation between $(\gamma_Y - \gamma_c)/\gamma_c$ and $K_{VFT}$. This clearly establishes the universal effect of fragility on the yielding transition in amorphous solids. Note that in some of the fragile cases, the relative change in the yield strain could be very large. \SKrev{In the inset of the same figure panel, we show the correlation between $(\gamma_Y - \gamma_c)/\gamma_c$ and $K_{VFT}$ for $T_{MCT}/T_p = 1.07$ such that we do not have to do any extrapolation. This also shows a strong correlation between the shift in the yield strain relative to the critical yield strain and increasing fragility.}
}

\vskip 0.1in
\noindent{\large Microscopic Understanding of Diverging Timescales \& Relation to Shear Bands:}
\rev{Till now, we have investigated the steady state properties of the system and the physics associated with the yielding diagram. We now focus on an important dynamical aspect of the systems, namely the time (or number of cycles) taken by a poorly annealed sample to reach the steady state. We refer to that time as $\tau$. We have fitted the zero-strain energy (stroboscopic energy) as a function of cycles using a stretched exponential function to find the timescale $\tau$ (see SI). It is known in the literature \cite{bhaumik2021role}, that the number of cycles required to reach steady state shows a power-law divergence as it approaches the critical yield strain, $\gamma_c$, as $\tau \sim |\gamma_{max} - \gamma_c|^{-\eta}$. In Fig. \ref{fig:4}(A), we show $\tau$ as a function of $\gamma_{max}/\gamma_Y$ for all the models studied in this work. The lines are the fits to the power-law expression. We see that all models universally show this power-law divergence of timescale while approaching the yielding strain from both sides. The divergence exponents are different while approaching the transition from below and above. The numerical values of the exponents for each glass-former are given in the SI. Interestingly, we see that strong glasses take a statistically larger number of cycles to reach a steady state than fragile glass-formers, and the power-law exponent for strong glass-formers is slightly larger than that of fragile ones.}

To understand the microscopic origin of this large timescale to reach absorbing steady states, we computed the number of plastic drops $N_{transient}$ that the system traverses through to reach the absorbing state. The total plastic drops in the transient states (before reaching the steady state) when plotted against normalized oscillatory shear strain amplitude $\gamma_{max}/\gamma_Y$ for \rev{all model systems,} one sees that strong glass-formers (see $\rho = 0.750$ \rev{for 3D HP and Silica}) go through a significantly large number of plastic drops at the same relative strain amplitude as shown in Fig. \ref{fig:4}(B). Next, to connect the total time to reach a steady absorbing state to the number of plastic drops the system has gone through, we plotted the correlation between timescale $\tau$ and $N_{transient}$ for all studied models in Fig. \ref{fig:4}(C). It is interesting to see that $\tau$ and $N_{transient}$ follow a universal power-law relation as $\tau \sim N_{transient}^{\alpha}$ for all the models\SKReRev{ irrespective of system sizes. The dotted line represents a power-law fit with exponent $\alpha \simeq 0.594 \pm 0.051$.} This observation is very similar to the observation made in \cite{adhikari2023encoding} for the 3D KA model. \SKrev{Note that the number of plastic drops depends on the system sizes in a power-law manner, but it was shown in \cite{adhikari2023encoding} that this power-law relationship is universally obeyed for all system sizes as well. Thus, we expect this universal relation to be valid for the studied model systems in this work at different system sizes. In \cite{C3SM53134A}, it was shown that the number of plastic drops increases as the system approaches the unjamming transition. For the 3D HP model, some of these microscopic reasons might also be at play. The effect of jamming transition on the yielding transition under oscillatory shear needs further exploration.}
In \cite{chatterjee2024effect}, it was also shown very convincingly that even in systems with random pinning, which alters the fragility of the parent unpinned system by a large factor with increasing pinning concentration, the universal relation between $\tau$ and $N_{transient}$ holds good both in two and three dimensions. Although we do not have a microscopic understanding of this universal relationship between $\tau$ and $N_{transient}$, this correlation certainly ensures that the timescale is connected with the number of plastic drops the system traverses through before reaching the steady absorbing states in a fundamental manner. A better understanding of this universal relation in these systems will undoubtedly have important implications for gaining deeper insights into the mechanical response of amorphous solids under oscillatory shear deformation and the memory formation in these materials.

\begin{figure}
    \centering
    \includegraphics[width=0.49\textwidth]{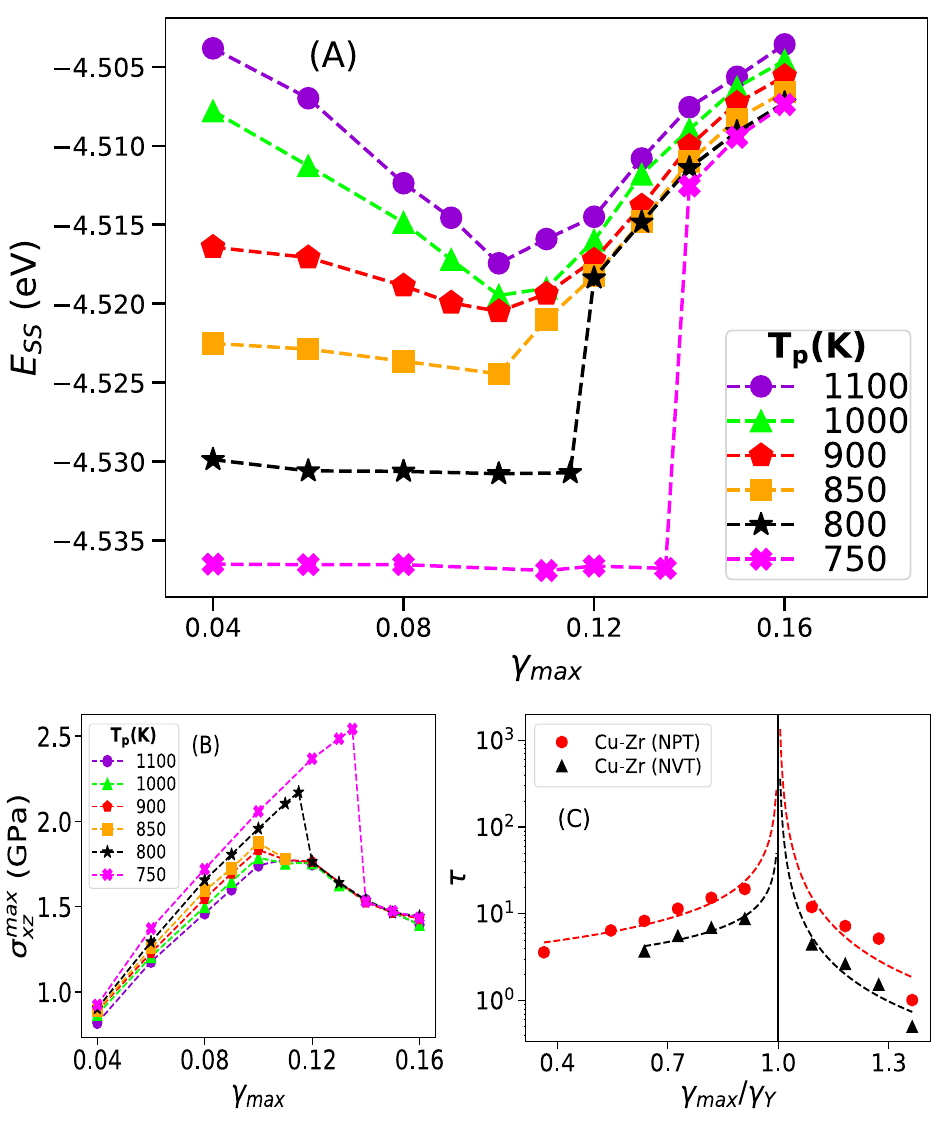}
    \caption{\rev{{\bf Ensemble Dependence.} (A) Shows the steady state yielding diagram of Cu-Zr metallic glass at zero pressure (NPT ensemble). The yielding diagram looks very similar to the one done in the NVT ensemble at $\rho = 7.4 gm/cm^3$, apart from the steady state energy profile beyond yield strain. In NPT, the steady states do not fall on top of each other for $\gamma_{max} > \gamma_c$. (B) Shows the corresponding $\sigma^{max}_{xz}$ vs $\gamma_{max}$ for all the parent temperatures. One clearly sees a large stress drop at the yield strain for well-annealed samples. It is interesting to see that the steady state stresses are the same for all the samples with different parent temperatures, although their steady state energies are different. The dashed lines are guides to the eye. (C) Shows the stark difference in the timescale, $\tau$, to reach the absorbing state in NPT and NVT. $\tau$ in the NPT ensemble is surprisingly much larger than in the NVT ensemble. The dashed lines represent fits to the data.}}
    \label{NPT_NVT}
\end{figure}

\SK{In \rev{the inset of } Figs. \ref{fig:1}(F) and \ref{allModel}(F), we have illustrated that for a similar degree of annealing, the stress jumps around the yielding strain gradually decrease with a decrease in fragility, indicating a more rounded, ductile-like response for strong liquids. To explore this behavior, we further investigated the formation of shear bands across the yielding transition. These investigations involved samples prepared at parent temperatures approximately $2\%$ below their respective Mode coupling temperatures ($T_{MCT}$). We had to perform simulations for a large system size ($N = 100000$) to study the formation of shear bands and their structure with the changing fragility of the corresponding glass-formers. In Fig. \ref{shearband}(A) and (B), we present typical horizontal and vertical bands, respectively, for a strong glass-former at a density $\rho = 0.750$, while similar shear bands for a fragile glass-former at a density $\rho = 0.943$ are depicted in  Fig. \ref{shearband}(C) and (D). It is evident that shear bands in strong glass-formers appear more diffusive than those in fragile glass-formers. \SKrev{Fig. \ref{shearband}(E) and (F) represent the typical horizontal shear bands in BKS silica ($T_p = 2700$K, $N = 48000$) and Cu-Zr metallic glass ($T_p = 750$K, $N = 50000$) respectively. Though the parent temperature of Silica is $15\%$ below $T_{MCT}$ we observe a diffused shear band. In case of metallic glass the parent temperature is only $8\%$ below $T_{MCT}$ but still we see a sharp localized shear band. Details are mentioned in SI. Earlier work \cite{bhaumik2022avalanches} also shows a diffused shear band in case of Silica even at very well annealed samples.} For a more quantitative comparison, we computed the local density ($\rho_{\text{local}}$) averaged in the $XY$ plane across the shear band (in the Z-direction) for the strong glass-former (Fig. \ref{shearband}(G)) and the fragile glass-former (Fig. \ref{shearband} (H)). Fig. \ref{shearband}(G) clearly shows that the density across the shear band in the strong glass-former remains nearly constant, with a slight decrease in the local density within the shear band region. The maximum percentage density change in the shear band region is approximately $1.6\%$. Conversely, for the fragile glass-former, as illustrated in panel (H), there is a sharp dip in the average local density profile, with the maximum percentage of density change in the shear band region being nearly $3.7\%$. These results strongly suggest that strain localization in the fragile glass former, with a similar degree of annealing, is much more significant than in the strong glass former. To directly compare the shear band width in these two glass-formers, we computed the mean squared displacement of particles for stroboscopic configurations ($\gamma = 0$) across the shear band, referred to as MSD($\gamma = 0$). In Fig. \ref{shearband}(I), we display the distributions of the MSD for strong and fragile glass-formers. The distribution clearly indicates that the shear band width in the strong glass-former is much larger ($\approx 1.6$ times) than in the fragile glass-former. \SKrev{We also observed a broader distribution of MSD $(\gamma = 0)$ in case of Silica than metallic glass in Fig. \ref{shearband}(J).} The detailed structural analysis of the shear band in strong and fragile glass-formers with similar degrees of annealing very well rationalizes the results obtained in our previous sections, namely, yielding in the strong glass-former is less brittle than in the fragile glass-former when compared at the same degree of annealing. \rev{It will be interesting to study a statistical correlation between the width of the shear band and fragility in the near future across model systems. This might be important for future applications in better material design.}}

\vskip 0.05in
\noindent{\large Ensemble Dependence:} 
\rev{In this section, we highlight some puzzling observations related to the dependence of our results on the details of the ensemble used for simulations. In Fig. \ref{NPT_NVT}(A), we have shown the steady state yielding diagram of Cu-Zr metallic glass performed at zero pressure in the NPT ensemble. The results are qualitatively similar to the results reported in Fig. \ref{allModel}(E) for the same model in the NVT ensemble, with the notable difference of steady state energy for $\gamma_{max} > \gamma_c$. One observes that the $E_{SS}$s are the same for $\gamma_{max} > \gamma_c$ for the NVT ensemble, while they are different for the NPT ensemble. This suggests that steady-state properties are not the same at a constant pressure, but rather they are the same at constant density. At the same time, one sees that the steady state stresses are the same for all the samples with different parent temperatures. A better understanding of this interesting observation will be crucial in understanding results from experiments, which are typically performed in a constant-pressure scenario. It is also important to note that the effect of fragility on the shift of $\gamma_Y$ compared to $\gamma_c$ is the same in both the NVT and NPT ensembles, as Cu-Zr has comparable fragility values in the NVT and NPT. This reconfirms that the effect of fragility on the yielding transition is universally applicable to different simulation conditions and thereby will be directly testable in experiments. Another interesting difference that draws our attention is the strong difference in the timescale required to reach steady states in these two ensembles. It seems that the number of cycles required to reach an absorbing state for $\gamma_{max} < \gamma_c$ or the steady state for $\gamma_{max} > \gamma_c$ in the NVT ensemble is much smaller than the same in the NPT ensemble. In Fig. \ref{NPT_NVT}(B), we have shown such a comparison. The stark difference is very puzzling, but we do not have any microscopic understanding of this observation, except for the fact that they follow the same power-law relation with the number of plastic drops in the transient states, $N_{transient}$. Also note that the 3D HP model shows significant variation in fragility with increasing pressure in the NPT ensemble, as shown in the SI; thus, one is nearly certain that the results for the 3D HP model in the NVT ensemble will be quantitatively similar even in the NPT ensemble.}

\vskip 0.1in
\noindent{\large Elastoplastic Mesomodel Results:}
\SK{We now consider whether the observed phenomena in simulation can be reproduced and the underlying mechanism could be understood using a simple mean-field mesoscopic model. While the understanding of yielding behavior under uniform shear deformations using a mesoscopic model has been extensively studied \cite{nicolas2018deformation,popovic2018elastoplastic,barlow2020ductile,talamali2012strain}, there is a growing interest in comprehending yielding behavior under cyclic shear deformation recently \cite{sastry2021models,parley2022mean, Maloney2021, rossi2022finite, kumar2022mapping,liu2022fate,cochran2024slow}. 
The main ingredient of many of these models is the assumption of a local energy landscape associated with a volume of material, often referred to as a meso block. However, it is not immediately clear how large a volume one needs to consider for the validity of the assumption. One assumes that each of these blocks of materials can be regarded as independent from the others, and a collection of such blocks can constitute a macroscopic body of materials.} 

The absence of interactions between blocks can be thought of as a mean-field approximation. Within this approximation, one further assumes that each of these meso-blocks evolves on the energy landscape when subjected to various deformations, such as oscillatory shear deformation. When the block reaches its local yield threshold, it topples to another energy minimum among all the accessible minima it has, and the process continues as long as the body is subjected to the deformation. In Ref. \cite{sastry2021models}, it was assumed that such a block is associated with an absolute energy minimum $E_0$ and a strain $\gamma_0$ (where energy is minimum as a function of strain). The yield threshold is modelled via a stability range of strain ($\gamma_{\pm}$), beyond which the system topples to a new energy minimum. 
Under strain, a quadratic variation of energy is considered, assuming Hooke's law of linear elasticity. It was further assumed that $E_0$ follows a Gaussian distribution, 
\begin{equation}
P\left(E_0\right) = \sqrt{\frac{2}{\pi {\sigma}^2}} \exp{\left(-\frac{{E_0}^2}{2 {\sigma}^2}\right)},
\end{equation}
where $-1 <E_0 <0$, $\sigma = 0.1$ and energy evolution with strain $\gamma$ is, 
\begin{equation}
E (\gamma,E_0,\gamma_0) = E_0 + \frac{\mu}{2} (\gamma - \gamma_0)^2,
\end{equation}
where $\mu = 1.1$ and the stability limit is, $\gamma_{\pm} = \gamma_0  \pm  \sqrt{|E_0|}$. 
The yielding diagram of this model closely resembles what we observed for fragile glass formers. However, neither this model nor others in the literature \cite{parley2022mean, Maloney2021, rossi2022finite, kumar2022mapping, liu2022fate, cochran2024slow} consider the impact of the initial glass former's fragility on the yielding diagram. However, our simulation results suggest a significant change in the yielding diagram when the fragility of the initial glass former varies. Therefore, we propose a new model similar to the concept discussed in Ref. \cite{sastry2021models} but with a crucial difference of energy barrier playing a crucial role in plasticity.
\begin{figure*}[htpb]
  \centering
  \includegraphics[width=0.98\textwidth]{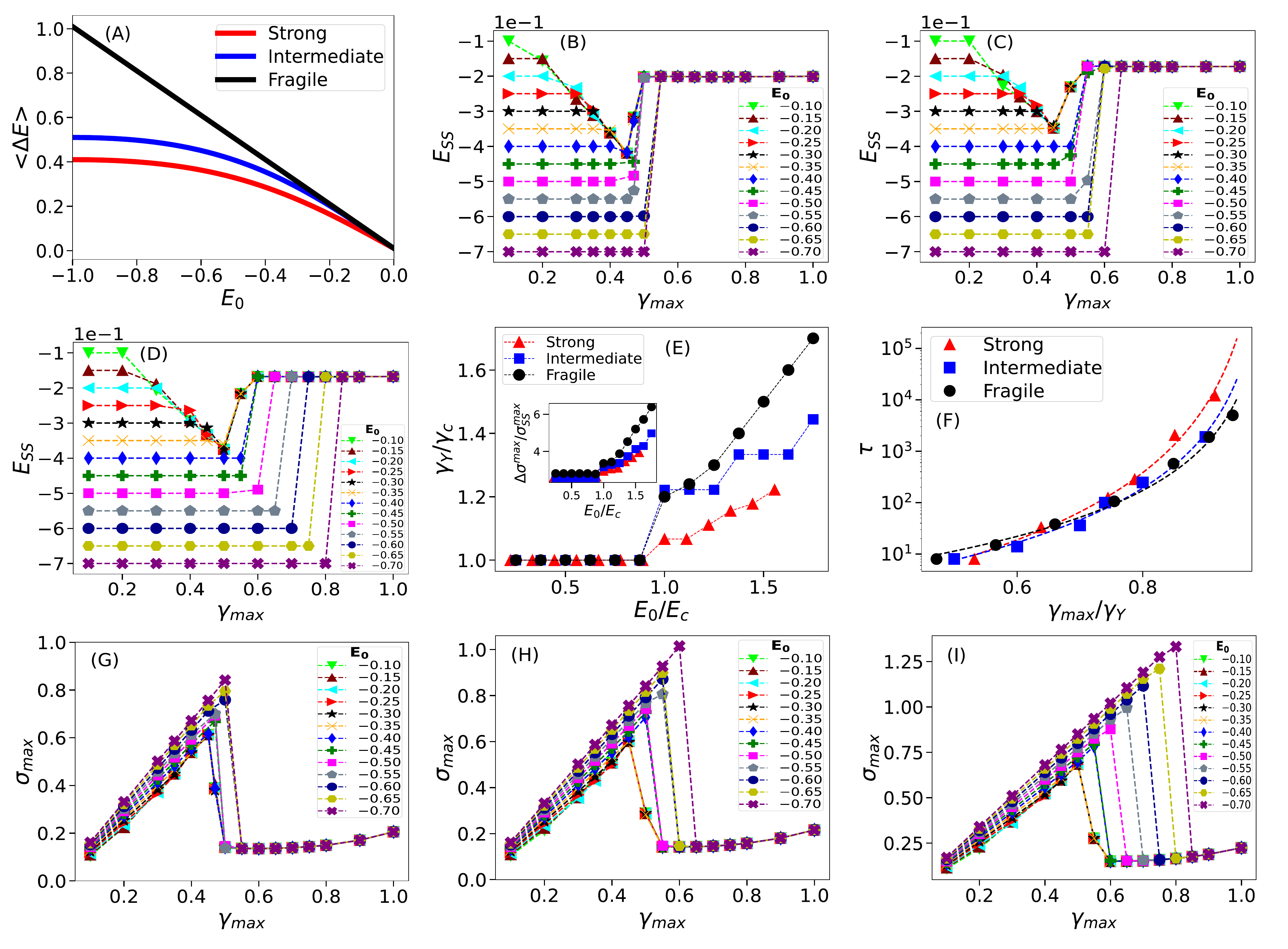} 
  \caption{{\bf Mesoscale Elastoplastic Model: } (A) Schematic of Mean barrier as a function inherent state energy $E_0$ following Eq.\ref{eqs_fragile} and Eq. \ref{eqs_strong}. (B) Steady-state energy vs $\gamma_{max}$ for a model strong glass former with energy barrier depicted by the red curve in panel (A). The energy barrier saturates with increasing annealing (equivalent to more negative inherent structure energy, $E_0$). Note that the poorly annealed states show further annealing with increasing $\gamma_{max}$, but for better-annealed states, there is no mechanical annealing, and yield strain does not increase. (C) A similar yielding diagram for a model system with intermediate fragility is depicted in panel (A) by a blue curve. In this case, the poorly annealed part of the diagram shows similar results to the strong glass former, but for the better-annealed part, yield strain increases with increasing annealing. (D) Shows the results for the most fragile case (depicted in panel (A) by a black curve). The striking difference is the better-annealed part of the yielding diagram. The yield strain increases significantly with increasing annealing in complete agreement with simulation results. (E) The shift in yield strain $\gamma_Y/\gamma_c$ is plotted with annealing $E_0/E_c$. The inset shows the stress drop with annealing. (F) Shows the number of cycles needed to reach steady state ($\tau$) vs $\gamma_{max}/\gamma_Y$ from the elastoplastic model. $\tau$ is the smallest for a given $\gamma_{max}$ for the fragile system, indicating the ease of reaching a steady state for fragile glass formers. Excellent qualitative match with the simulation results is indeed very encouraging. (G) $\sigma_{max}$ vs $\gamma_{max}$ plot for strong glass-former. (H) and (I) show a stress-strain curve for intermediate and fragile scenarios. Again, the qualitative match with the simulation results shown in Fig. \ref{fig:1}(H)-(J) is excellent. See the text for more discussions. The dashed lines are guides to the eye.}
\label{fig:3}
\end{figure*}

Previous studies \cite{tah2022kinetic,tah2022fragility} have shown that the energy barrier ($\Delta E$) of the fragile glass-formers increases rapidly with decreasing temperature or increasing annealing, but for strong glass-formers, the energy barrier increases rather modestly. We have introduced this crucial aspect of variation of the energy barrier with parent temperature or annealing in our version of the elastoplastic model. The energy barrier distribution is assumed to be Gaussian, with the mean energy barrier, $\langle\Delta E\rangle$, varying with $E_0$ or parent temperature, $T_p$. We have chosen two trial functions for $\langle\Delta E\rangle$ variation with annealing ($E_0$). For fragile glass-formers,
\begin{equation}
\langle \Delta E\rangle = a + bc |E_0|,
\label{eqs_fragile}
\end{equation}
and for strong glass-formers 
\begin{equation}
\langle \Delta E\rangle = a + bc\left(\frac{E_0^2}{1 + E_0^2}\right),
\label{eqs_strong}
\end{equation}
with $a = 0.01$, $b = 0.80$ and c is a tuning parameter. We have chosen $c = 1$ and 1.25 for strong and fragile glass formers, respectively. We have used Eq. \ref{eqs_strong} with $c = 1.25$ to produce an intermediate case between strong and fragile. A schematic is shown in Fig. \ref{fig:3}(A). We propose that the stability limit should depend on the energy barrier ($\Delta E$) as
\begin{equation}
\gamma_\pm = \gamma_0  \pm  k {[\Delta E]}^{2/3},
\label{barrierEq}
\end{equation}

\begin{figure*}[htpb]
    \centering
    \includegraphics[width=0.985\textwidth]{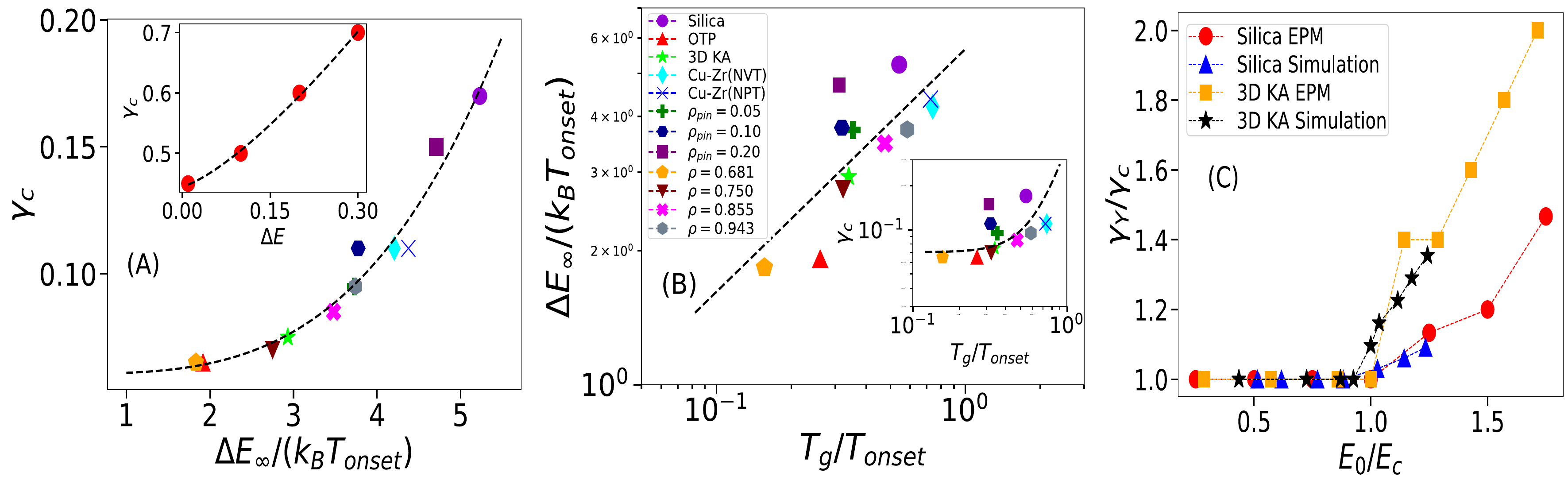}
    \caption{\rev{{\bf Direct Comparison between Simulation and EPM.} (A) The cross correlation between the critical yield strain $\gamma_c$ and high temperature energy barrier (see text for details) suggests \SKrev{a power-law dependence with an exponent close to $3.43$}. The inset shows a similar \SKrev{ cross-correlation from EPM with understandably different power-law exponent due to the mean field nature of the EPM. The exponent is $1.33$ in this case. (B) Cross-correlation between the scaled energy barrier and scaled $T_g$. They seem to be proportional to each other. The inset shows a similar cross-correlation between $\gamma_c$ and $T_g$. The line is the power-law fit with exponent $3.43$. Although there is scatter in the data, it broadly obeys the power-law relation.} (C) $\gamma_Y/\gamma_c$ is plotted against inherent structure energy scaled by the critical energy $E_0/E_c$ (a proxy for $T_p/T_{MCT}$). Predicted $\gamma_Y/\gamma_c$ from the elastoplastic model (EPM) using the energy barrier information from $\tau_{\alpha}$ vs $T$ data compares very well with the oscillatory shear simulation results. This suggests that the energy barrier directly controls the shift in the yield strain. Comparison in both the Silica and the 3D KA models is observed to be very good. The dashed lines are guides to the eye.}}
    \label{gammacVsE}
\end{figure*}
where $k = 1$. The exponent $2/3$ is used instead of $1/2$ as it was already shown that the energy barrier vanishes as a power-law with exponent $2/3$ as the plastic strain is approached \cite{karmakar2010statistical}. \rev{Note that an exponent $1/2$ will also produce qualitatively similar results, and the effect of a different exponent is expected to be relevant when one studies the statistics of plastic drops and avalanches and their size distribution, etc}. We also considered uniform distribution of $\gamma_0 \in (-1,1)$ as in \cite{sastry2021models}. The cyclic shear in this model goes as follows. 
\begin{enumerate}[label=\Roman*.]
    \item Start with a initial value of $E_0$ and $\gamma_0 = 0$. 
    \item For this $E_0$, get $\langle\Delta E\rangle$ from the Eq. \ref{eqs_fragile} or Eq. \ref{eqs_strong} and the corresponding $\Delta E$ is chosen from Gaussian distribution whose width $\sigma_{\Delta E}$ decreases linearly from 10\% to 0.001\% of $\langle\Delta E\rangle$ with annealing. 
    \item Define the stability limit using Eq.\ref{barrierEq}. 
    \item Energy is then varied with strain as $E(\gamma, E_0,\gamma_0) = E_0 + \frac{\mu}{2} (\gamma - \gamma_0)^2$ until the stability limit is reached. $\mu$ is increased linearly from $1$ to $2$ with annealing ($E_0$).
    \item Whenever the system crosses its stability limit $\gamma_\pm$, it jumps to any lower energy mesostate at the transition strain, and a new set of $E_0$ and $\gamma_0$ is chosen from the respective distributions, and we repeat the process from (ii). 
\end{enumerate}
 The strain $\gamma$ is varied cyclically with amplitude $\gamma_{max}$, and the number of cycles is around $10000$. If the zero strain energy of the system doesn't change with successive cycles, the procedure is terminated. The average zero-strain energies of the last $500$ cycles are calculated to obtain $E_{SS}$. We took 100 independent samples for better averaging. We studied the model within the same range of $E_0 \in (-0.10, -0.70)$ for all the samples. $E_{SS}$ vs $\gamma_{max}$ is plotted for all the cases for this elastoplastic model in Fig. \ref{fig:3}(B) - (D). As we can see, the yielding point for the strong model barely shifts below the critical energy. In contrast, for the fragile model, the shift in the yielding point with annealing is very large, in complete agreement with simulation results. \rev{In Fig. \ref{fig:3}(E), we have shown the normalised yield strain $\gamma_Y/\gamma_c$ as a function of $E_0/E_c$ (a proxy for the parent temperature) to highlight the larger shift in yield strain in fragile glass-formers than in strong glass-formers.}

From our modified elastoplastic model, we get additional information about timescales (the number of cycles needed to reach a steady state) for both the strong and fragile glass formers.  The timescales are found to show a power-law-like divergence at their respective yield strain $\gamma_{max} \approx \gamma_Y$. In Fig. \ref{fig:3}(F), we show the timescale obtained from our simulations of the elastoplastic models for strong, intermediate, and fragile glass-formers. One can clearly see that the number of cycles needed to reach a steady state for a strong glass-former is way more significant than for fragile glass-formers for a given $\gamma_{max}/\gamma_Y$. \rev{We found that the power-law divergence exponent is higher for strong glass than for the fragile one at a given $\gamma_{max}/\gamma_Y$, in agreement with the MD simulation results. These results suggest the crucial role played by the energy barrier in modeling these systems even at the mean-field level.} A detailed characterisation of this divergence is shown in the SI.

In Fig. \ref{fig:3}(G)-(I), we show $\sigma_{max}$ vs $\gamma_{max}$ as obtained from our elastoplastic model (EPM) to draw a parallel comparison with the simulation results in Fig. \ref{fig:1}(G)-(I). One also sees that for strong liquids, the yield strain does not change much, and the stress jump is found to be small compared to the fragile ones. In fragile glass-formers, the yield strain changes significantly with annealing, and the corresponding stress drops at the yielding also increase, as seen in the simulations. Excellent agreement between the mean-field elastoplastic model and the simulation results is really encouraging, suggesting that much of the oscillatory shear yielding in a wide variety of amorphous solids might be possible to understand within the mean-field framework. In the future, we plan to extend this model to include spatial dimensions to see whether some of the missing physics can be incorporated into the elastoplastic model.

\rev{Till now, we have discussed about the relation between fragility and the relative shift in the yield strain with respect to poorly annealed values and convincingly demonstrated that changes in the energy barrier with annealing play the most important role in determining this. Our mean-field EPM calculations reconfirmed this. To extend this further, we computed the actual changes in the energy barrier with parent temperature in the BKS Silica and 3D KA model (see SI for details). We then feed this formation into our mean-field EPM. We readjusted the absolute value of the energy barrier by a scale factor without changing its parent-temperature dependence such that the yield strain, $\gamma_c$, in the EPM calculation matches the MD simulation results. This is done without any loss of generality, as we discovered that $\gamma_c$ is strongly correlated with the value of the high-temperature energy barrier $\Delta E_{\infty}$ as demonstrated in Fig. \ref{gammacVsE}(A). \SKrev{The data suggests a power-law dependence of $\gamma_c$ on $\Delta E_{\infty}$ scaled by the onset temperature $T_{onset}$ as 
\begin{equation}
\gamma_c \sim \left(\frac{\Delta E_{\infty}}{k_B T_{onset}}\right)^\delta,
\end{equation} 
where $\Delta E_{\infty}$ is computed by fitting the $\tau_{\alpha}$ vs $1/T$ data using Arrhenius relation at high temperatures as shown in the SI. The exponent $\delta\sim3.43$. In Fig.\ref{gammacVsE}(B), we show the correlation between the scaled energy barrier and $T_g$. They seem to be proportional to each other within the range of our data. Thus, we expect $\gamma_c$ to have a power-law relation with $T_g$ as shown in the inset of the same figure. The line is a power-law fit with the same exponent $\delta \sim 3.43$. There is some scatter in the data, but it broadly follows the relation; more data are needed to draw definitive conclusions.}} 

\rev{The expectation of a stronger correlation between $\gamma_c$ and $\Delta E_{\infty}$ came from our mean-field EPM calculations, which showed a similar power-law dependence on the chosen energy barrier, as shown in the inset of Fig. \ref{gammacVsE}(A) \SKrev{with $\delta \simeq 1.33$}. The difference between the simulation and EPM results on the particular dependence of $\gamma_c$ on $\Delta E_{\infty}$ can be associated with the fact that the EPM is at a mean-field level, and it will be interesting to see if, when one incorporates spatial information, how the dependence changes. At this point, we do not have a microscopic understanding of this observed correlation between $\gamma_c$ and $\Delta E_{\infty}$, but it is very encouraging to see how the high-temperature energy barrier correlates with the yield strain. One could potentially predict the yield strain of poorly annealed samples based on the relaxation time data from their high-temperature liquid states. This hypothesis is worth investigating in the near future.} 

\SK{In Fig. \ref{gammacVsE}(C), we show $\gamma_Y/\gamma_c$ as a function of inherent structure energy scaled by the critical energy $E_c$. This parameter can be thought of as a proxy for the parent temperature or degree of annealing. The direct comparison between the predicted $\gamma_Y/\gamma_c$ from EPM using the energy barrier information from the relaxation time data and the oscillatory shear simulation results clearly suggests that the energy barrier solely controls the shift in the yield strain. Comparison in both the Silica and the 3D KA models is observed to be very good. These results once again reconfirm the universal role of fragility via the energy barrier in controlling the yielding transition in amorphous solids.}

\rev{We believe that our discovery of the energy barrier universally controlling the shift in yield strain represents a significant advancement in understanding the physics of the yielding transition for a broad range of amorphous solids encompassing granular, colloidal, molecular and metallic glasses. In fact, our results allow us to rationalise the striking differences in the yielding behaviour observed in the 3D KA and Silica models from a microscopic perspective. Recent investigations \cite{sharma2023activeannealing,goswami2023yielding} on the shear deformation of active glasses indicate that the yielding point changes little with increasing annealing. Our results indicate that one can rationalise that all these apparent changes in yielding behaviour are closely related to the system's fragility across all models and irrespective of deformation conditions.}

\vskip +0.1in
\noindent{\large\bf Discussion: }
\SK{\rev{Through extensive molecular dynamics simulations of a wide class of model amorphous solids having a large variation in their respective fragilities,} we show that the initial glass former's fragility strongly influences the nature of the yielding transition under cyclic shear deformation \rev{in a universal manner}. The fragility of a glass former, which indicates how quickly the viscosity or relaxation time \rev{and thereby their energy barriers} increases with decreasing temperature in the supercooled liquid temperature regime, \rev{dictates both the critical yield strain for poorly annealed glasses and changes in yield strain with respect to it in well-annealed glasses}. Our findings are particularly noteworthy because they reveal how the kinetic properties of liquids can strongly influence the mechanical properties of amorphous solids, providing new insights into the behavior of these materials.} 

\SK{Our results reveal that, while the yielding response of strong and fragile liquids in their poorly annealed states is similar, a striking difference emerges when they are better annealed. We found that fragile glass formers show a large increase in yield strain \rev{relative to their poorly annealed critical yield strain} with increasing annealing. In contrast, strong glass formers show almost no change in their yielding strain with better annealing. This difference is also apparent in their stress-strain relationship. Strong glass-formers show nearly ductile yielding with no significant change in their characteristic with increasing annealing. In contrast, fragile glass-formers show increasingly brittle yielding behavior with increasing annealing, characterized by a very large stress jump at the corresponding yield strain. Additionally, we observe that strong glass formers statistically take more shear cycles to reach an absorbing state when the strain amplitude is below their yield strain, while fragile glass formers reach their absorbing state much faster at the same strain amplitude. We also demonstrated that the number of oscillatory cycles needed to reach the absorbing steady state universally depends on the number of plastic events the system has to go through before reaching the absorbing state. This suggests a fundamental connection between the plasticity and the absorbing states in these disordered solids.}

\rev{The striking differences in the yielding diagrams of fragile and strong glass-formers} led us to propose a modified mean-field elastoplastic model that considers the crucial role of energy barriers in amorphous plasticity, which was previously missing in other works. By introducing a temperature-dependent growth of energy barrier, particularly a weakly growing energy barrier for strong glass-formers and a strongly growing energy barrier for fragile glass-formers with parent temperature \rev{(equivalently, the degree of annealing)}, we could accurately explain the observations from the molecular dynamics simulation. We also found that the yielding diagram under oscillatory shear can be useful for estimating the growth of energy barriers in various glass-formers and may help us better understand their liquid properties. 

\rev{We found that the critical yield strain for poorly annealed samples is strongly correlated with the high-temperature energy barrier that governs the Arrhenius temperature dependence of relaxation time or viscosity in the liquid regime. This correlation appears to be power-law in nature, suggesting that it may not merely be coincidental but rather a crucial factor in determining the yield strain in amorphous solids, as indicated by mean-field elastoplastic models. While the dependence of critical yield strain on the energy barrier in the mean-field elastoplastic model is also power-law-like, albeit with a different exponent, we believe this difference arises from the model's lack of spatial information. Conducting a systematic study in this area is essential, as it could provide predictive insights for assessing failure strain and, consequently, failure load for materials based solely on their dynamic characteristics. This knowledge would also facilitate the engineering of materials with desirable mechanical properties and improved load-bearing capacities.} 

\SKrev{Although we have studied a wide range of model systems interacting via both binary and multi-body potential (metallic glass), it will be interesting to extend our analysis to other models with anisotropic interaction as well as systems with three-body angular interactions \cite{weber1985local} to understand how microscopic properties play roles in the oscillatory shear response, especially shear band formation. It will also be interesting to study oscillatory shear yielding transition in biological systems like epithelial tissues using models like the vertex model or deformable particle model, which are minimal physics-based models of these systems \cite{fletcher2014vertex,nayak2025glassy,bose2025understanding,li2018role,gnan2019microscopic}.}

Finally, our results offer valuable insights into how variations in energy barriers due to annealing influence the mechanical behavior of amorphous solids under oscillatory shear in a universal manner. These findings will inspire future research aimed at developing a unified understanding of the yielding transition in various amorphous materials, while emphasizing the critical role that energy barriers play in plasticity.

\section*{Methods}\label{sec:methods}

\noindent{\large Model and Simulation details:}\label{sec:model}
\rev{We have investigated a number of model systems which qualitatively represent granular, colloidal, molecular, metallic and network-forming glasses. We have given brief details of these models below.} First model is a binary mixture ($50:50$) of particles that interact via a Harmonic potential given by \cite{durian1995foam,adhikari2021spatial}: 
\begin{eqnarray}\label{eqs_1}
\nonumber
V_{\alpha \beta}(r) &=& \frac{\epsilon_{\alpha \beta}}{2}\left(1-\frac{r}{\sigma_{\alpha \beta}}\right)^{2}, \hspace{1.58cm} r_{\alpha \beta} \leq \sigma_{\alpha \beta}\\ 
&=&0 ,\hspace{3.94cm} r_{\alpha \beta } > \sigma_{\alpha \beta}
\end{eqnarray}
where $\alpha, \;  \beta$ $\in$ (A, B) indicates the type of particle. The two types of particles differ in their size, with $\sigma_{BB} = 1.4 \sigma_{AA}$ (and the diameters are additive), with the interaction strengths the same for all pairs ($\epsilon_{AA} = \epsilon_{AB} = \epsilon_{BB}$). We have performed simulations at different packing fractions, $\phi$, which are larger than $\phi_J$ as well as $\phi_0$ \cite{adhikari2023dependence}. The relation between number density $\rho$ and packing fraction $\phi$ for binary mixture is,
\begin{eqnarray}\label{eqs_2}
\phi &=& \rho 2^{-d} \frac{\pi^{d/2}}{\Gamma (1 + \frac{d}{2})} (c_A \sigma^d_{AA} + c_B \sigma^d_{BB}),
\end{eqnarray}
where $\rho = N/V$, N is the number of particles, V is the simulation box's volume, and $c_A = c_B = 1/2$. We have performed NVT (constant particle number N, volume V, temperature T) Molecular Dynamics (MD) simulation using Nose-Hover thermostat in cubic box maintaining periodic boundary condition for $N = 5000$ particles at four packing fractions $\phi = \left[0.667, 0.735, 0.837, 0.925\right]$ and the corresponding number densities (from Eq. \ref{eqs_2}) $\rho = \left[0.681, 0.750, 0.855, 0.943\right]$. The time step of molecular dynamics is chosen $dt = 0.01$.

\rev{Next model is the well-known van Beest-Kramer-van Santen (BKS) model \cite{van1990force} for Silica (SiO$_2$) as implemented in \cite{saika2004free}. We prepared the samples via NVT molecular dynamics at density $\rho = 2.8$ gm/cm$^3$ with system size $N = 4800$ ions and integration time step $dt = 1$fs. The interaction potential is given by,
\begin{equation}
V(r_{ij}) = \frac{q_i q_j}{4\pi \epsilon_0 r_{ij}} + A_{ij}exp(-B_{ij}r_{ij}) + C_{ij}r^{-6}_{ij} + \phi(r_{ij})
\end{equation}
where,
\begin{equation}
\phi(r_{ij}) = 4\epsilon_{ij}\left[\left(\sigma_{ij}/r_{ij}\right)^{30} - (\sigma_{ij}/r_{ij})^6\right]
\end{equation}}
\rev{The potential cut-off is at $r_c = 10\AA$ and the Ewald parameter $\alpha = 2.5\AA^{-1}$. \SKrev{To study the shear band formation we simulated $N = 48000$ at $T_p = 2700$K.}}

\rev{Third model that we have simulated is the model of ortho-terphenyl (OTP) molecule \cite{lewis1993relaxation,lewis1994molecular,lewis1994rotational} which is a well-known fragile glass-former. OTP is a rigid molecule with a three-site planar isosceles triangle structure, where each site represents a whole phenyl ring. The bond length of the two sides of the triangle is $\sigma = 4.83\AA$ and the bond angle is $75\degree$. The interaction potential,
\begin{equation}
V(r) = 4\epsilon\left[\left(\frac{\sigma}{r}\right)^{12} - \left(\frac{\sigma}{r}\right)^6\right]
+ \lambda_1 + \lambda_2r
\end{equation}
where $\epsilon = 5.276$ kJ/mol, $\lambda_1 = 0.461$ kJ/mol, $\lambda_2 = -0.313$ kJ/(mol.nm), the interaction cut off $r_c = 12.616\AA$  as used by Mossa et all in \cite{mossa2002dynamics}. In addition, in our model, the three sites within
a single molecule interacts via a stiff harmonic potential,
\begin{equation}
V_{bond}(r,\theta) = \frac{1}{2}k(r_{ij} - r_0)^2 + \frac{1}{2}k(\theta_{ij} - \theta_0)^2
\end{equation}
where for rigid bond, $r_0 = 4.83\AA$, $\theta_0 = 75\degree$ and $k = 10^4$ kcal.mol$^{-1}.\AA^{-2}$. We have prepared samples via NVT MD simulation with system size $N = 4800$ atoms, i.e. $N_m = 1600$ molecules at density $\rho = 1.135$ gm/cm$^3$ with integration time step $dt = 5$fs.}

\rev{Fourth model is the well-known model of metallic glass Cu$_{64.5}$Zr$_{35.5}$ interacting via embedded atom model (EAM) potential as in \cite{mendelev2019development} via NPT (constant particle number $N$, pressure $P$, temperature $T$) MD simulation at zero pressure with system size $N = 5000$ and integration time step $dt = 1$fs. We have also performed NVT MD for this model at density $7.40 gm/cm^3$. \SKrev{To study the shear band formation we also simulated $N = 50000$ at $T_p = 750$K.}}

\rev{Finally, we have simulated the well-known Kob-Andersen binary mixture (80:20) in three dimensions (3D) \cite{kob1994scaling,kob1995testing}. The interaction potential is given by
\begin{equation}
V_{\alpha\beta}(r)=4\epsilon_{\alpha\beta}\left[\left(\frac{\sigma_{\alpha\beta}}{r}\right)^{12}-\left(\frac{\sigma_{\alpha\beta}}{r}\right)^6+C_0+C_2\left(\frac{r}{\sigma_{\alpha\beta}}\right)^2\right]
\end{equation}
where $\alpha,\beta \in {A,B}$; interaction strength of pairs $\epsilon_{AB}/\epsilon_{AA} = 1.5, \epsilon_{BB}/\epsilon_{AA} = 0.5$; and diameters $\sigma_{AB}/\sigma_{AA} = 0.8, \sigma_{BB}/\sigma_{AA} = 0.88$. The interaction cut-off is at $r_c = 2.5 \sigma_{\alpha\beta}$ and number density $\rho = 1.2$. We have prepared samples via NVT MD simulation with system size $N = 5000$ and the integration time step $dt = 0.005$.}

\vskip +0.1in
\noindent{\large Initial states:} For each density, we equilibrated the liquids at a wide range of temperatures, including temperatures well below the $T_{MCT}$ (Mode Coupling Temperature) to study the effect of annealing on yielding. $T_{MCT}$ are given in SI for each density. Within this temperature range, the relaxation time $\tau_\alpha$ reaches $10^7$ for all densities. $\tau_\alpha$ is defined as the time when the two-point correlation function becomes $q(t = \tau_\alpha) = 1/e$ (the details are shown in the SI). We have used the conjugate gradient (CG) method on these equilibrated configurations to obtain the energy-minimized or inherent structure configurations (IS configurations). 

\vskip +0.1in
\noindent{\large Oscillatory Shear Protocol:} We performed cyclic shear on IS configurations in xz direction at different strain amplitude $\gamma_{max}$ via the athermal quasistatic (AQS) shear protocol where (i) each particle is displaced by affine transformation (${r_x}^{\prime} \rightarrow r_x + r_z \times d\gamma$, ${r_y}^{\prime} \rightarrow r_y$, ${r_z}^{\prime} \rightarrow r_z$) and then (ii) energy minimization using the CG method. The strain $\gamma$ varies as ($0 \rightarrow{\gamma_{max} \rightarrow{-\gamma_{max} \rightarrow 0}}$) indicating a complete cycle, where elementary strain increment $d\gamma = 2\times10^{-4}$.  We prepared 12 independent samples at each temperature and density. See SI for further details.

\vskip 0.1in
\noindent\textbf{Data availability:}
All data generated and analyzed for this study are included in the manuscript and its Supplementary Information.
The source data underlying the figures are publicly available on Figshare at
https://doi.org/10.6084/m9.figshare.31146514 \cite{Chatterjee2026Figshare}.

\vskip 0.1in
\noindent\textbf{Code availability:}
All simulations were performed using the open-source molecular dynamics package
LAMMPS. Codes are publicly available on Figshare at
https://doi.org/10.6084/m9.figshare.31146514 \cite{Chatterjee2026Figshare}.

\newpage
\makeatletter
\def\bibsection{}
\makeatother
\noindent{\bf \large References }\\
\vskip 0.1in

\bibliographystyle{ieeetr}
\bibliography{Effectoffragility}

\vskip +0.4in
\noindent{\bf Acknowledgments:}
\SK{We want to thank Srikanth Sastry, Pinaki Chaudhuri, Viswas Venkatesh, and Shiladitya Sengupta for many useful discussions. We acknowledge funding by intramural funds at TIFR Hyderabad from the Department of Atomic Energy (DAE) under Project Identification No. RTI 4007. SK acknowledges the Swarna Jayanti Fellowship grants DST/SJF/PSA01/2018-19 and SB/SFJ/2019-20/05 from the Science and Engineering Research Board (SERB) and Department of Science and Technology (DST) and the National Super Computing Mission (NSM) grant $\mathrm{DST/NSM/R\&D\_HPC\_Applications/2021/29}$ for generous funding. Most of the computations are done using the HPC clusters procured using Swarna Jayanti Fellowship grants DST/SJF/PSA01/2018-19, SB/SFJ/2019-20/05 and Core Research Grant CRG/2019/005373. SK also acknowledges research support from MATRICES Grant MTR/2023/000079 from SERB. MA acknowledges support from NSM grant $\mathrm{DST/NSM/R\&D\_HPC\_Applications/2021/29}$ for financial support.}

\vskip +0.1in
\noindent{\bf Author Contributions: }
SK conceived the project. SK supervised the project. RC performed the research and simulations. RC, MA, and SK designed analysis methods. RC performed all the analyses. RC, MA, and SK wrote the paper jointly.

\end{document}




\title[]{\Large \bf Role of Fragility of the Glass Formers in the Yielding Transition under Oscillatory Shear - Supplementary Information}

\author{Roni Chatterjee}
\affiliation{Tata Institute of Fundamental Research, 36/P, Gopanpally Village, Serilingampally Mandal, Ranga Reddy District, Hyderabad 500046, Telangana, India}
\author{Monoj Adhikari }
\affiliation{Tata Institute of Fundamental Research, 36/P, Gopanpally Village, Serilingampally Mandal, Ranga Reddy District, Hyderabad 500046, Telangana, India}
\author{Smarajit Karmakar}
\affiliation{Tata Institute of Fundamental Research, 36/P, Gopanpally Village, Serilingampally Mandal, Ranga Reddy District, Hyderabad 500046, Telangana, India}

\maketitle

\section{Equilibration of the liquid}
As discussed in the main text, our initial configuration before shear deformation starts is obtained by equilibrating the liquid at a given temperature. Below, we show the mean squared displacement (MSD) and overlap function $q(t)$ as functions of time to demonstrate that the liquid has reached to equilibrium.

\subsection*{Mean squared displacement}
Mean squared displacement (considering the B-type particles only) is defined as :
\begin{equation}
    \Delta r^2 (t) = \frac{1}{N_B} \sum_{i=1}^{N_B} |\mathbf{r_i}(t + t_0)-\mathbf{r_i}(t_0)|^2,
\end{equation}
\begin{figure}[htp]
    \centering
    \includegraphics[width=0.4\textwidth]{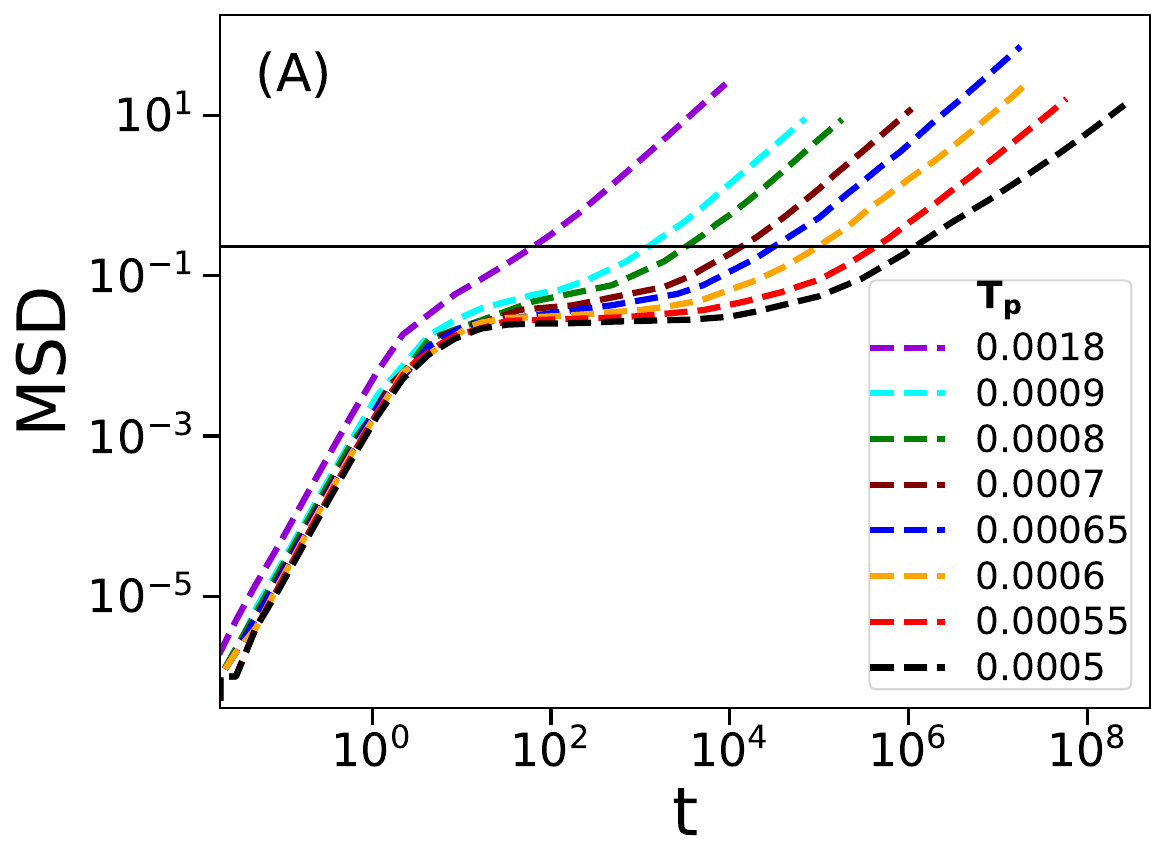}
    \includegraphics[width=0.4\textwidth]{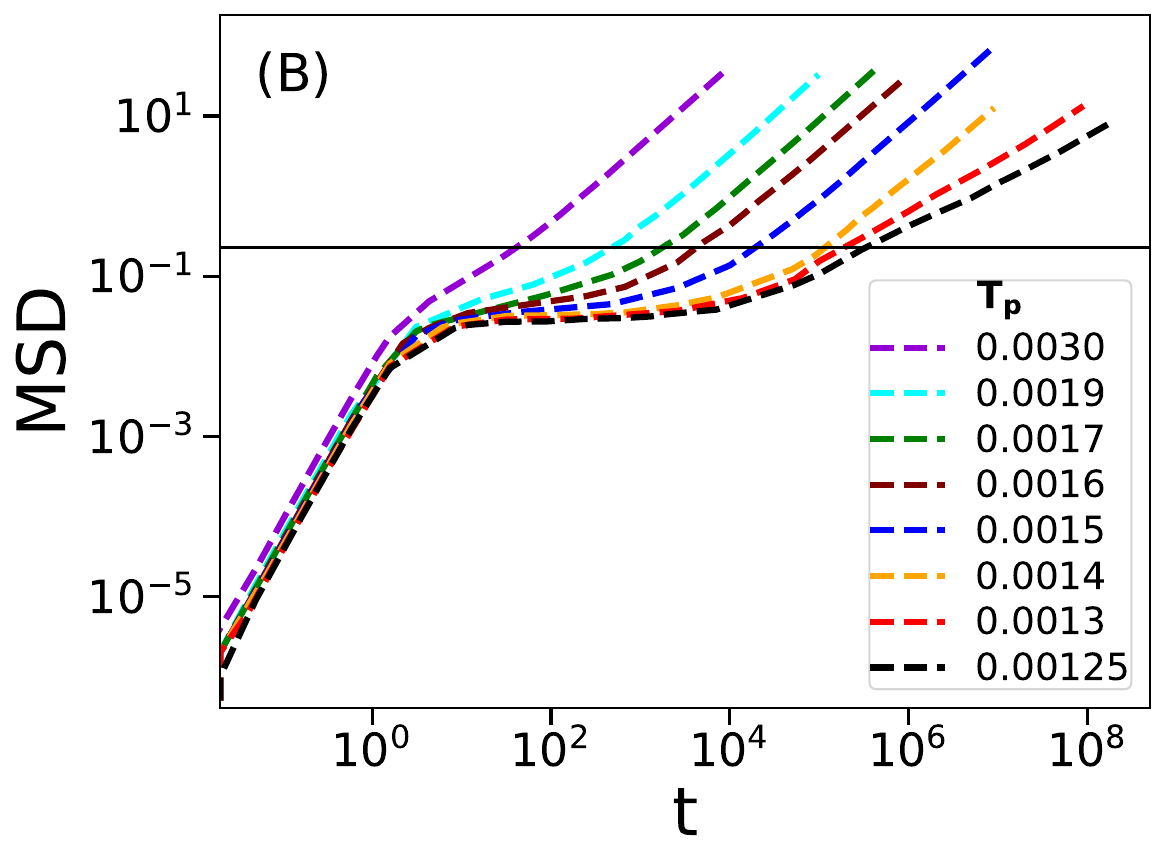}
    \includegraphics[width=0.4\textwidth]{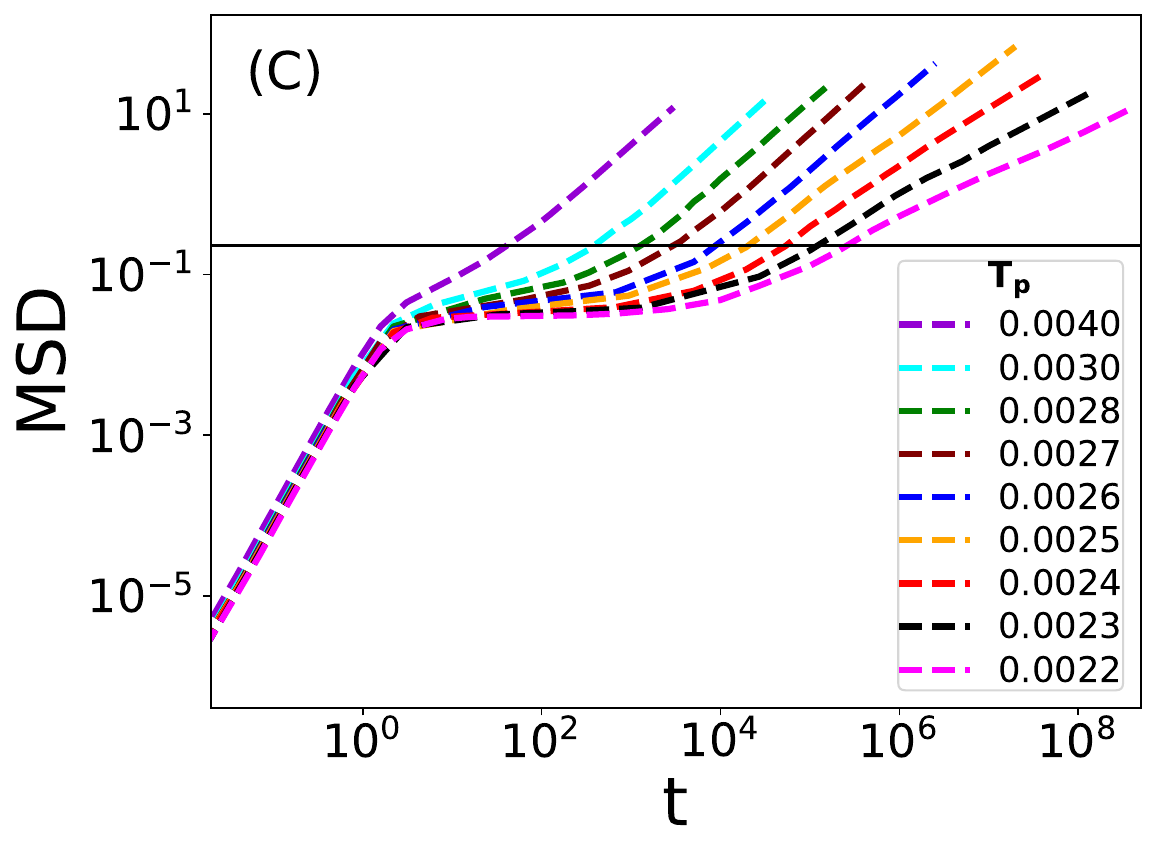}
    \includegraphics[width=0.4\textwidth]{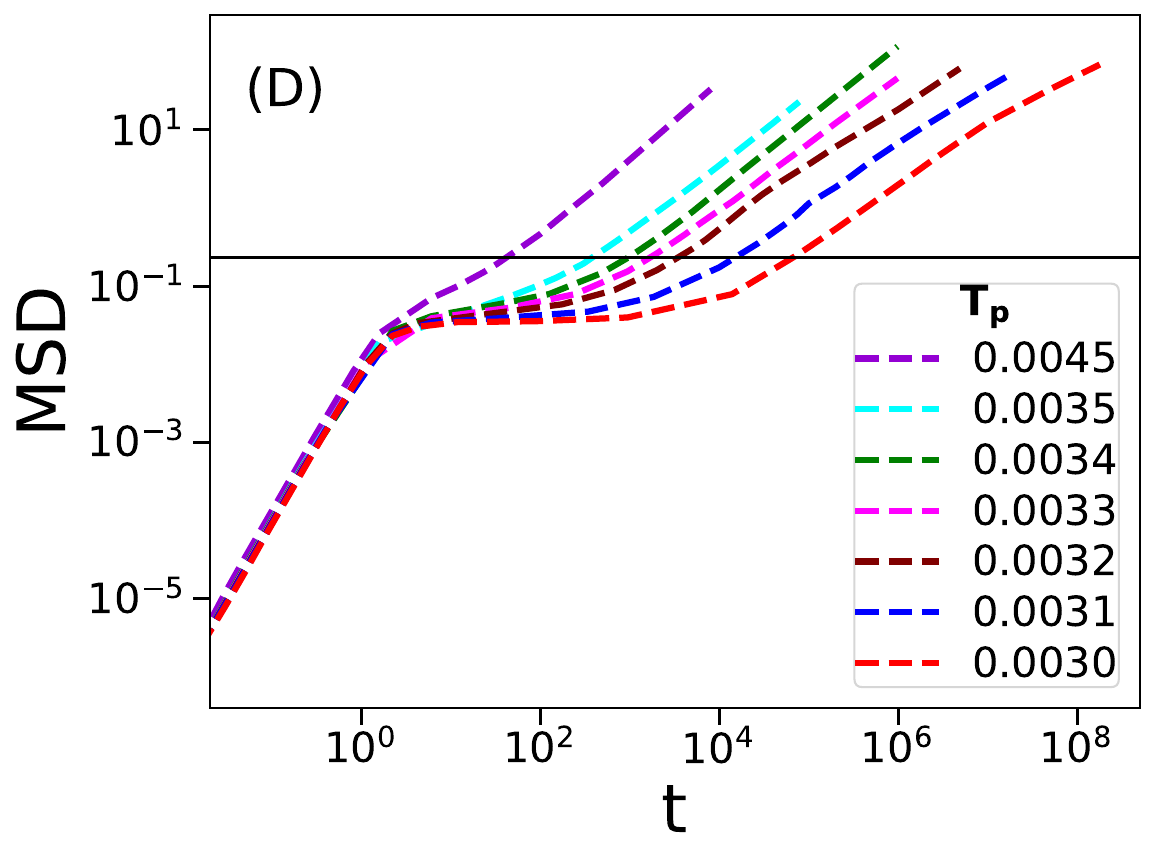}
    \caption{\textbf{Mean squared displacement for 3D HP.} Plot of mean squared displacement (MSD) with time t for system size N = 5000 at different parent temperatures ($T_p$) at densities (A) $\rho = 0.681$, (B) $\rho = 0.750$, (C) $\rho = 0.855$, (D) $\rho = 0.943$. The solid black line corresponds to the cut-off parameter $a^2 = 0.2304$.}
    \label{SI_fig:1}
\end{figure}
where $\mathbf{r_i}(t)$ is the position of the i-th particle at time t. The Mean Squared Displacement (MSD) is calculated by averaging over 12 samples (Fig. \ref{SI_fig:1}). We have conducted long MD simulation runs (with a duration of approximately $10^8$ ) to ensure that the MSD reaches the diffusive region for each temperature. The solid black line represents the value of the cut-off parameter $a^2 = 0.2304$, which we use to compute the overlap function, which is shown next.

\subsection*{Overlap function}
\begin{figure}[htp]
    \centering
    \includegraphics[width=0.49\textwidth,height=0.40\textwidth]{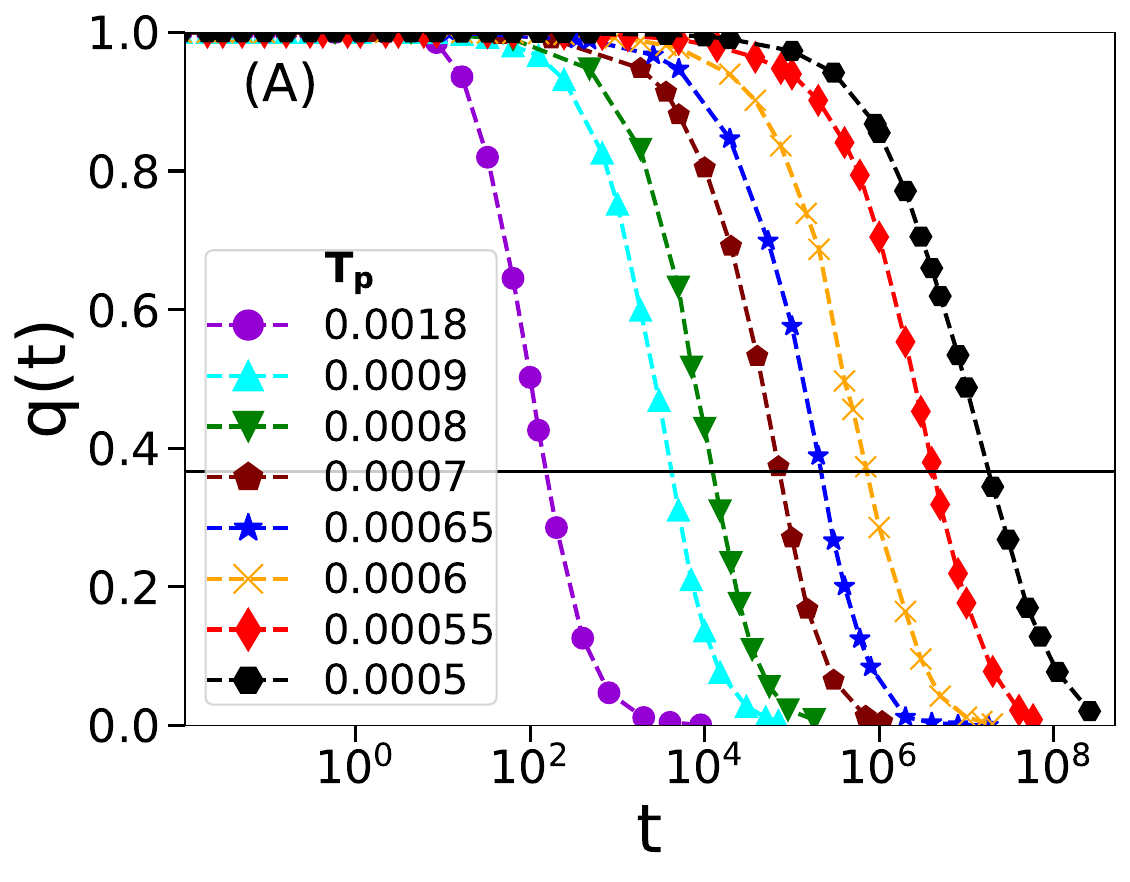}
    \includegraphics[width=0.49\textwidth,height=0.40\textwidth]{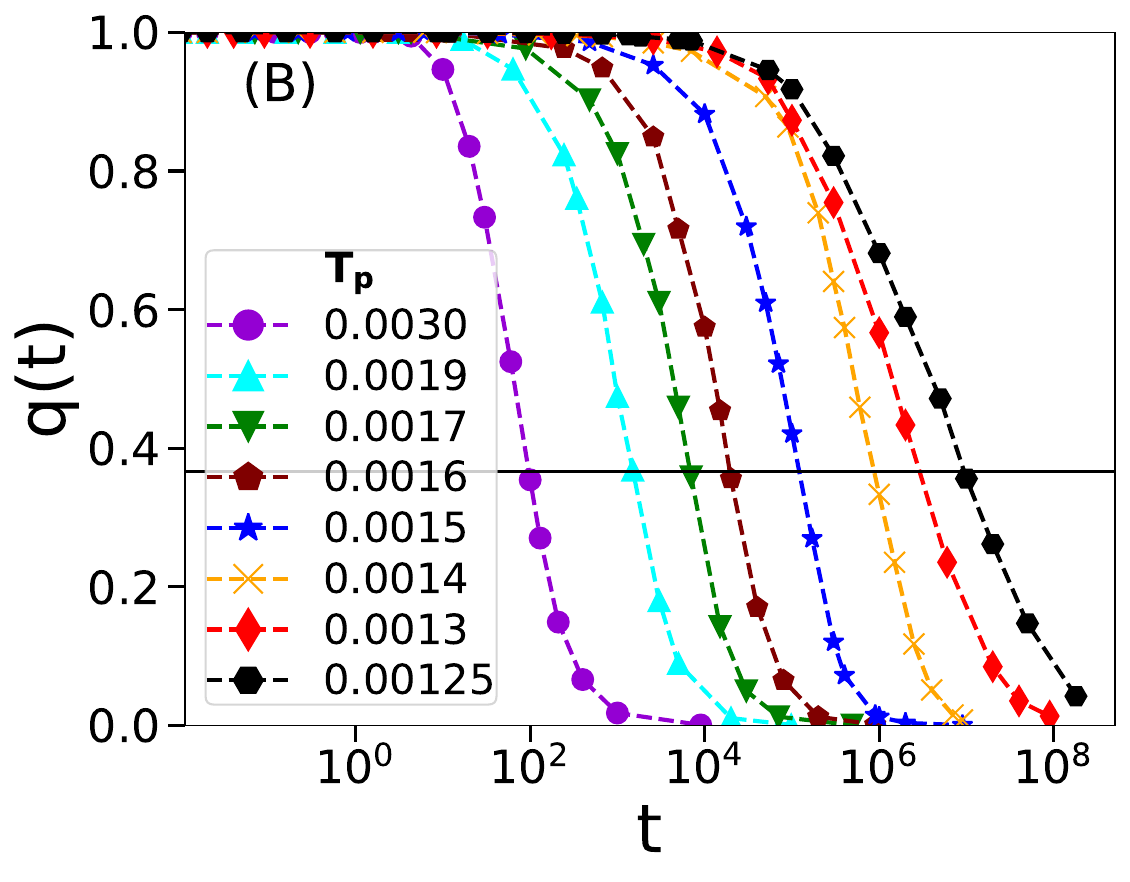}
    \includegraphics[width=0.49\textwidth,height=0.40\textwidth]{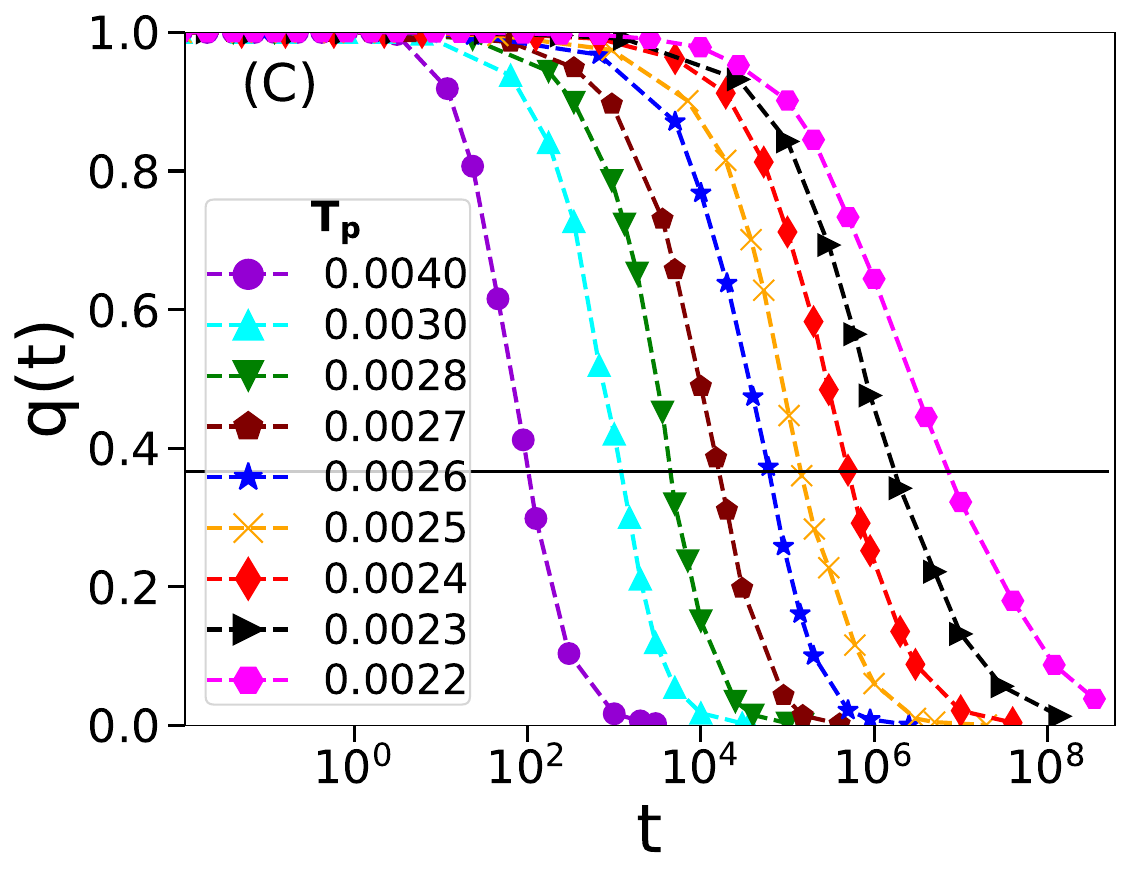}
    \includegraphics[width=0.49\textwidth,height=0.40\textwidth]{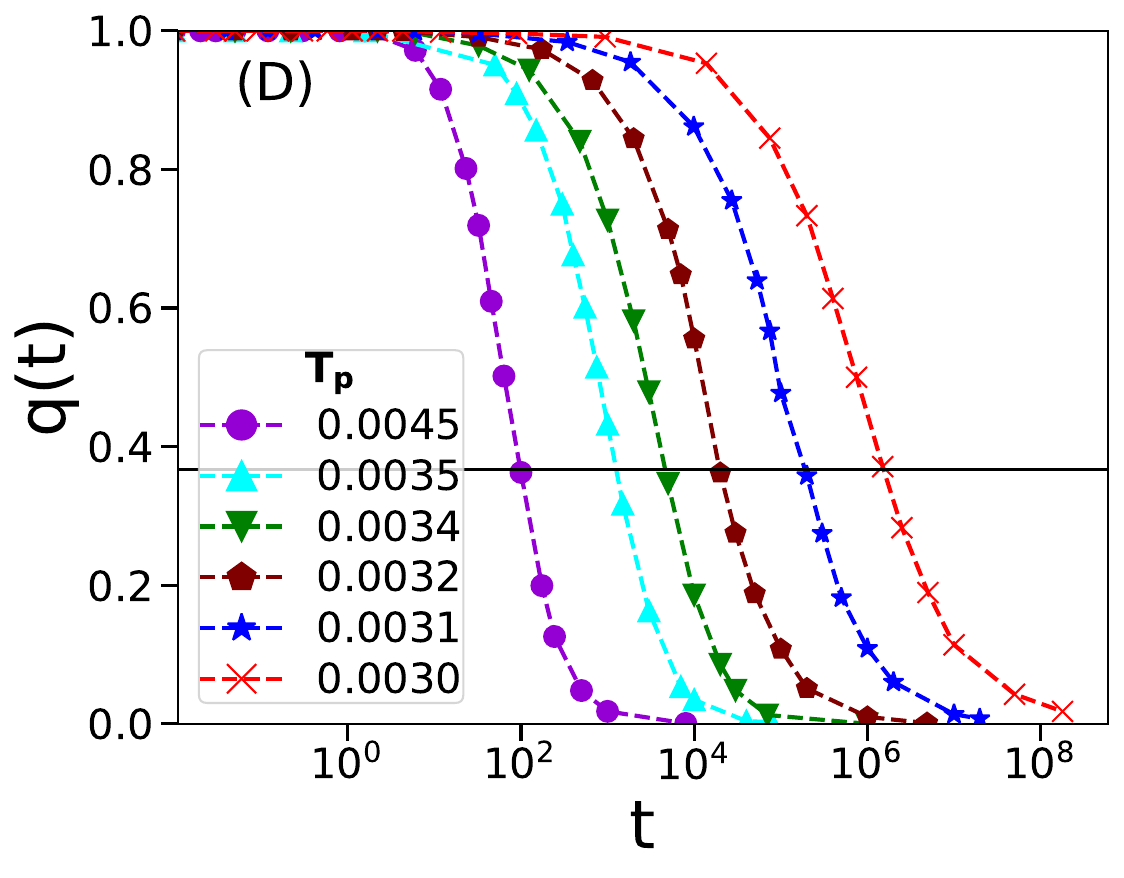}
    \caption{\textbf{Overlap function for 3D HP.} Plot of overlap function q(t) with time t at wide range of parent temperatures ($T_p$) at densities (A) $\rho = 0.681$, (B) $\rho = 0.750$, (C) $\rho = 0.855$, (D) $\rho = 0.943$. The solid black line corresponds to $q(t=\tau_\alpha) = \frac{1}{e}$. System size $N = 5000$ for all panels.}
    \label{SI_fig:2}
\end{figure}

The overlap function q(t) (considering the B-type particles only) is defined as :
\begin{equation}
    q(t) = \frac{1}{N_B} \sum_{i=1}^{N_B} w\left(|\mathbf{r_i}(t_0)-\mathbf{r_i}(t + t_0)|\right)
\end{equation}
where $w(x) = 1.0$ if $x \leq a$ and $=0$ otherwise. Here, $a = 0.48$ is chosen from the plateau value ($a^2 = 0.2304$) of the MSD curves. $q(t)$ is calculated by averaging over $12$ samples. Again, the decay of $q(t)$ to zero ensures that we have equilibrated all the liquids at their respective temperatures (Fig. \ref{SI_fig:2}). The relaxation time ($\tau_\alpha$) is calculated at the time where $q(t=\tau_\alpha) = \frac{1}{e}$.

\section{Inherent structure energy}
\begin{figure}[htp]
    \centering
    \includegraphics[width=0.49\textwidth,height=0.40\textwidth]{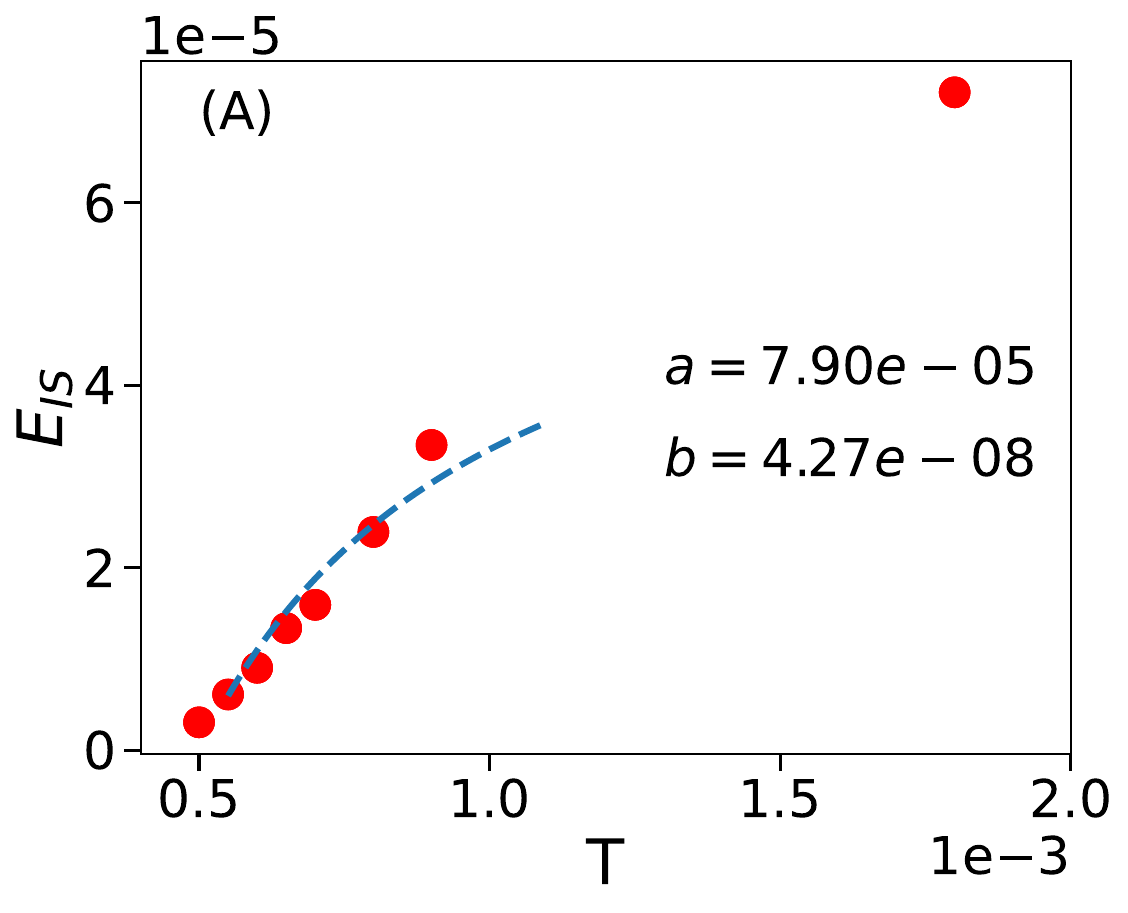}
    \includegraphics[width=0.49\textwidth,height=0.40\textwidth]{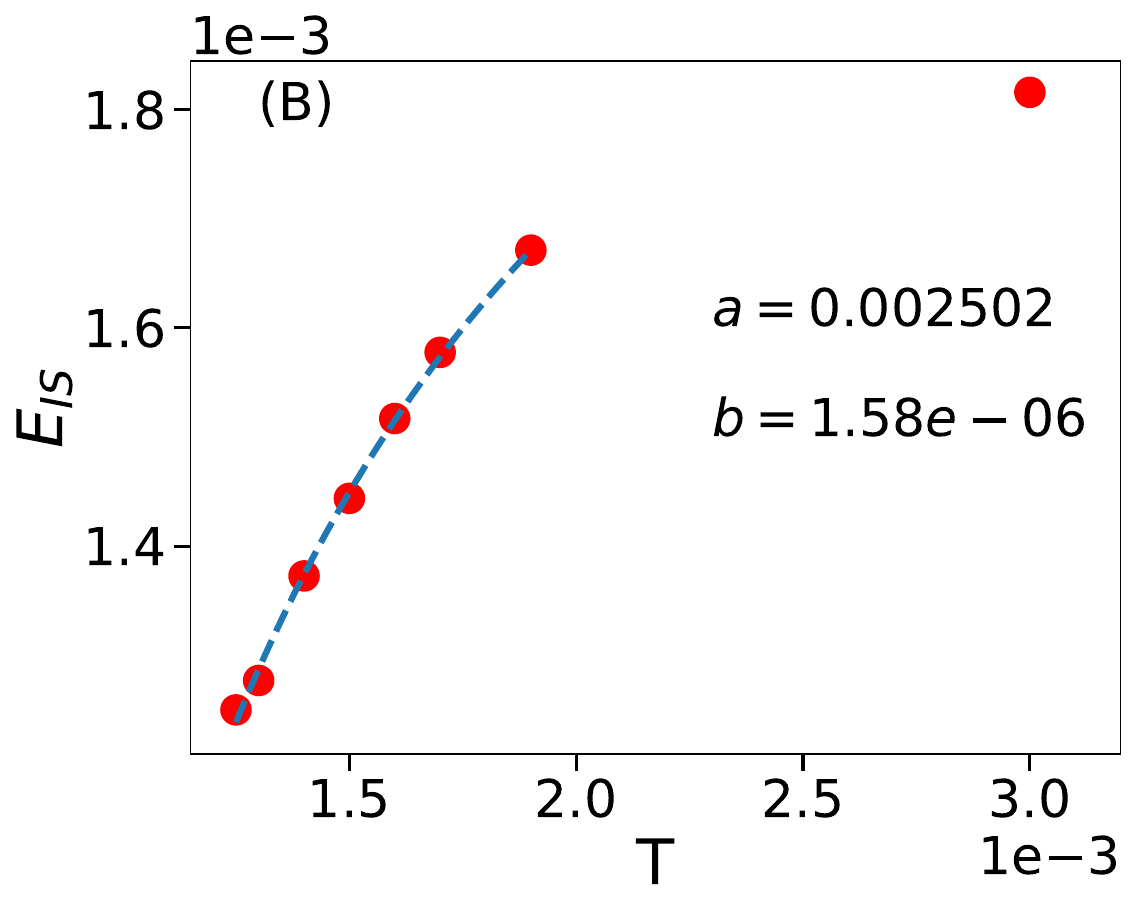}
    \includegraphics[width=0.49\textwidth,height=0.40\textwidth]{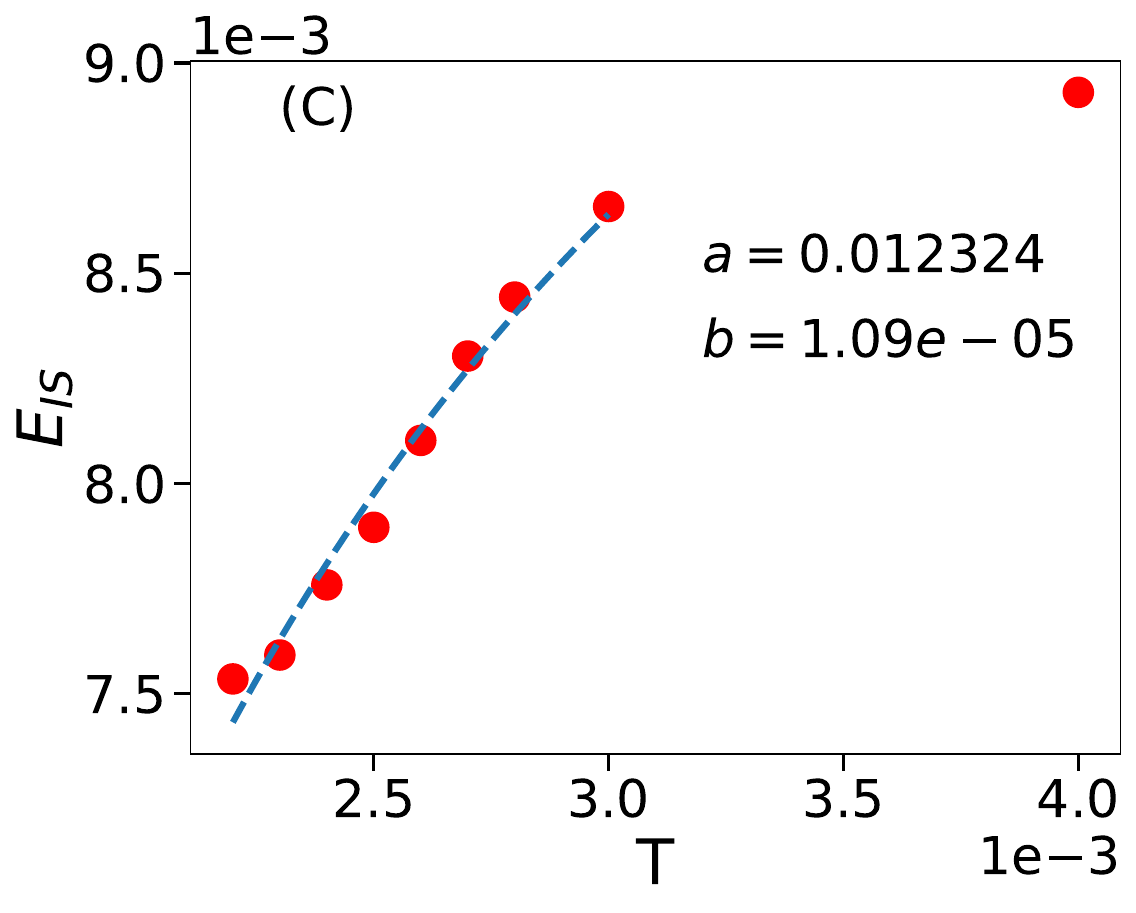}
    \includegraphics[width=0.49\textwidth,height=0.40\textwidth]{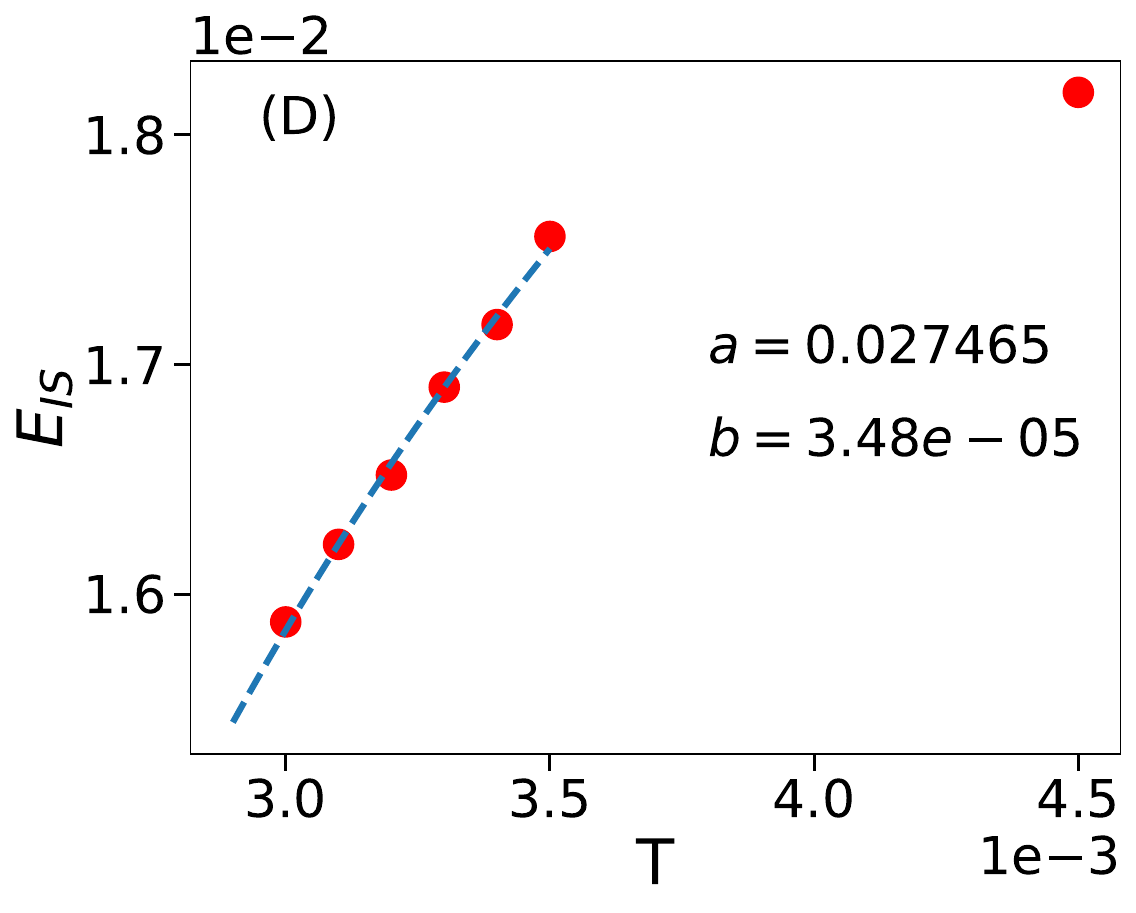}
    \caption{\textbf{Inherent structure energy for 3D HP.} Plot of inherent structure energy $E_{IS}$ against temperature T for densities (A) $\rho = 0.681$, (B) $\rho = 0.750$, (C) $\rho = 0.855$, (D) $\rho = 0.943$. Each curve is fitted at low temperatures via the equation: $E_{IS} = a -b/T$. Only for the lowest density $\rho = 0.681$, $E_{IS}$ shows deviation from $1/T$ behaviour.}
    \label{SI_fig:3}
\end{figure}

After equilibration, we achieve the amorphous solid state by minimizing the liquid. We employ the Conjugate Gradient method (CG) to minimize the liquid. These minimized configurations are referred to as inherent structure configurations. At low temperature region inherent structure energy ($E_{IS}$) varies with temperature (T) as : $E_{IS} = a - \frac{b}{T}$ (Fig. \ref{SI_fig:3}).  Although this relation holds well at high density, this nature deviates for the lowest density $\rho = 0.681$. The fitted data (along with fitting parameters) are plotted for all densities. $E_{IS}$ is calculated by averaging over $12$ samples at each temperature and density.

\section{$T_{MCT}$ for 3D HP Model at different densities}
\begin{figure}[H]
    \centering
    \includegraphics[width=0.49\textwidth,height=0.40\textwidth]{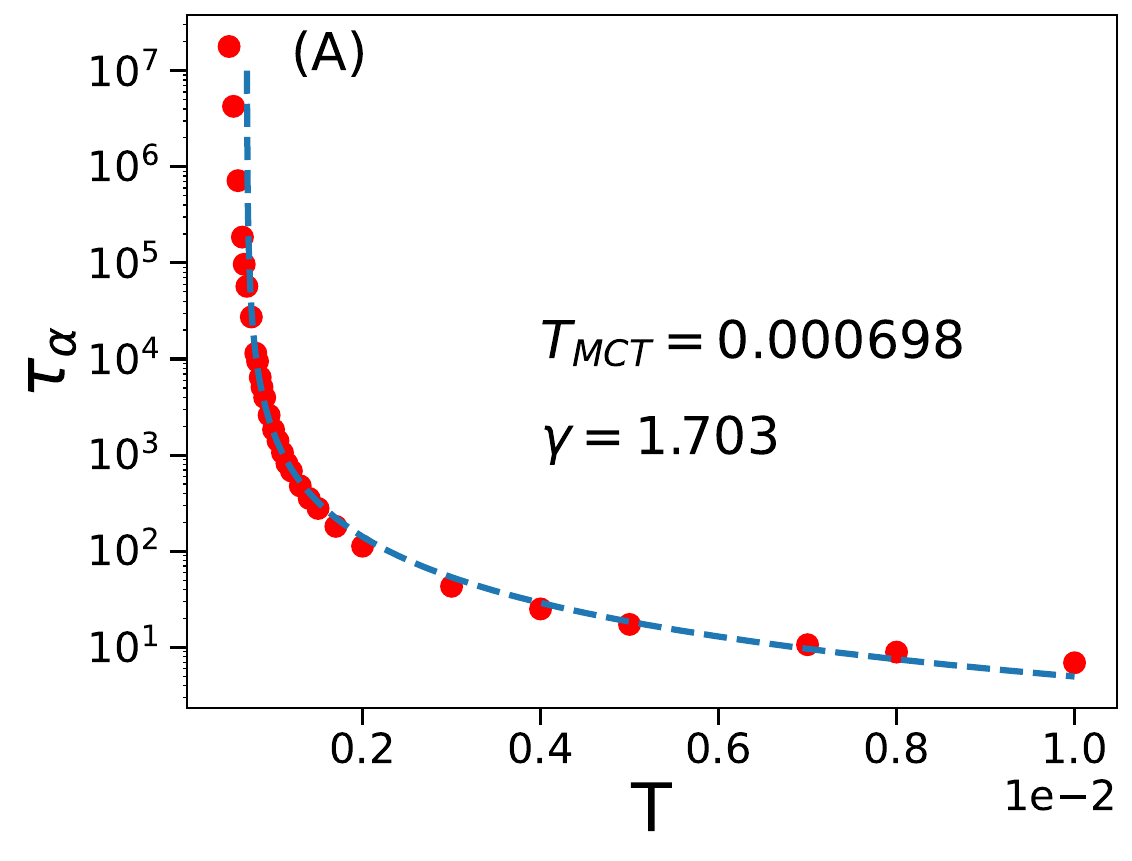}
    \includegraphics[width=0.49\textwidth,height=0.40\textwidth]{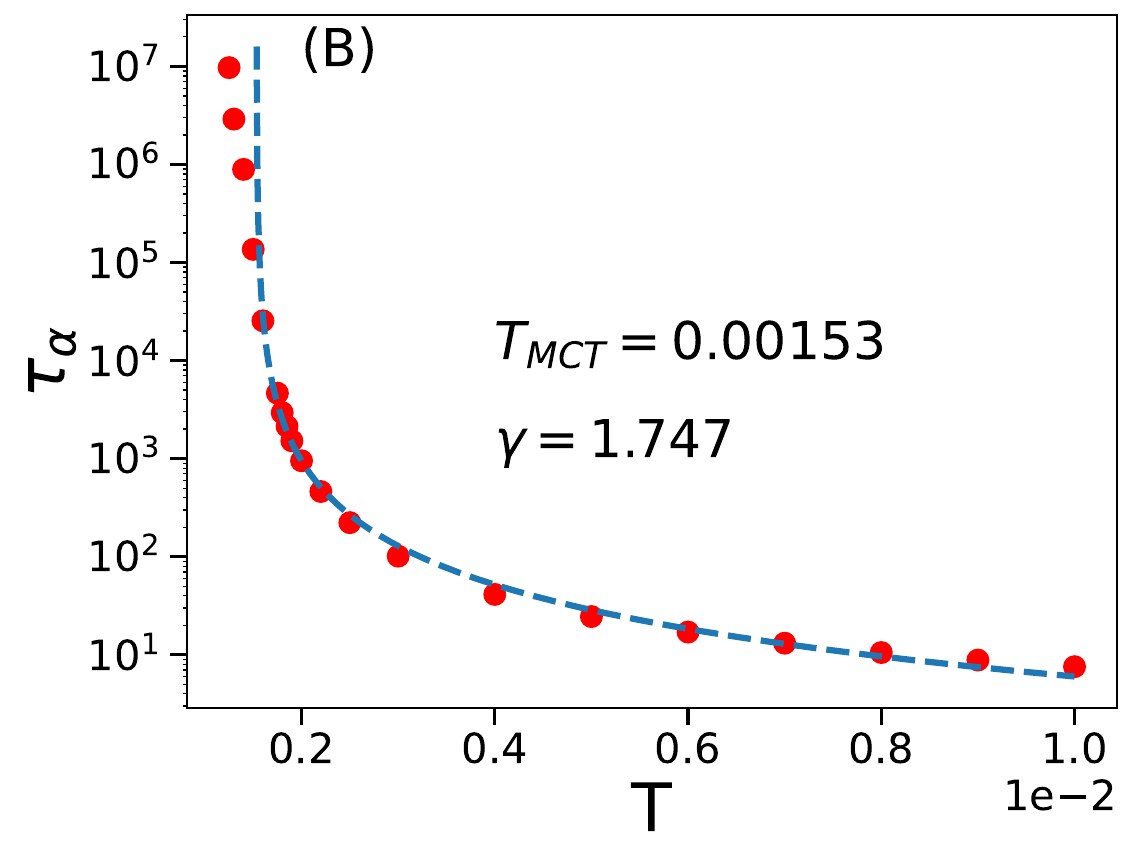}
    \includegraphics[width=0.49\textwidth,height=0.40\textwidth]{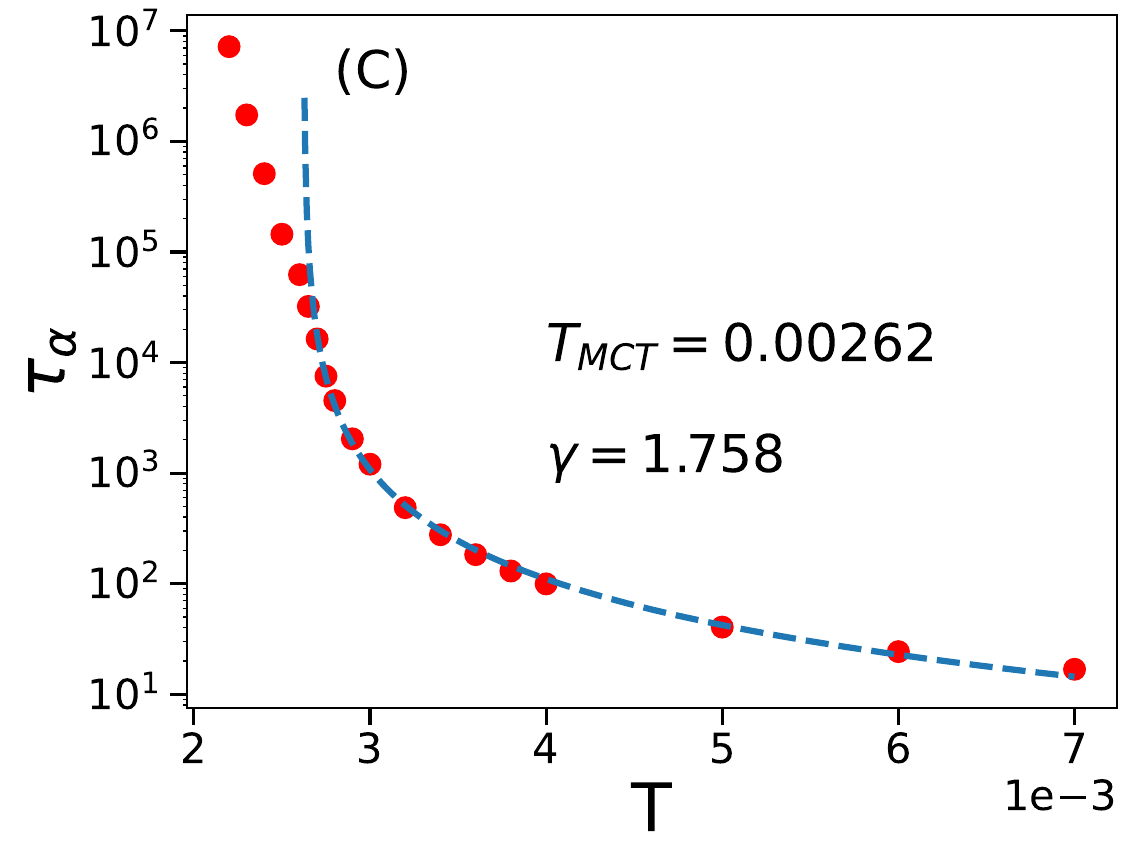}
    \includegraphics[width=0.49\textwidth,height=0.40\textwidth]{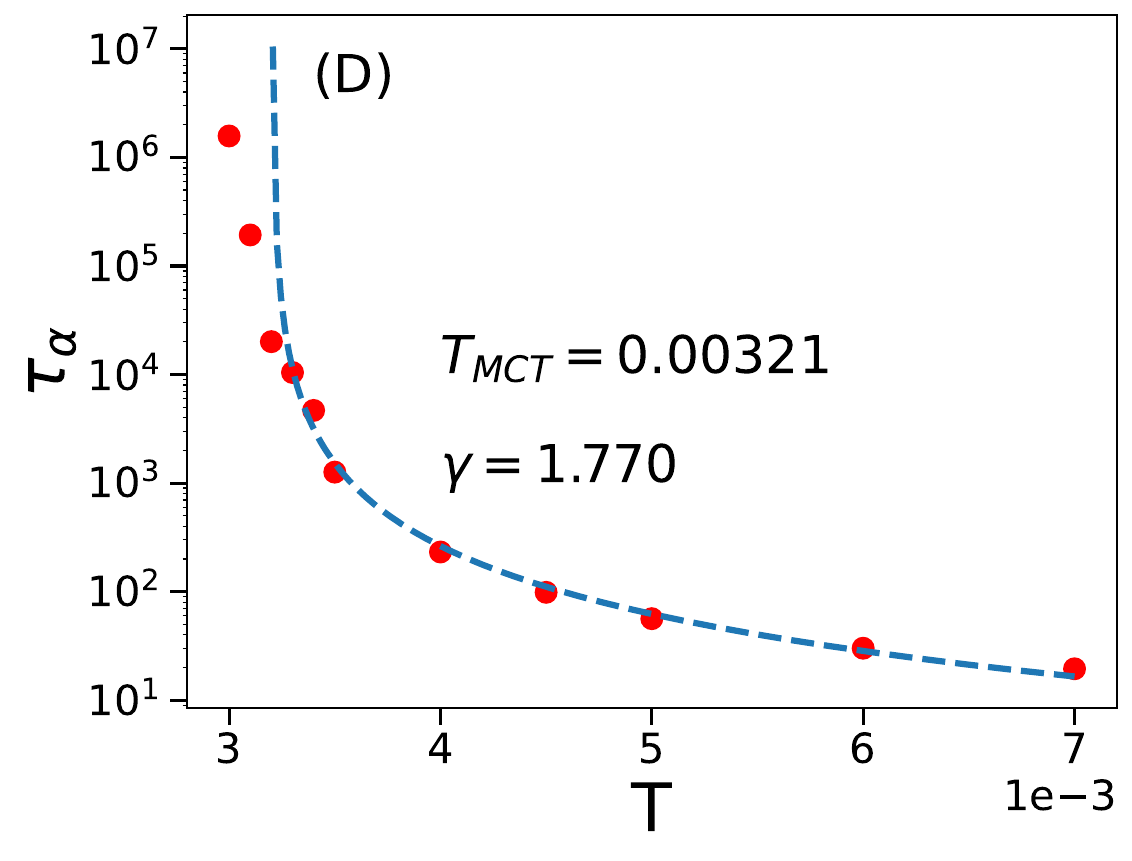}
    \caption{\textbf{Mode coupling temperature for 3D HP.} Plots of $\tau_\alpha$ extracted from the overlap function q(t), shown as a function of the temperature T on a semi-logarithmic scale, for densities (A) $\rho = 0.681$, (B) $\rho = 0.750$, (C) $\rho = 0.855$, (D) $\rho = 0.943$. Each curve is fitted by equation : $\tau_\alpha = \tau_0 (T - T_{MCT})^{-\gamma}$. $T_{MCT}$ is shown for all $\rho$ for system size N = 5000.}
    \label{SI_fig:4}
\end{figure}

As discussed in the main text, we observe a change in shear response behavior around the Mode Coupling temperature. Here, we show how we compute $T_{MCT}$ for different densities. According to the Mode-Coupling theory, variation of relaxation time ($\tau_\alpha$) with temperatures (T) is well described by a power-law form: $\tau_\alpha = \tau_0 (T - T_{MCT})^{-\gamma}$, where $\tau_\alpha$ diverges at a critical temperature $T_{MCT}$. In Fig. \ref{SI_fig:4}, we illustrate $\tau_{\alpha}$ of the liquids as a function of $T$ at different densities $\rho$ along with MCT fits, and the corresponding fitted $T_{MCT}$ is also indicated within the figures.

\section{Stroboscopic energy with cycles at different densities for 3D HP Model}
\begin{figure}[htp]
    \centering
    \includegraphics[width=0.49\textwidth]{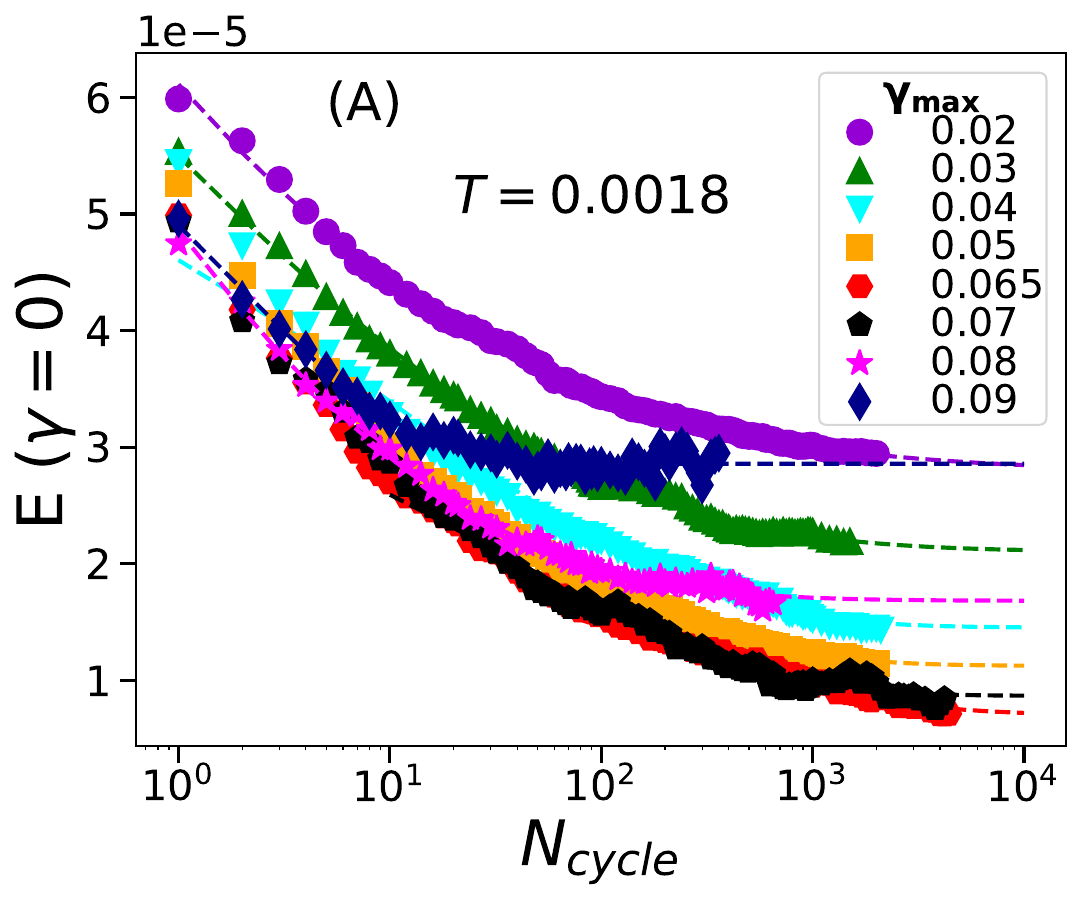}
    \includegraphics[width=0.49\textwidth]{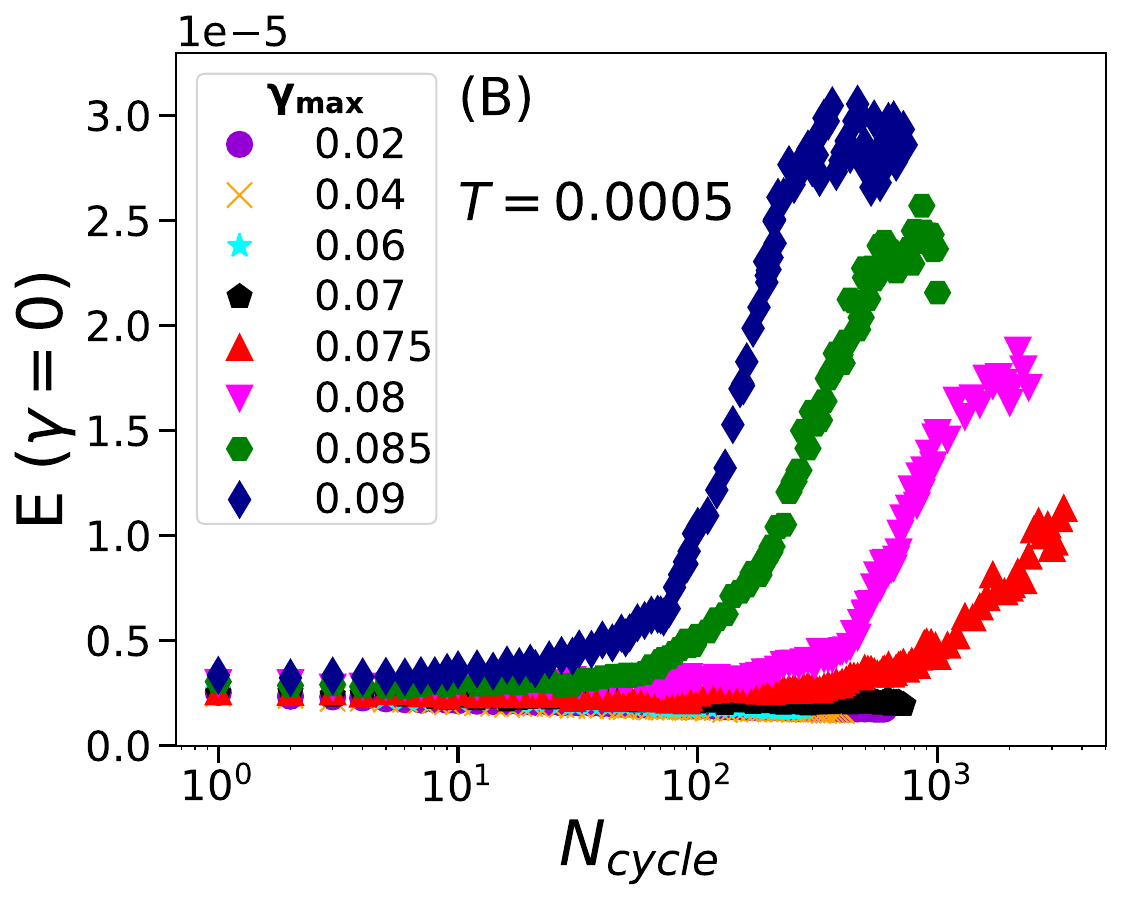}
    \caption{\textbf{Stroboscopic energy vs cycle for $\rho = 0.681$.} Plot of stroboscopic energies with cycles at different $\gamma_{max}$ at density $\rho = 0.681$ for (A) poorly annealed ($T = 0.0018$) and (B) well annealed ($T = 0.0005$). The data are averaged over 12 samples for the system size N = 5000. Dashed lines through the data set are fits to a stretched exponential form. From fitting, we get the yield point is $\gamma_Y \approx 0.065$ for the poorly annealed case. The yielding transition of the well-annealed sample is at $\gamma_Y \approx 0.075$.}
    \label{SI_fig:5}
\end{figure}

\begin{figure}[htp]
    \centering
    \includegraphics[width=0.49\textwidth]{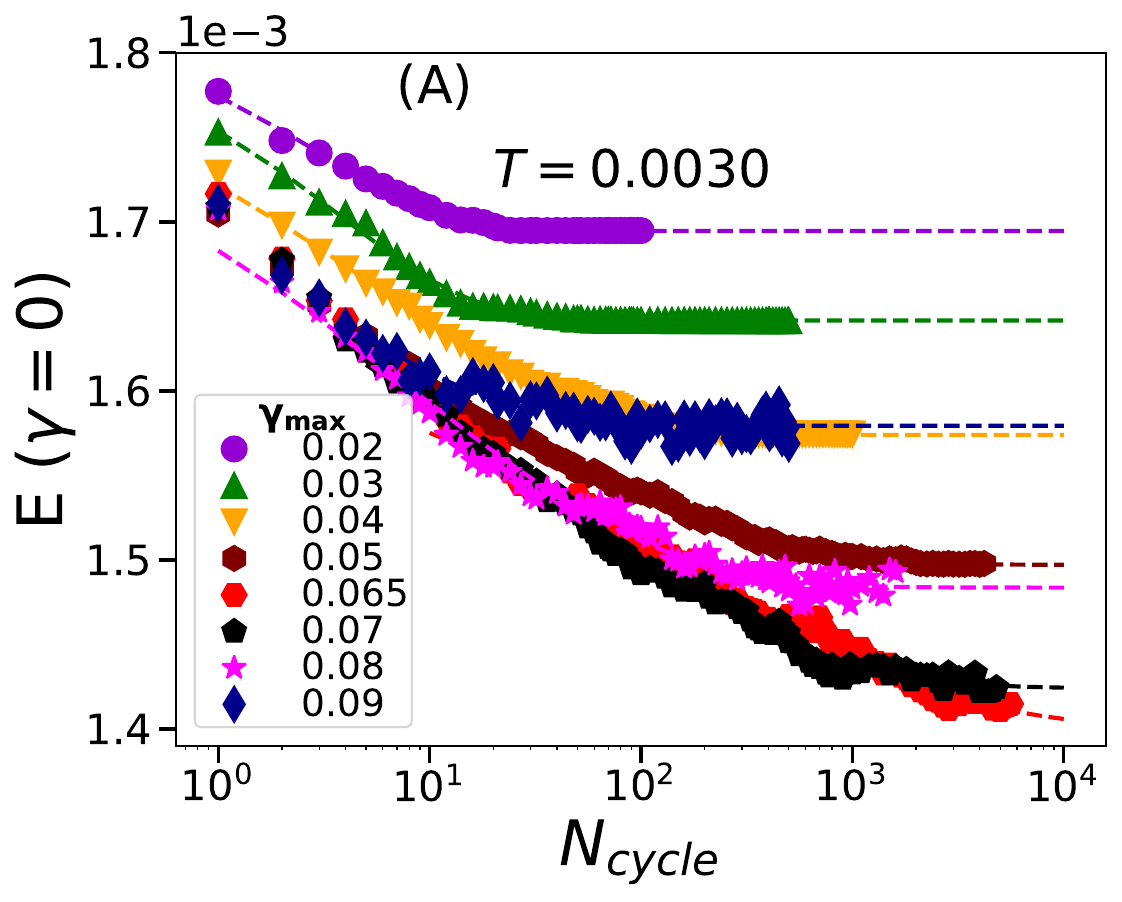}
    \includegraphics[width=0.49\textwidth]{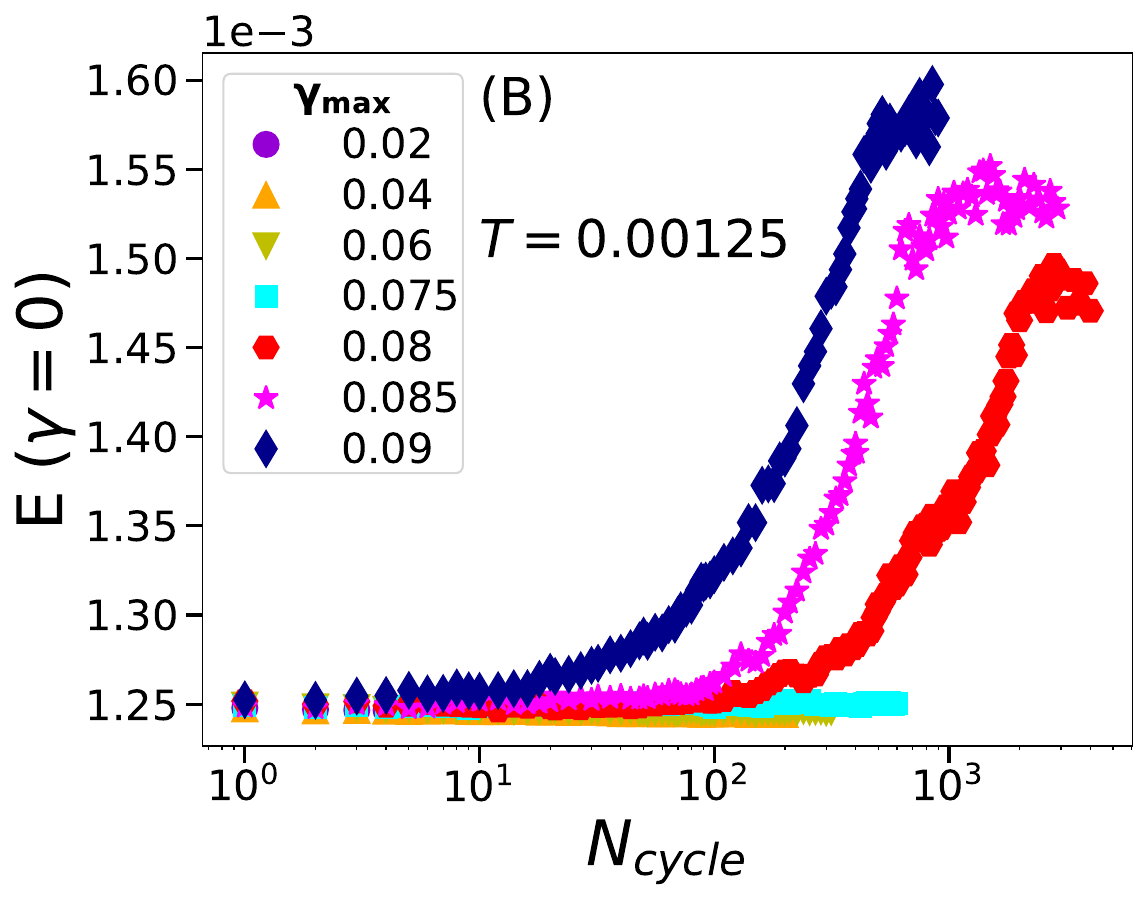}
    \caption{\textbf{Stroboscopic energy vs cycle for $\rho = 0.750$.} Plot of stroboscopic energies with cycles at different $\gamma_{max}$ at density $\rho = 0.750$ for (A) poorly annealed ($T = 0.0030$) and (B) well annealed ($T = 0.00125$).  From fitting, we get the yield point of poorly annealed is at $\gamma_Y \approx 0.065$. The yielding is at $\gamma_Y \approx 0.08$ for the well-annealed case.}
    \label{SI_fig:6}
\end{figure}

\begin{figure}[htp]
    \centering
    \includegraphics[width=0.49\textwidth]{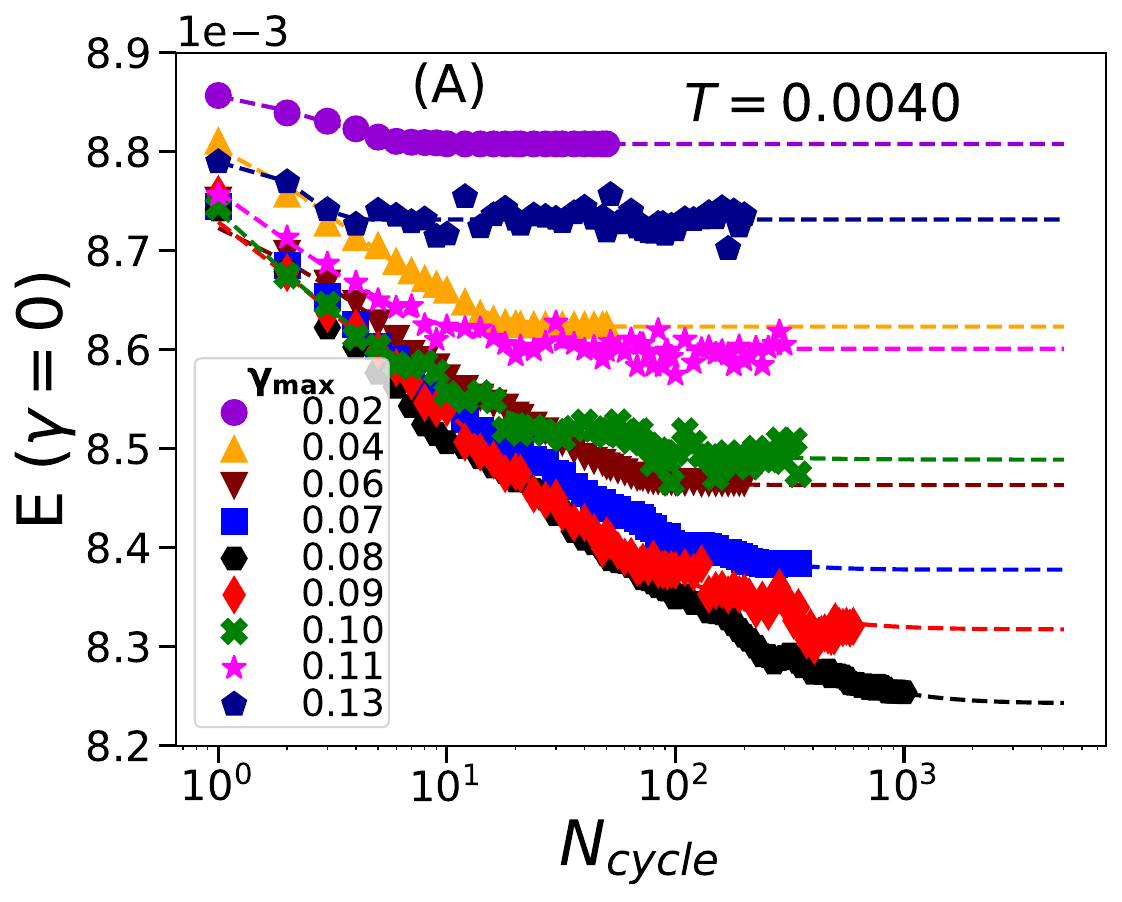}
    \includegraphics[width=0.49\textwidth]{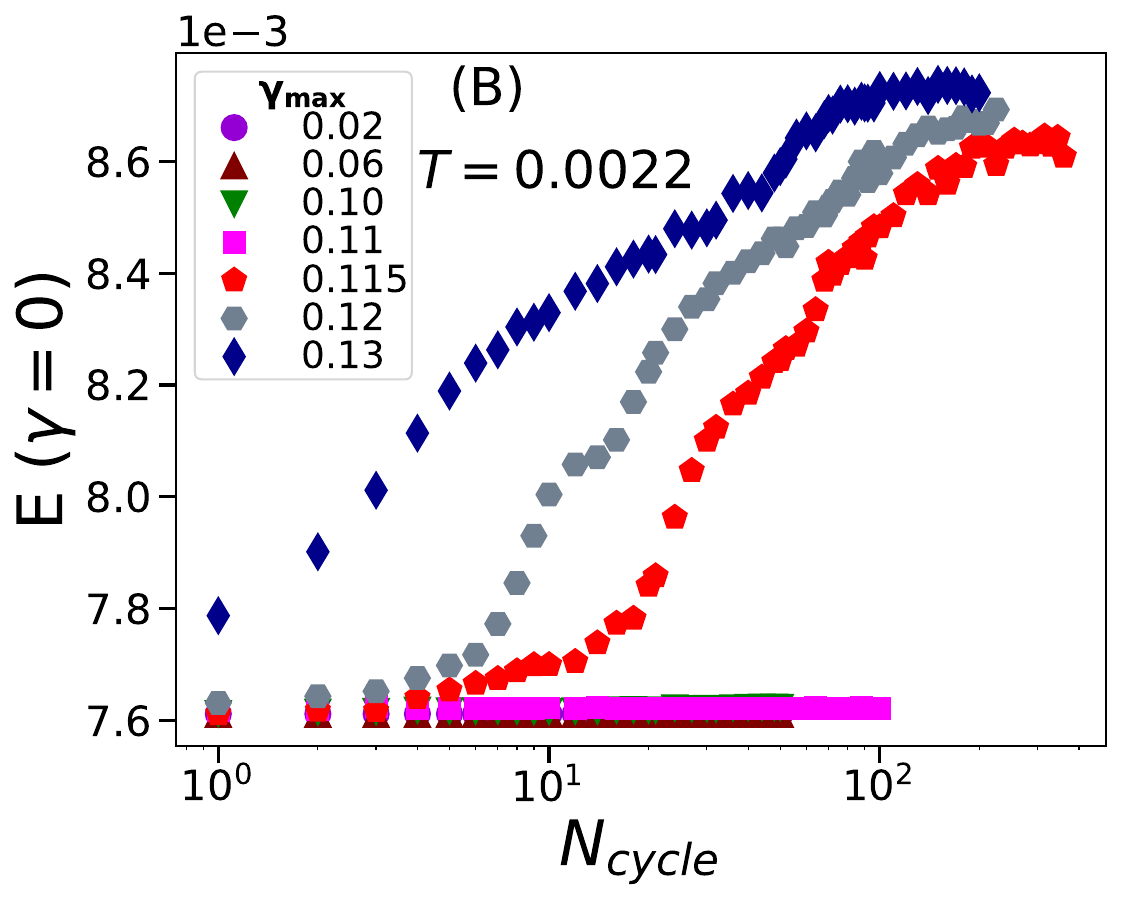}
    \caption{\textbf{Stroboscopic energy vs cycle for $\rho = 0.855$.} Plot of stroboscopic energies with cycles at different $\gamma_{max}$ at density $\rho = 0.855$ for (A) poorly annealed ($T = 0.0040$) and (B) well annealed ($T = 0.0022$) samples. The yield points of the poorly annealed and well-annealed samples are at $\gamma_Y \approx 0.085$ and $\gamma_Y \approx 0.115$, respectively.}
    \label{SI_fig:7}
\end{figure}

\begin{figure}[htp]
    \centering
    \includegraphics[width=0.49\textwidth]{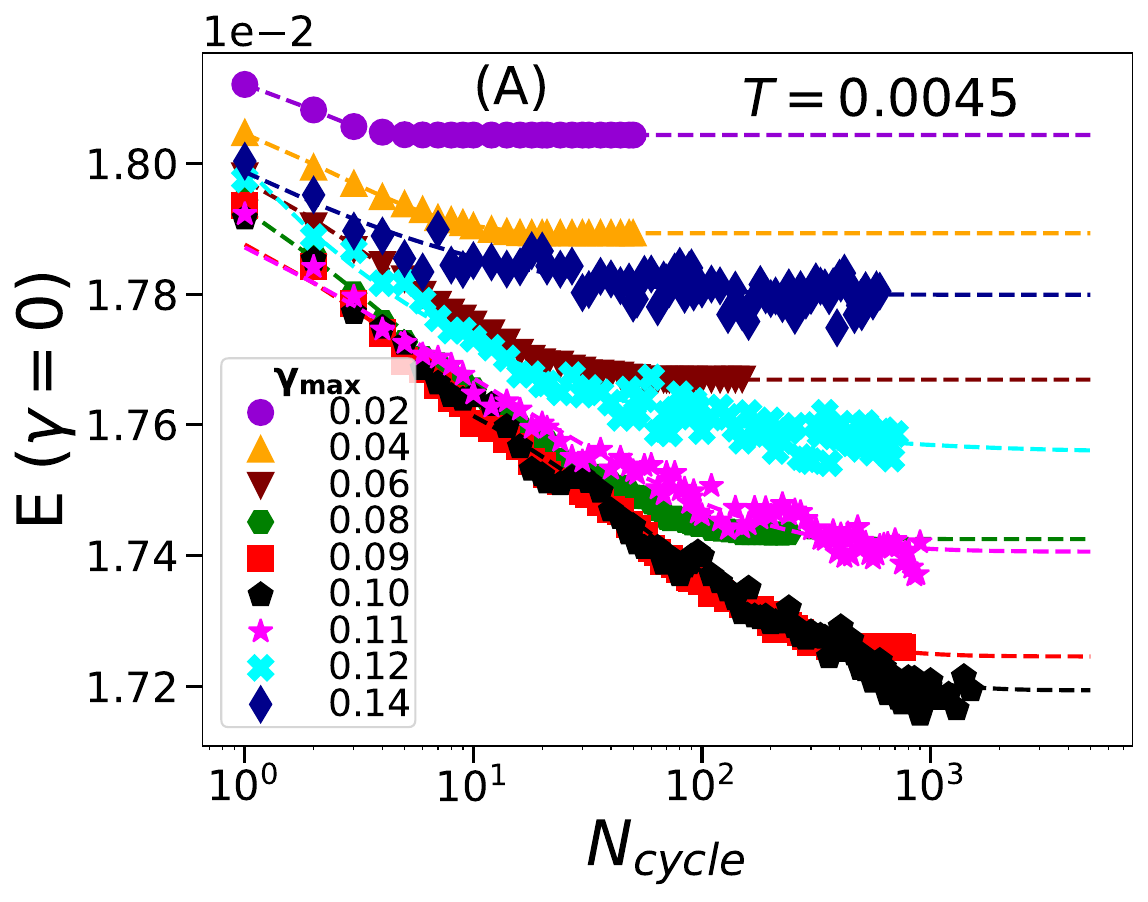}
    \includegraphics[width=0.49\textwidth]{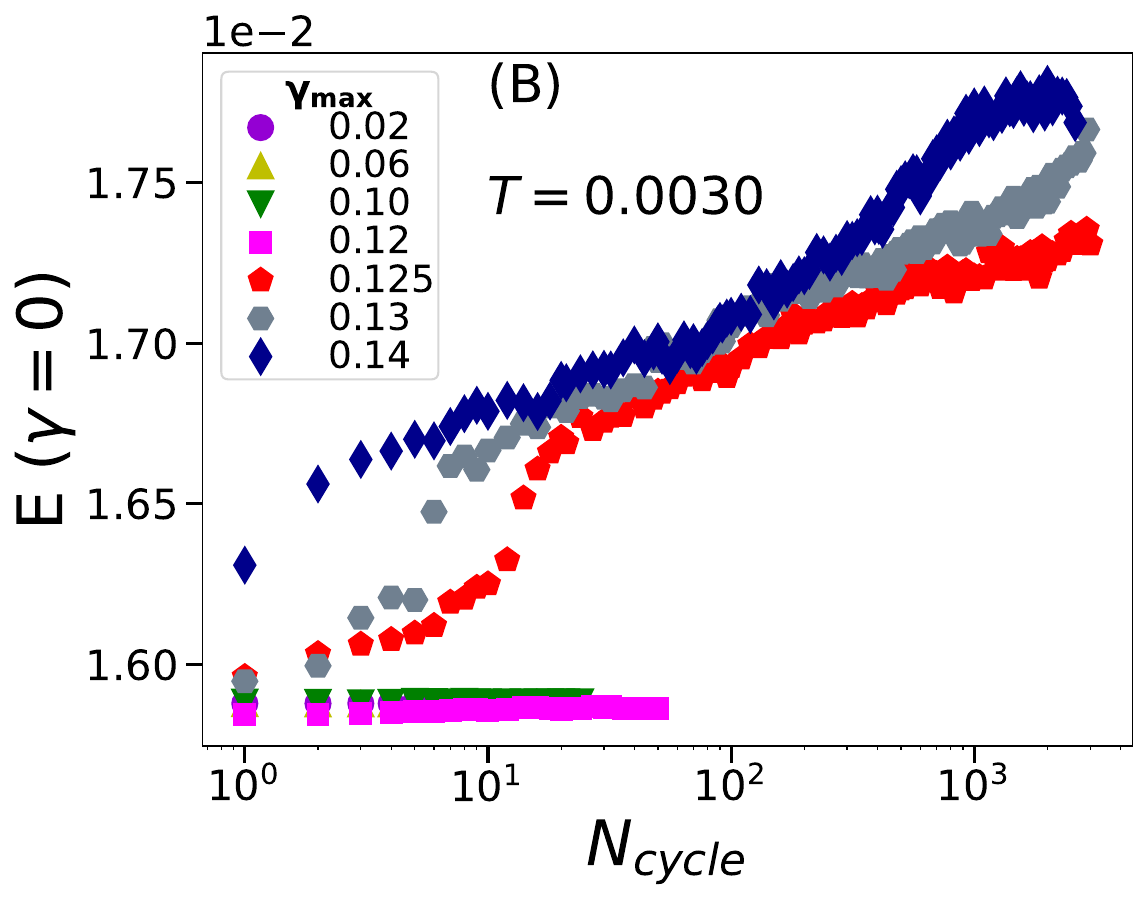}
    \caption{\textbf{Stroboscopic energy vs cycle for $\rho = 0.943$.} Plot of stroboscopic energy vs cycles at different $\gamma_{max}$ at density $\rho = 0.943$ for (A) poorly annealed ($T = 0.0045$) and (B) well annealed ($T = 0.0030$) samples. The yield points of the poorly annealed and well-annealed samples are at $\gamma_Y \approx 0.095$ and $\gamma_Y \approx 0.125$, respectively.}
    \label{SI_fig:8}
\end{figure}

Next, we show how the stroboscopic energy $E(\gamma = 0$) evolves with the number of oscillatory shear cycles. Two extreme temperature cases — the highest T (representing poorly annealed glass) and the lowest T (representing well-annealed glass)—are considered within the studied temperature range for each $\rho$. Stroboscopic energy obtained from $T = 0.0018$ and $T = 0.0005$ at $\rho = 0.681$ are plotted for different values of strain amplitude ($\gamma_{max}$) in Fig. \ref{SI_fig:5}. At the lowest density ($\rho = 0.681$), stroboscopic energies do not reach a perfect steady state even after thousands of cycles for a few $\gamma_{max}$ in the poorly annealed case. Therefore, we fitted the E($\gamma=0$) vs $N_{cycle}$ data using the stretched exponential relation: $E(N_{cycle}) = E_0 + b\exp\left(-\left(\frac{N_{cycle}}{\tau}\right)^{\beta}\right)$ to extract the steady-state energies as well as the relaxation time, $\tau$, representing the number of cycles to reach the steady state. The dashed lines represent the fitted data. From the fitting, we can extract $\tau$, the relaxation time shown in the main text. 
Similarly, the variation of stroboscopic energies with cycles at different $\gamma_{max}$ for the poorly annealed and well-annealed cases is also shown for $\rho = 0.750, 0.8556, 0.943$ in Fig. \ref{SI_fig:6}, \ref{SI_fig:7}, \ref{SI_fig:8} respectively.
\newpage

\newpage

\section{Steady state maximum stress $\sigma^{\text{max}}_{xz}$ vs $\gamma_{\text{max}}$ for $\rho = 0.681$ (3D HP)}
\begin{figure}[htp]
    \centering
    \includegraphics[width=0.40\textwidth]{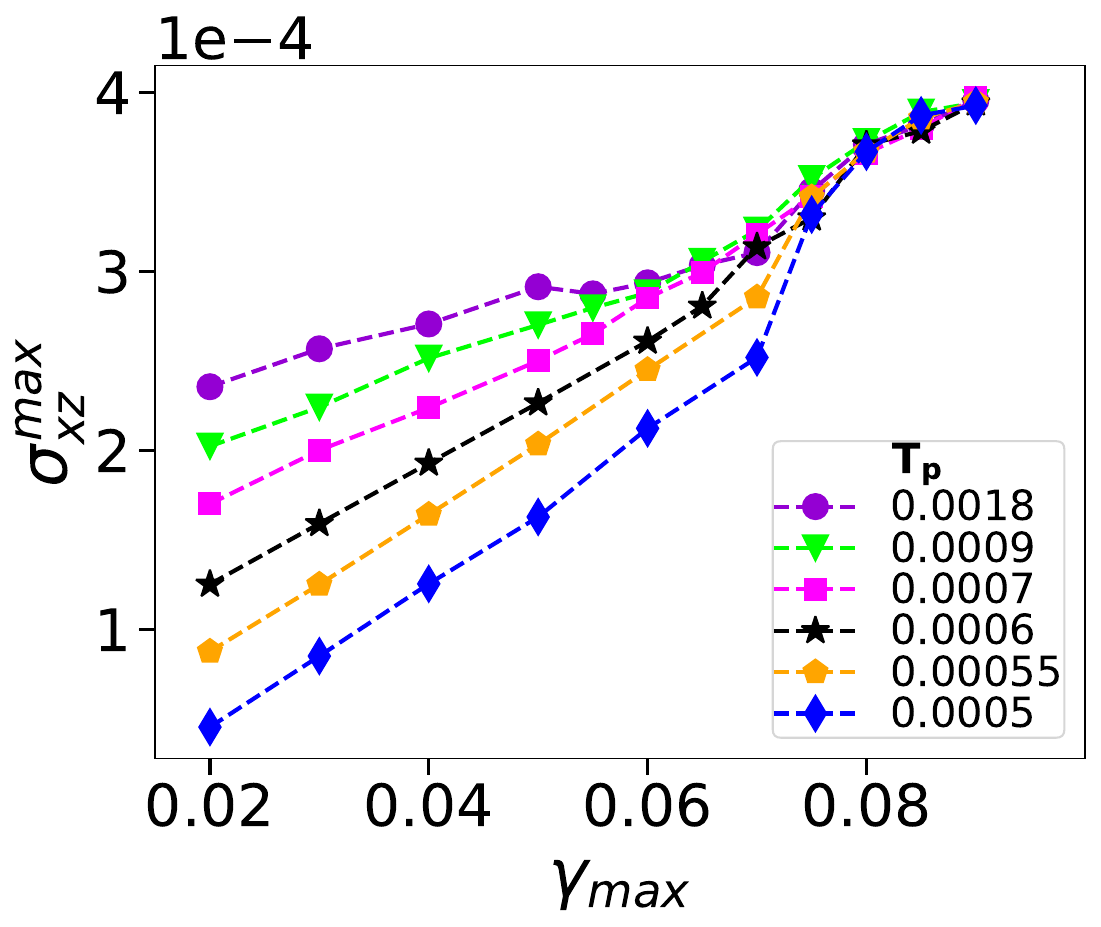}
    \caption{\textbf{Maximum steady state stress for $\rho = 0.681$.} The maximum steady state stress, $\sigma^{\text{max}}_{xz}$, is plotted as a function of strain amplitude $\gamma_{\text{max}}$ for different parent temperatures $T_p$ at the lowest density $\rho = 0.681$ of the 3D HP model. Each data point is averaged over 12 independent ensembles. System size $N = 5000$.}
    \label{SI_fig:22}
\end{figure}

In the main text, we presented the steady-state $\sigma^{\text{max}}_{xz}$ as a function of $\gamma{\text{max}}$ under oscillatory shear for densities $0.750$, $0.855$, and $0.943$. Here, we show the corresponding $\sigma^{\text{max}}_{xz}$ vs. strain data for the lowest density studied, $0.681$ for 3dHP model. As seen in Fig. \ref{SI_fig:22}, $\sigma^{\text{max}}_{xz}$ exhibits a markedly different behavior compared to all other densities. In particular, it appears to show shear hardening below yielding, as indicated by the slope of the $\sigma^{\text{max}}_{xz}$ vs. $\gamma_{\text{max}}$ curve. We emphasize that the results at this low density differ from all other cases, and whether this behavior is related to the unjamming/jamming transition remains an interesting open question for future study. Importantly, this observation does not affect the main conclusion of this work, namely that the shift in yielding behavior with the degree of annealing differs for liquids of different fragilities.  

\section{Timescale to reach the steady state} 
\begin{figure}[htp]
    \centering
    \includegraphics[width=0.30\textwidth]{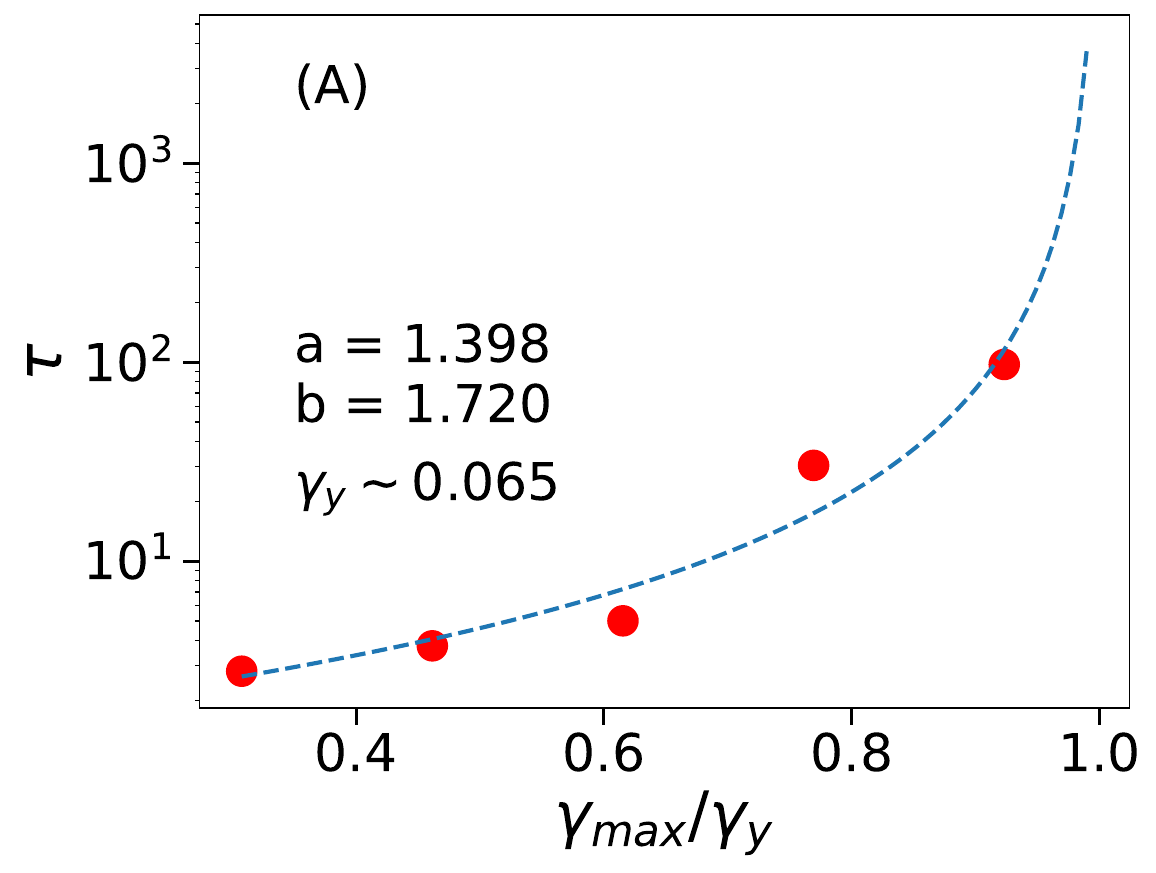}
    \includegraphics[width=0.30\textwidth]{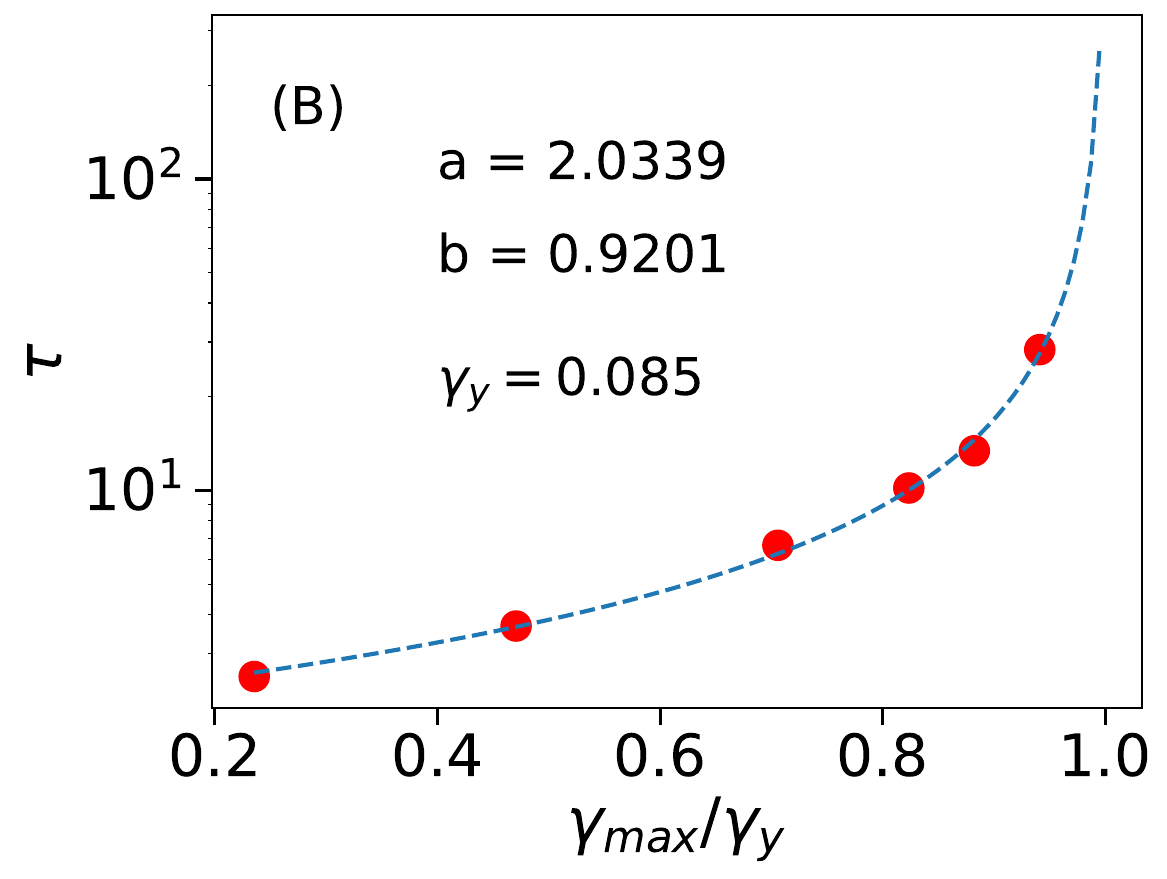}
    \includegraphics[width=0.30\textwidth]{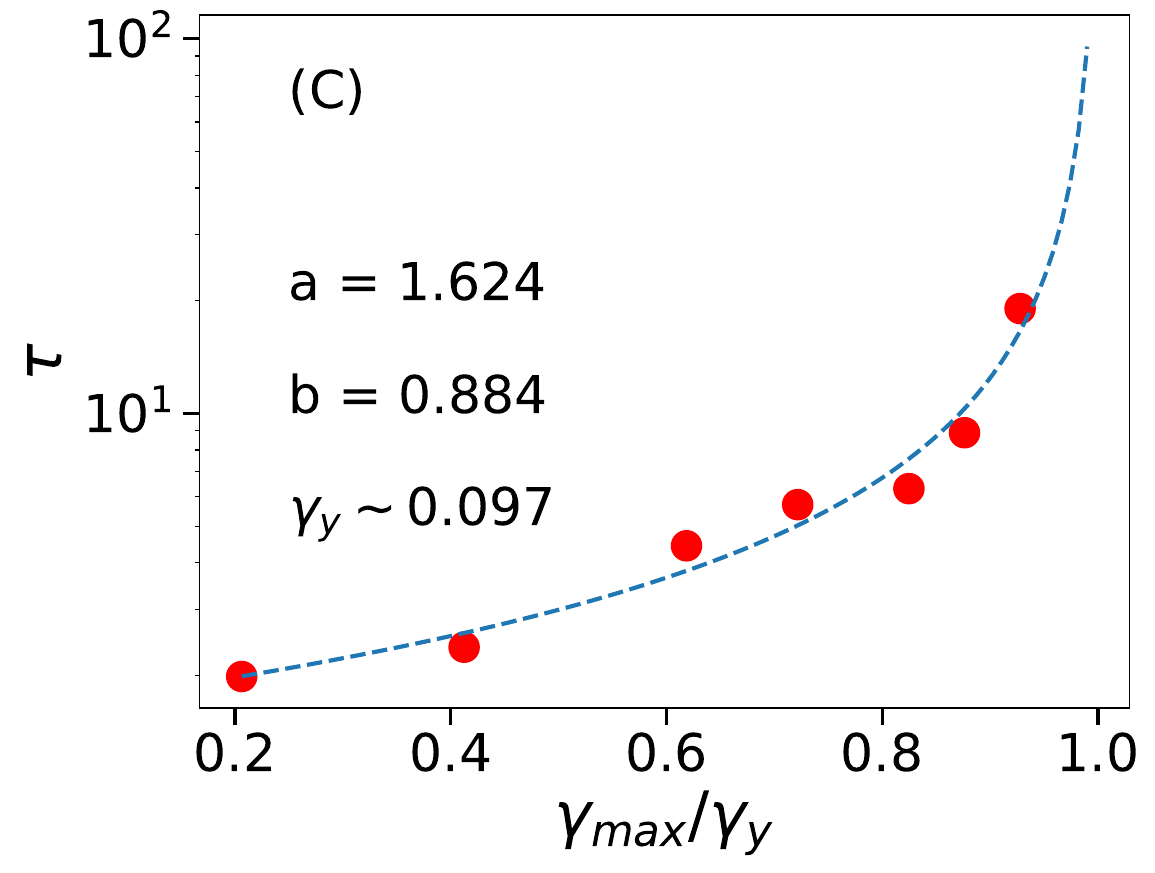}
    \caption{\textbf{Simulation timescale for 3D HP.} Relaxation time $\tau$ to reach the steady state for different strain amplitude $\gamma_{max}$ for (A) $\rho = 0.750$, (B) $\rho = 0.855$, (C) $\rho = 0.943$ in 3D HP model. Red dots correspond to actual data, and the dashed lines are the fit to the data points. Each data point is averaged over 12 independent ensembles.}
    \label{SI_fig:9}
\end{figure}

\begin{figure}[htp]
    \centering
    \includegraphics[width=0.30\textwidth,height=0.25\textwidth]{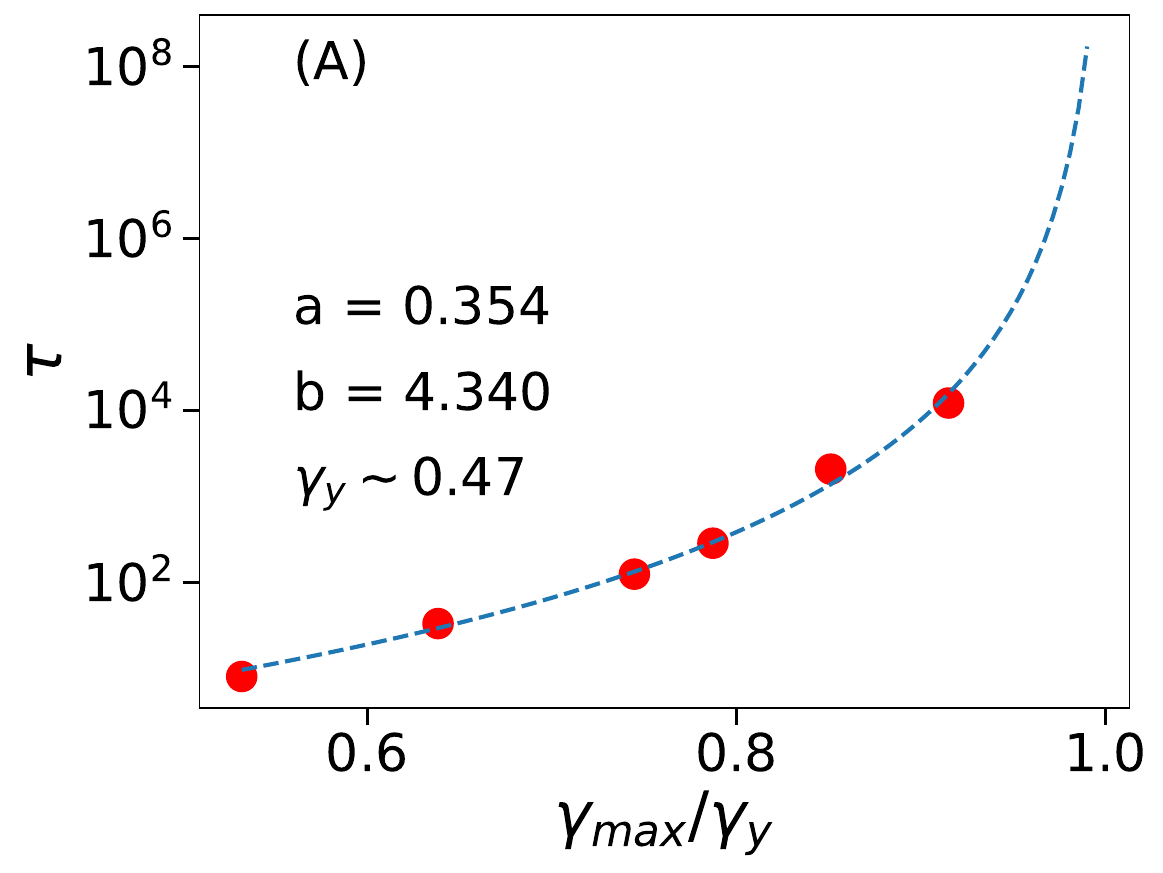}
    \includegraphics[width=0.30\textwidth,height=0.25\textwidth]{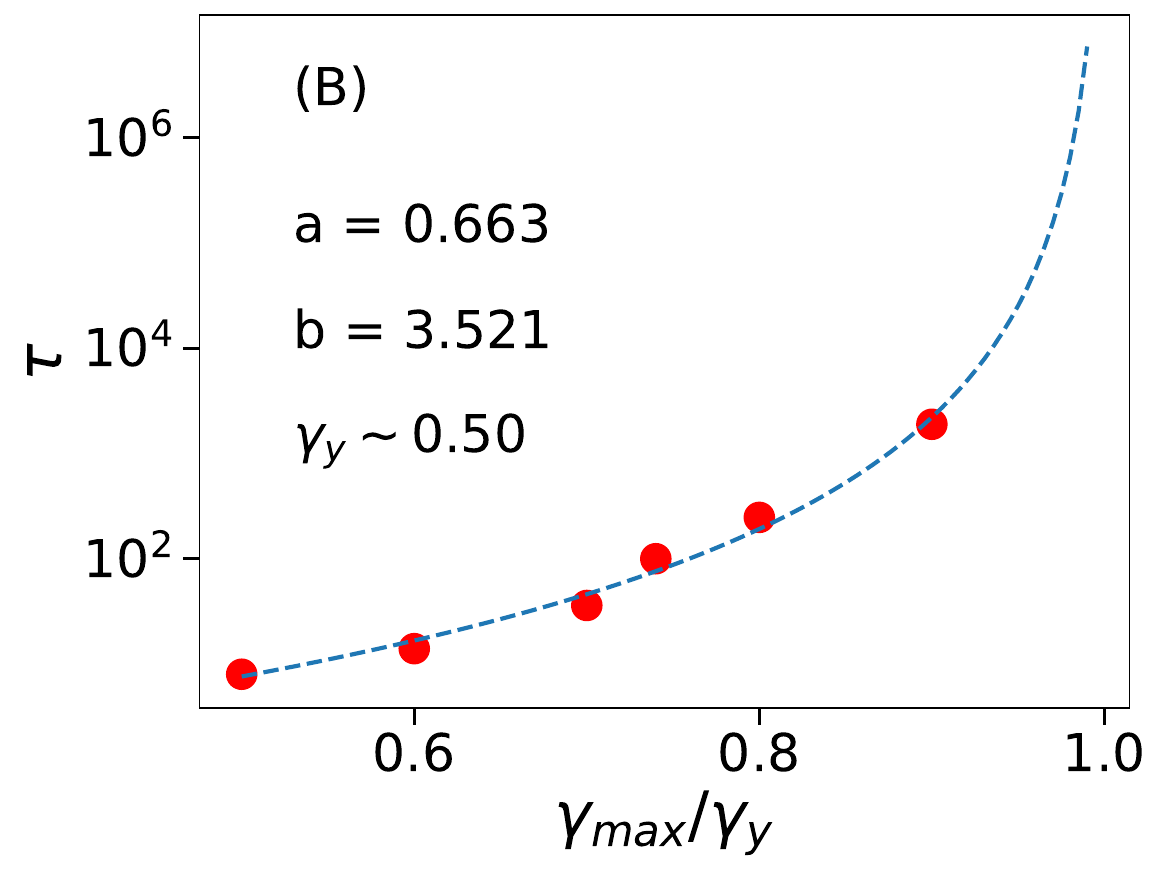}
    \includegraphics[width=0.30\textwidth,height=0.25\textwidth]{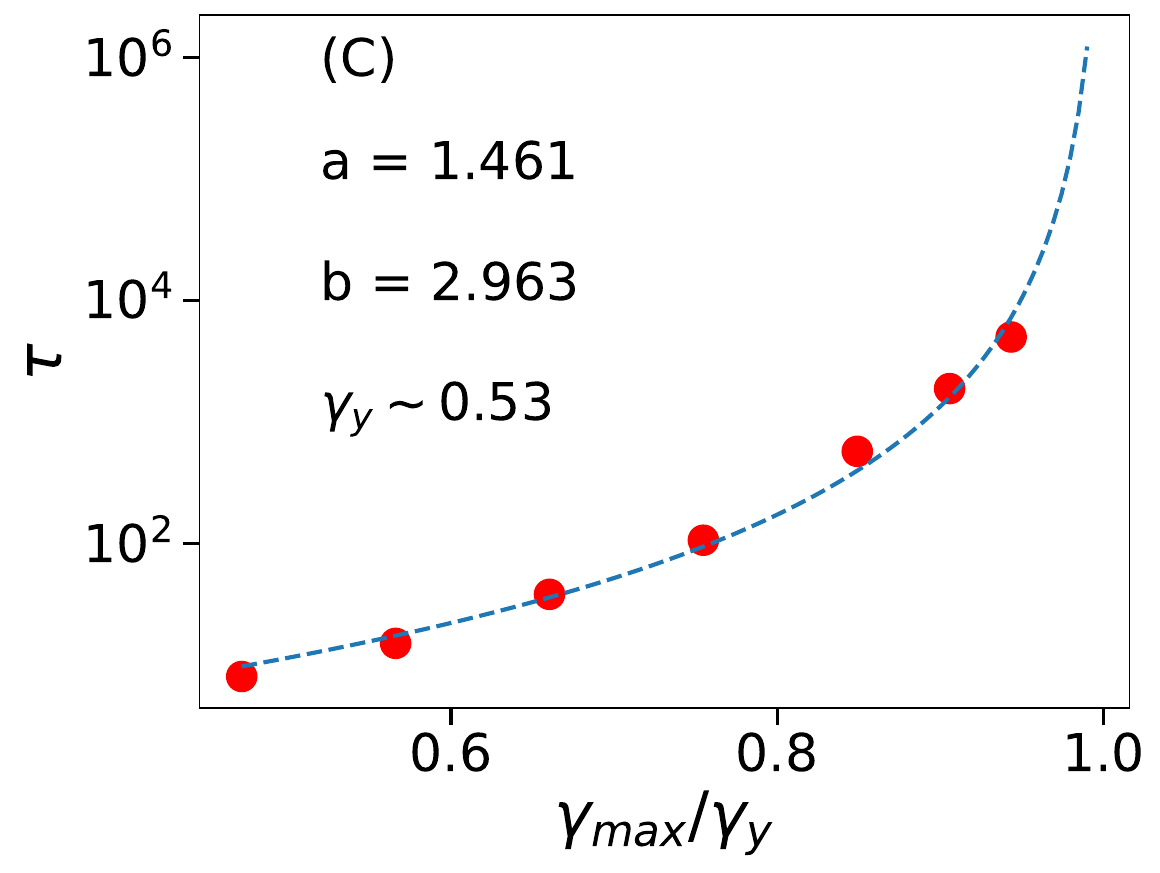}
    \caption{\textbf{Elastoplastic Model timescale.} Number of cycles $\tau$ required to reach the steady state for different strain amplitude $\gamma_{max}$ for (A) strong, (B) intermediate, (C) fragile case. Red dots correspond to actual data, and the dashed lines are the fit to the data points. Each data point is averaged over 100 independent ensembles.}
    \label{SI_fig:10}
\end{figure}

\newpage

Now, we show the relaxation time, $\tau$, obtained from the fitting described above as a function of $\gamma_{\text{max}}$. As $\gamma_{{max}}$ approaches the yield point $\gamma_Y$, the relaxation time to reach the steady state diverges. Although various mechanisms have been proposed in the literature to explain how $\tau$ increases with $\gamma_{\text{max}}$, we have observed that $\tau = a (\gamma_Y - \gamma_{\text{max}})^{-b}$ describes our data well. In Fig. \ref{SI_fig:9}, we show the relaxation time as a function of $\gamma_{max}$ for different densities, where the dashed line represents the fit and the points represent the actual data from the simulation. Similar behaviour is also observed for the elastoplastic model we studied. The corresponding data is shown in Fig. \ref{SI_fig:10}. In the table below (Table S1), we show all the fit parameters for different densities in 3D HP as well as for other models. 

\begin{table}[htbp]
\centering
\renewcommand{\arraystretch}{1.25}   
\setlength{\tabcolsep}{10pt}          
\begin{tabular}{|l|c|c|c|c|}
\hline
System & $a_1$ & $b_1$ & $a_2$ & $b_2$ \\
\hline
$\rho = 0.681$ & 1.485 & 1.896 & 0.390 & 2.194 \\
\hline
$\rho = 0.750$ & 1.398 & 1.720 & 0.203 & 2.118 \\
\hline
$\rho = 0.855$ & 2.033 & 0.920 & 0.425 & 1.007 \\
\hline
$\rho = 0.943$ & 1.624 & 0.884 & 1.064 & 0.996 \\
\hline
Silica & 1.463 & 1.318 & 0.058 & 1.962 \\
\hline
OTP & 1.795 & 0.925 & 0.173 & 1.801 \\
\hline
3D KA & 1.816 & 0.898 & 0.041 & 1.663 \\
\hline
Cu--Zr (NVT) & 2.358 & 0.576 & 0.391 & 1.536 \\
\hline
Cu--Zr (NPT) & 3.912 & 0.896 & 0.170 & 1.444 \\
\hline
\end{tabular}
\caption{Fit parameters for the timescale $\tau$ to reach the steady state for different glass-forming systems before and after yielding.
The timescale follows $\tau = a \left|1 - \gamma_{\max}/\gamma_Y\right|^{-b}$.}
\label{table:1A}
\end{table}

\section{Response under uniform shear deformation for 3D HP}
\subsection{Stress vs strain Responses under Uniform shear Deformation}
\begin{figure}[htp]
    \centering
    \includegraphics[width=0.46\textwidth]{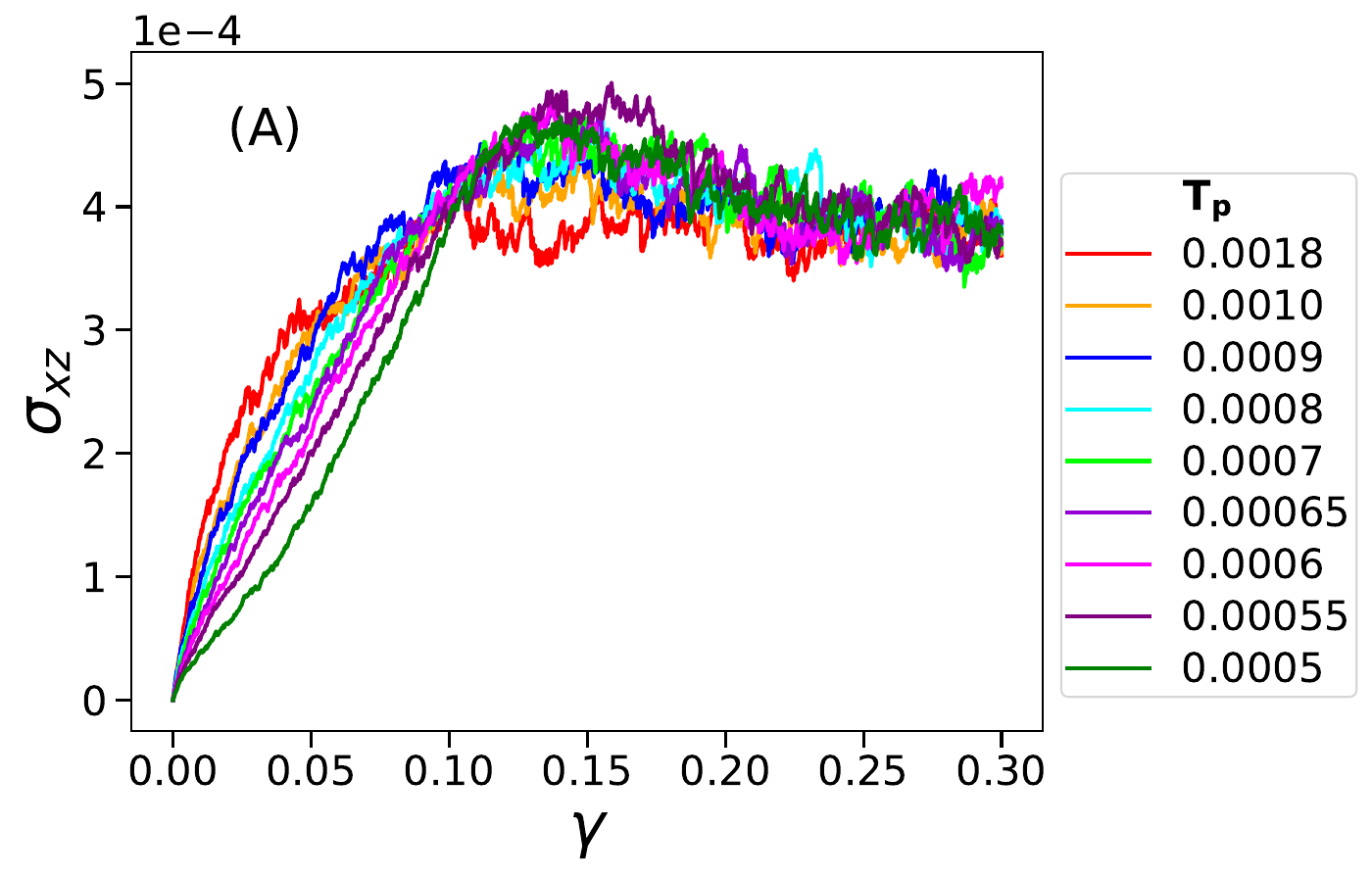}
    \includegraphics[width=0.46\textwidth]{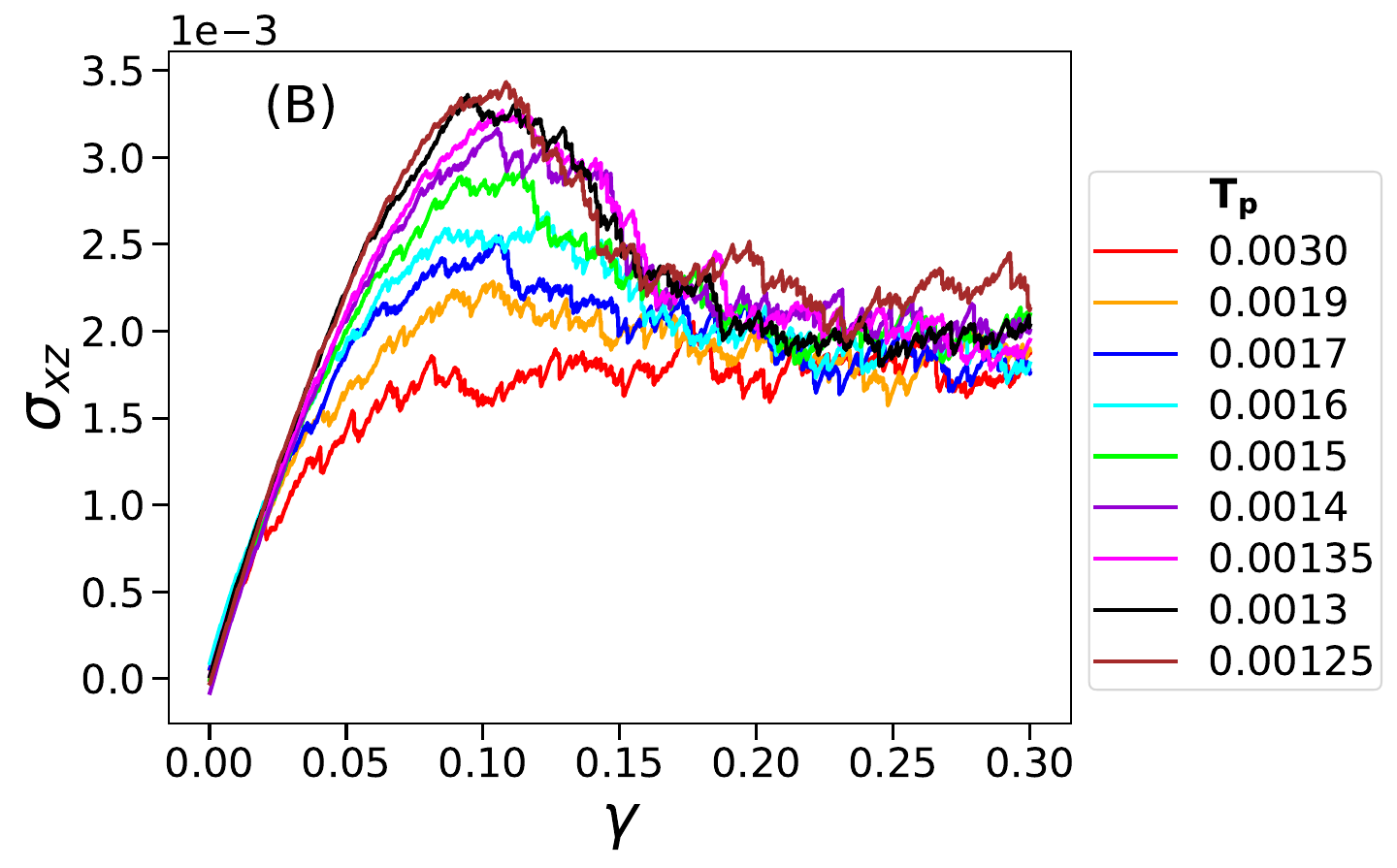}
    \includegraphics[width=0.46\textwidth]{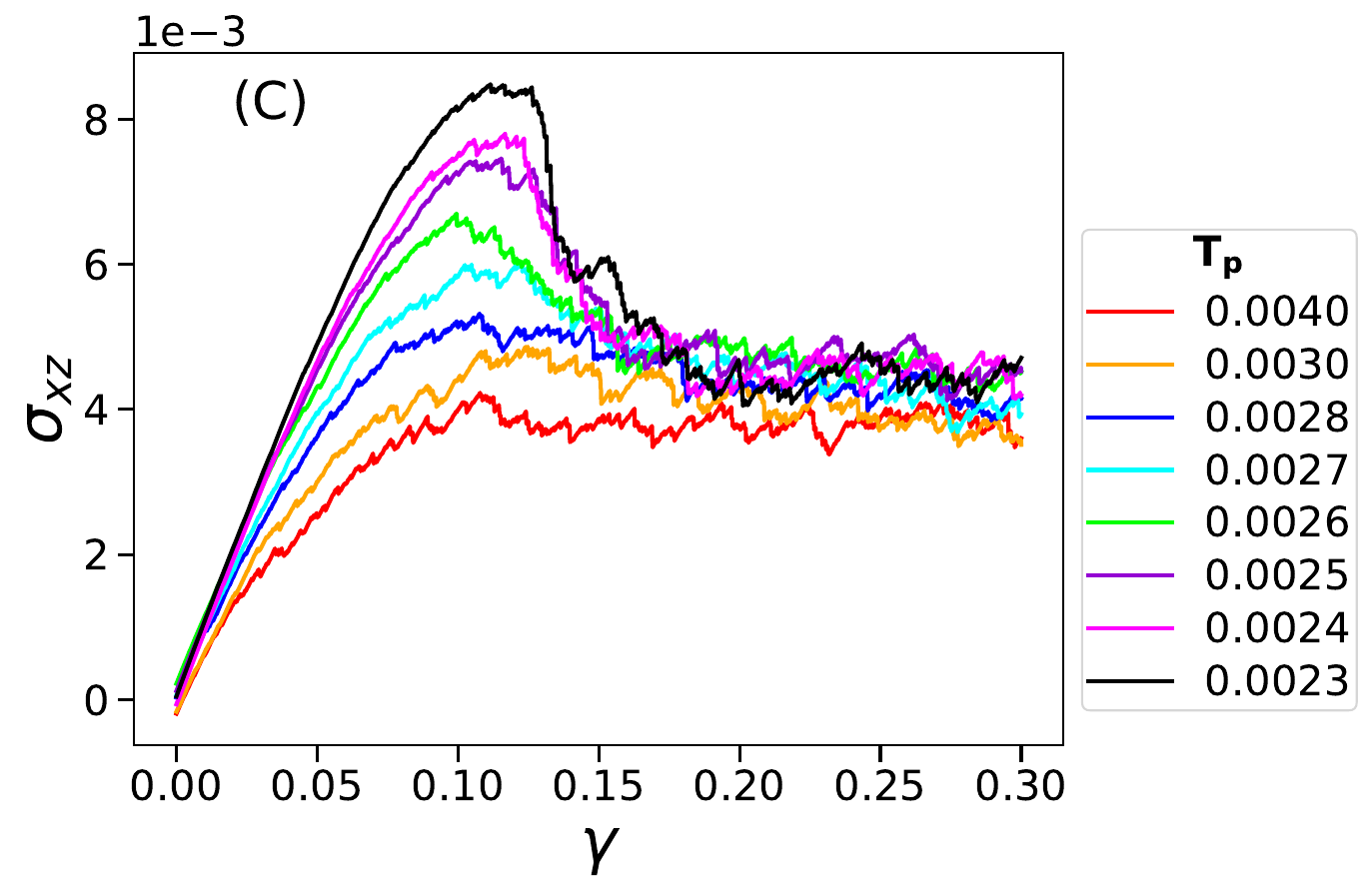}
    \includegraphics[width=0.46\textwidth]{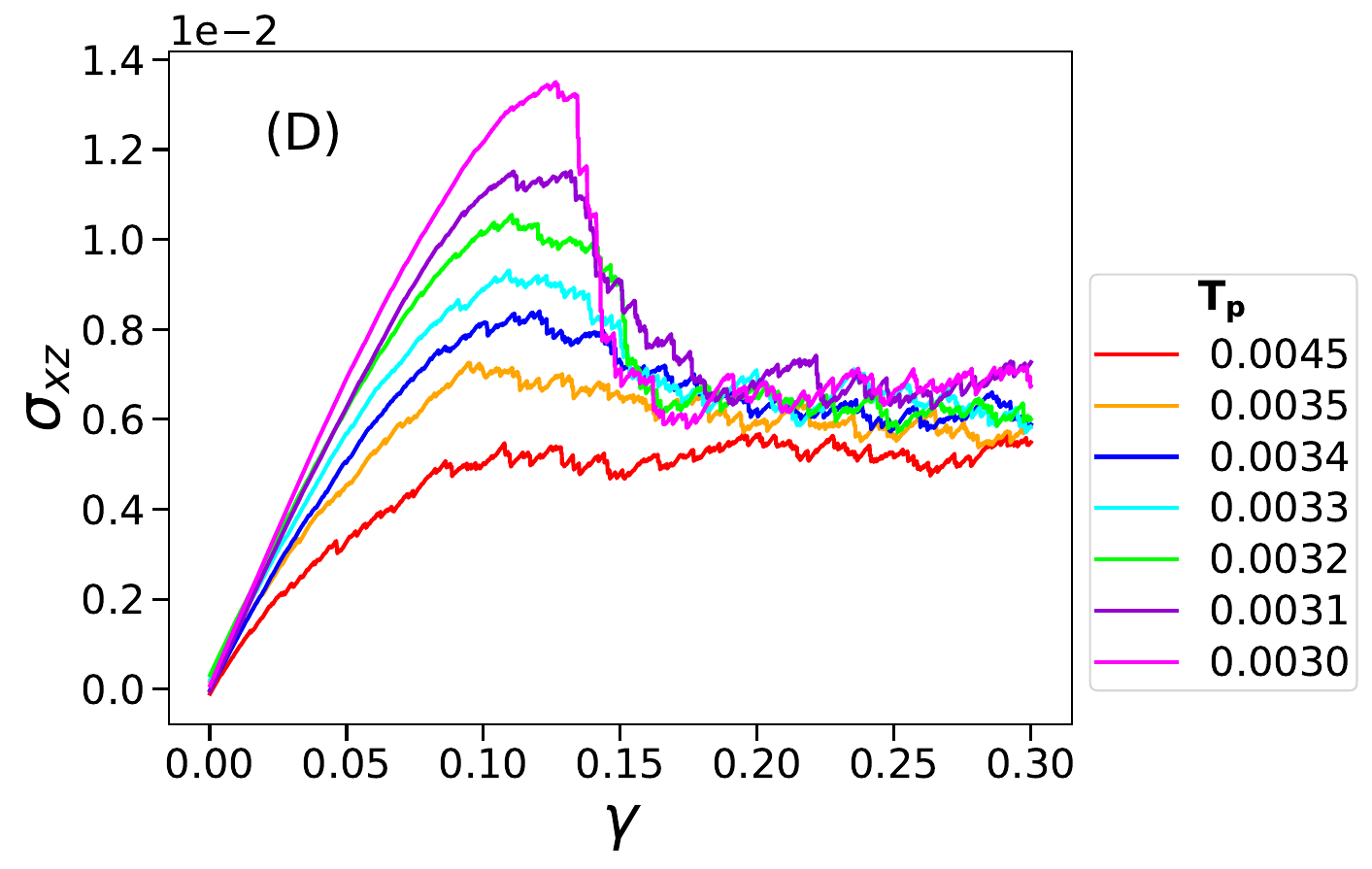}
    \caption{\textbf{Stress-strain curve for 3D HP under uniform shear.} The shear stress $\sigma_{xz}$ is plotted as a function of strain $\gamma$ under uniform shear for (A) $\rho = 0.681$, (B) $\rho = 0.750$, (C) $\rho = 0.855$, and (D) $\rho = 0.943$. The lowest-density system exhibits a more ductile response, whereas increasing density leads to progressively more brittle behavior. Data points are averaged over 8 independent samples. System size $N = 5000$ for each panel.}
    \label{SI_fig:16}
\end{figure}
In the main text, we presented the cyclic shear response for 3D HP. For completeness, here we also show the results for uniform shear deformation. In Fig. \ref{SI_fig:16}, we show stress vs strain during uniform shear deformation for different densities.  As mentioned before, the stress–strain relation at the lowest density ($\rho = 0.681$) exhibits behavior distinct from the other densities: the shear modulus shows a significant variation with annealing, unlike at higher densities. The possible origin of this anomalous behavior requires further investigation in future work.
For all other densities, when the system is poorly annealed, the stress increases continuously (without overshoot) from the elastic branch to the steady state. However, as the degree of annealing increases, an overshoot emerges. At the highest density, the most annealed systems display a sharp stress drop, resembling brittle failure. In contrast, at $\rho = 0.750$, even the most annealed glasses exhibit a smoother stress drop, characteristic of ductile behavior.

By comparing uniform shear and cyclic shear (discussed in the main text), we conclude that cyclic shear is a more effective protocol for identifying both the yield strain and the nature of yielding, consistent with previous observations.
\subsection{Energy vs strain for Uniform shear}
\begin{figure}[htp]
    \centering
    \includegraphics[width=0.49\textwidth,height=0.35\textwidth]{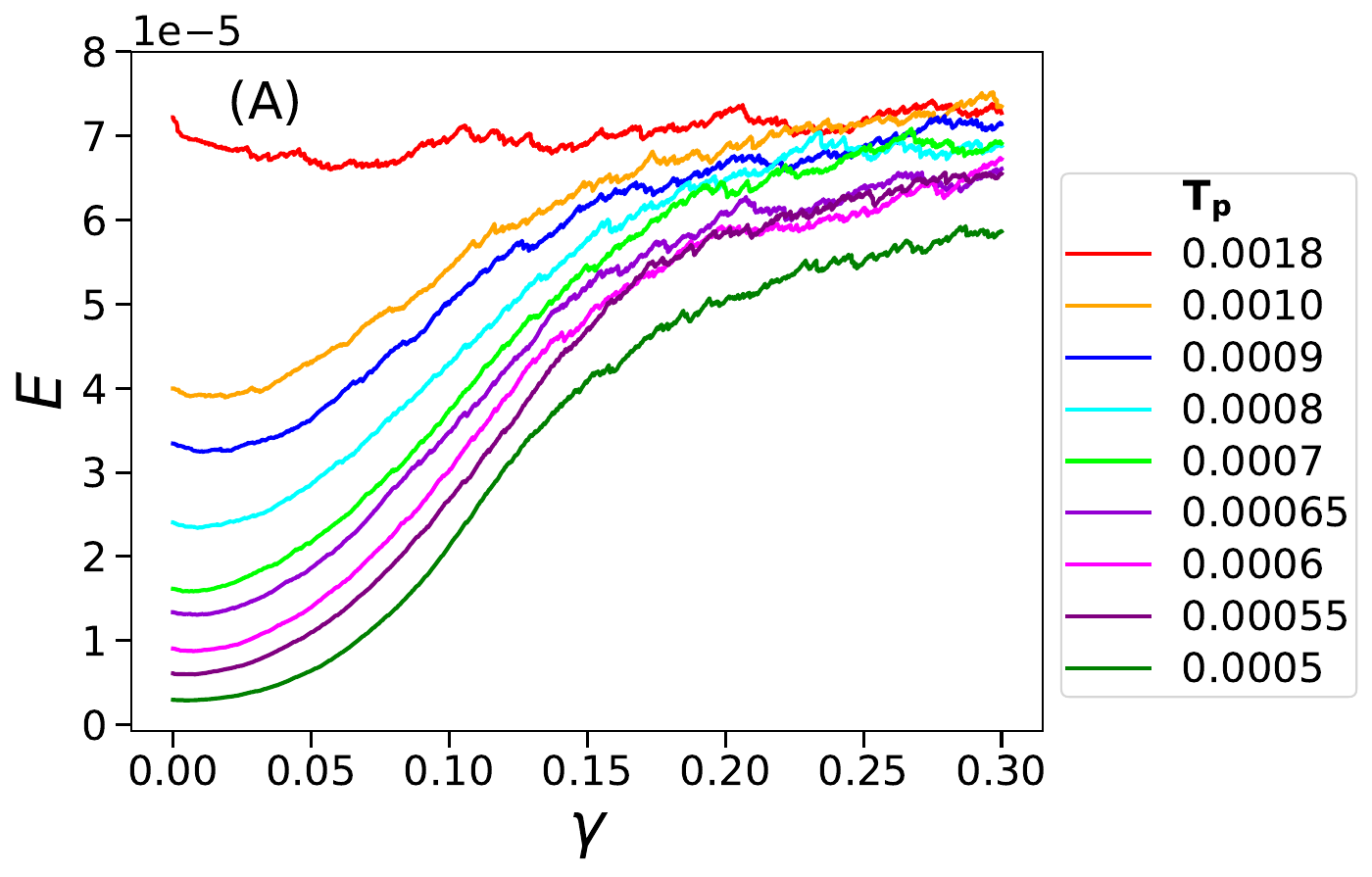}
    \includegraphics[width=0.49\textwidth,height=0.35\textwidth]{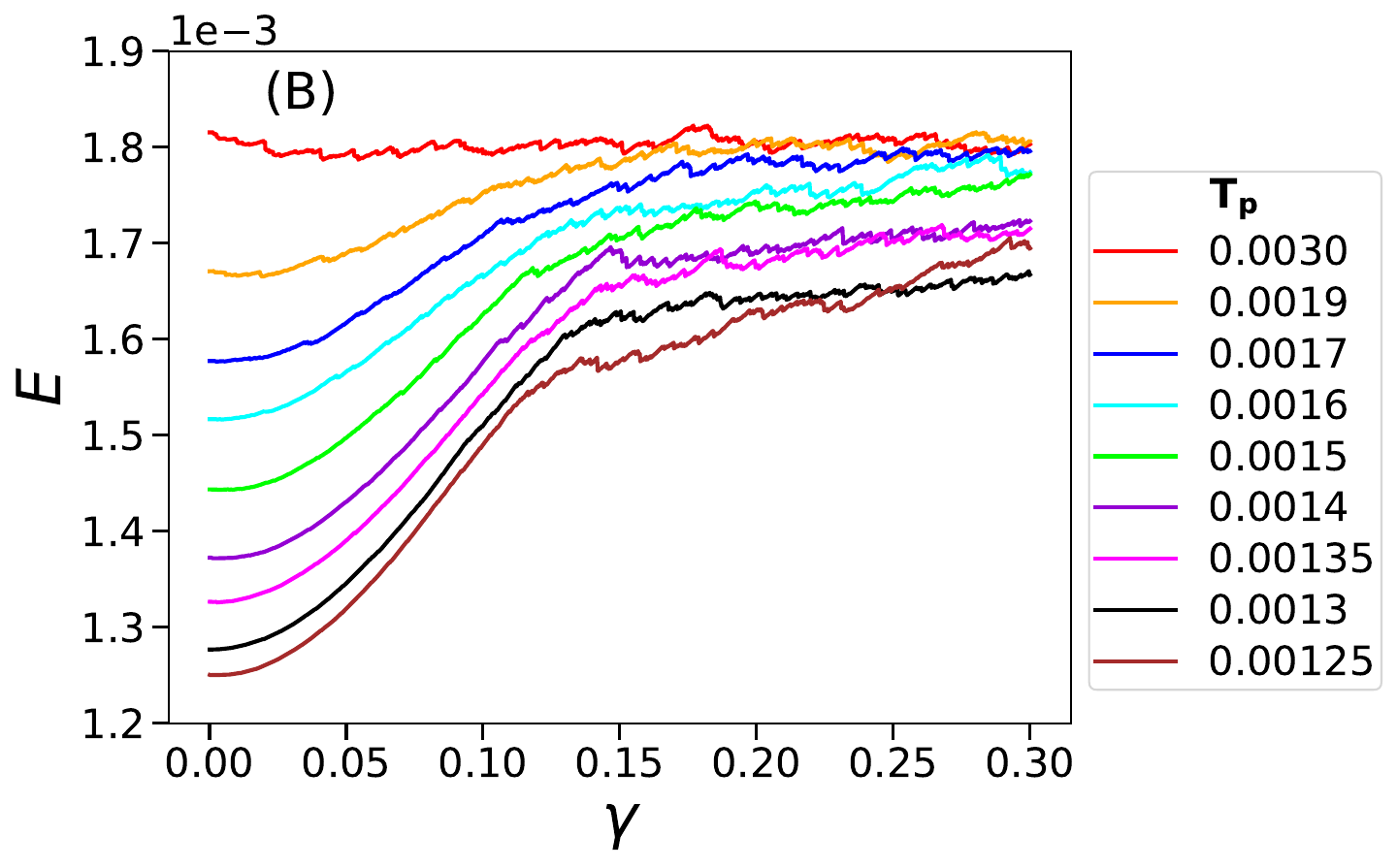}
    \includegraphics[width=0.49\textwidth,height=0.35\textwidth]{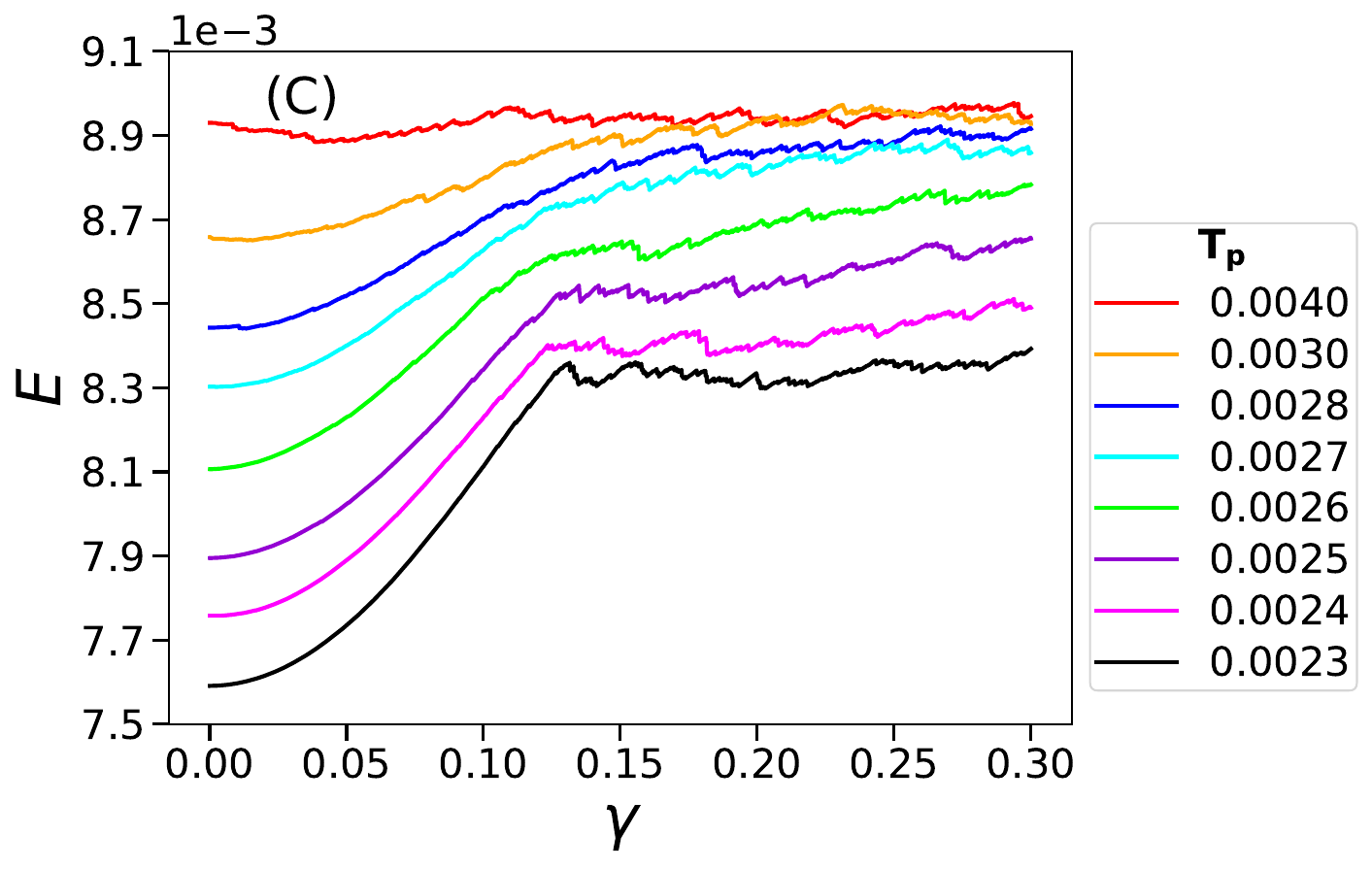}
    \includegraphics[width=0.49\textwidth,height=0.35\textwidth]{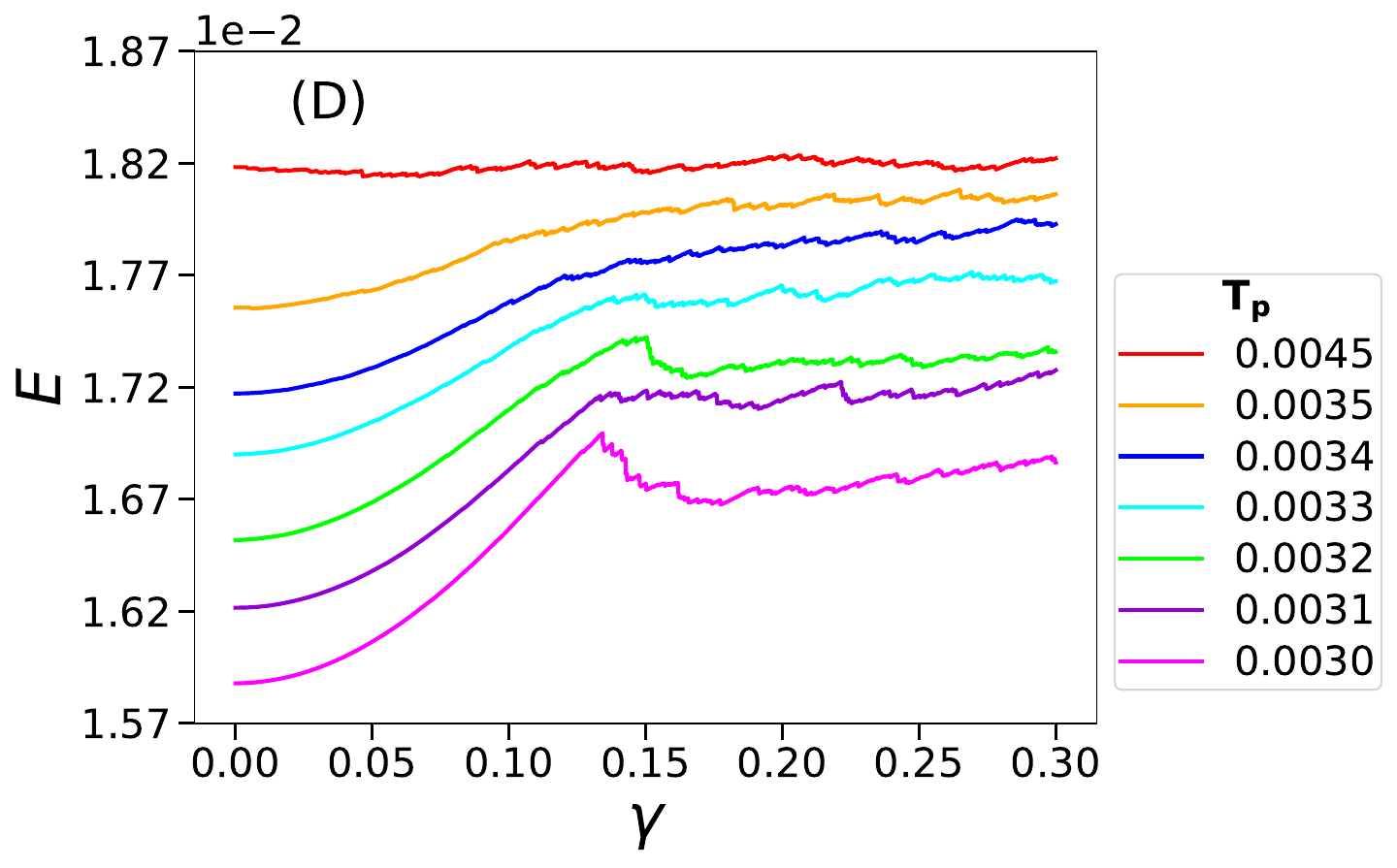}
    \caption{\textbf{Energy-strain curve for 3D HP under uniform shear.} The potential energy $E$ is plotted as a function of strain $\gamma$ under uniform shear for densities (A) $\rho = 0.681$, (B) $\rho = 0.750$, (C) $\rho = 0.855$, and (D) $\rho = 0.943$. Data points are averaged over 8 independent samples. System size $N = 5000$ for each panel.}
    \label{SI_fig:15}
\end{figure}

Here, we present the energy as a function of strain during uniform shear deformation (Fig. \ref{SI_fig:15}). For poorly annealed glasses, the energy shows little change with increasing strain. However, as the degree of annealing increases, a distinct change in behavior appears at the yield strain. Thus, we observed that, although energy—particularly the stroboscopic energy—is a central quantity for understanding cyclic shear behavior, it does not exhibit any dramatic change during uniform shear, especially in poorly annealed glasses.  

\newpage

\section{Tuning fragility with changing pressure}
\begin{figure}[H]
    \centering
    \includegraphics[width=0.45\textwidth]{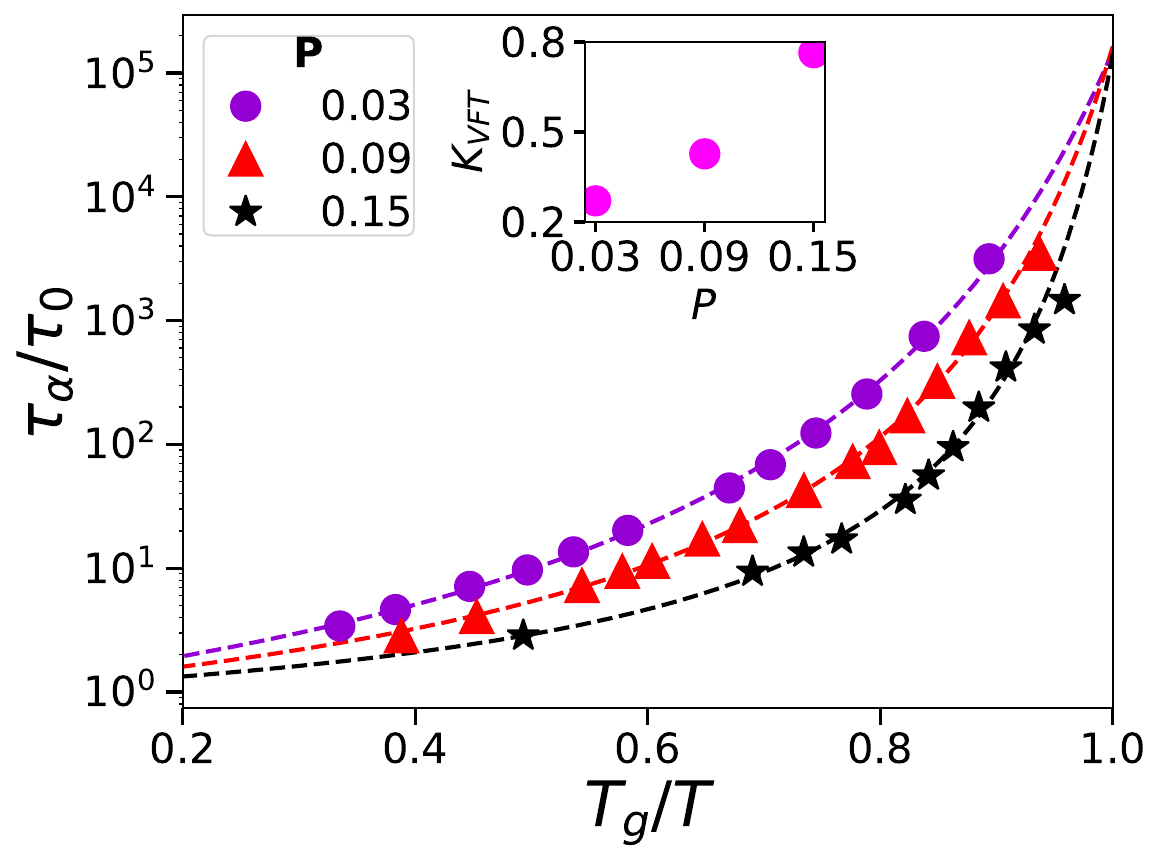}
    \caption{\textbf{Angell plot at different pressures for 3D HP.} The structural relaxation time $\tau_\alpha$ is plotted as a function of the normalized temperature $T_g/T$, where $T_g$ denotes the glass transition temperature. Results show that the fragility of the system increases with increasing pressure. The inset displays the fragility $K_{\mathrm{VFT}}$ as a function of pressure $P$.}
    \label{SI_fig:21}
\end{figure}

So far, we have discussed how fragility varies with density. In Fig. \ref{SI_fig:21}, we show that fragility can also be tuned by changing pressure. Thus, our result—that fragility governs the yielding behavior—is expected to be independent of the ensemble used. (We explicitly demonstrate this for the 3D HP model below.)

\section{Angell Plot for four different models}
\begin{figure}[htp]
    \centering
    \includegraphics[width=0.45\textwidth]{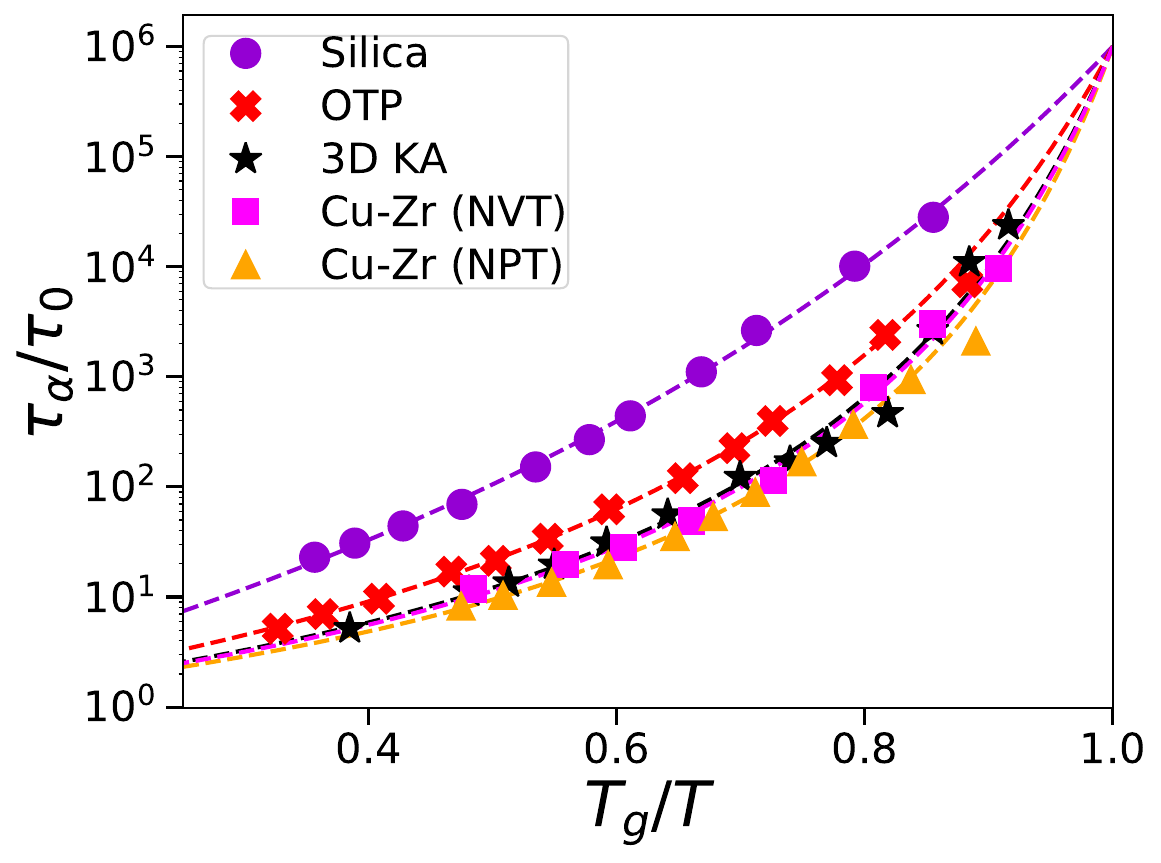}
    \caption{\textbf{Angell plot for microscopically different glass-forming models.} The structural relaxation time $\tau_\alpha$ is plotted as a function of the normalized temperature $T_g/T$, where $T_g$ denotes the glass transition temperature. Silica corresponds to the most strong glass, whereas Cu-Zr metallic glass is the most fragile in this study. Fitting parameters are mentioned in TABLE S2.}
    \label{SI_fig:11}
\end{figure}

In Fig. \ref{SI_fig:11}, we present the scaled relaxation time data as a function of scaled temperature. We also show the VFT fitting using the full range of data, and the corresponding fit parameters are listed in the table below. In the main text, we report the best-fit results obtained by adjusting the fitting range. Qualitatively, the behaviour remains the same: among all four different models, silica is the strongest glass former with the lowest fragility, while Cu–Zr is the most fragile glass. For Cu–Zr, we present data from both the NVT and NPT ensembles, and the results are qualitatively similar in both cases.In the following table (Table :S2), we listed $\tau_0$, $T_{VFT}$ and $K_{VFT}$ for all four models.  

\begin{table}[htbp]
\centering
\renewcommand{\arraystretch}{1.25}   
\setlength{\tabcolsep}{10pt}          
\begin{tabular}{|l|c|c|c|}
\hline
System & $\ln \tau_0$ & $T_{\mathrm{VFT}}$ & $K_{\mathrm{VFT}}$ \\
\hline
Silica & 2.859 & 1050.346 & 0.0698 \\
\hline
OTP & 6.743 & 233.253 & 0.181 \\
\hline
3D KA & -0.972 & 0.299 & 0.256 \\
\hline
Cu-Zr (NVT) & -2.358 & 571.121 & 0.266 \\
\hline
Cu-Zr (NPT) & -2.117 & 573.502 & 0.301 \\
\hline
\end{tabular}
\caption{Fit parameters obtained from Vogel--Fulcher--Tammann (VFT) fits to the relaxation-time data as a function of temperature.}
\label{table:VFT}
\end{table}

\newpage
\section{$T_{MCT}$ for four different models} 
\begin{figure}[H]
    \centering
    \includegraphics[width=0.49\textwidth,height=0.40\textwidth]{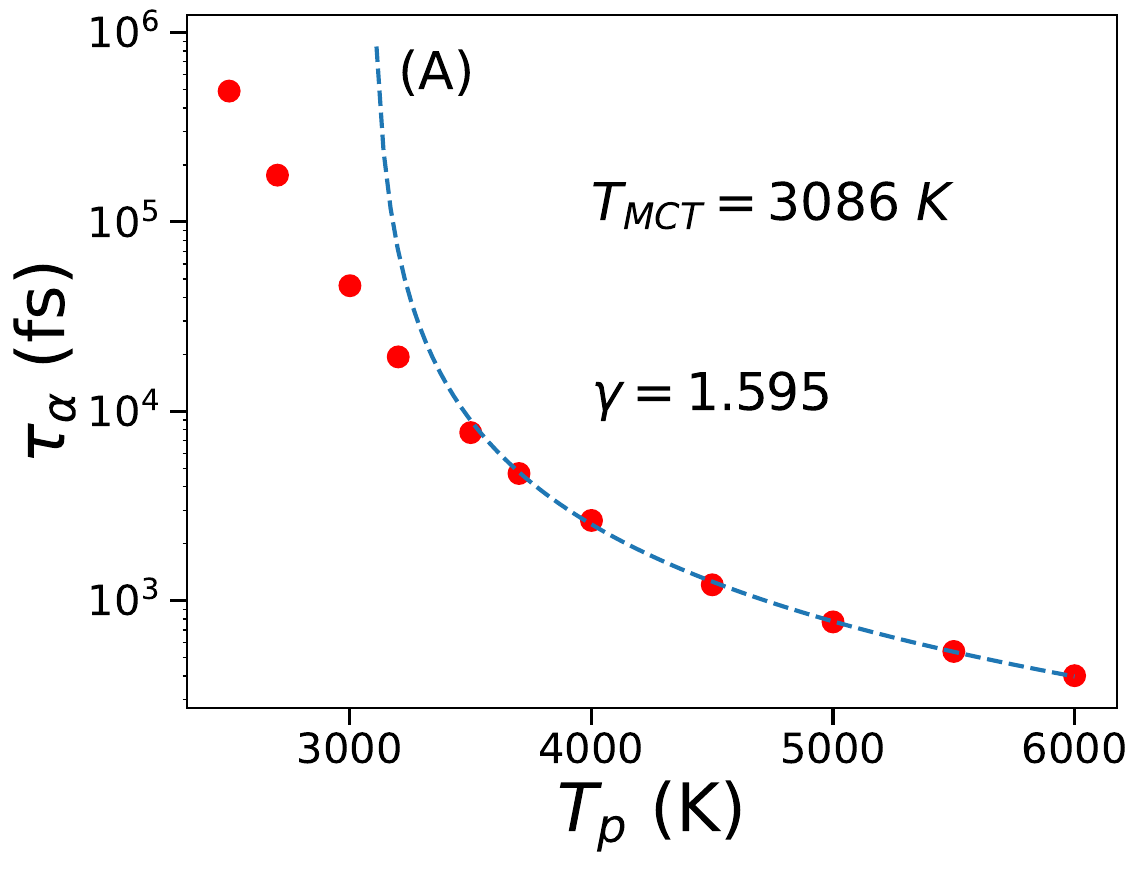}
    \includegraphics[width=0.49\textwidth,height=0.40\textwidth]{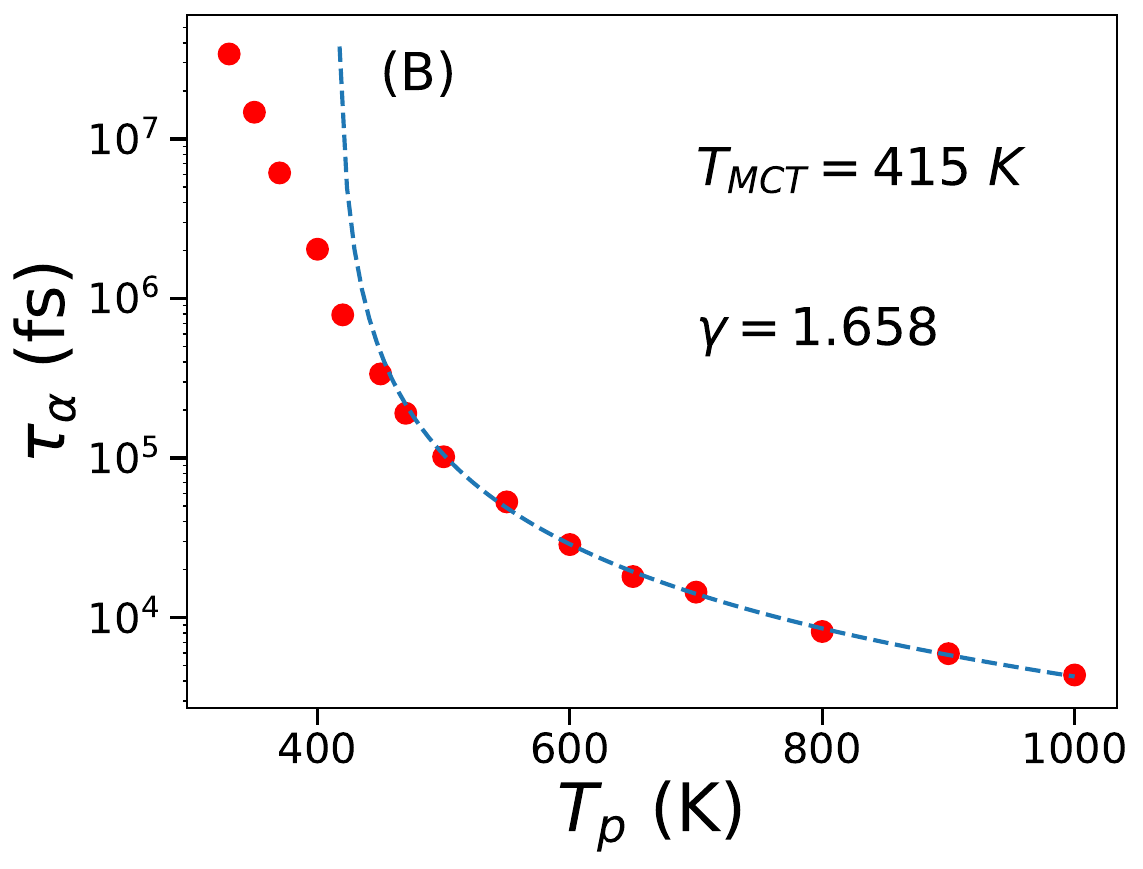}
    \includegraphics[width=0.49\textwidth,height=0.40\textwidth]{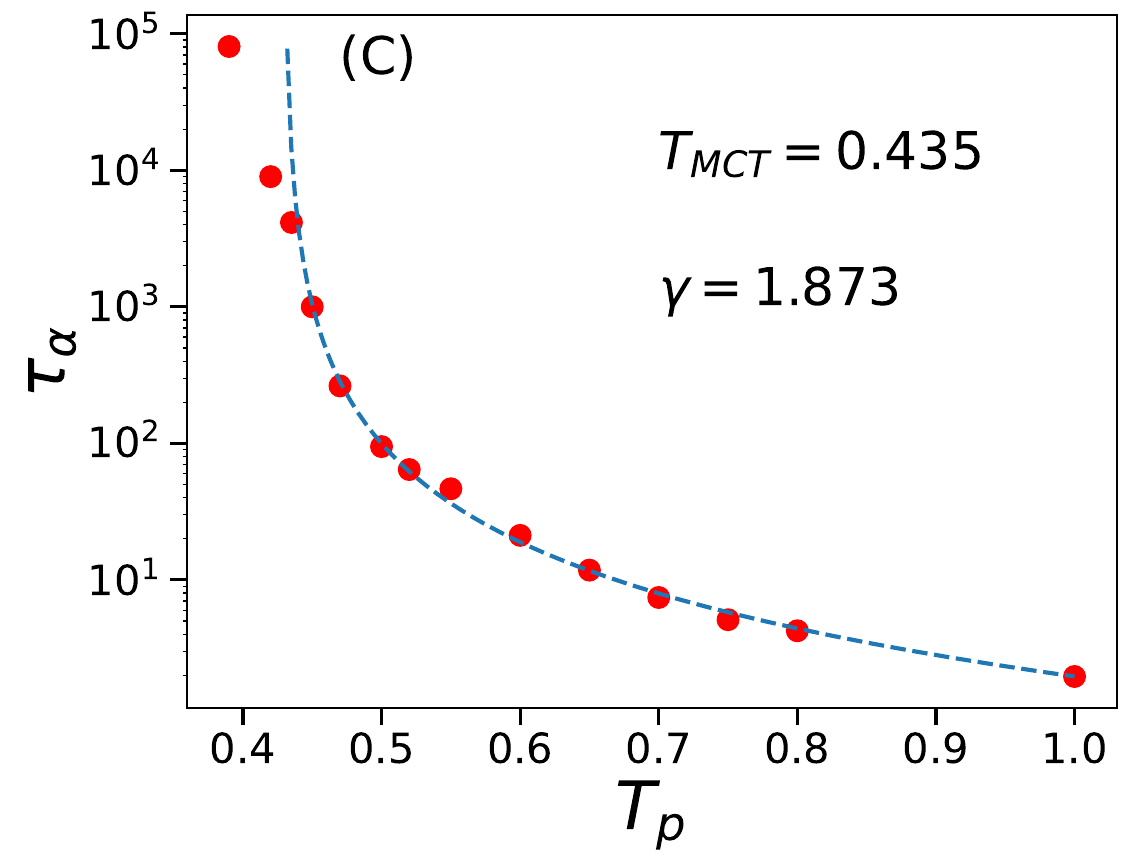}
    \includegraphics[width=0.49\textwidth,height=0.40\textwidth]{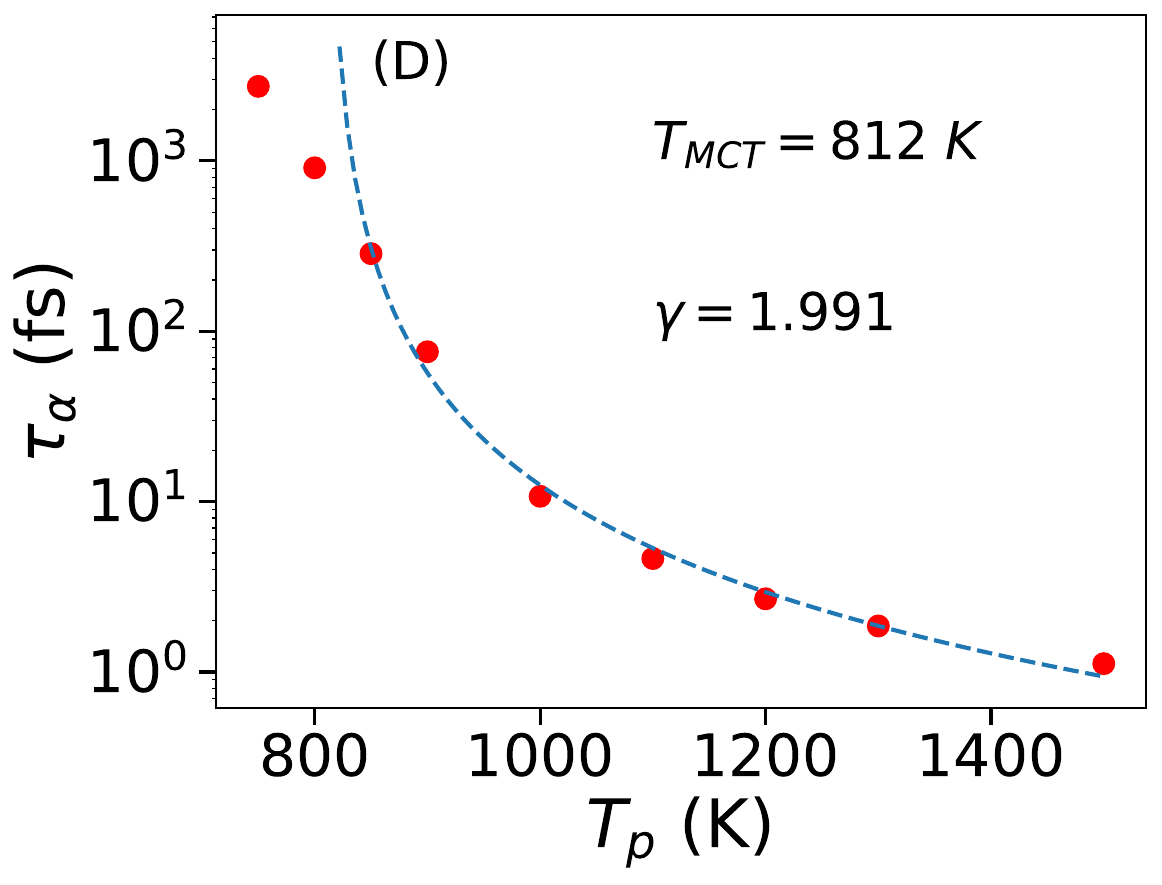}
    \caption{\textbf{Mode coupling temperature for microscopically different glass-forming models.} We show relaxation time $\tau_{\alpha}$ vs. parent temperature $T_p$ along with the corresponding MCT fits in panel (A) BKS Silica, (B) OTP, (C) 3D KA, and (D) Cu-Zr glass. Data points are fitted to the MCT equation, $\tau_\alpha = \tau_0 (T - T_{MCT})^{-\gamma}$. The extracted values of $T_{MCT}$ and $\gamma$ are indicated within the figures.}
    \label{SI_fig:28}
\end{figure}

We have discussed in the main text that, similar to 3D HP, we also observe a change in yielding behavior for all four other studied models around $T_{MCT}$. In Fig. \ref{SI_fig:28}, we show how $T_{MCT}$ is computed for the different models. The $\tau_{\alpha}$ vs. $T_p$ data are fitted to the MCT power-law equation, $\tau_\alpha = \tau_0 (T - T_{MCT})^{-\gamma}$, where $\tau_\alpha$ diverges at a critical temperature $T_{MCT}$.  The corresponding $T_{MCT}$ is shown inside each figure. 
\newpage
\section{Stroboscopic energy with cycles for different glasses}
\begin{figure}[H]
    \centering
    \includegraphics[width=0.49\textwidth]{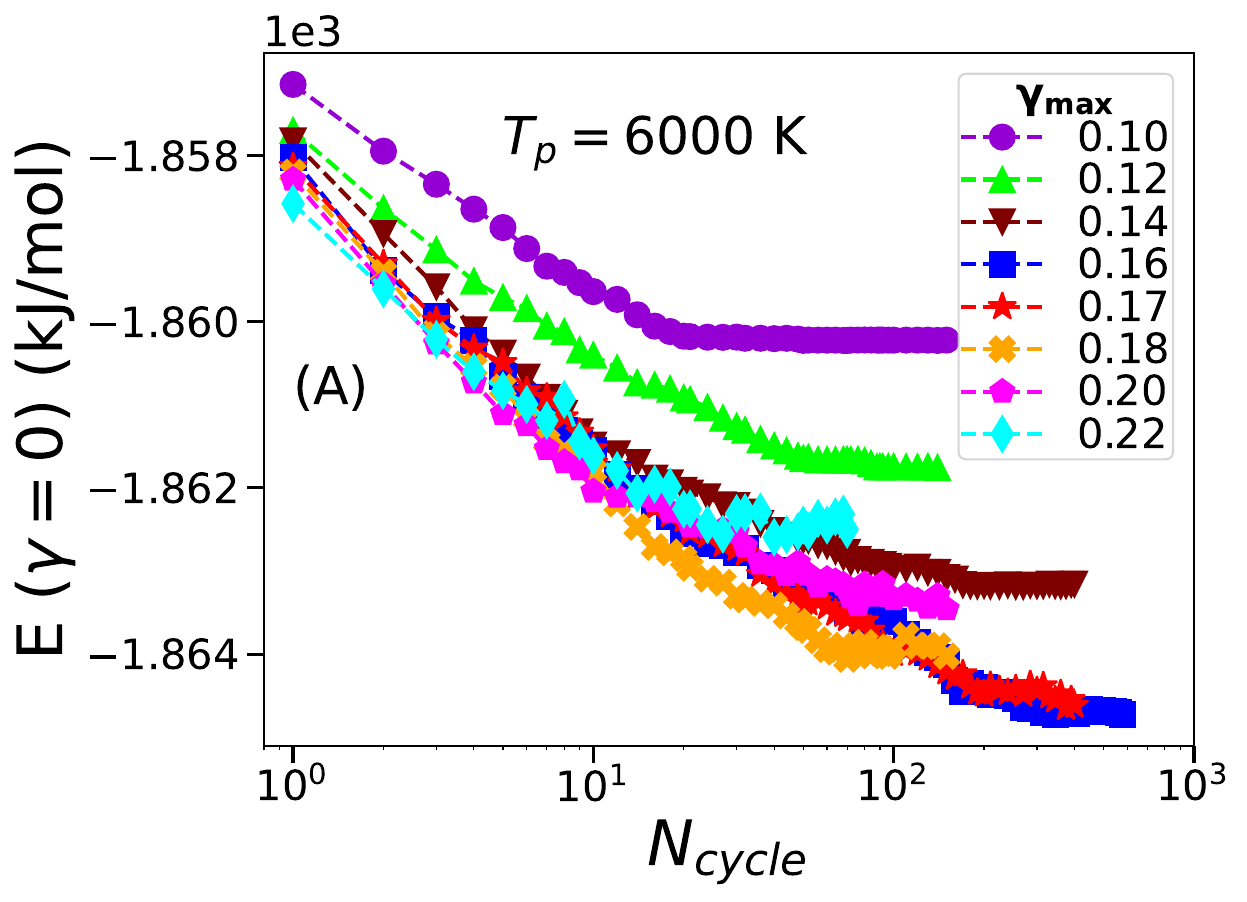}
    \includegraphics[width=0.49\textwidth]{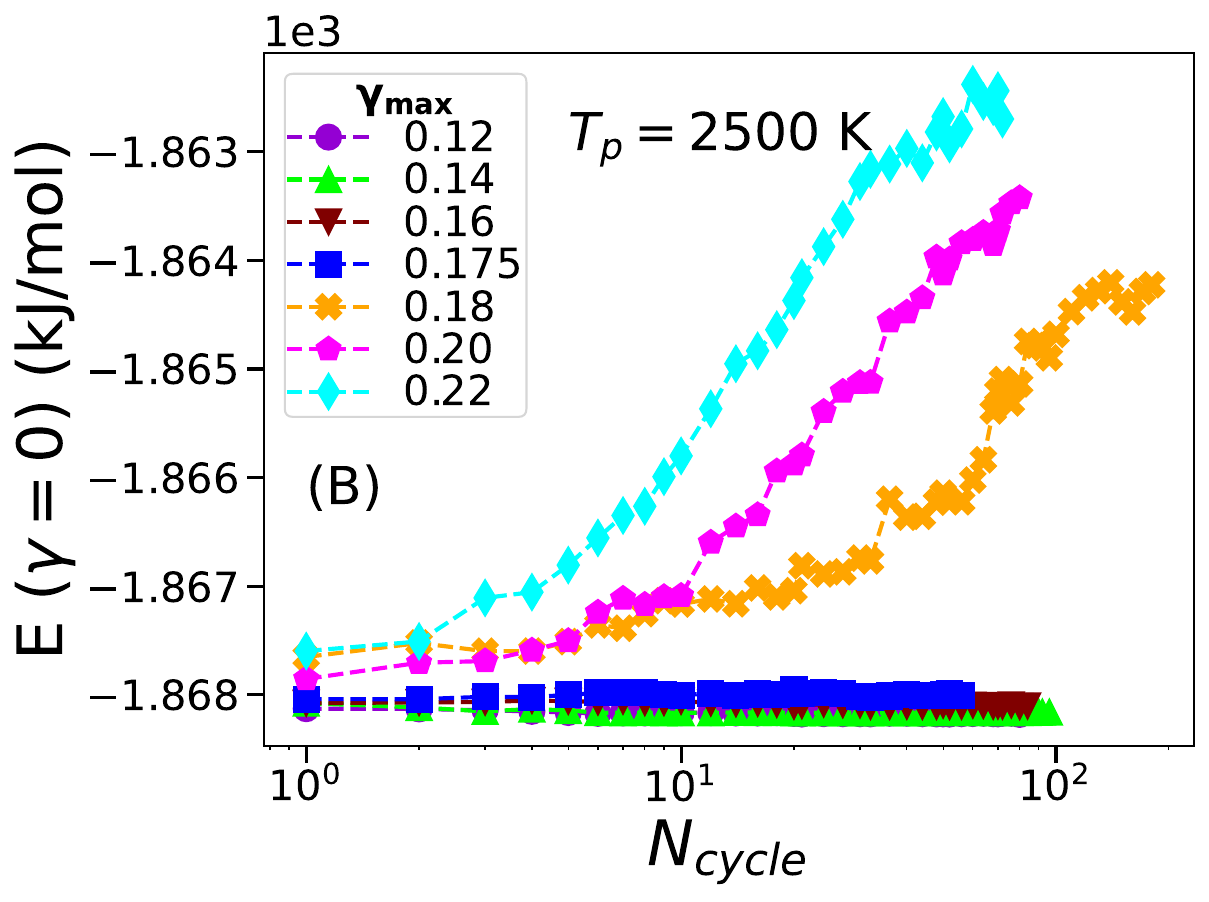}
    \caption{\textbf{Stroboscopic energy vs cycle for BKS Silica.} Variation of the stroboscopic energy, $E(\gamma = 0)$, with number of cycles for (A) poorly annealed ($T_p = 6000\,\mathrm{K}$) and (B) well-annealed ($T_p = 2500\,\mathrm{K}$) samples in the BKS silica model. The yielding strain is $\gamma_Y \approx 0.165$ for the poorly annealed sample and $\gamma_Y \approx 0.18$ for the well-annealed sample.}
    \label{SI_fig:50}
\end{figure}

\begin{figure}[H]
    \centering
    \includegraphics[width=0.49\textwidth]{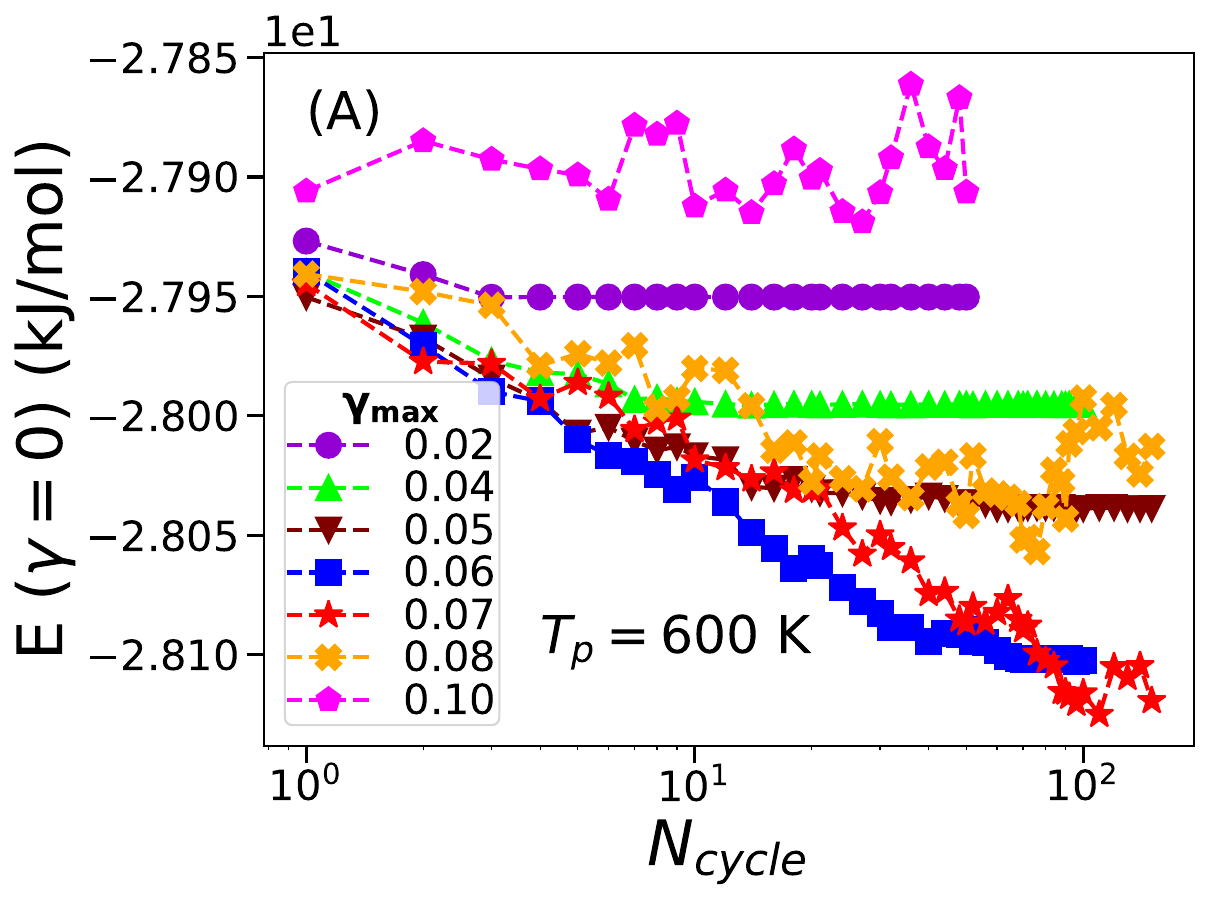}
    \includegraphics[width=0.49\textwidth]{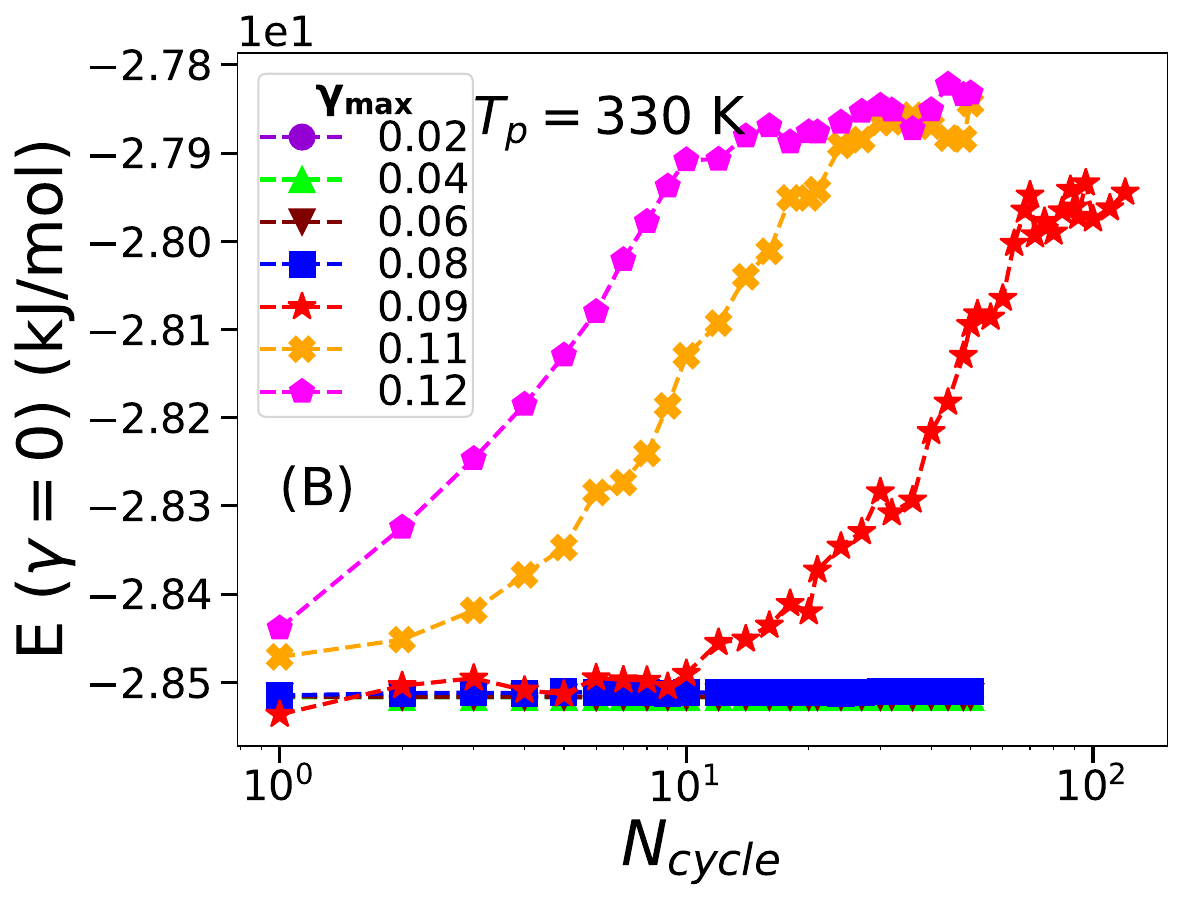}
    \caption{\textbf{Stroboscopic energy vs cycle for OTP glass.} Variation of stroboscopic energy E($\gamma = 0$) with cycles is plotted for (A) poorly annealed ($T_p = 600$K) and (B) well annealed ($T_p = 330$K) samples in the OTP model. The yielding strain is $\gamma_Y \approx 0.07$ for the poorly annealed sample and $\gamma_Y \approx 0.09$ for the well-annealed sample.}
    \label{SI_fig:51}
\end{figure}

\begin{figure}[H]
    \centering
    \includegraphics[width=0.49\textwidth]{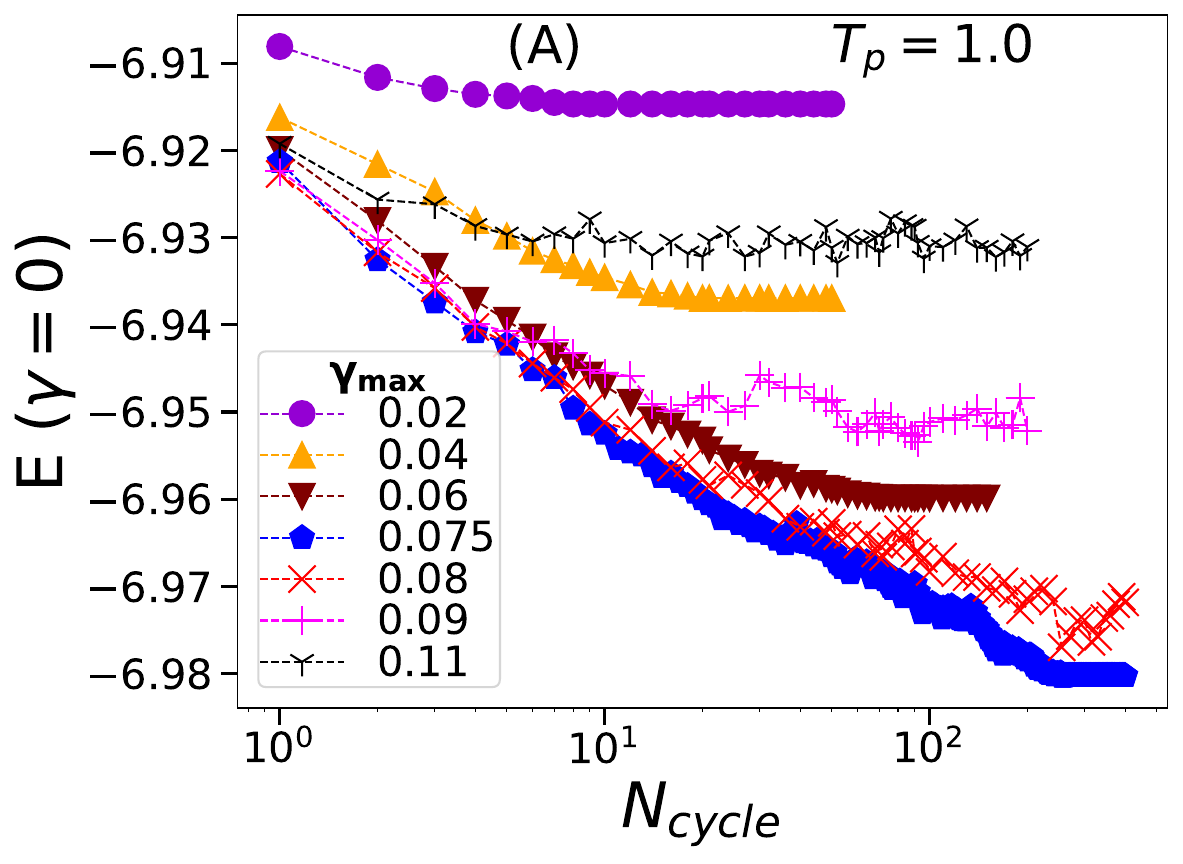}
    \includegraphics[width=0.49\textwidth]{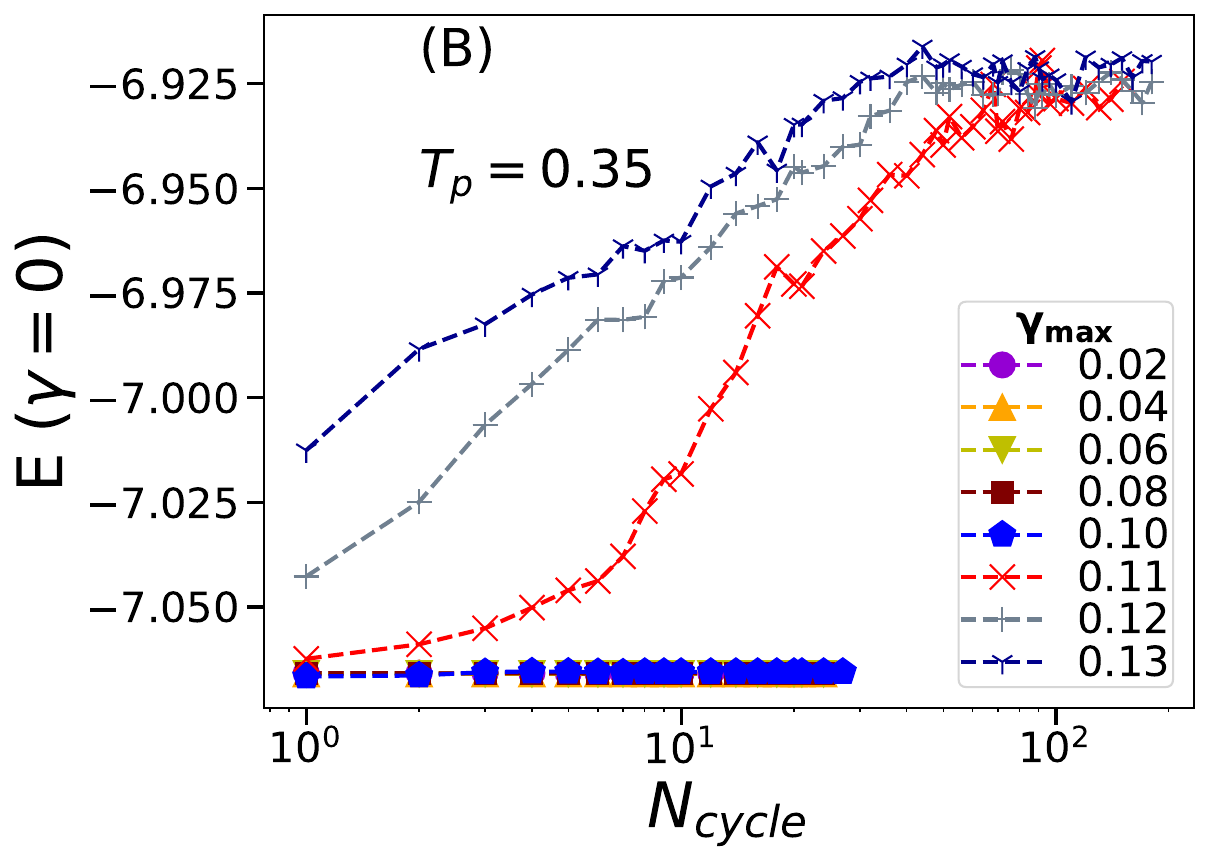}
    \caption{\textbf{Stroboscopic energy vs cycle for 3D KA.} Variation of stroboscopic energy E($\gamma = 0$) with cycles is plotted for (A) poorly annealed ($T_p = 1.0$) and (B) well annealed ($T_p = 0.35$) samples in the 3D Kob-Andersen model. The yielding strain is $\gamma_Y \approx 0.075$ for the poorly annealed sample and $\gamma_Y \approx 0.11$ for the well-annealed sample.}
    \label{SI_fig:52}
\end{figure}

\begin{figure}[H]
    \centering
    \includegraphics[width=0.49\textwidth]{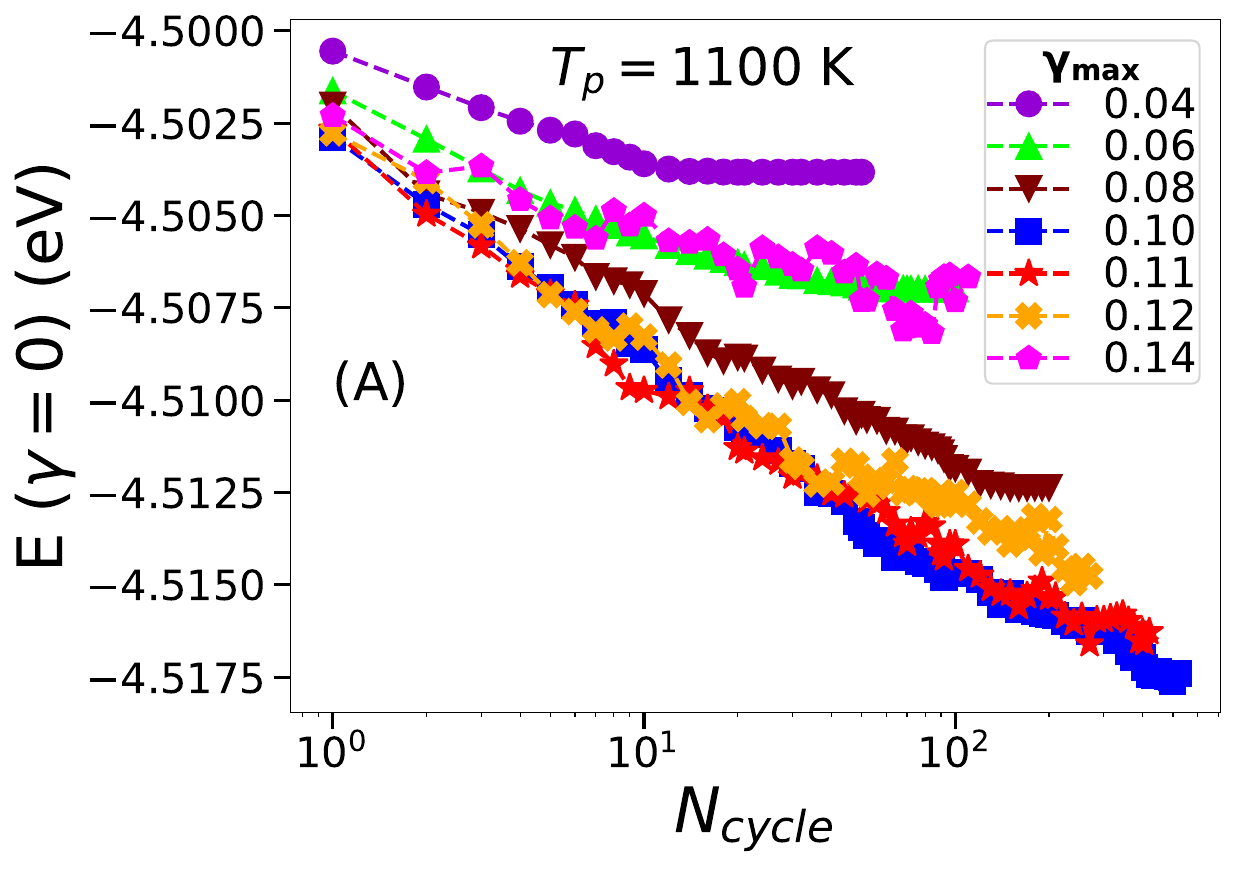}
    \includegraphics[width=0.49\textwidth]{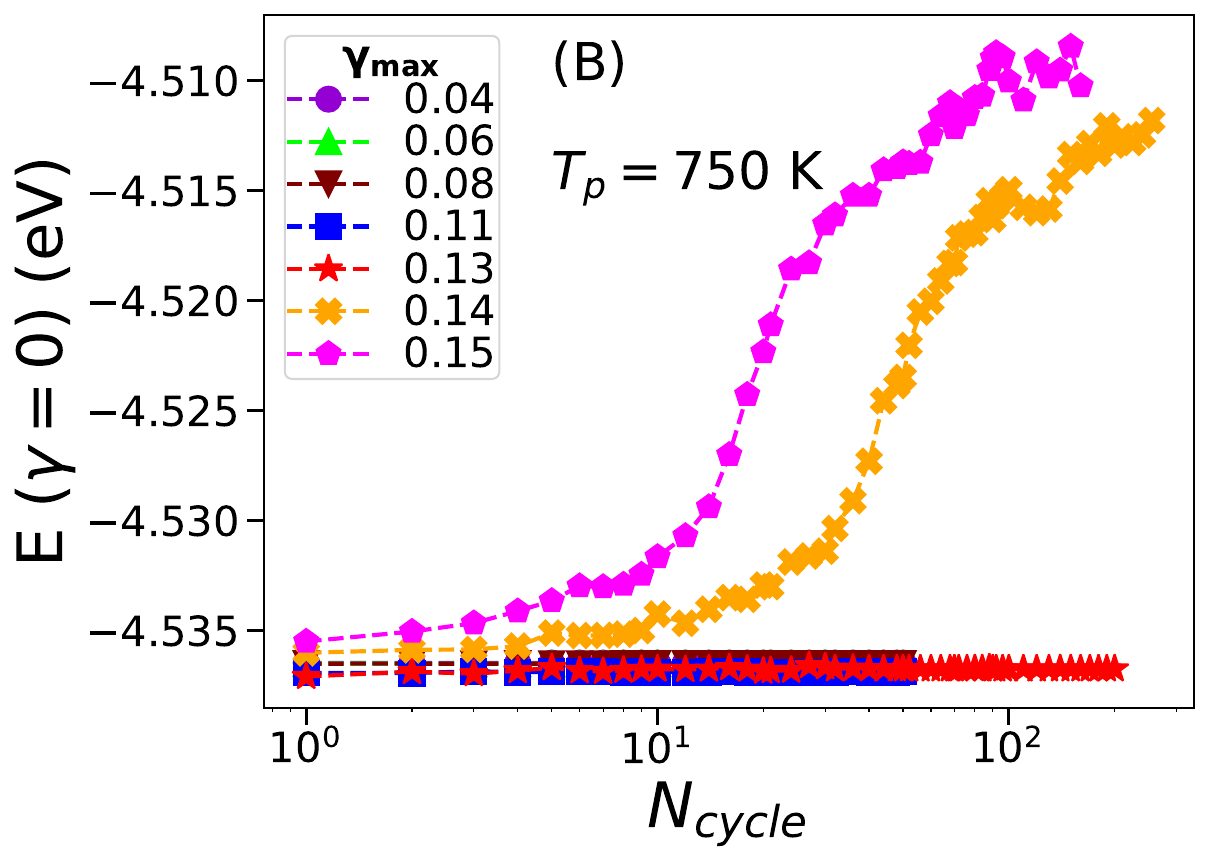}
    \caption{\textbf{Stroboscopic energy vs cycle for Cu-Zr metallic (NPT) glass.} Variation of stroboscopic energy E($\gamma = 0$) with cycles is plotted for (A) poorly annealed ($T_p = 1100$K) and (B) well annealed ($T_p = 750$K) samples in the Cu-Zr (NPT) model. The yield points of the poorly annealed and well-annealed samples are at $\gamma_Y \approx 0.105$ and $\gamma_Y \approx 0.14$, respectively.}
    \label{SI_fig:53}
\end{figure}

\begin{figure}[H]
    \centering
    \includegraphics[width=0.49\textwidth]{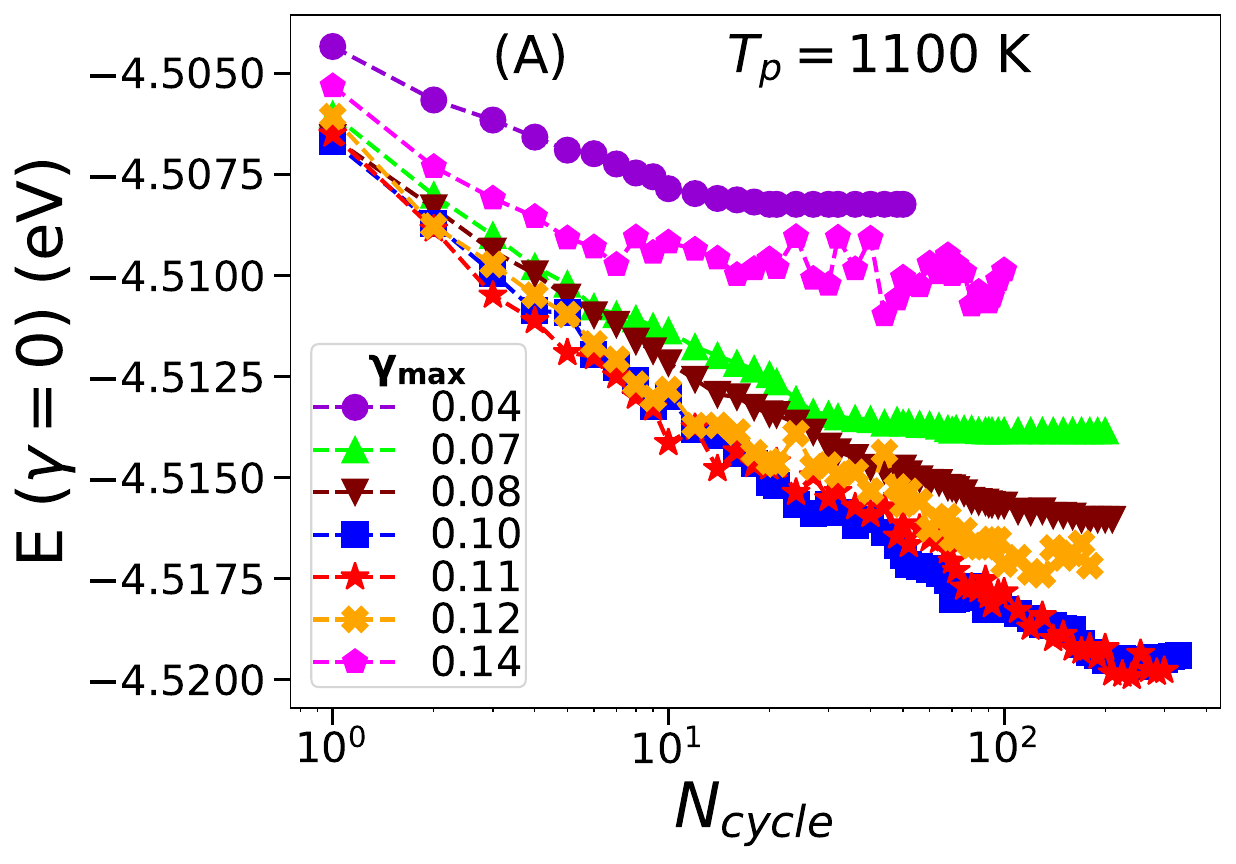}
    \includegraphics[width=0.49\textwidth]{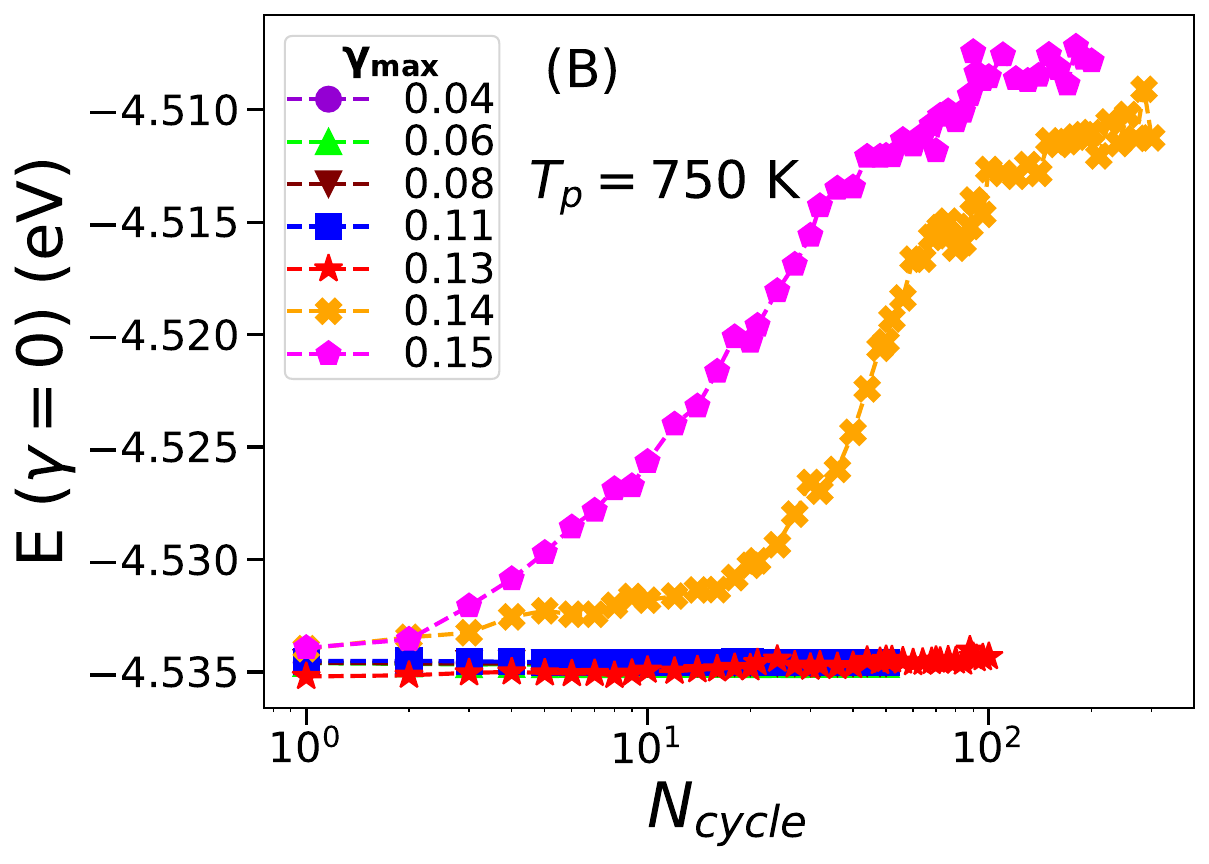}
    \caption{\textbf{Stroboscopic energy vs cycle for Cu-Zr metallic (NVT) glass.} Variation of stroboscopic energy E($\gamma = 0$) with cycles is plotted for (A) poorly annealed ($T_p = 1100$K) and (B) well annealed ($T_p = 750$K) samples in the Cu-Zr (NVT) model. The yield points of the poorly annealed and well-annealed samples are at $\gamma_Y \approx 0.105$ and $\gamma_Y \approx 0.14$, respectively.}
    \label{SI_fig:54}
\end{figure}

In the main text, we present the stroboscopic energy in the steady state as a function of different $\gamma_{max}$ values and temperatures. Here, we show the evolution of stroboscopic energy as a function of the number of cycles for different $\gamma_{max}$, focusing on two extreme cases of annealing in Fig. \ref{SI_fig:50}, Fig. \ref{SI_fig:51}, Fig. \ref{SI_fig:52}, Fig. \ref{SI_fig:53} and Fig. \ref{SI_fig:54} for BKS Silica, OTP, 3DKA, Cu-Zr (NPT) and Cu-Zr (NVT) respectively. The data clearly demonstrate that we performed a sufficient number of shear cycles for the system to reach steady states, where the energy either remains constant (absorbing states) or fluctuates around a mean value (diffusive states). 

\newpage

\section{Uniform shear response of four different models}
\begin{figure}[H]
    \centering
    \includegraphics[width=0.38\textwidth]{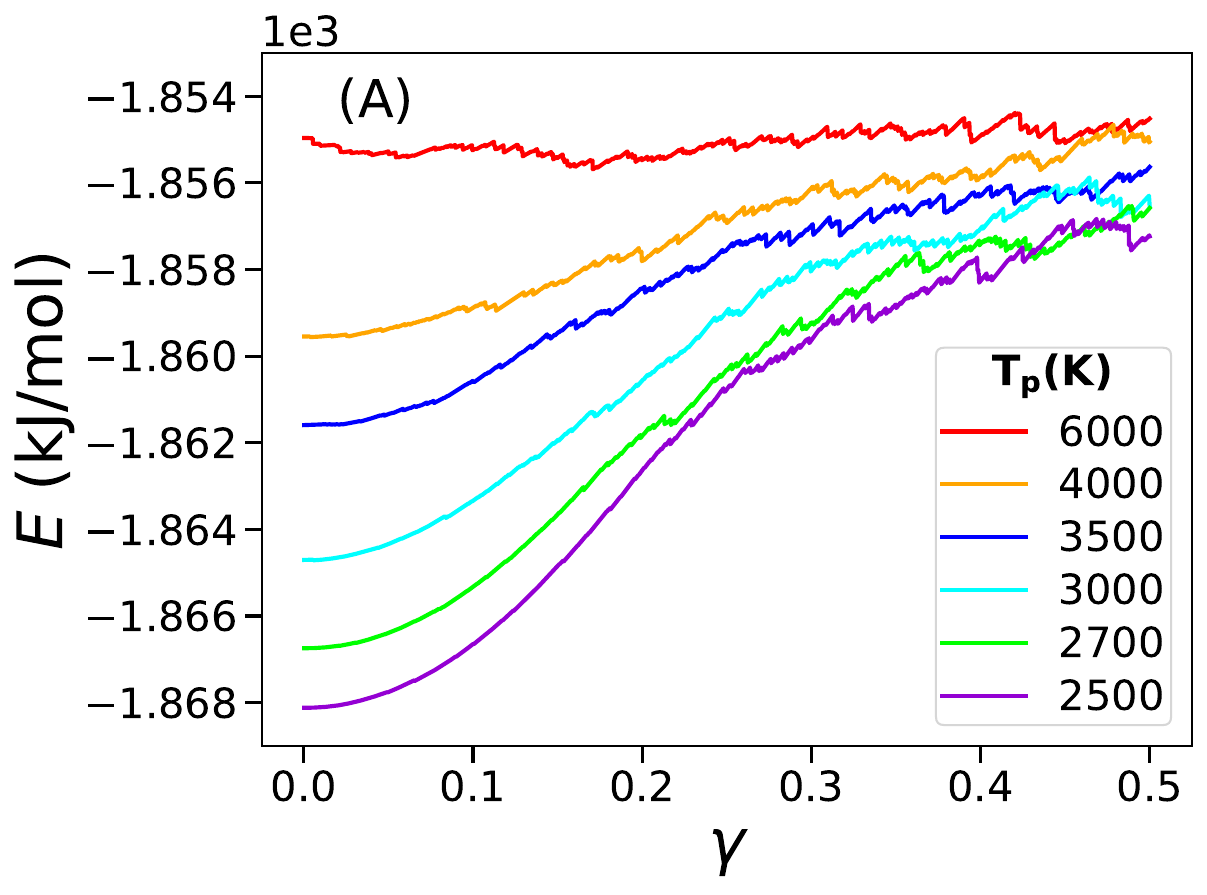}
    \includegraphics[width=0.36\textwidth]{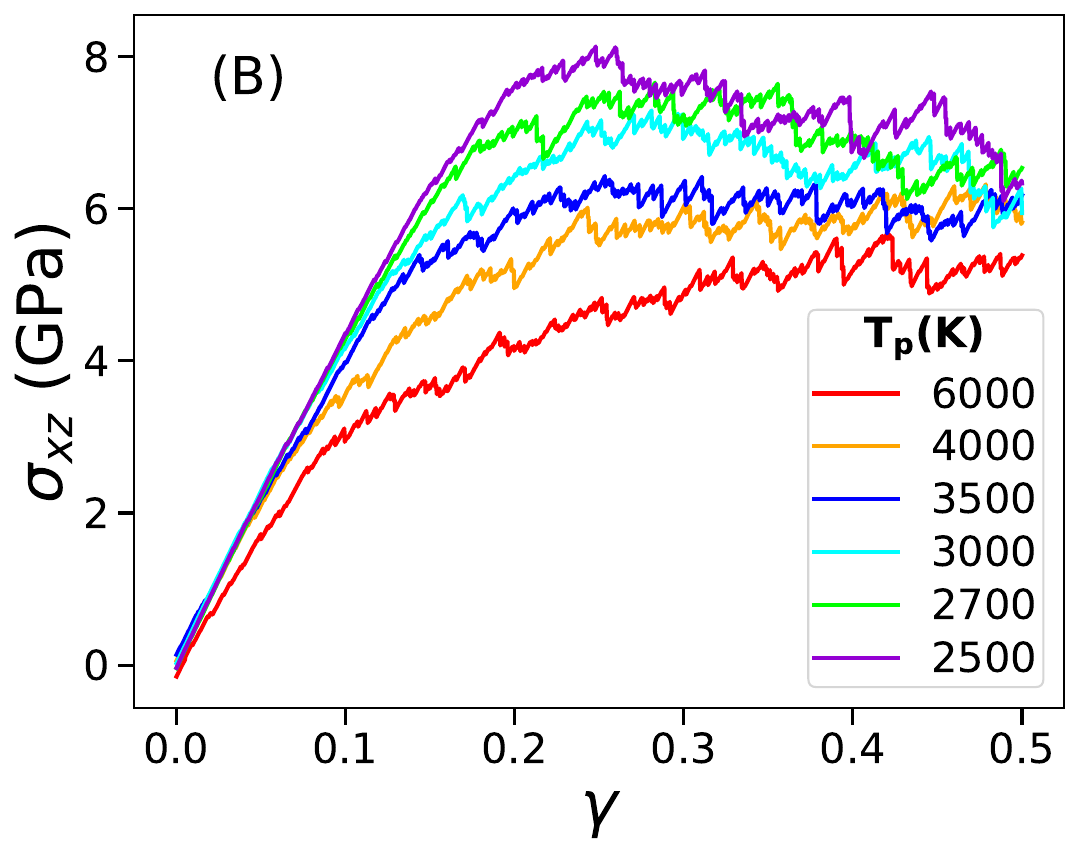}
    \caption{\textbf{Uniform shear response of BKS Silica.} Variation of (A) energy E and (B) stress $\sigma_{xz}$ with strain $\gamma$ is plotted for uniform shear in BKS Silica for different parent temperatures as indicated in the legend. Data sets are averaged over 8 independent samples. System size $N = 4800$.}
    \label{SI_fig:17}
\end{figure}

\begin{figure}[H]
    \centering
    \includegraphics[width=0.38\textwidth]{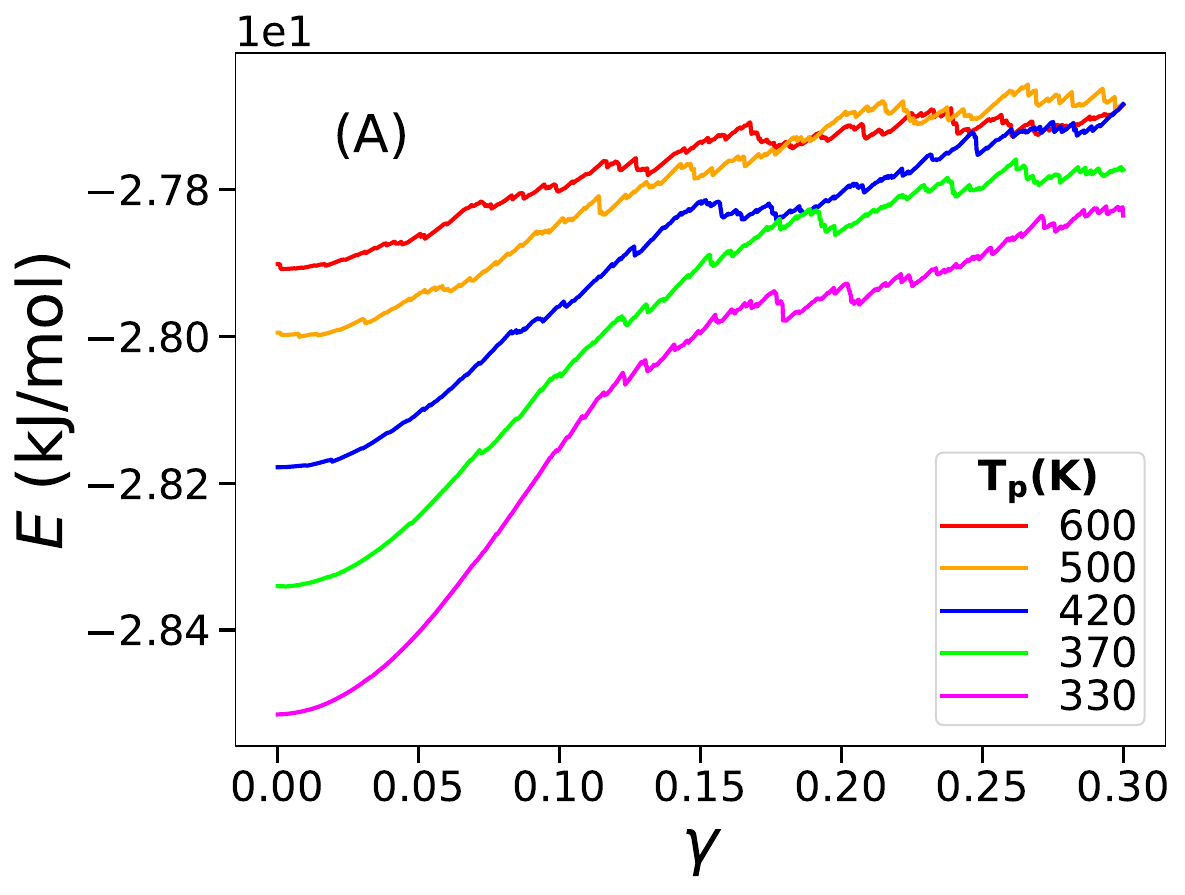}
    \includegraphics[width=0.36\textwidth]{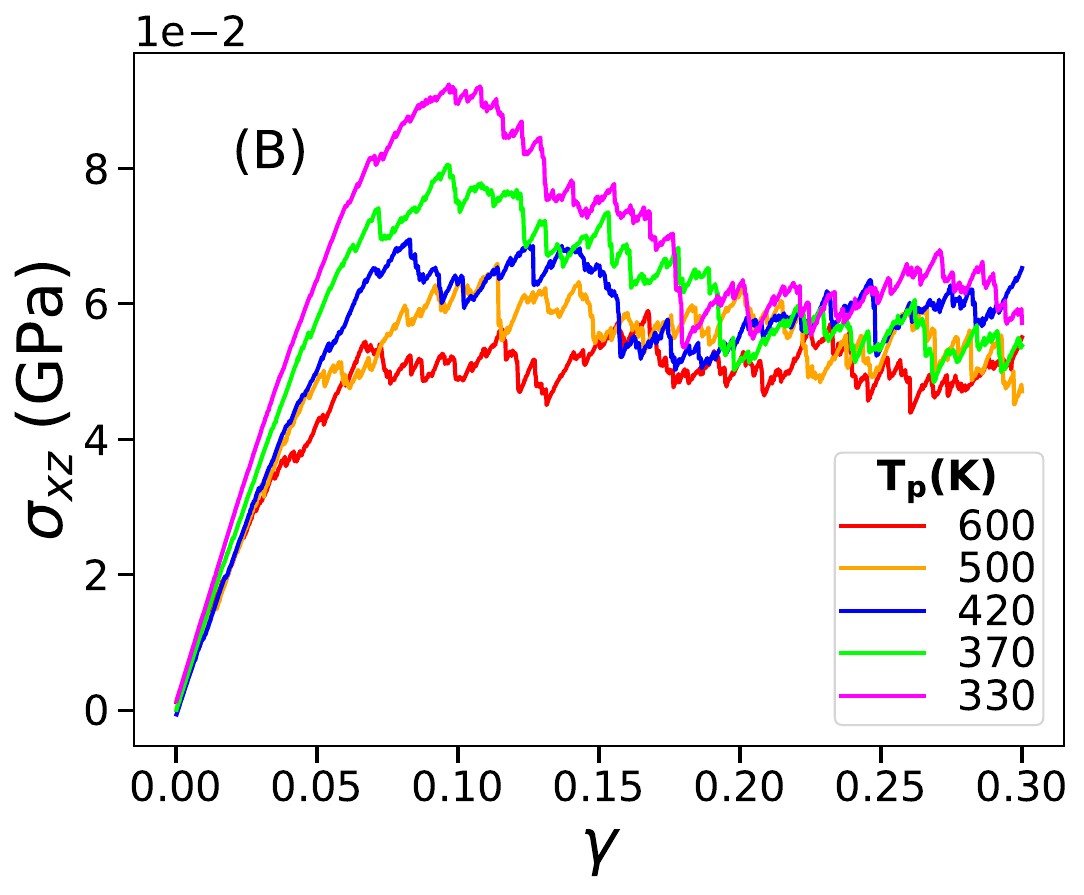}
    \caption{\textbf{Uniform shear response of OTP glass.} Variation of (A) energy, E and (B) stress, $\sigma_{xz}$, with strain $\gamma$ is plotted for uniform shear in OTP for different parent temperatures as indicated in the legend. Data sets are averaged over 8 independent samples. System size $N = 4800$.}
    \label{SI_fig:18}
\end{figure}

\begin{figure}[H]
    \centering
    \includegraphics[width=0.38\textwidth]{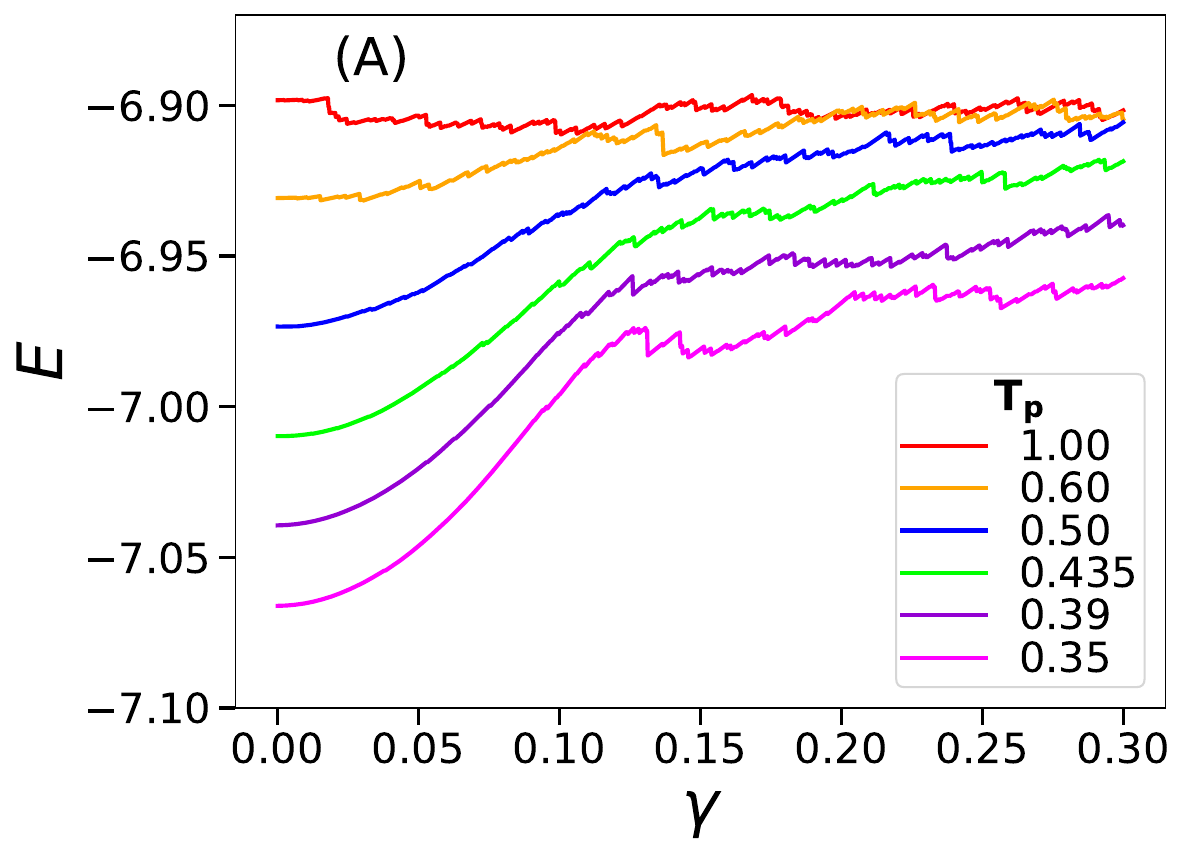}
    \includegraphics[width=0.36\textwidth]{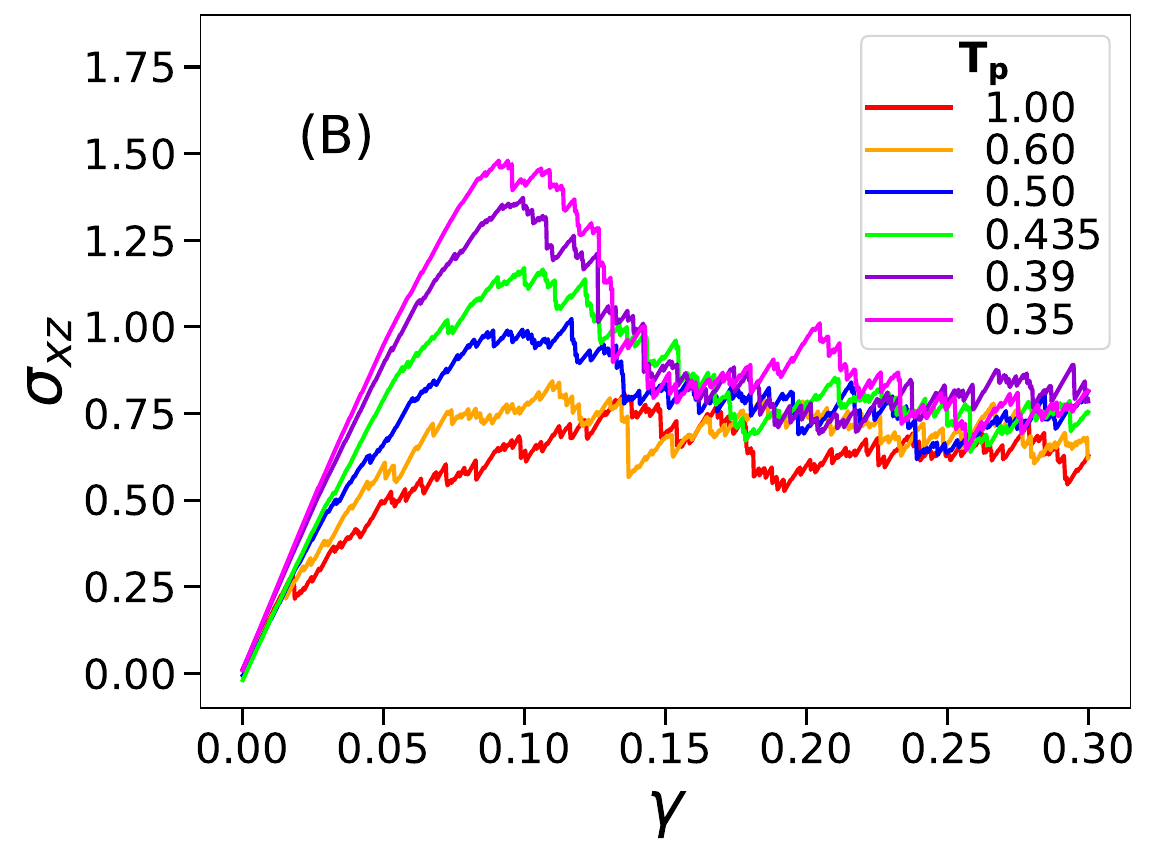}
    \caption{\textbf{Uniform shear response of 3D KA.} Variation of (A) energy E and (B) stress $\sigma_{xz}$ with strain $\gamma$ is plotted for uniform shear in 3D KA for different parent temperatures as indicated in the legend. Data sets are averaged over 8 independent samples. System size $N = 5000$.}
    \label{SI_fig:19}
\end{figure}

\begin{figure}[H]
    \centering
    \includegraphics[width=0.39\textwidth]{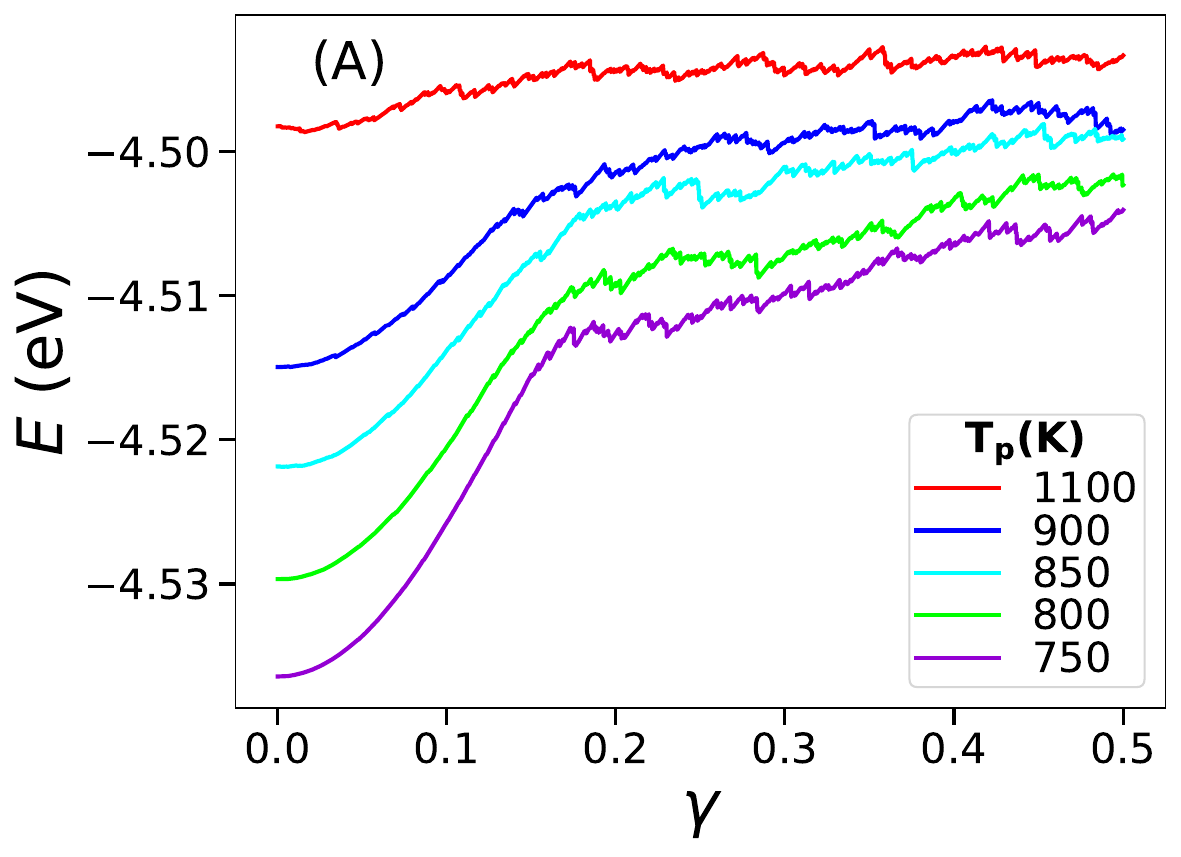}
    \includegraphics[width=0.37\textwidth]{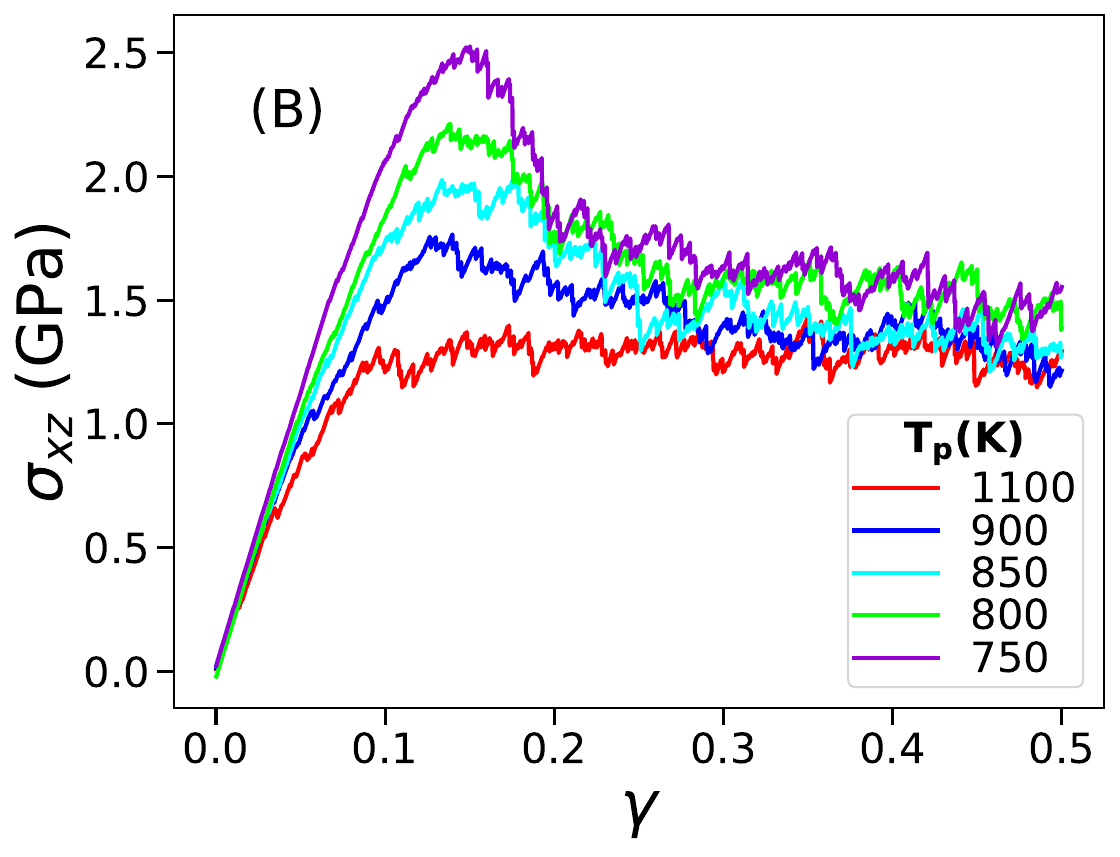}
    \caption{\textbf{Uniform shear response of Cu-Zr metallic glass.} Variation of (A) energy E and (B) stress $\sigma_{xz}$ with strain $\gamma$ is plotted for uniform shear in Cu-Zr for different parent temperatures as indicated in the legend. Data sets are averaged over 8 independent samples. System size $N = 5000$.}
    \label{SI_fig:20}
\end{figure}

For completeness, we also show the response of the four different model systems under uniform shear in Fig. \ref{SI_fig:17}, Fig. \ref{SI_fig:18}, Fig. \ref{SI_fig:19}, and Fig. \ref{SI_fig:20}, corresponding to silica, OTP, 3D KA and Cu-Zr glasses, respectively. We plot both the energy and stress as functions of strain for differently annealed samples. Similar to the behavior observed at different densities in the 3D HP model, well-annealed glasses exhibit stress overshoot, which becomes more pronounced with increasing fragility. In contrast, for poorly annealed glasses, the stress evolves continuously from the elastic branch to the steady state without overshoot.


\newpage
\section{Energy Barrier}
\subsection{3D HP: different density}
\begin{figure}[H]
    \centering
    \includegraphics[width=0.44\textwidth]{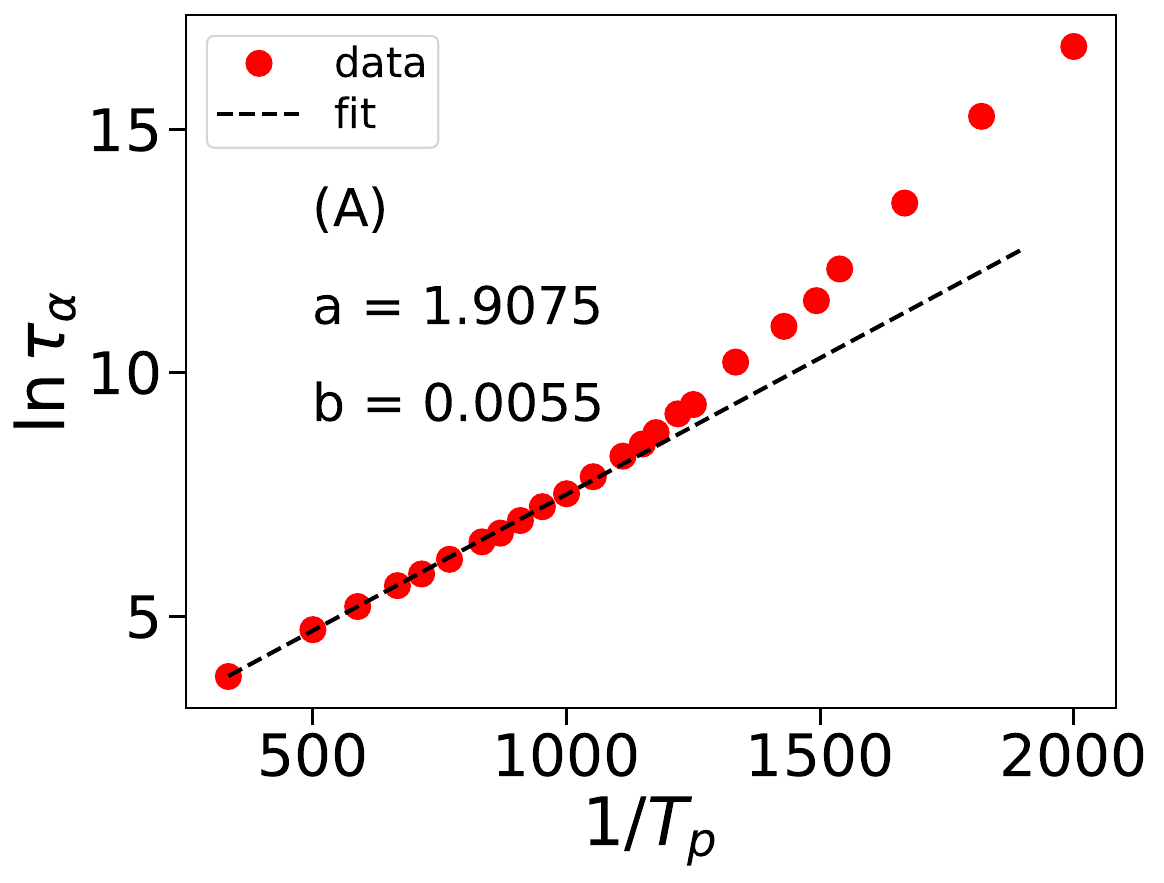}
    \includegraphics[width=0.44\textwidth]{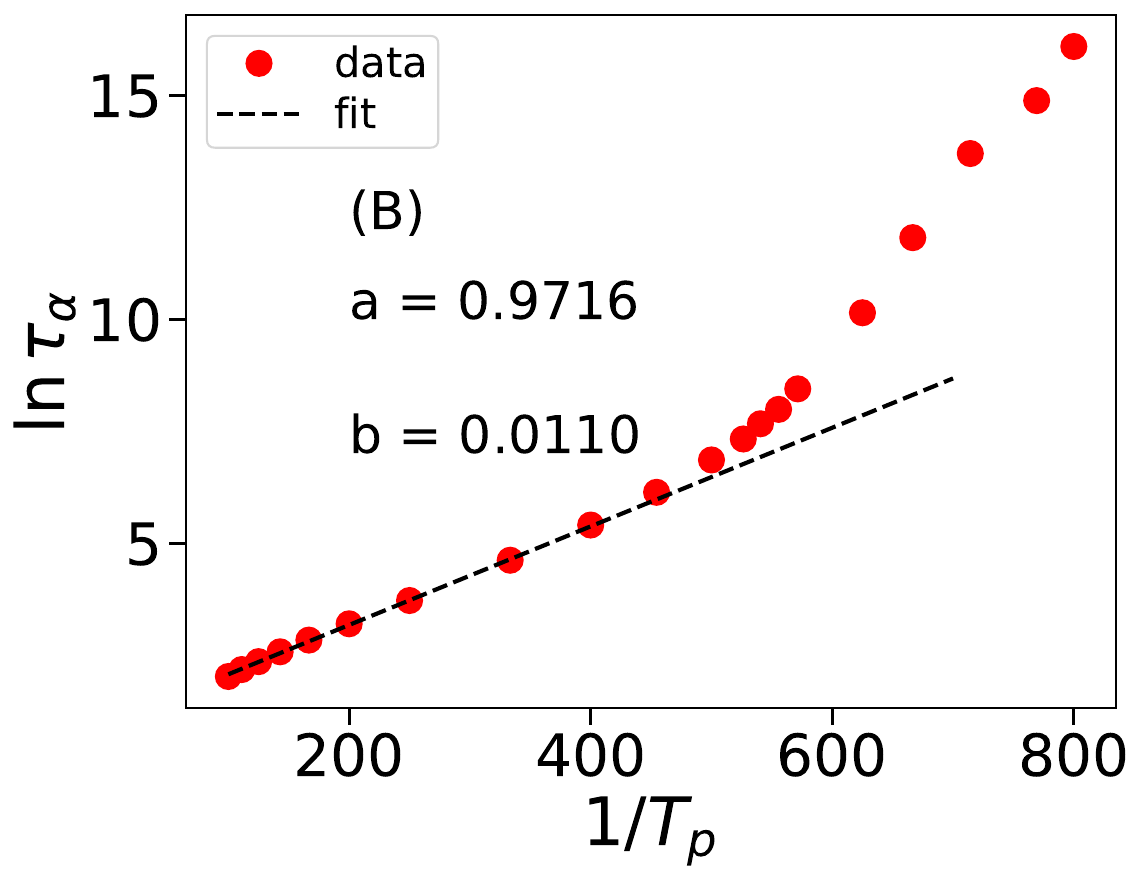}
    \includegraphics[width=0.44\textwidth]{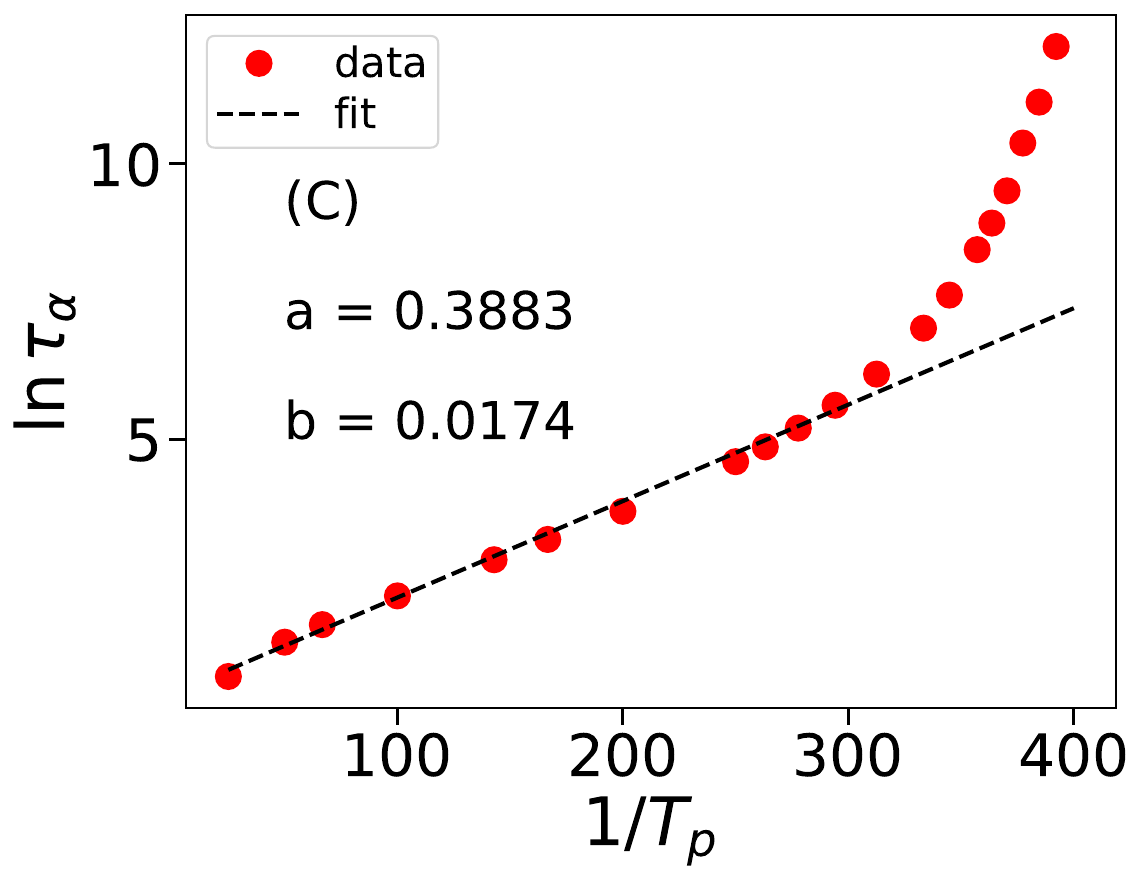}
    \includegraphics[width=0.44\textwidth]{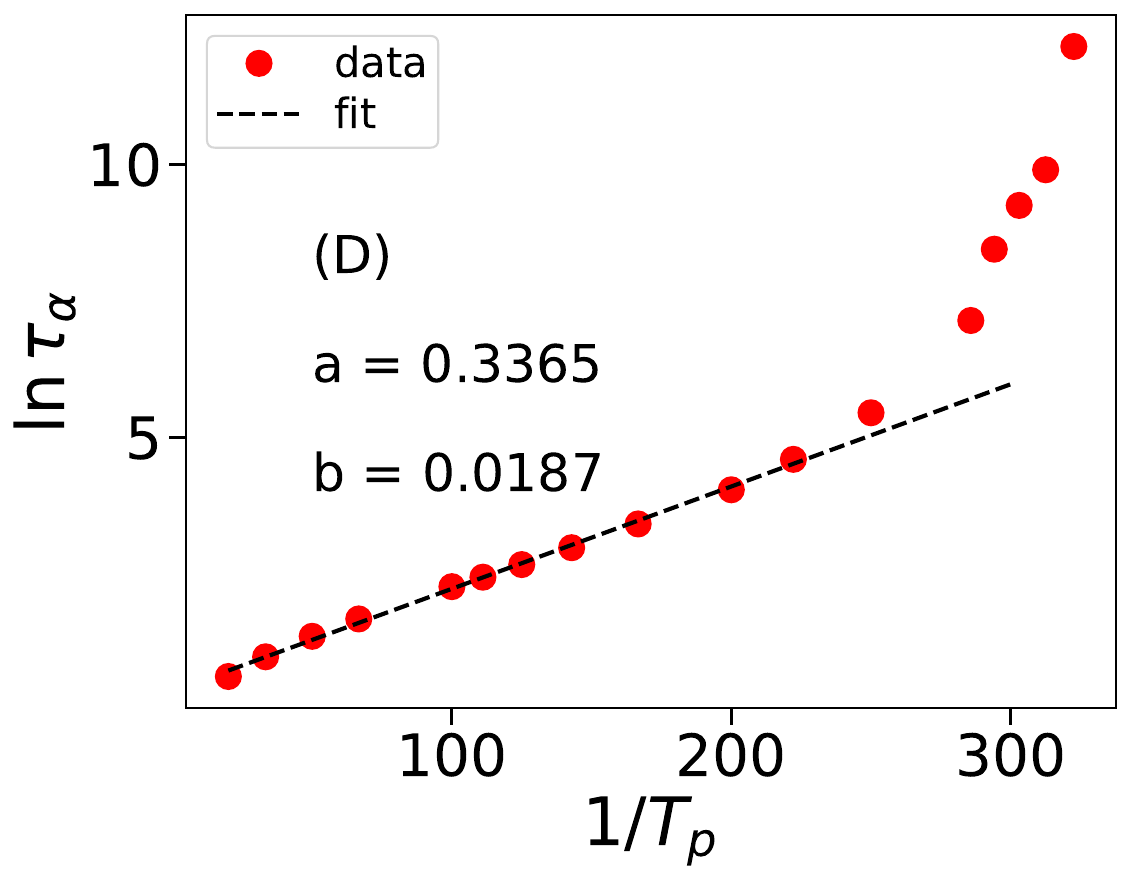}
    \caption{\textbf{Energy barrier for 3D HP model.} Relaxation time $\tau_\alpha$ is plotted with $1/T_p$ for very high parent temperatures $T_p$ at (A) $\rho = 0.681$, (B) $\rho = 0.750$, (C) $\rho = 0.855$ and (D) $\rho = 0.943$. $\tau_\alpha$ is fitted via equation : $ln \tau_\alpha = a + \frac{b}{T_p}$, where $a = ln \tau_0$ and $b = \frac{\Delta E_{\infty}}{k_B}$. Fitting parameters are mentioned in the plots.}
    \label{SI_fig:23A}
\end{figure}
\begin{figure}[H]
    \centering
    \includegraphics[width=0.49\textwidth,height=0.35\textwidth]{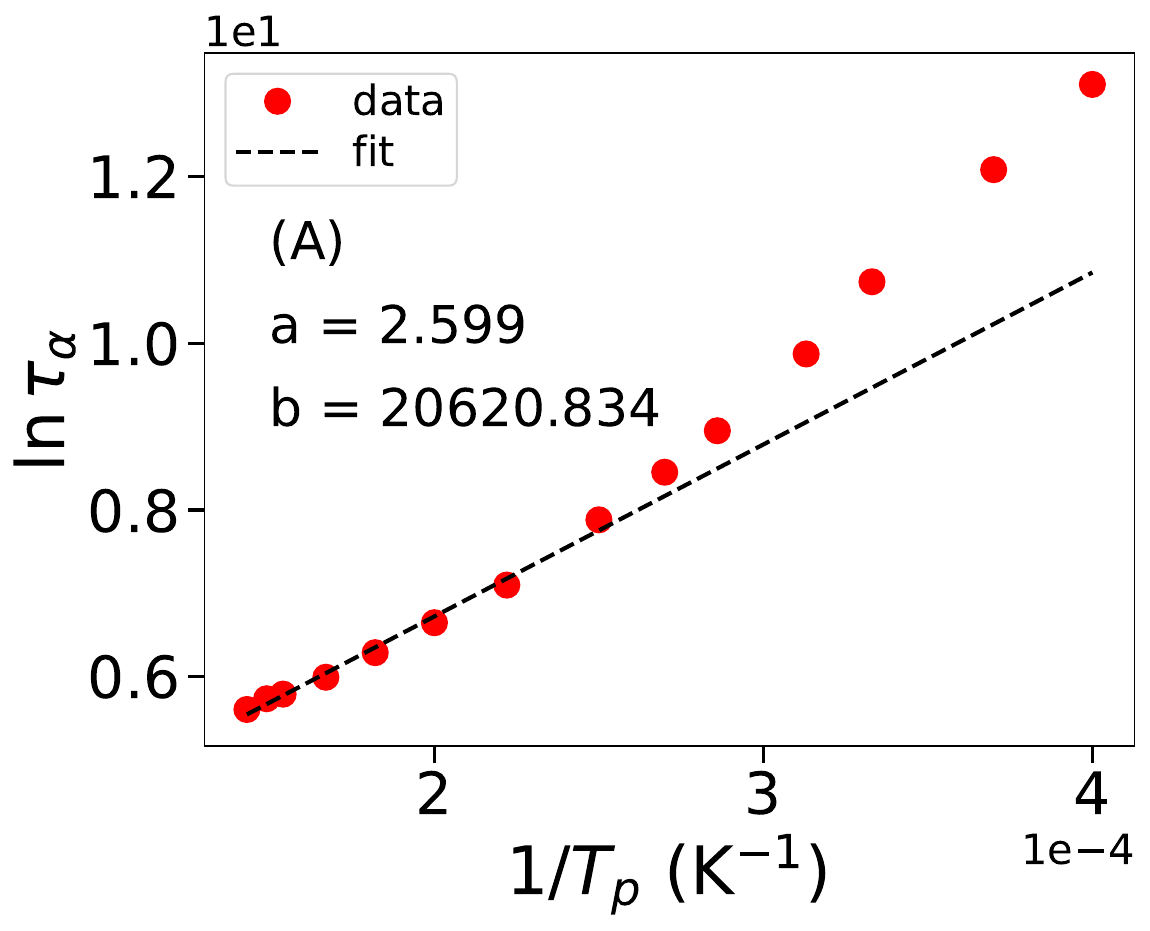}
    \includegraphics[width=0.49\textwidth,height=0.35\textwidth]{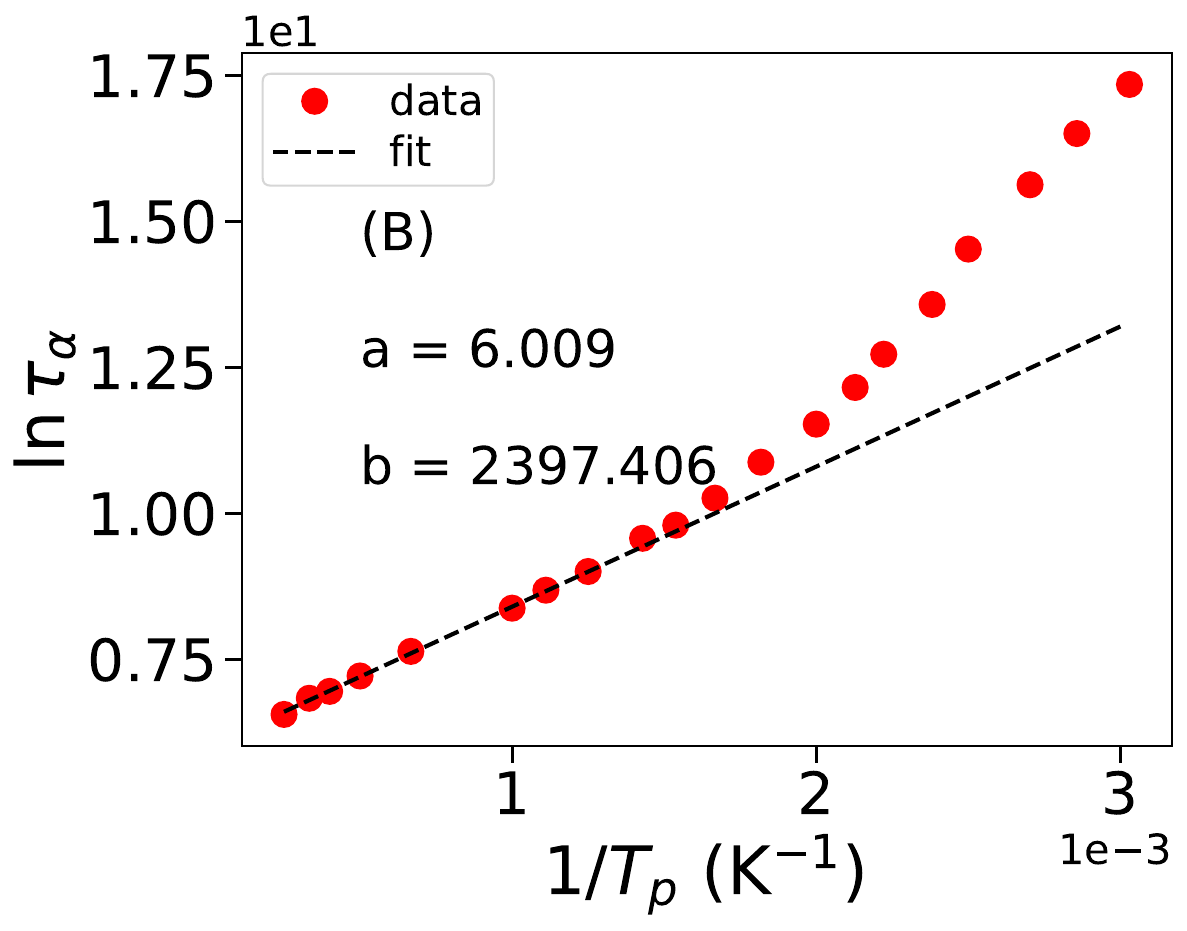}
    \includegraphics[width=0.49\textwidth,height=0.35\textwidth]{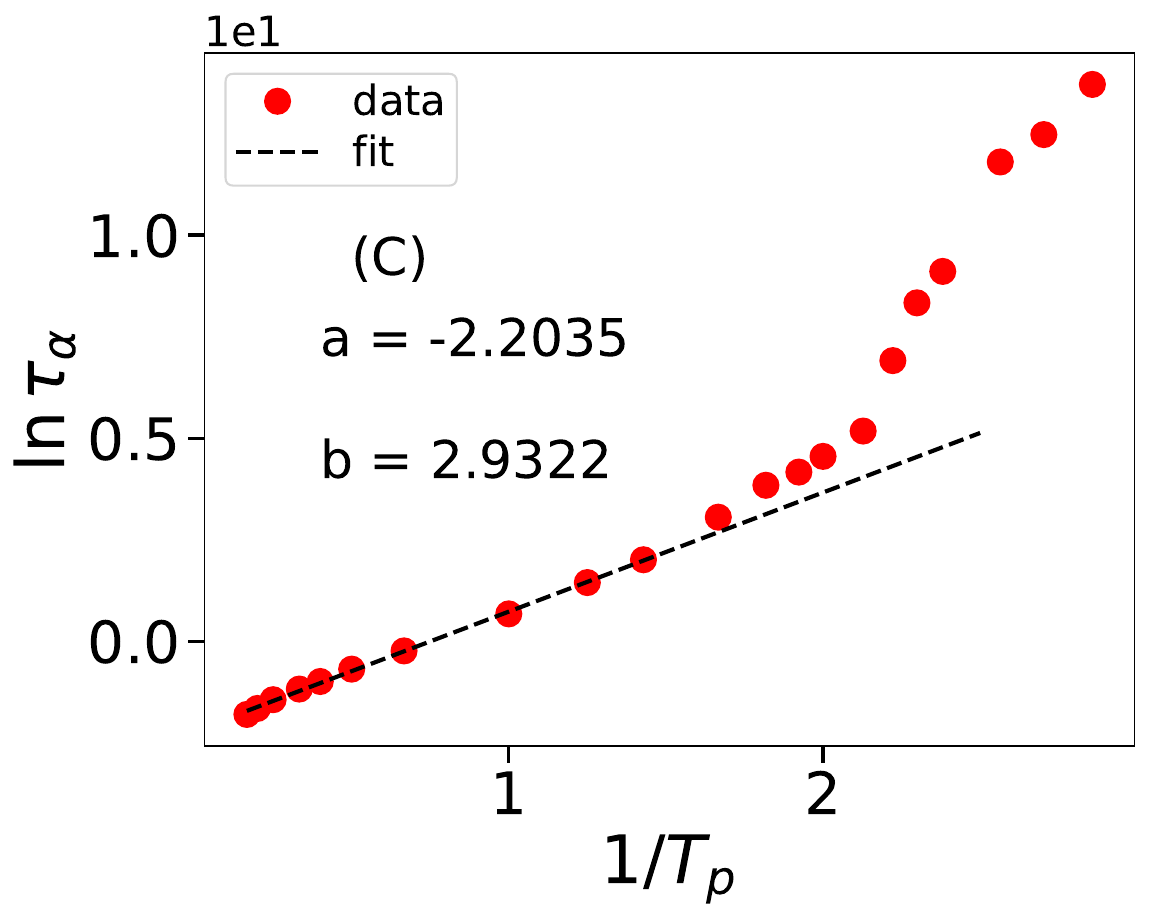}
    \includegraphics[width=0.49\textwidth,height=0.35\textwidth]{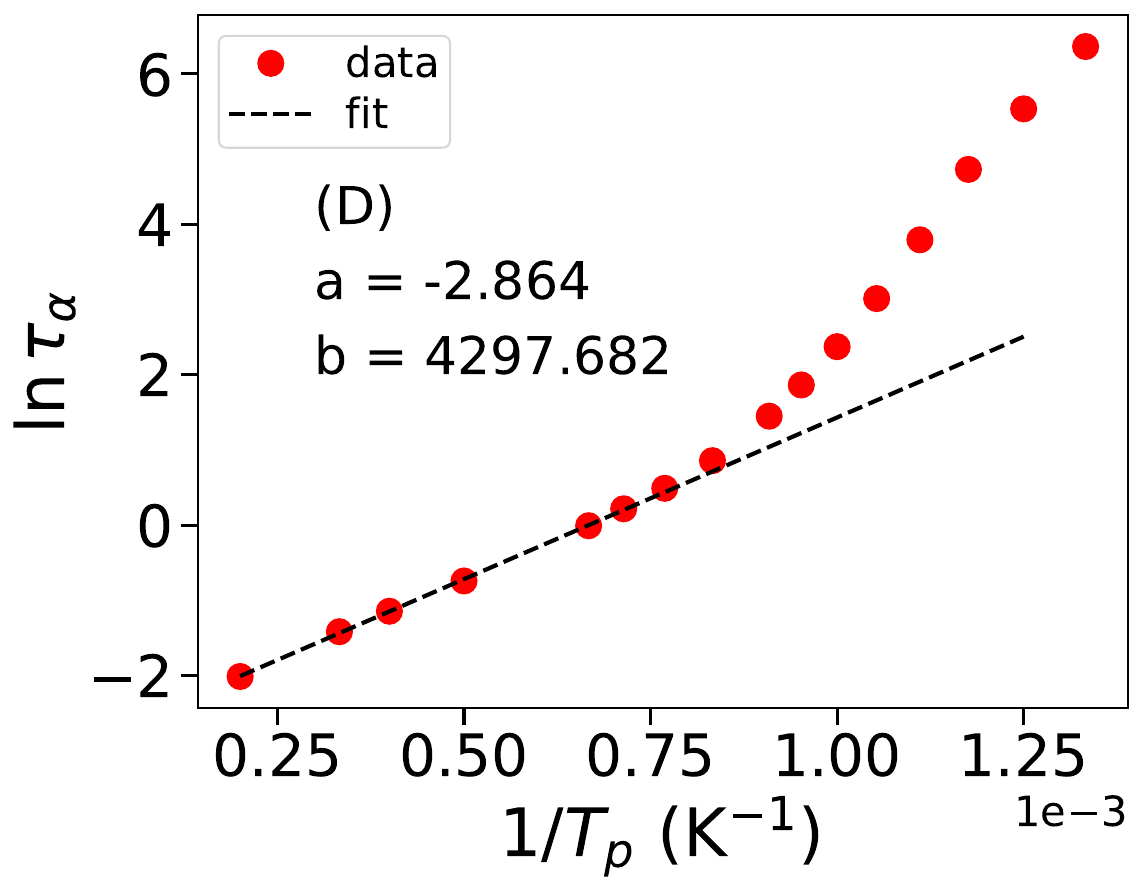}
    \caption{\textbf{Energy barrier for microscopically different glass models.} Relaxation time $\tau_\alpha$ is plotted with $1/T_p$ for very high parent temperatures $T_p$ at (A) SiO$_2$, (B) OTP, (C) 3D KA and (D) Cu-Zr (NPT) . $\tau_\alpha$ is fitted via equation : $ln \tau_\alpha = a + \frac{b}{T_p}$, where $a = ln \tau_0$ and $b = \frac{\Delta E}{k_B}$. Fitting parameters are mentioned in the plots.}
    \label{SI_fig:24}
\end{figure}
\begin{figure}[H]
    \centering
    \includegraphics[width=0.49\textwidth]{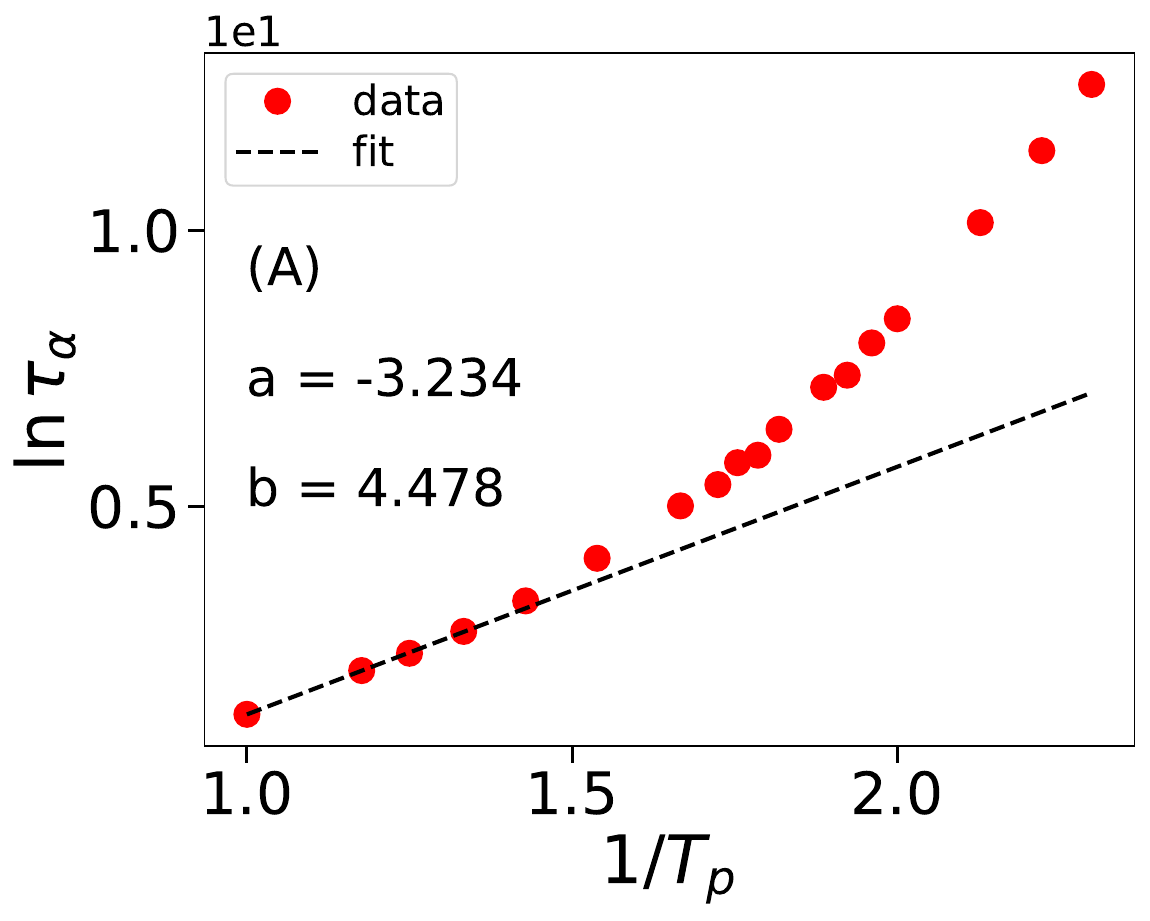}
    \includegraphics[width=0.49\textwidth]{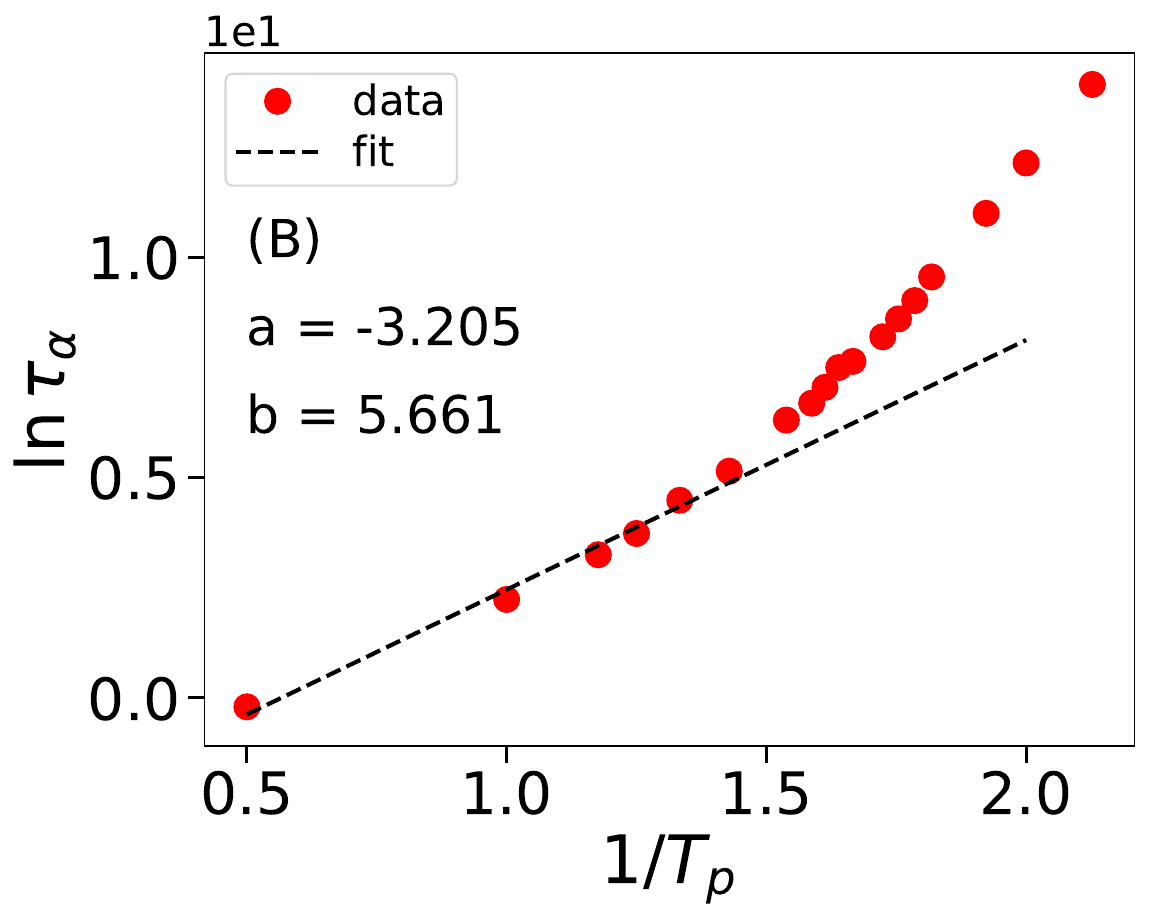}
    \includegraphics[width=0.49\textwidth]{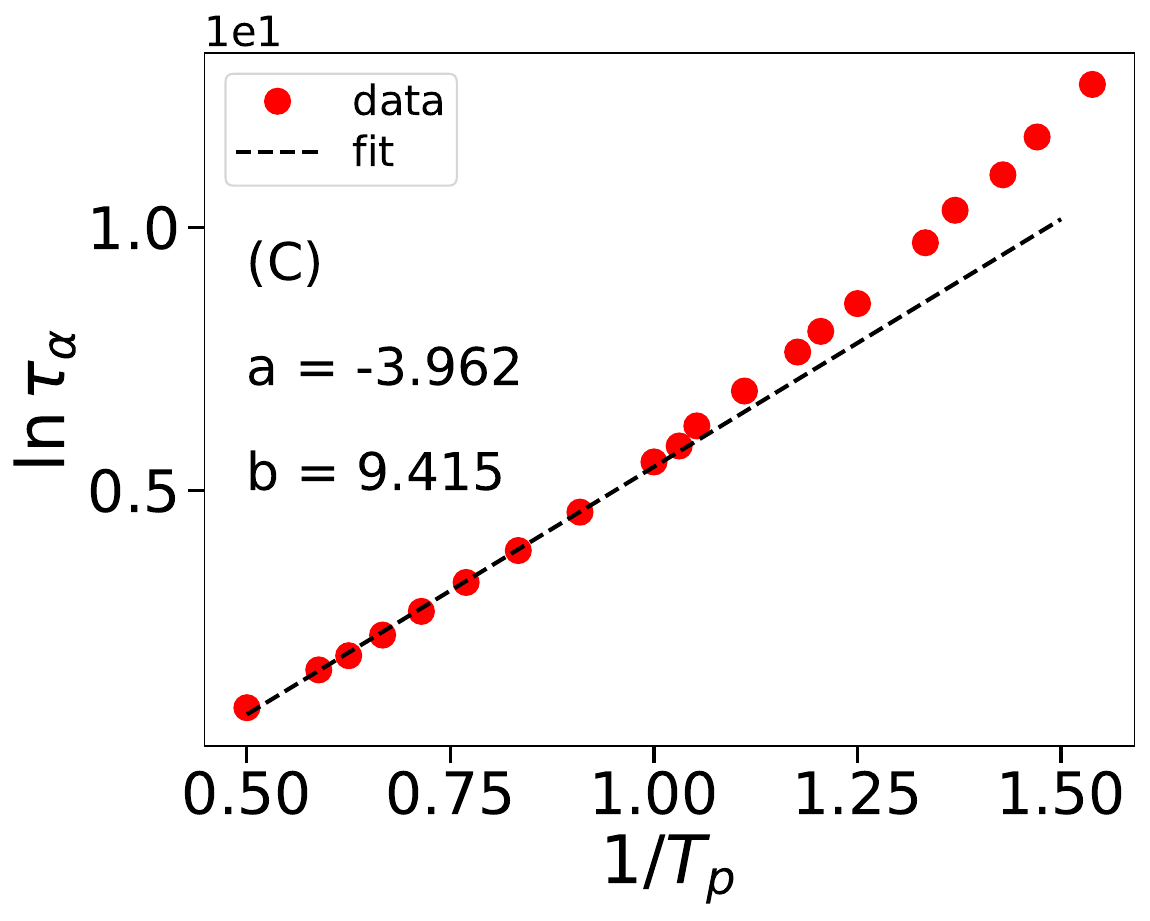}
    \includegraphics[width=0.49\textwidth]{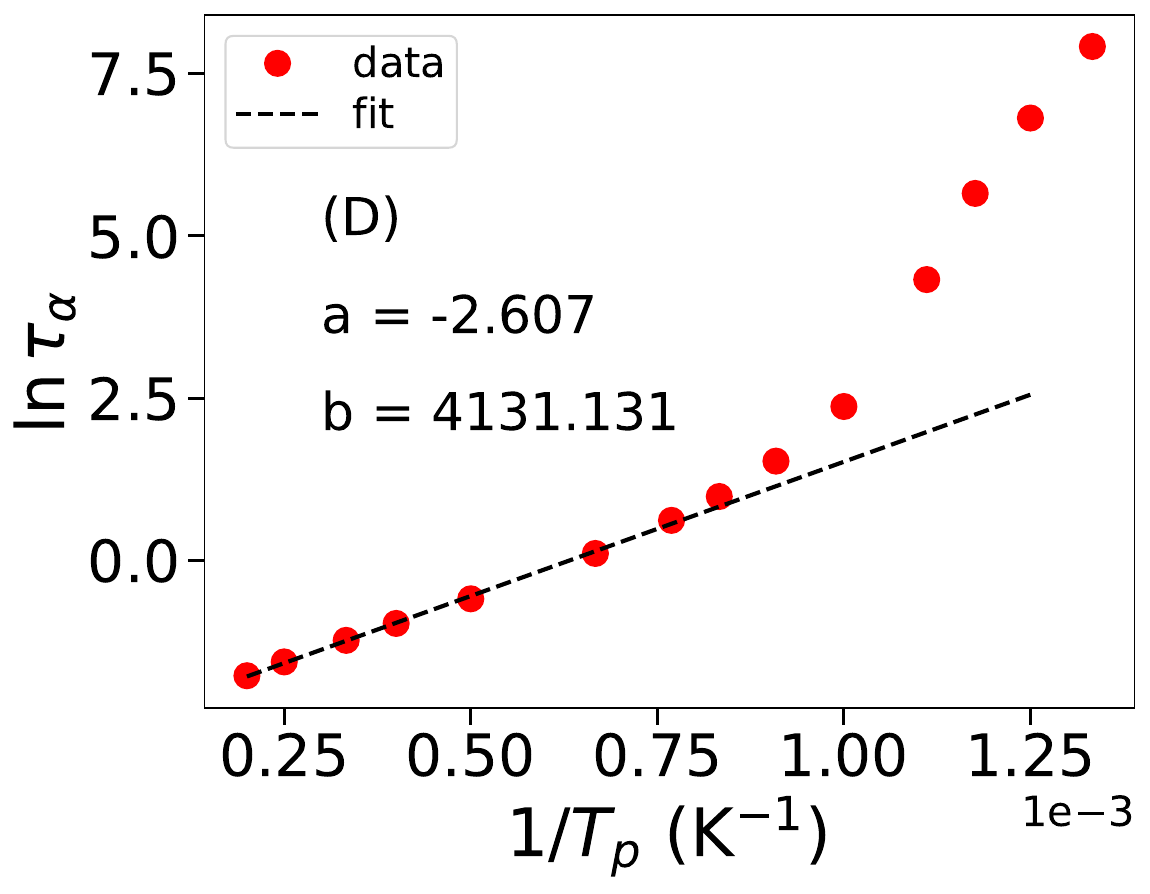}
    \caption{\textbf{Energy barrier for pinned systems.} Relaxation time $\tau_\alpha$ is plotted with $1/T_p$ for very high parent temperatures $T_p$ at (A) $\rho_{pin} = 0.05$, (B) $\rho_{pin} = 0.10$, (C) $\rho_{pin} = 0.20$ and (D) Cu-Zr (NVT) system. $\tau_\alpha$ is fitted via equation : $ln \tau_\alpha = a + \frac{b}{T_p}$, where $a = ln \tau_0$ and $b = \frac{\Delta E}{k_B}$. Fitting parameters are mentioned in the plots. The pinning data is taken from the ref [61] of the main manuscript.}
    \label{SI_fig:23}
\end{figure}

As discussed in the main text, previous studies have attempted to connect fragility to the dynamical properties of liquids. In this study, we demonstrate a connection between fragility and yielding behavior, which raises the interesting question of how yielding is linked to other dynamical quantities of the liquid. Our results from the elastoplastic model indicate that the energy barrier plays an important role in yielding. This naturally leads to the question: Is there a connection between yielding and the energy barrier in our simulation models as well? We indeed find such a connection. Specifically, we observe a correlation between the high-temperature energy barrier and the yielding point ($\gamma_c$) for poorly annealed glasses. In the figures Fig. \ref{SI_fig:23A}, we show how the energy barrier is extracted for different densities in the 3D HP model, in Fig. \ref{SI_fig:24} for the other four models and in Fig. \ref{SI_fig:23} for 3D KA with pinning (this data is taken from ref. [61] of the main text). Arrhenius fits ($\tau_{\alpha} = \tau_0 \exp (\Delta E_{\infty}/k_BT) $) of the relaxation time using high-temperature data are used to obtain the corresponding high-temperature energy barriers ($\Delta E_{\infty}$).
\subsection{Energy Barrier for different models}
\newpage
\subsection{Energy Barrier for 3DKA Model with Random Pinning}

\section{Energy Barrier controlling critical strain $\gamma_c$:  Elastoplastic Model}
\begin{figure}[H]
    \centering
    \includegraphics[width=0.49\textwidth,height=0.35\textwidth]{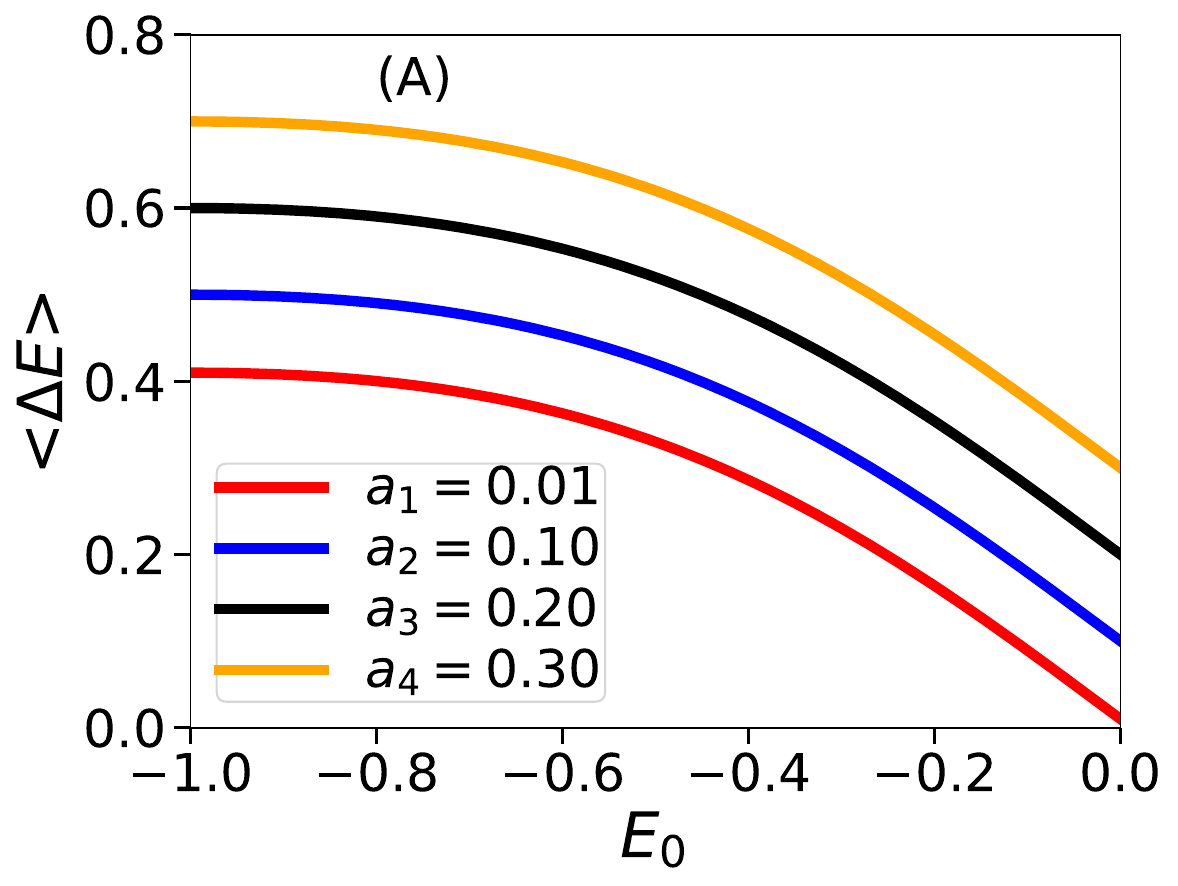}
    \includegraphics[width=0.49\textwidth,height=0.35\textwidth]{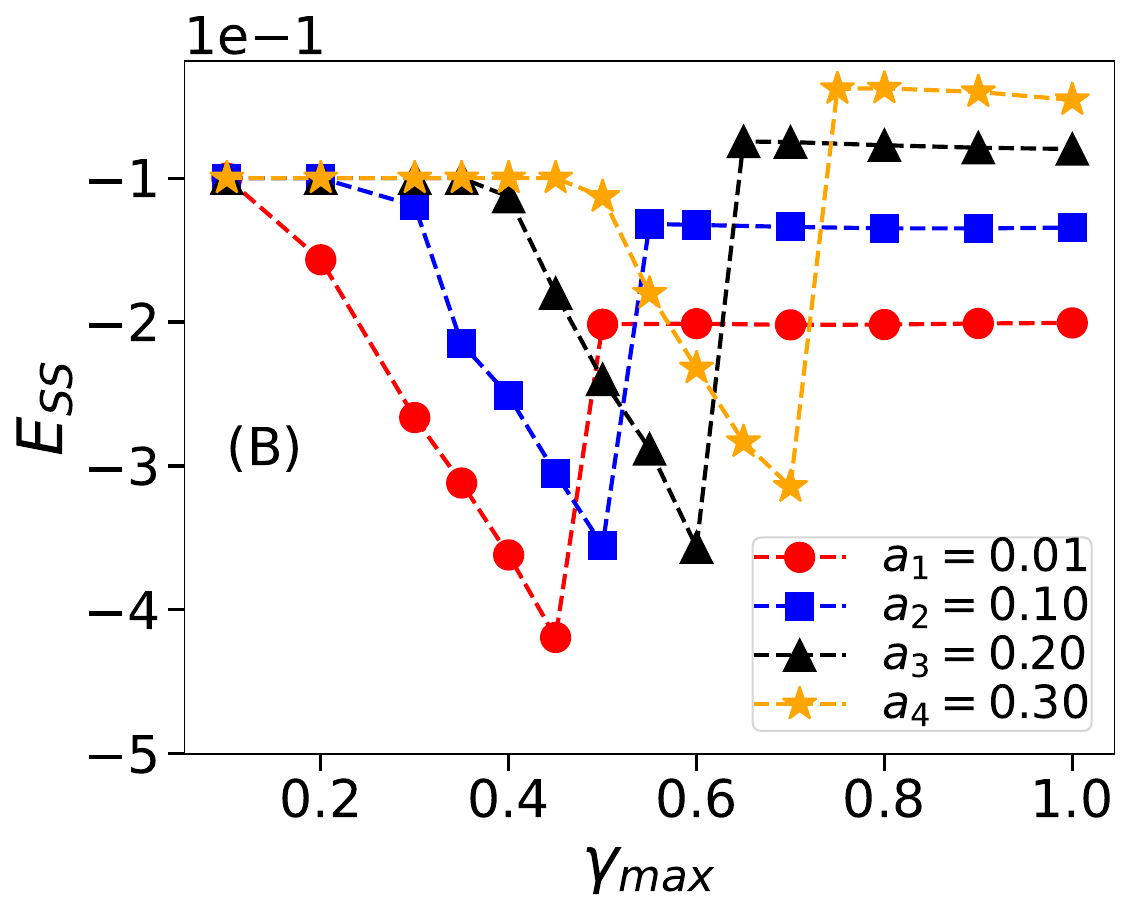}
    \includegraphics[width=0.49\textwidth,height=0.35\textwidth]{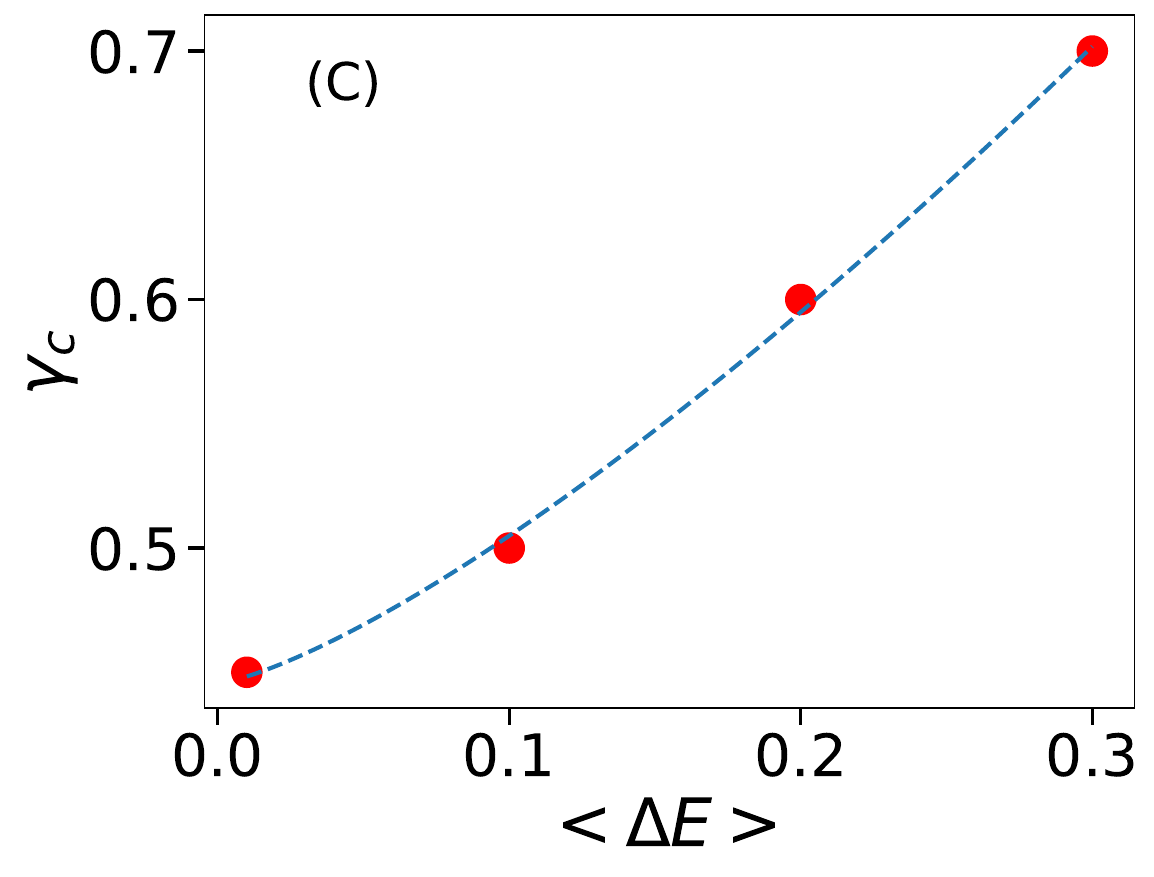}
    \includegraphics[width=0.49\textwidth,height=0.35\textwidth]{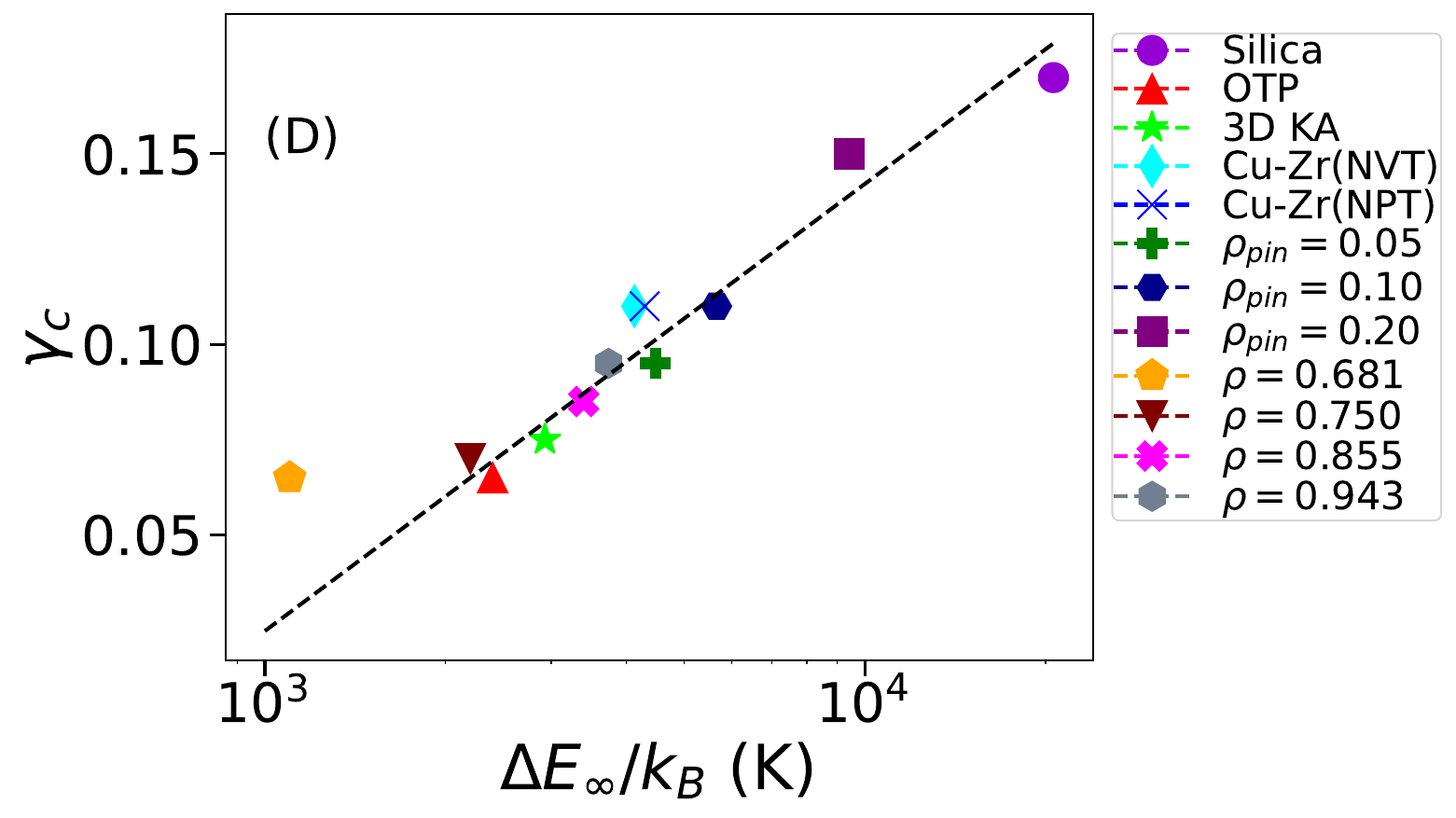}
    \caption{\textbf{Dependence of critical strain $\gamma_c$ on energy barrier from Elastoplastic model.} (A) Mean Energy barrier $<\Delta E>$ is plotted with annealing $E_0$ for different offset a$_1$,a$_2$,a$_3$,a$_4$ in $<\Delta E>$. (B) Steady state energy $E_{SS}$ is plotted with strain amplitude $\gamma_{max}$ for annealing $E_0 = -0.10$ in the Elastoplastic model for different offsets in the barrier. Variation of critical strain $\gamma_c$ is plotted with energy barrier from (C) EPM data and (D) simulation data. For Kob-Andersen model we converted the reduced unit to real unit by mimicking Ni$_{80}$P$_{20}$ metallic glass. In case of soft sphere model we multiplied the energy scale by a factor of $2\times10^5$ to match the data. }
    \label{SI_fig:25}
\end{figure}

Similar to the simulations, we now test whether the elastoplastic model introduced here can also reproduce the relation between $\gamma_c$ and $\Delta E$. To this end, we vary the parameter ``a" in Eq. (4) of the main text. Changing ``a" modifies $\Delta E$, as shown in Fig. \ref{SI_fig:25}(A). Using these different ``a" values, we then generate the yielding diagram for a poorly annealed case and indeed observe distinct yielding behavior [Fig. \ref{SI_fig:25}(B)]. In Fig. \ref{SI_fig:25}(C), we explicitly show $\gamma_c$ as a function of $\Delta E$. For comparison, the corresponding data for all other models are also shown in Fig. \ref{SI_fig:25}(D). Interestingly, while the elastoplastic model exhibits a linear relationship between $\gamma_c$ and $\Delta E$, the particle-based models display a logarithmic dependence. A possible explanation is that the elastoplastic model, being a mean-field, single-site description, lacks spatial fluctuations. Exploring these behaviors in a multi-site version of the model would be an interesting direction for future work. 

\section{Onset Temperature ($\text{T}_{onset}$)}
\subsection{3D HP: different density}
\begin{figure}[H]
    \centering
    \includegraphics[width=0.44\textwidth]{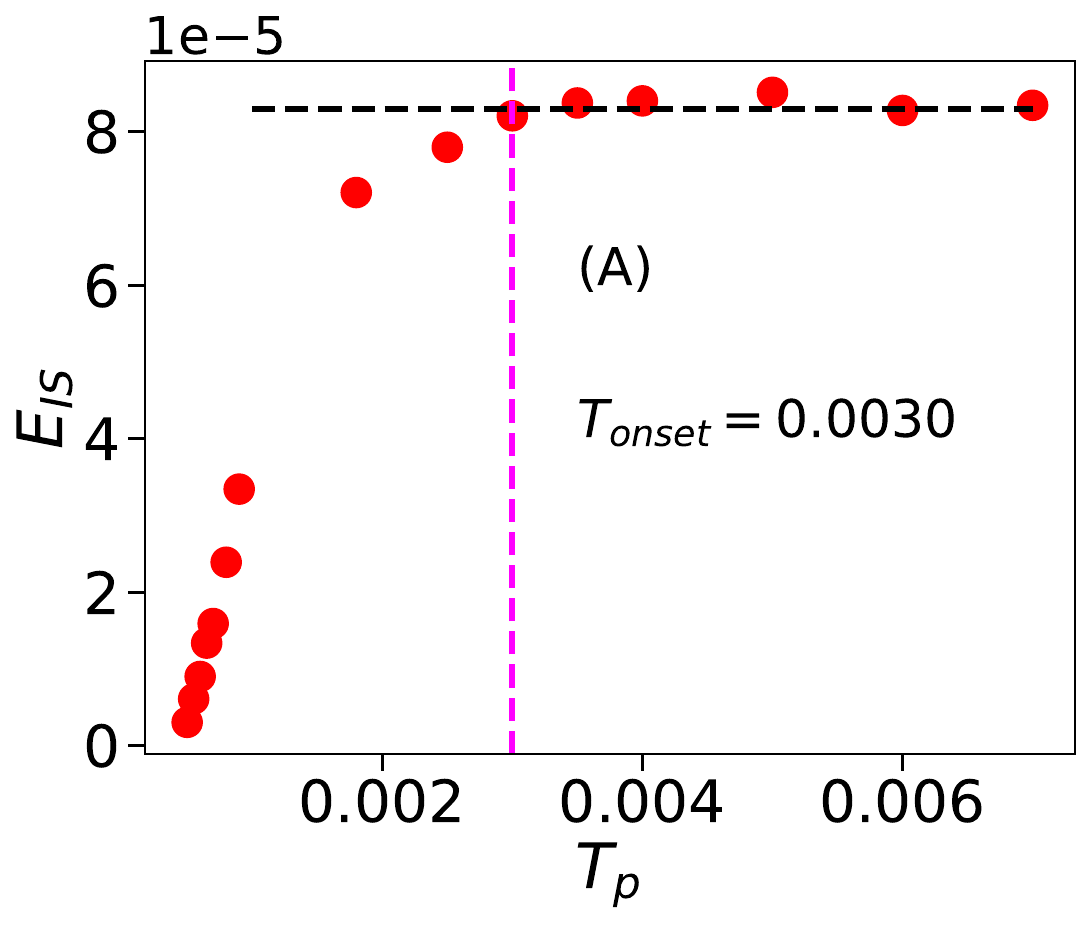}
    \includegraphics[width=0.44\textwidth]{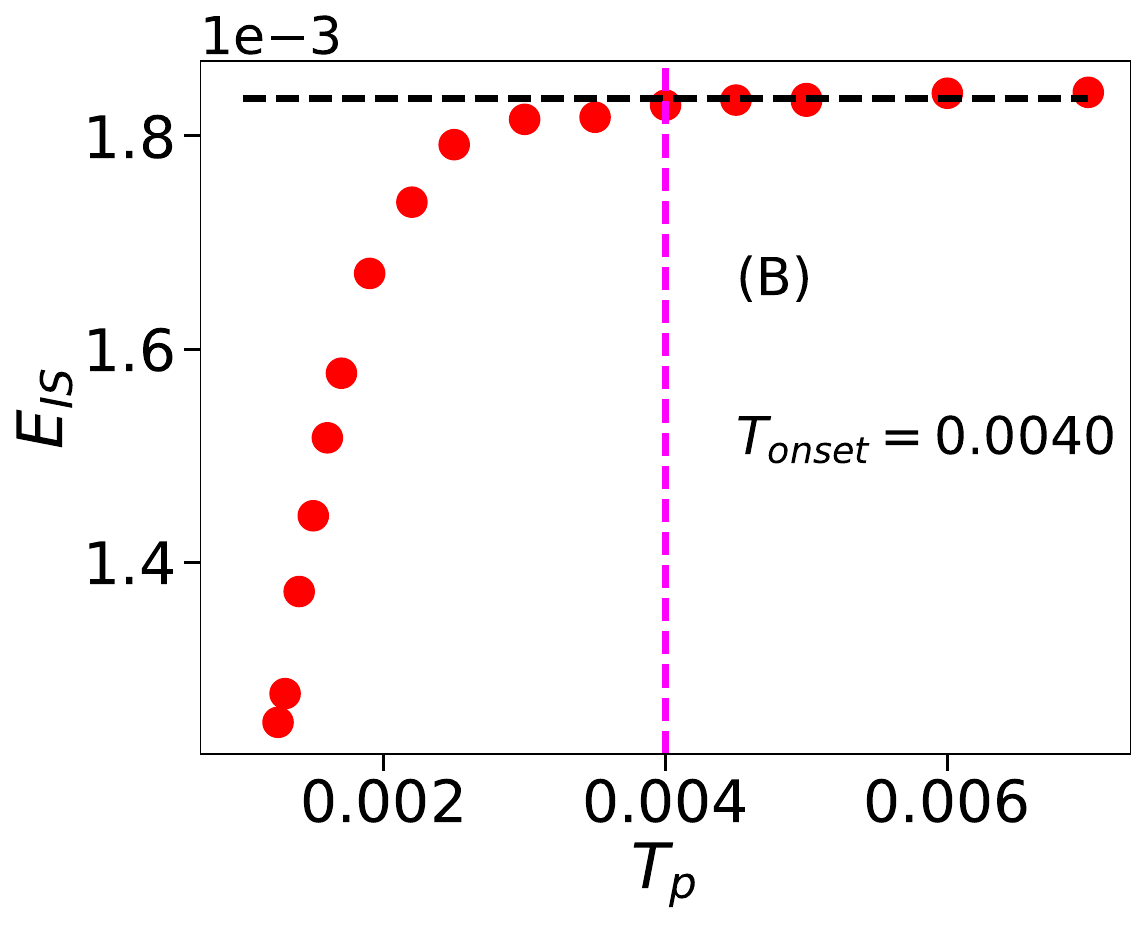}
    \includegraphics[width=0.44\textwidth]{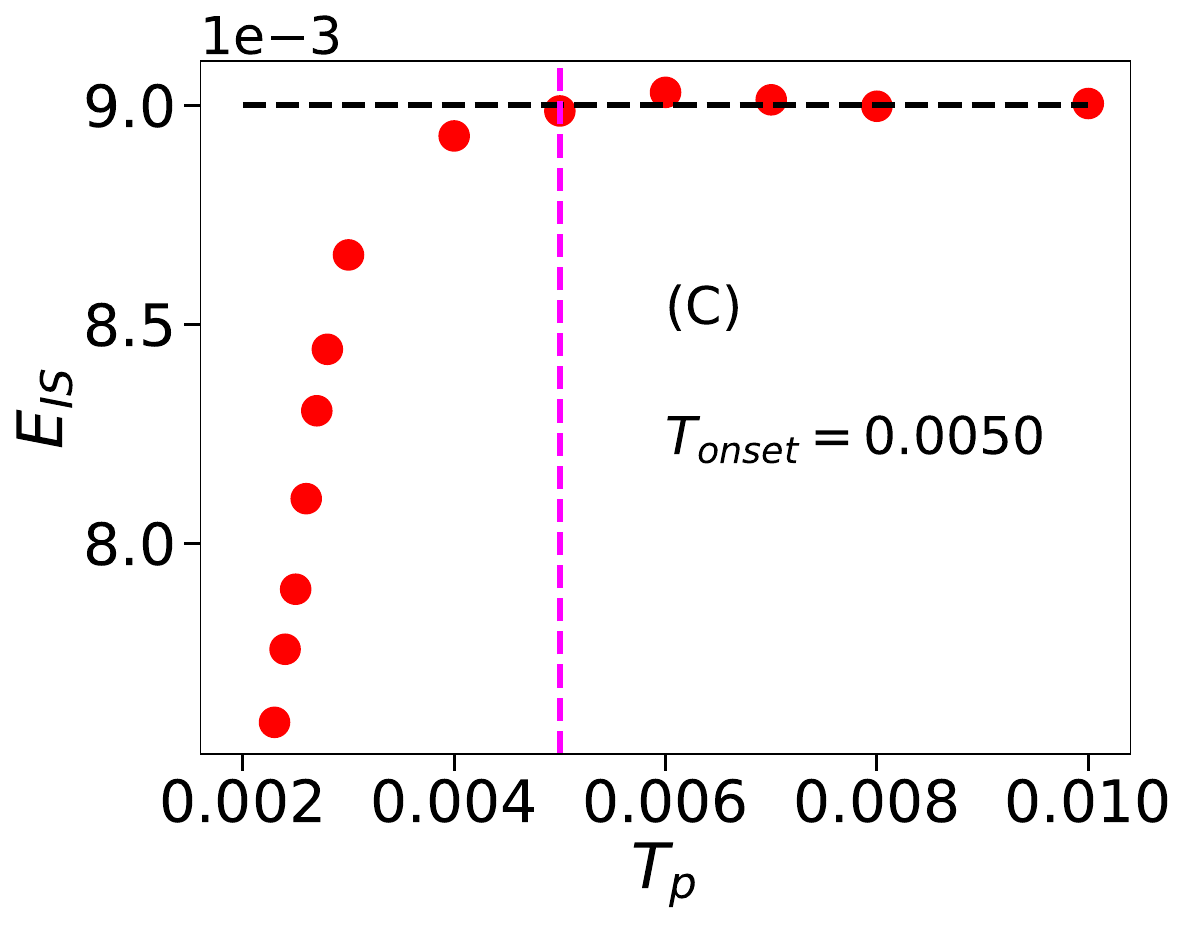}
    \includegraphics[width=0.44\textwidth]{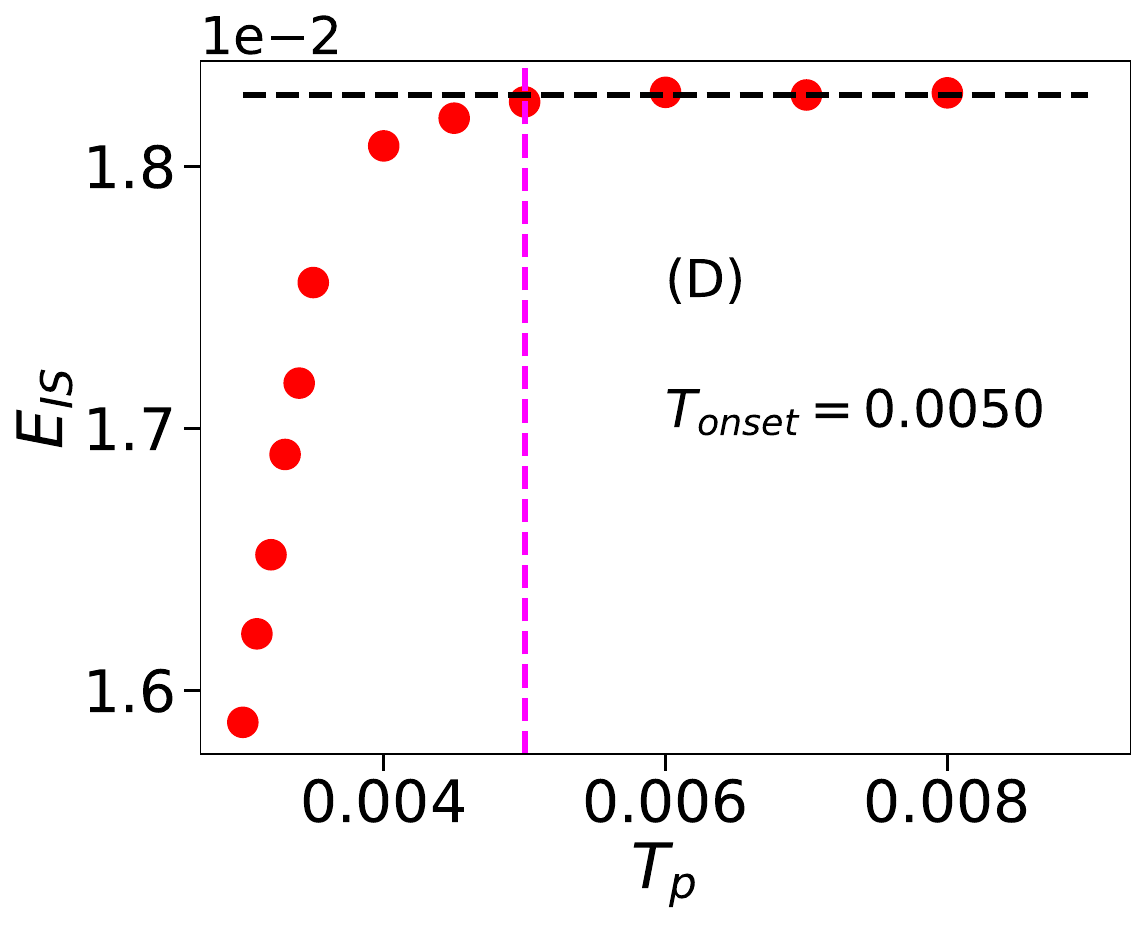}
    \caption{\textbf{Onset temperature $T_{onset}$ for 3D HP model.} Onset temperature ($T_{onset}$) of glassy dynamics is calculated where inherent structure energy $E_{IS}$ starts to deviate from high temperature steady value. $T_{onset}$ is shown for (A) $\rho = 0.681$, (B) $\rho = 0.750$, (C) $\rho = 0.855$ and (D) $\rho = 0.943$. Each data point is averaged over 12 independent samples.}
    \label{SI_fig:55}
\end{figure}

\subsection{Onset Temperature for different models}
\begin{figure}[H]
    \centering
    \includegraphics[width=0.49\textwidth,height=0.35\textwidth]{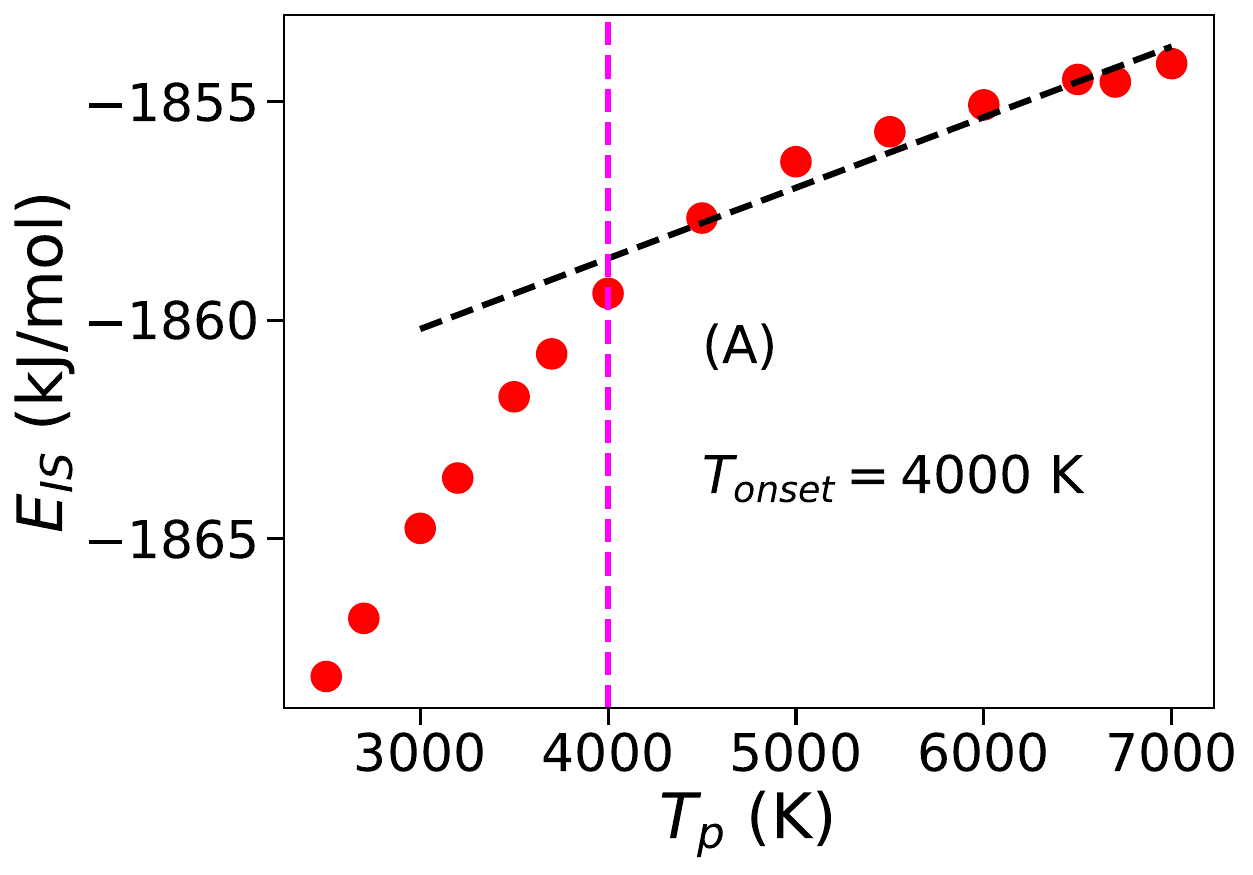}
    \includegraphics[width=0.49\textwidth,height=0.35\textwidth]{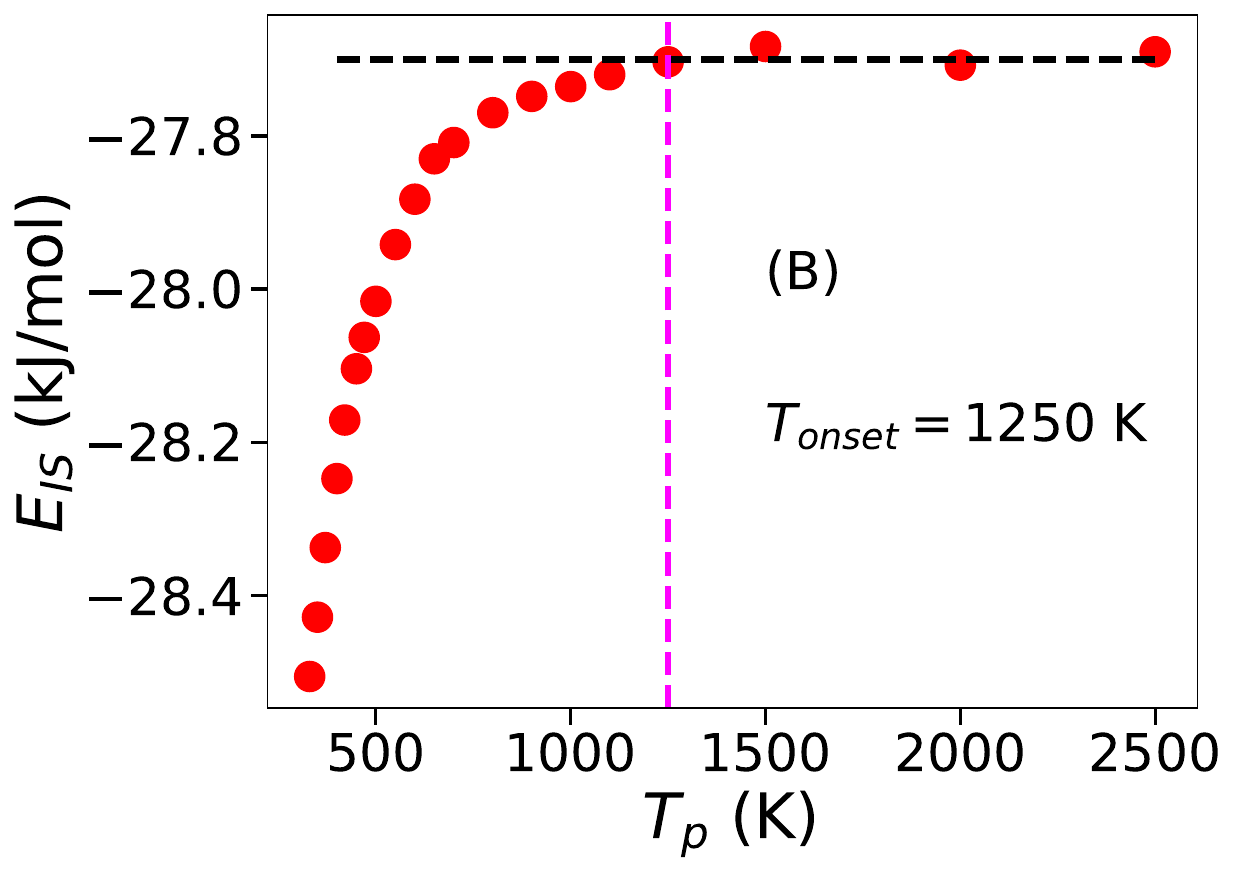}
    \includegraphics[width=0.49\textwidth,height=0.35\textwidth]{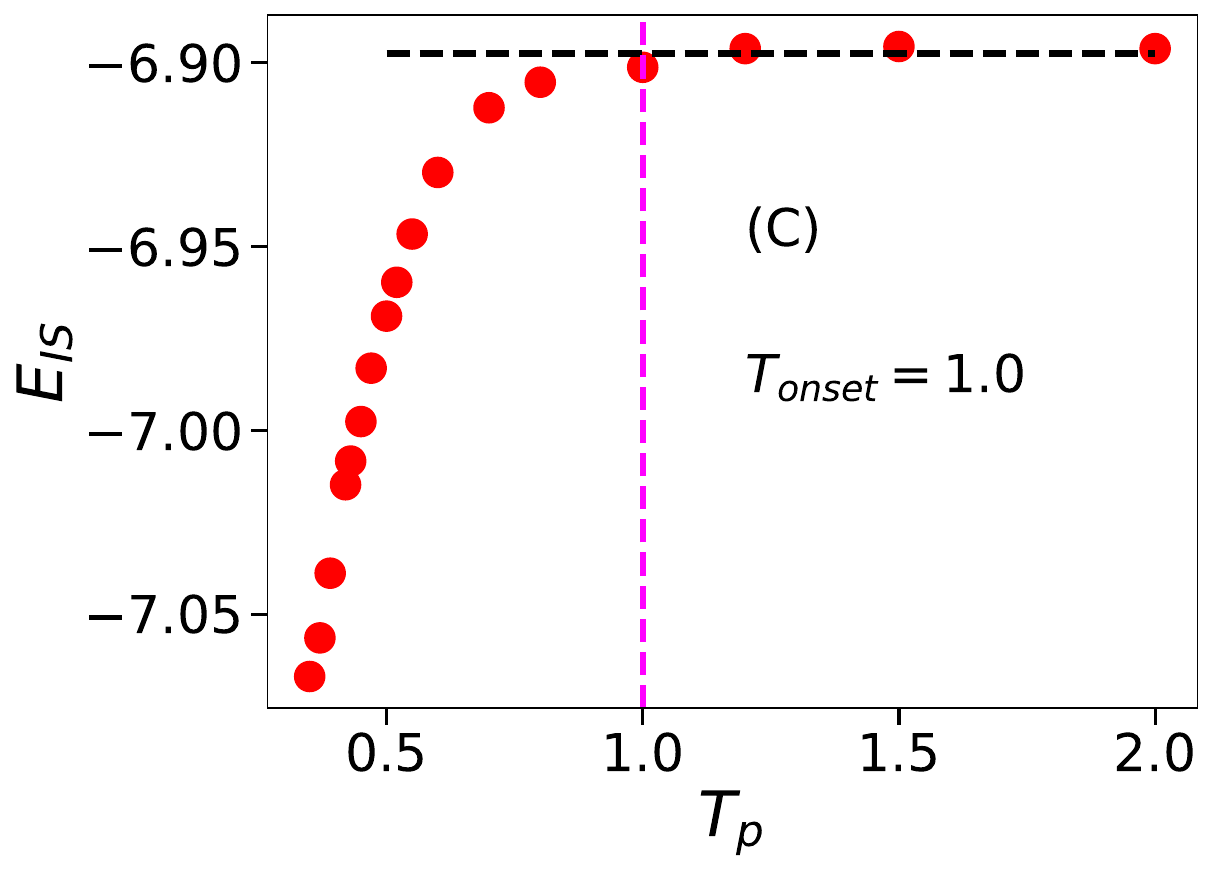}
    \includegraphics[width=0.49\textwidth,height=0.35\textwidth]{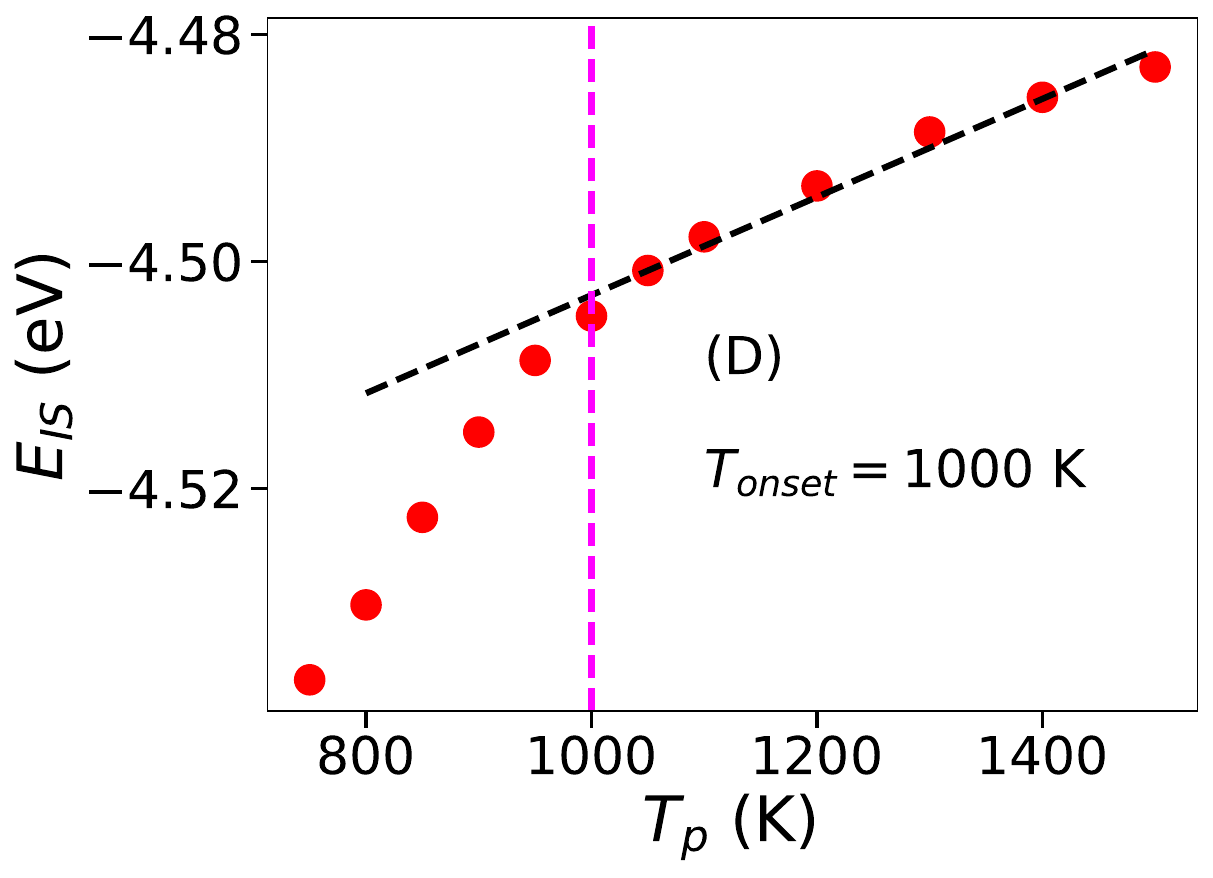}
    \caption{\textbf{Onset temperature $T_{onset}$ for microscopically different glass models.} Onset temperature ($T_{onset}$) of glassy dynamics is calculated where inherent structure energy $E_{IS}$ starts to deviate from high temperature steady value. $T_{onset}$ is shown for (A) SiO$_2$, (B) OTP, (C) 3D KA and (D) Cu-Zr (NPT). Each data point is averaged over 12 independent samples.}
    \label{SI_fig:56}
\end{figure}

\subsection{Critical yield strain $\gamma_c$ vs energy barrier in reduced unit}
\begin{figure}[H]
    \centering
    \includegraphics[width=0.85\textwidth,height=0.5\textwidth]{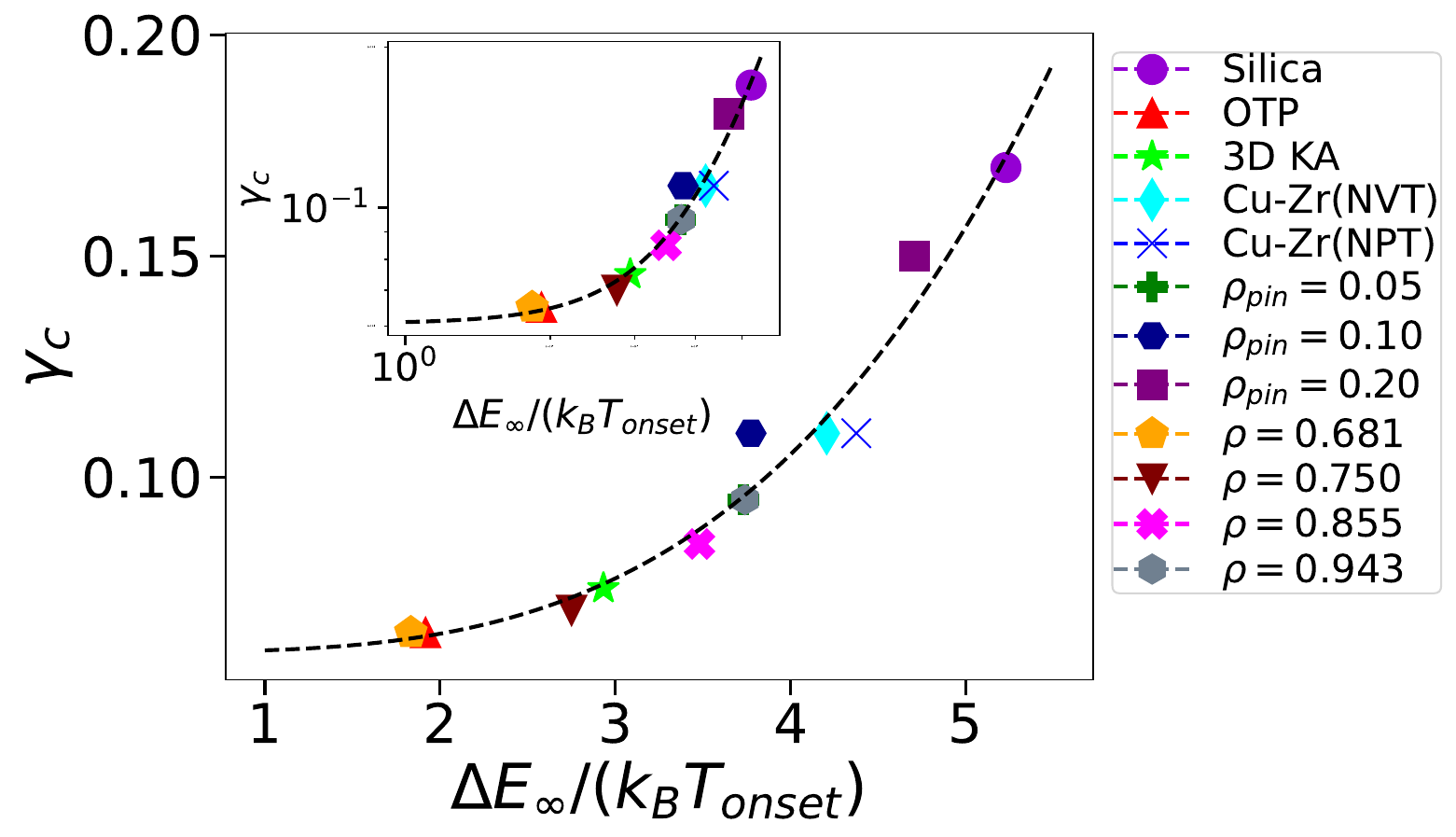}
    \caption{\textbf{Dependence of critical strain $\gamma_c$ on energy barrier from simulation models.} Plot for $\gamma_c$ vs high temperature energy barrier in reduced unit (normalised by $T_{onset}$). Data points are fitted via power law, $y = ax^b+c$, where $a = 0.00038$, $b = 3.432$ and $c = 0.061$. Same plot is shown in log scale in the inset.}
    \label{SI_fig:57}
\end{figure}

\section{Comparison between Simulation and EPM }
As discussed in the main text, the energy barrier plays an important role in the yielding transition. The EPM model introduced in the main text explicitly incorporates the energy barrier. We also demonstrated how $\Delta E$ can be computed for particle-based models. Here, we test whether a quantitative comparison can be made by extracting $\Delta E$ from particle-based simulations and using it as input to the EPM.  

We calculated the energy barrier for silica glass and 3D KA at each parent temperature $T_p$ using the expression $\Delta E = k_B T_p \ln\left(\tau_\alpha/\tau_0\right)$ and scaled it by the shear modulus $\mu$. The scaled quantity $\Delta E/\mu$ is plotted against the annealing parameter $E_{IS} - E_{\infty}$, where $E_{\infty}$ denotes the high-temperature limit of the inherent structure energy. We used the form $E_{IS} = a - b/T_p$ with $a = E_\infty$. We find that as $E_{IS} \to E_\infty$, the scaled energy barrier $\Delta E/\mu$ saturates. The data were fitted with $\Delta E/\mu = ax^2 + bx + c$, where $x = E_{IS} - E_\infty$, and this fitted expression was then used as input in our elastoplastic model.  

\begin{figure}[H]
    \centering
    \includegraphics[width=0.49\textwidth,height=0.35\textwidth]{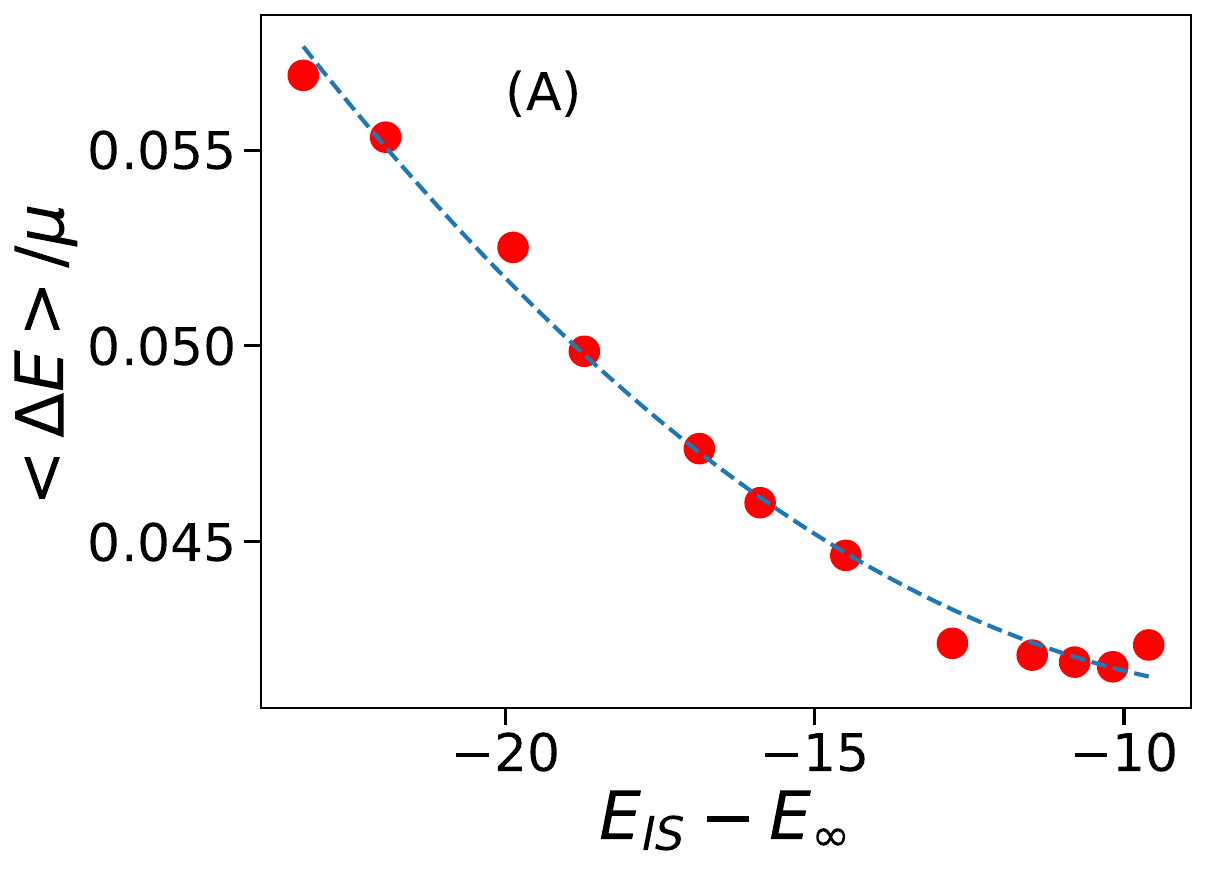}
    \includegraphics[width=0.49\textwidth,height=0.35\textwidth]{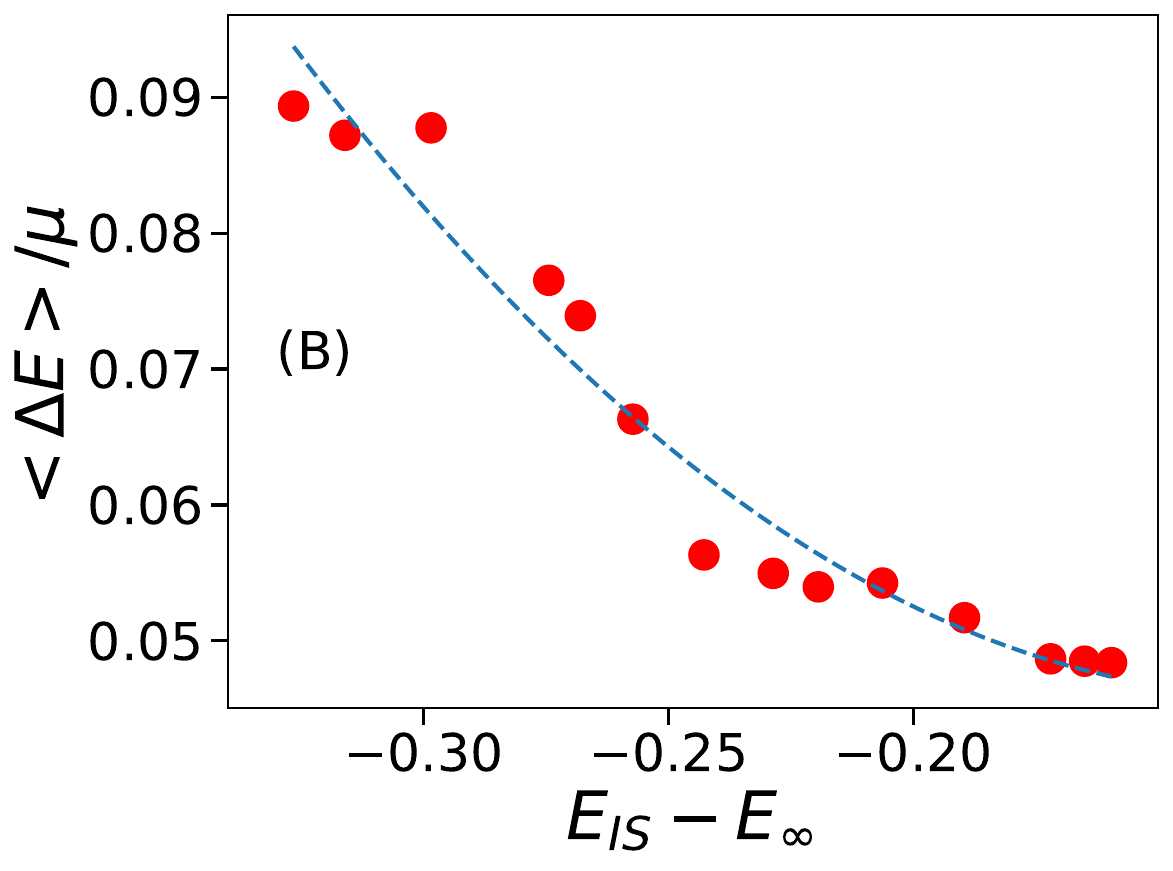}
    \includegraphics[width=0.49\textwidth,height=0.35\textwidth]{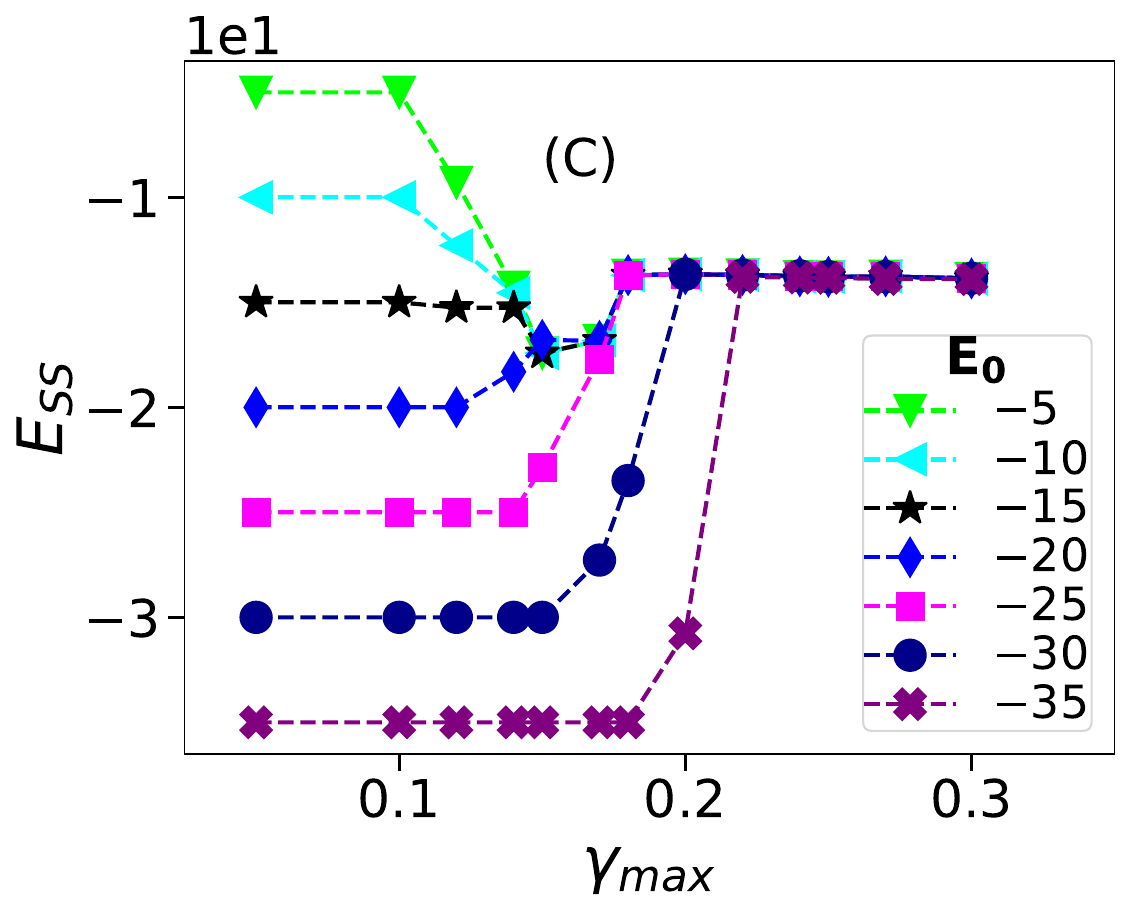}
    \includegraphics[width=0.49\textwidth,height=0.35\textwidth]{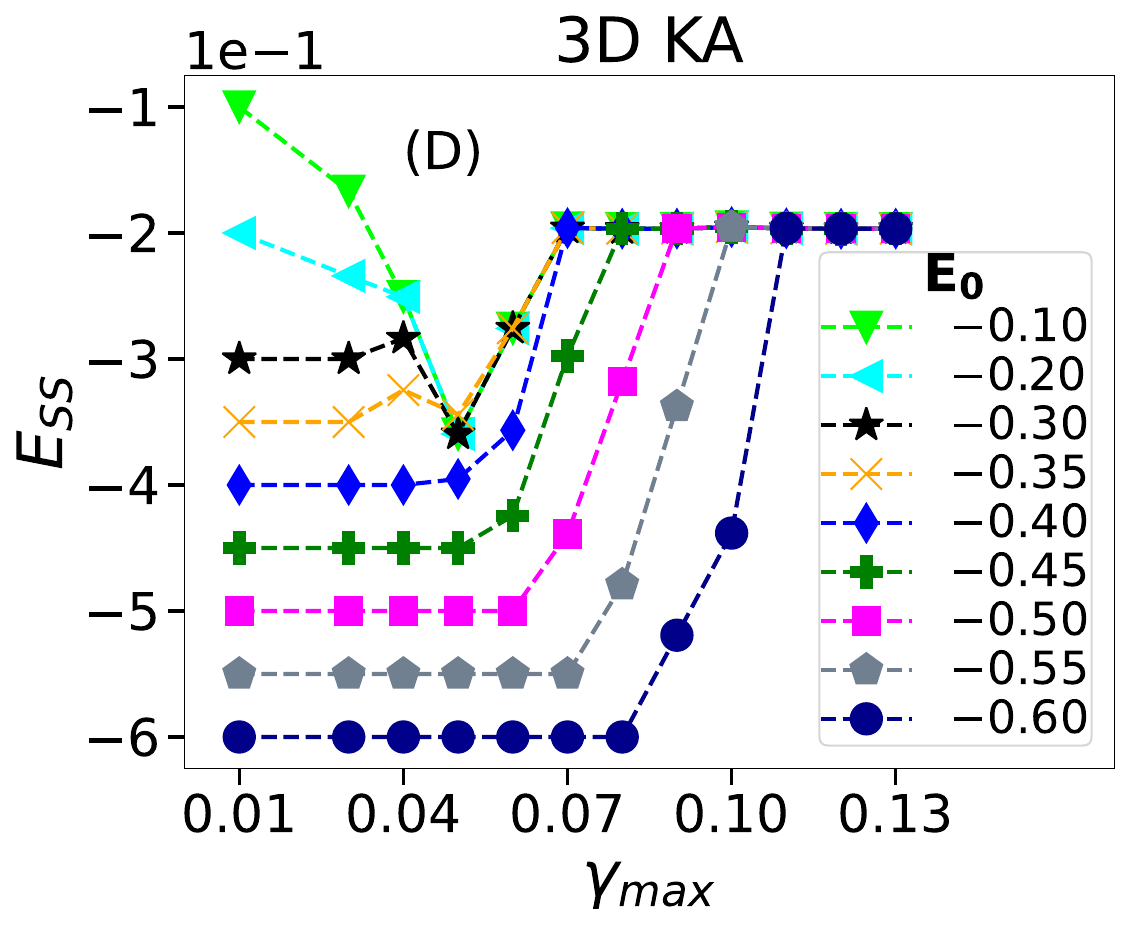}
    \caption{\textbf{Simulation-derived energy-barrier with annealing and its implementation in the EPM.} Mean energy barrier $<\Delta E>$ scaled by modulus $\mu$ is plotted with annealing $E_{IS}$ (reference taken as very high temperature energy) for (A) Silica and (B) 3D KA. The data set are fitted via equation $y = ax^2 + bx +c$, where $y = \Delta E/\mu$ and $x = E_{IS} - E_\infty$. We feed these fitted equations in our EPM, and the corresponding steady state energy $E_{SS}$ vs strain amplitude $\gamma_{max}$ is plotted for (C) Silica and (D) 3DKA.}
    \label{SI_fig:26}
\end{figure}

For silica, we chose $E_0 \in (0, -50)$ with $\sigma = 5.0$. To match the yield point with the simulation results, we set $k = 1.0$ in Eq.~(6) of the main text. The shear modulus $\mu$ in Eq.~(3) of the main text was increased linearly from 200 to 500 with annealing. The barrier width $\sigma_{\Delta E}$ was decreased linearly from 30\% to 1\% with annealing. Using these parameters, we generated the yielding diagram shown in  Fig.~\ref{SI_fig:26}(C).  

For 3D KA, we chose $E_0 \in (0, -1)$ with $\sigma = 0.1$. We set $k = 0.2$, increased $\mu$ linearly from 100 to 200 with annealing, and decreased the barrier width $\sigma_{\Delta E}$ from 50\% to 1\% linearly with annealing. Using these parameters, we generated the yielding diagram shown in Fig.~\ref{SI_fig:26}(D).

\begin{figure}[htp]
    \centering
    \includegraphics[width=0.85\textwidth,height=0.6\textwidth]{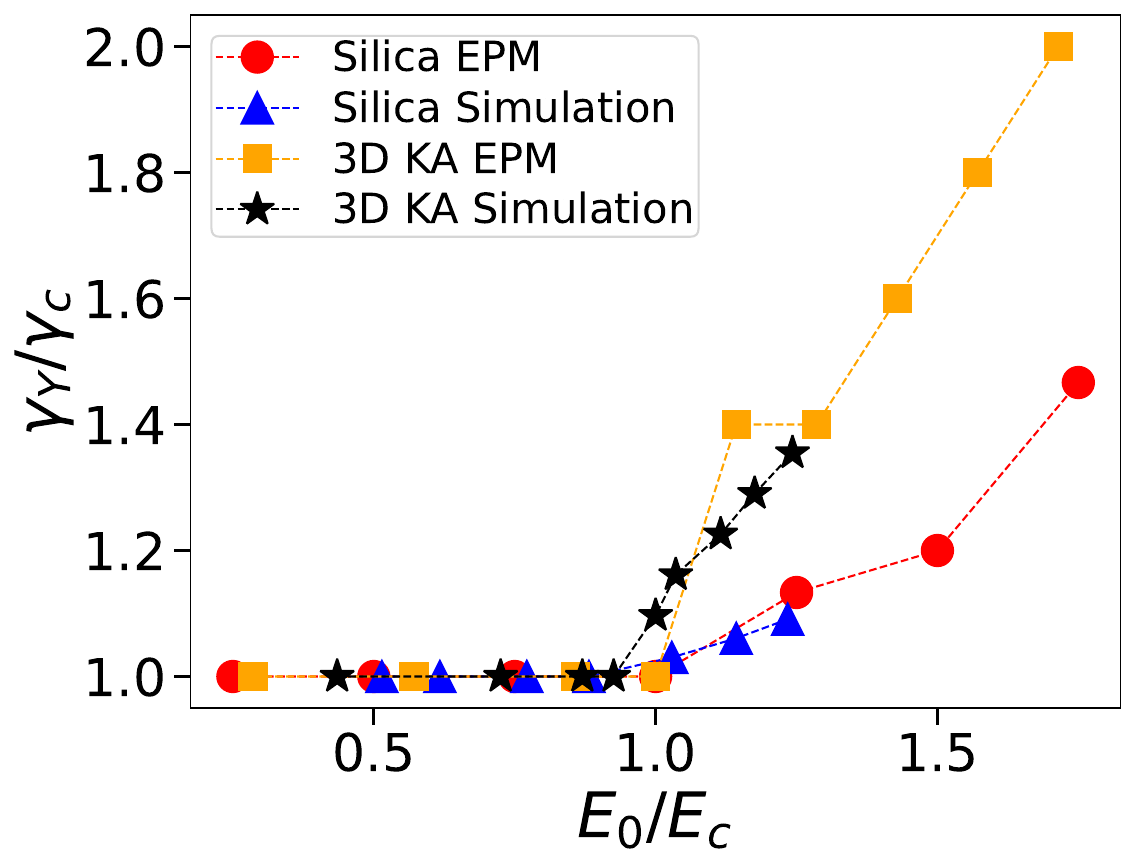}
    \caption{\textbf{Comparison between Simulation and Elastoplastic model.} $\gamma_Y/\gamma_c$ is plotted against inherent structure energy scaled by the critical energy $E_0/E_c$ (a proxy for $T_p/T_{MCT}$). Predicted $\gamma_Y/\gamma_c$ from the elastoplastic model (EPM) using the energy barrier information from $\tau_{\alpha}$ vs $T$ data compares very well with the oscillatory shear simulation results. This suggests that the energy barrier directly controls the shift in the yield strain. Comparison in both the Silica and the 3D KA models is observed to be very good.}
    \label{SI_fig:27}
\end{figure}

In Fig.~\ref{SI_fig:27}, we show the scaled yield strain ($\gamma_Y/\gamma_c$) as a function of scaled annealing (used here as a proxy for parent temperature) for the EPM model, which takes input from particle-based simulations, alongside the actual simulation data. We observe a strikingly similar quantitative behavior in both cases. 


\newpage

\section{Shear Band Analysis}

To analysis shear band we have prepared samples at parent temperatures below respective Mode coupling temperature $T_{MCT}$. We prapared BKS silica samples at $T_p = 2700$K with system size $N = 48000$. For comparison with fragile glass, we have also prepared Cu-Zr metallic glass samples at $T_p = 750$K with system size $N = 50000$. We sheared the glass samples cyclically at their respective yielding amplitude $\gamma_{max} \approx \gamma_Y$ ($\gamma_{max} = 0.18$ for BKS Silica and $\gamma_{max} = 0.14$ for metallic glass). We have measured the stroboscopic mean squared displacement $MSD$ $(\gamma = 0)$ between two successive cycles in the steady state. We have reached steady state within $N_{cycle} = 60$ for metallic glass. For Silica it takes much larger cycles to reach steady state. We have seen sharp shear band for metallic glass (fragile glass) but in case of Silica the shear band is diffusive.

\begin{figure}[H]
    \centering
    \includegraphics[width=0.49\textwidth,height=0.35\textwidth]{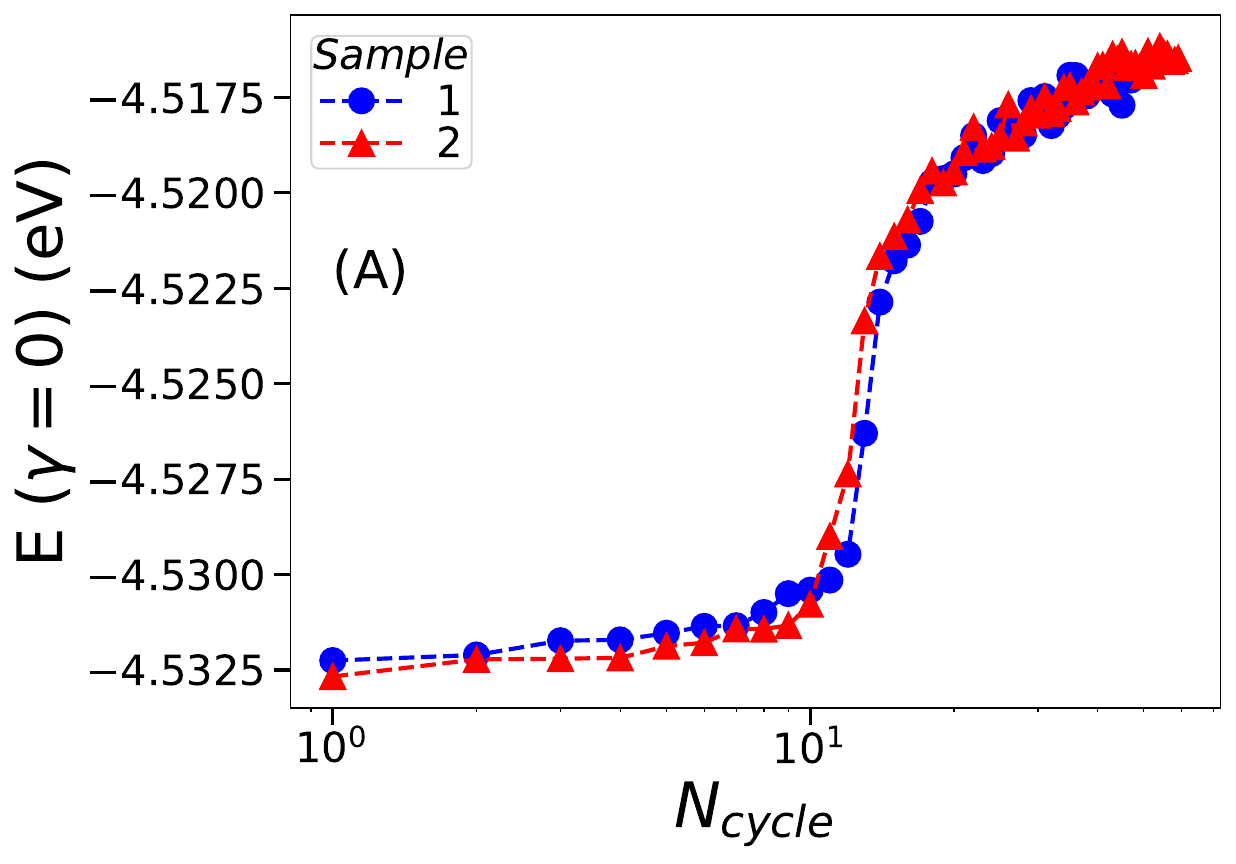}
    \includegraphics[width=0.49\textwidth,height=0.35\textwidth]{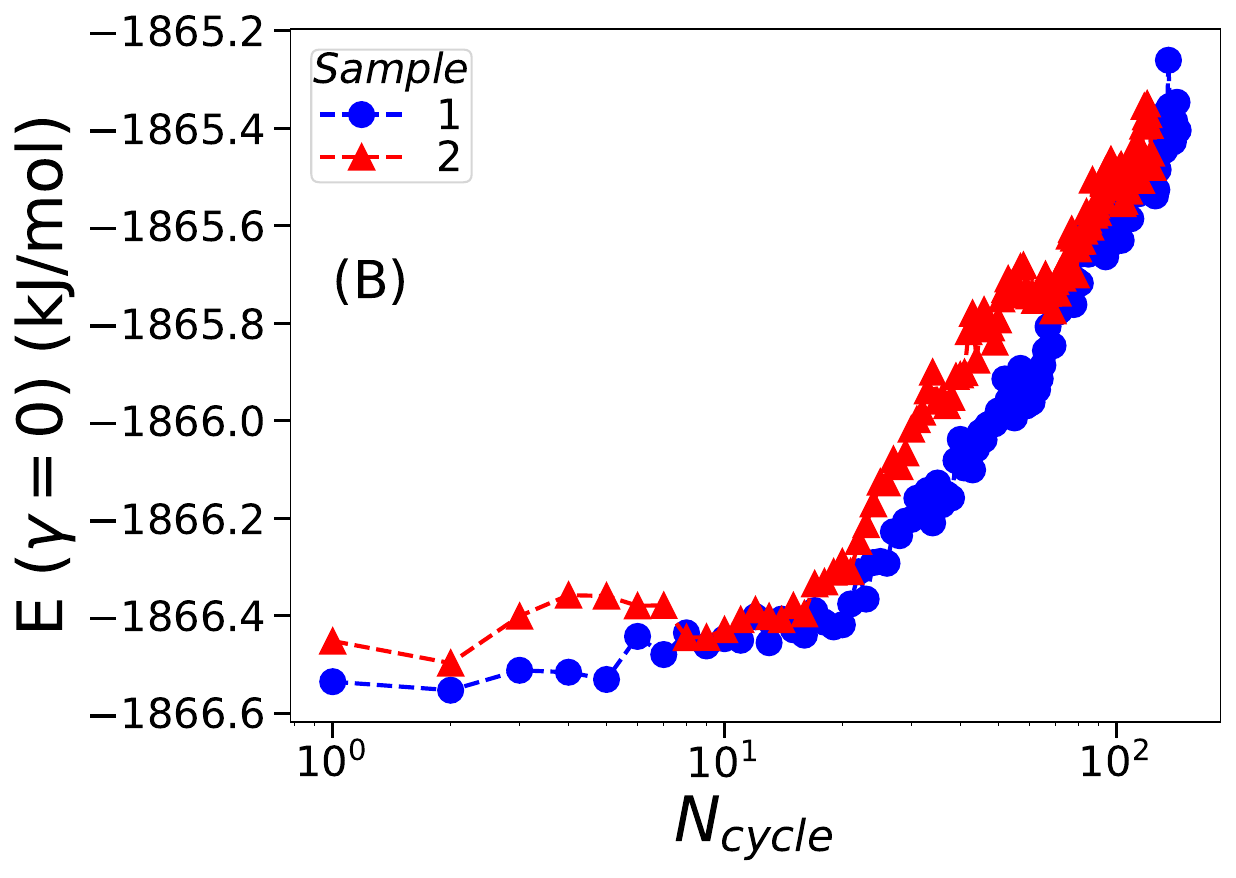}

    \caption{\textbf{Stroboscopic energy vs cycle.} Variation of stroboscopic energy E($\gamma = 0$) with $N_{cycle}$ is plotted for (A) Cu-Zr metallic glass ($T_p = 750$K and $N = 50000$) and (B) BKS Silica model ($T_p = 2700$K and $N = 48000$). The parent temperature of the respective model is below the Mode coupling temperature $T_{MCT}$. Samples are sheared at yielding amplitude $\gamma_{max} \approx \gamma_Y$. $\gamma_{max} = 0.18$ for BKS Silica and $\gamma_{max} = 0.14$ in case of metallic glass.}
    \label{SI_fig:100}
\end{figure}